%% file: main.tex
\documentclass[12pt,vi,a4paper,twoside]{mitthesis}

\usepackage[T1]{fontenc}
\usepackage[utf8]{inputenc}
\usepackage{lmodern}

\usepackage{aas_macros}

\usepackage{caption}
\usepackage{subcaption}
\usepackage{graphicx}
\graphicspath{{figures/}}
\usepackage{tabularx}

\usepackage{amsmath}
\usepackage{bm}

\usepackage{enumitem}

\usepackage{multirow}
\usepackage{booktabs}

\usepackage[figuresright]{rotating}

\usepackage{pdfpages}

\usepackage{emptypage}

\usepackage[
    pdftitle = {PhD Thesis},
    pdfauthor = {Filipe Pereira},
    colorlinks = true,
    linkcolor = blue,
    citecolor = blue,
    urlcolor = blue,
    bookmarks,
    bookmarksopen,
    bookmarksdepth=1,
]{hyperref}
\usepackage{bookmark}

\usepackage[
    backend=biber,
    style=authoryear,
    citestyle=authoryear,
    uniquename=false,
    uniquelist=false,
    doi=true,
    dashed=false,
    backref=true,
    backrefstyle=two,
    mincitenames=1,
    maxcitenames=2,
    minbibnames=3,
    maxbibnames=3,
    giveninits=true                 
]{biblatex}
\DefineBibliographyStrings{english}{
    backrefpage  = {see page},
    backrefpages = {see pages},
}
\DeclareNameAlias{sortname}{last-first}

\begin{document}

\pagenumbering{roman}
\include{preliminary/preliminary}

\include{preliminary/contents}

\pagenumbering{arabic}

\include{chapters/introduction/main}

\include{chapters/transits/main}

\include{chapters/characterization/main}

\include{chapters/pipeline/main}
\include{chapters/search/main}


\include{chapters/conclusions}

\appendix
\include{chapters/appendix}

\printbibliography[heading=bibintoc, title=References]

\end{document}

%% file: preliminary/preliminary.tex
\includepdf{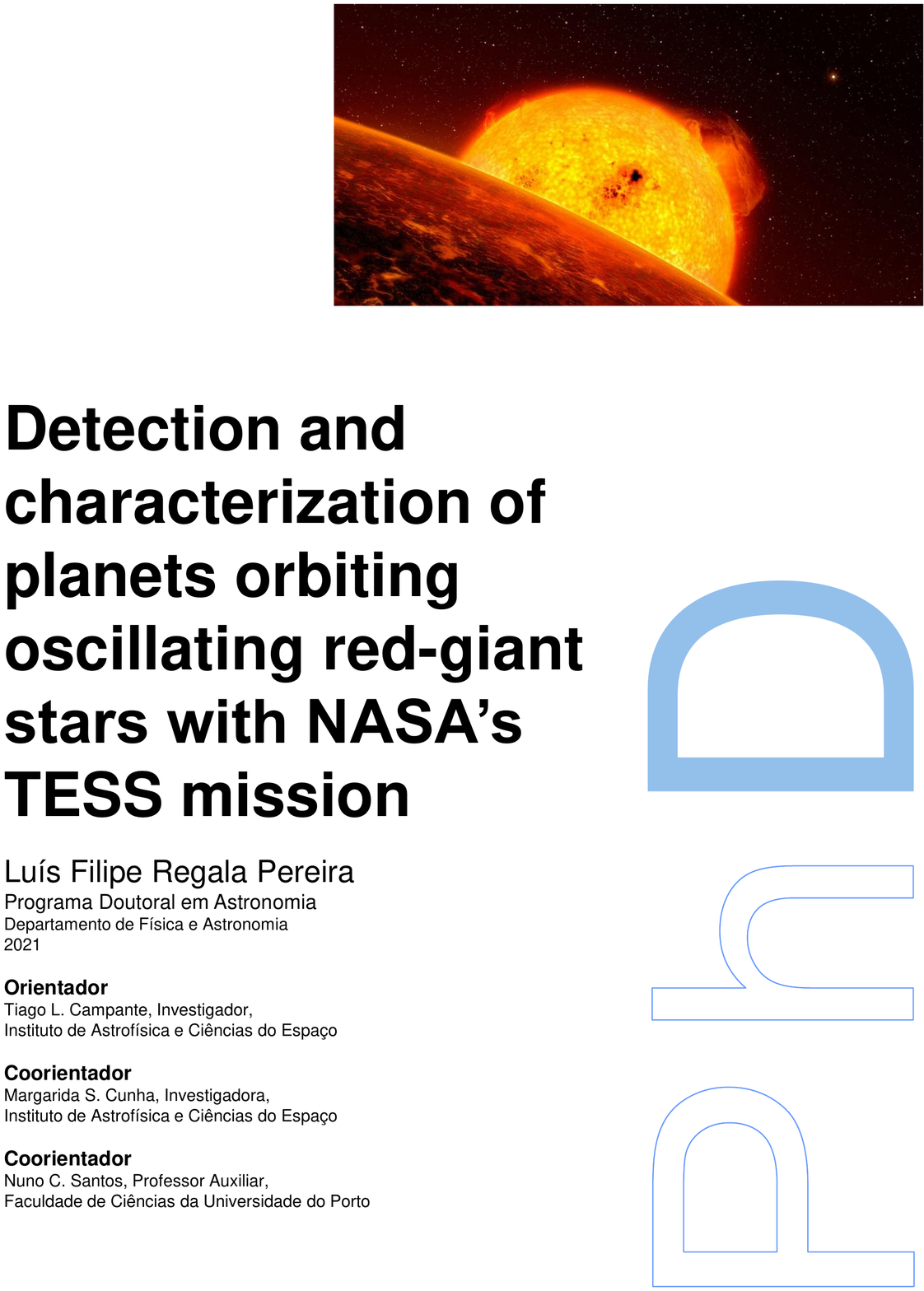}

\newpage
\vspace*{\fill}
Cover illustration source: ESO/L. Calçada

\cleardoublepage

\chapter*{Acknowledgments}
\input{preliminary/acknowledgements}

\cleardoublepage

\pdfbookmark[chapter]{Abstract}{abstract}
\chapter*{Abstract}
\input{preliminary/abstract}

\cleardoublepage

\pdfbookmark[chapter]{Resumo}{resumo}
\chapter*{Resumo}
\input{preliminary/resumo}

%% file: preliminary/acknowledgements.tex
I'd like to start more formally by thanking all the projects and fellowships that made it possible for me to work on and write this thesis.
I acknowledge support from fellowship PD/BD/135227/2017 funded by FCT - Fundação para a Ciência e Tecnologia (Portugal) and POPH/FSE - Programa Operacional Potencial Humano (EC).
This work was also supported by FCT through national funds and by FEDER - Fundo Europeu de Desenvolvimento Regional funds through COMPETE2020 - Programa Operacional Competitividade e Internacionalização (POCI) by these grants: UID/FIS/04434/2019; UIDB/04434/2020; UIDP/04434/2020; PTDC/FIS-AST/30389/2017 \& POCI-01-0145-FEDER-030389; PTDC/FIS-AST/32113/2017 \& POCI-01-0145-FEDER-032113 and PTDC/FIS-AST/28953/2017 \& POCI-01-0145-FEDER-028953.

The remaining acknowledgements, more personal, will be in portuguese.

Em primeiro lugar, gostava de agradecer aos meus orientadores por toda a ajuda dada ao longo deste doutoramento.
Ao Tiago, obrigado pela minuciosidade e diligência em todos os aspetos da orientação.
Tenho a certeza de apenas ter conseguido o que consegui devido a toda a dedicação que deste a este projeto.
À Margarida, obrigado pelo incessante interesse na ciência, que facilmente se espalha por quem trabalha contigo.
Ao Nuno, obrigado pelo imenso conhecimento partilhado nesta procura por outros planetas, mesmo tendo uma agenda sempre a abarrotar.

Estou também agradecido a todos com que me cruzei em reuniões, almoços, seminários e muitas outras atividades durante todos estes anos no Caup.

Aos meus amigos de Aveiro: Luciano, Mariana, Xavi, Miguel, Kiko, Bia, Sacchetti e Susy, obrigado pela companhia ao longo destes 5 anos, protagonizada um pouco por todos os cafés da nossa cidade... até mandarem toda a gente para casa.
Que as waffles voltem ao menu em pouco tempo.
Queria também adicionar a nota de que a Mariana é a melhor do mundo e faz anos a 24 de Julho.

À Liana, obrigado pela boa disposição e espírito aventureiro.
Que continuemos a encontrar mais saltos pouco recomendados para fazer na vida.

À Sara, obrigado pela companhia, por todas as conversas, das mais parvas as mais sérias e pelo tubérculo mais memorável da História.
Apesar das interrupções não covidadas nos terem atrasado os planos, espero voltar à nossa procura do sushi mais bem tempurado do país.

Aos meus amigos (e primos) dos campos de arroz: Leugim, Lança, Américo and Carlos, obrigado pelas longas conversas sobre tudo e mais alguma coisa, pelas noitadas de futebol de pára-choques e pela colaboração na exploração do limiar da parvoíce humana.

Gostava também de agradecer aos meus tios e tias de Famalicão pela hospitalidade, à minha tia Ana por toda a ajuda e pelos peixinhos grelhados, ao meu pai pelas conversas animadas e por me acompanhar nos voleys semanais e a minha irmã pelo suporte a carregar os stresses familiares e pelos passeios culturais.

Finalmente, queria agradecer à minha mãe, por tudo, seja a presença constante, em bons ou maus momentos, ou a ajuda inquestionável.
Certamente que não seria possível chegar onde cheguei, ou fazer esta tese, sem ti.

%% file: preliminary/abstract.tex
Driven largely by multiple ground-based radial-velocity (RV) surveys and photometric space missions such as \textit{Kepler} and \textit{K2}, the discovery of new exoplanets has increased rapidly since the early 2000s.
However, due to a target selection bias in favor of main-sequence stars, only a handful of transiting planets have been found orbiting evolved hosts.
These planets, most of which are giants, hold important information regarding the formation and evolution of planetary systems.

Correlations between stellar mass or metallicity and giant-planet occurrence contain evidence pointing to the mechanisms responsible for giant-planet formation.
Occurrence rate studies of close-in giant planets around dwarf and giant stars can inform on the impact of stellar evolution on planetary systems.
For instance, planetary radius characterization in evolved systems can shed light on the processes responsible for giant-planet radius inflation.
However, to learn about these and other processes, it is necessary to have a statistical sample of well characterized giant planets orbiting evolved stars.

In this thesis, I sought to increase the sample of known giant planets orbiting red-giant stars, focusing on data from NASA's \textit{Transiting Exoplanet Survey Satellite} (\textit{TESS}) mission, and to improve their characterization.
Specifically, I focused on close-in giant planets orbiting (preferably) oscillating low-luminosity red-giant branch (LLRGB) stars.

To improve characterization, I developed a method to model planetary transits and stellar signals simultaneously, implementing Gaussian processes to model stellar granulation and the oscillations envelope in the time domain using equivalent expressions to the ones commonly used in frequency-domain models from the literature.
Tests with \textit{TESS} simulated light curves and \textit{Kepler} light curves show that this model is capable of recovering the underlying stellar signals present in light curves and that it agrees, within uncertainties, with current power-spectrum fitting methods.
In particular, results show that the model enables time-domain asteroseismology, inferring the frequency of maximum oscillation amplitude, $\nu_\text{max}$, to within 1\%.

Additionally, when modeling stellar signals and transits simultaneously, estimated planetary parameters have higher accuracies and lower uncertainties overall, when compared to simpler transit models, common in the literature.
The method's implementation is open-source and available to the community.

Regarding the planet search, I assembled a pipeline, mostly comprised of third-party open-source software, for the extraction and correction of light curves from the \textit{TESS} full frame images, and subsequent transit search and validation.
With this pipeline, I explored a sample of $\sim$40,000 bright LLRGB stars in the southern hemisphere of \textit{TESS}'s field of view.
The sample was limited to stars with radii above 3 $\rm \ \rm R_\odot$, to ensure that oscillations would be present below the Nyquist frequency of \textit{TESS} long-cadence data, and below 8 $\rm R_\odot$, to ensure that transits of Jupiter-sized planets would be visible in the light curve.
Targets selected were also constrained to \textit{TESS} magnitudes lower than 10 to increase the probability of detecting oscillations.

Overall, I identified four planet candidates, two of which are not currently known planets and orbit red-giant stars.
Radial-velocity follow-up observations of both these candidates have tentatively confirmed their planetary nature.
The first candidate is a hot Jupiter with a possibly inflated radius of $1.24 \ R_\text{J}$ on a 6.20-day-period orbit.
The second candidate has a radius of $1.02 \ R_\text{J}$ and an orbital period of 9.96 days.

Finally, I also confirmed the planetary nature of an additional candidate, not part of the above sample, through RV observations.
This planet is also a possibly inflated $1.28 \ R_\text{J}$ giant planet on a 4.38-day-period orbit around a red-giant star.

The formal confirmation of all candidates, in particular the two found in the above planet search, should allow for an occurrence rate study of close-in giants orbiting giants with \textit{TESS}.
Furthermore, two of the planets have potentially inflated radii, and can be useful as testbeds to determine the origins and efficiency of planetary radius inflation.


%% file: preliminary/resumo.tex
Múltiplos rasterios de velocidades radiais e missões espaciais de fotometria como o \textit{Kepler} e o \textit{K2} impulsionaram consideravelmente o número de novos exoplanetas descobertos desde o início deste século.
No entanto, dado um enviesamento da seleção de alvos para estrelas de sequência principal, poucos planetas são conhecidos que transitem estrelas evoluídas.
Estes, cuja maioria são planetas gigantes, contêm informação importante para desvendar os mistérios da formação e evolução dos sistemas planetários.

Correlações entre a metalicidade e/ou a massa de estrelas e a presença de planetas gigantes fornecem pistas acerca dos mecanismos responsáveis pela formação de planetas gigantes.
Já a comparação de taxas de ocorrência de planetas gigantes de curto período entre estrelas anãs e gigantes permite inferir o impacto que a evolução estelar tem nos sistemas planetários.
Finalmente, a caracterização detalhada do raio de planetas gigantes que orbitem estrelas gigantes pode conter informação única para perceber as origens da inflação dos raios observada em planetas gigantes.
Por estas razões, é necessário encontrar e caracterizar detalhadamente mais planetas gigantes que orbitem estrelas evoluídas.

Nesta tese procurei aumentar a população de planetas gigantes conhecidos que orbitam gigantes vermelhas e melhorar a sua caracterização, tendo como foco os dados da missão \textit{TESS}, recentemente lançada.
Especificamente, concentrei-me em planetas gigantes de curto período que orbitem gigantes vermelhas de baixa luminosidade, preferencialmente com oscilações.

Para melhorar a caracterização destes sistemas planetários, desenvolvi um modelo para descrever os sinais do trânsito do planeta e da estrela simultaneamente.
Para isso, utilizei processos Gaussianos para capturar os sinais da granulação e oscilações estelares no domínio temporal, através de expressões equivalentes às do domínio das frequências, encontradas na literatura.

Testes deste modelo com curvas de luz simuladas do \textit{TESS} e curvas de luz do \textit{Kepler} demonstram que o modelo é capaz de recuperar os sinais verdadeiros presentes nos dados e que obtém resultados iguais (dentro das incertezas) aos dos obtidos através do espectro de potências.
Em particular, os resultados demonstram a capacidade do modelo de fazer asterosismologia no domínio temporal, estimando $\nu_\text{max}$ a 1\% do valor real, desde que as oscilações sejam detetáveis.

Para além disso, quando são considerados trânsitos e sinais estelares simultaneamente, os parâmetros planetários estimados pelo nosso modelo têm maior exatidão e menores incertezas, quando comparados com os de um modelo mais simples de trânsito, comum na literatura.

No contexto da procura de planetas, desenvolvi uma \textit{pipeline}, constituída maioritariamente por \textit{software} aberto, para extrair e corrigir curvas de luz das imagens de campo largo do \textit{TESS}, fazendo posteriormente uma procura e validação de trânsitos nas mesmas.
Usando esta \textit{pipeline}, explorei uma amostra de gigantes vermelhas brilhantes, de baixa luminosidade, no hemisfério sul do \textit{TESS}.
A amostra foi limitada a estrelas com raios superiores a 3 $\rm R_\odot$, de forma a garantir que as oscilações estelares se encontrassem abaixo da frequência de Nyquist dos dados de longa cadência do \textit{TESS}, e com raios inferiores a 8 $\rm R_\odot$, para garantir que trânsitos de planetas do tamanho de Júpiter fossem visíveis nas curvas de luz.
As estrelas selecionadas foram também restringidas a magnitudes \textit{TESS} inferiores a 10 para que houvesse maior probabilidade de detetar oscilações.

Como resultado da procura, identifiquei quatro planetas candidatos, dois dos quais não são planetas conhecidos e orbitam gigantes vermelhas.
Observações de velocidades radiais de ambos apontam provisoriamente para a sua natureza planetária.
O primeiro candidato é um Júpiter quente com um raio possivelmente inflacionado de $1.24 \ R_\text{J}$, numa órbita de 6.20 dias.
Já o segundo candidato tem um raio de $1.02 \ R_\text{J}$ e um período orbital de 9.96 dias.

Finalmente, confirmei também a natureza de outro planeta candidato, que não fazia parte da amostra anterior, através de observações de velocidades radiais.
Este planeta é gigante, tem um raio que poderá estar inflacionado de $1.28 \ R_\text{J}$, e orbita uma gigante vermelha a cada 4.38 dias.

A confirmação formal de todos os candidatos, em particular os dois encontrados na procura acima mencionada, possibilitarão um estudo da taxa de ocorrência de gigantes de curto período que orbitam gigantes com o \textit{TESS}.
Para além disso, dois dos planetas encontrados têm raios potencialmente inflacionados, o que poderá permitir estudar as origens e eficiência da inflação de raios planetários.


%% file: preliminary/contents.tex
\cleardoublepage

\pdfbookmark[chapter]{\contentsname}{toc}
\tableofcontents

\cleardoublepage

\pdfbookmark[chapter]{\listfigurename}{lof}
\listoffigures

\cleardoublepage

\pdfbookmark[chapter]{\listtablename}{lot}
\listoftables

\cleardoublepage

%% file: chapters/introduction/main.tex
\chapter{Introduction}
\label{cha:introduction}

The field of exoplanets has had a meteoric rise in research ever since its recent inception.
In the 26 years since the first confirmed exoplanet discovery orbiting a solar-like star, by \textcite{Mayor_1995}, the number of known exoplanets has grown to over 4000 (from \textit{The Extrasolar Planets Encyclopaedia}\footnote{\href{http://www.exoplanet.eu}{exoplanet.eu}}; \cite{Schneider_2011}), driven largely by radial-velocity (RV) surveys and photometric space missions.
Most of these systems, however, orbit main-sequence stars, with the population of planets around evolved stars being considerably smaller.
Of these planets orbiting evolved stars, the majority has been found through radial-velocity surveys, with only a handful of known transiting planets.

In this introduction, I try to motivate the search and characterization of more of these transiting systems, in particular with the advent of NASA's \textit{Transiting Exoplanet Survey Satellite} (\textit{TESS}) mission.
I start, in Section~\ref{sec:exoplanet_history}, with some highlights of this short yet highly packed history of exoplanetary science to arrive at the \textit{TESS} mission \parencite{Ricker_2015}, one of the latest photometric space missions to launch and promise new discoveries, and the one at the forefront of this thesis' work.
I then expand on the ways in which the current understanding of planets orbiting evolved stars has lagged behind that of main-sequence hosts and how \textit{TESS} presents a great opportunity to advance this particular topic.
Section~\ref{sec:synergy_asteroseismology} then discusses the synergies that exoplanet studies have with asteroseismology, underlining some of the methodologies from the latter field that enhance our understanding of planetary systems.
In Section~\ref{sec:giant_planets}, I discuss the state of the art with respect to giant exoplanets orbiting evolved stars, exploring open questions and weighing in on the potential of \textit{TESS} to provide answers.
Finally, Section~\ref{sec:objectives} lays out our objectives and details the structure of this thesis.

\input{chapters/introduction/exoplanet_history}

\input{chapters/introduction/synergies}

\input{chapters/introduction/scientific_rationale}

\input{chapters/introduction/objectives}

%% file: chapters/introduction/exoplanet_history.tex
\section{Exoplanets: An historical perspective}
\label{sec:exoplanet_history}

Although initial detections of planets would still take a few more decades, in 1952 Otto Struve wrote a paper outlining two techniques that could lead to the detection of planets around other stars \parencite{Struve_1952}, which have proven to be the two major techniques in the discovery of new exoplanets today.

The first technique involved detecting the gravitational influence a planetary companion has on a host star. 
This influence can be detected by determining the line of sight velocity of the star through Doppler shifts in the star's spectral lines over the companion's orbital period. 
This technique is referred to as radial velocity (RV), and the velocity semi-amplitude, $K$, is proportional to the ratio between the masses of the planet, $M_\text{p}$, and the star, $M_\star$,
\begin{equation}
    K \propto \frac{M_\text{p}}{M_\star^{2/3}}.
    \label{eq:struve_rv}
\end{equation}
For inclined orbits, $M_\text{p}$ is replaced by $M_\text{p} \sin i$, where $i$ is the orbital inclination ($i = 90^\circ$ for edge-on orbits and $i = 0^\circ$ for face-on orbits).

The second technique involved the dimming of the host star's brightness, $F$, in the event that a planetary companion crossed in front of it from an observer's line-of-sight. 
This dimming is proportional to the fraction of the stellar disk blocked, and consequently proportional to the squared ratio between the radii of the planet, $R_\text{p}$, and the star, $R_\star$,
\begin{equation}
    \Delta F \propto \left( \frac{R_\text{p}}{R_\star} \right)^2.
    \label{eq:struve_transit}
\end{equation}
Through the two techniques, a measure of both a planet's mass and radius can be estimated, which, when combined, can provide a measure of the planet's mean density.
Both these techniques are responsible for most of the confirmed exoplanets known today, with the \textit{The Extrasolar Planets Encyclopaedia} citing $\sim$91.8\% of all discoveries.
For a discussion on the multiple existing methods for exoplanet detection, including the two mentioned before, see the work of \textcite{Santos_2020} and chapter~1 of \textcite{Perryman_2018}.

\subsection{First detections}

The first confident detection of a planet-mass object outside the Solar System was not found orbiting a solar-like star but a pulsar \parencite{Wolszczan_1992}.
For main-sequence stars, early suggestions of planetary-mass object detections started in the late 1980s with the works of \textcite{Campbell_1988,Latham_1989,Hatzes_1993}, using the RV method. 
But it was the discovery of 51 Pegasi b, by \textcite{Mayor_1995}, that sparked the rise of exoplanet studies.

Besides being the first unambiguous detection of a planet orbiting a solar-like star, 51 Pegasi b was also a surprising discovery due to the unexpected anatomy of its planetary system.
This planet had a minimum mass of 0.47 Jupiter masses ($M_\text{J}$), but was orbiting its star every 4.2 days at a distance of 0.05 AU (astronomical units), much closer than was expected, given what was known from the Solar System (for reference, the elliptical orbit of Mercury is never closer than 0.3 AU from the Sun, with Jupiter orbiting at about 5.2 AU).
These close-in giant planets took on the name of \textit{hot Jupiters} and were the most common type of planet found in the early days \parencite{Marcy_1996,Butler_1996b} due to their large sizes and masses, and very short orbital periods.
The discovery of these short-period giants also led to a revolution in planet formation theories, which until then could only rely on the Solar System's planets to inform their hypotheses (see Section~\ref{sec:giant_planets}).

The first confirmed detection of a transiting planet would take a few extra years, with \textcite{Charbonneau_2000} and \textcite{Henry_2000} both announcing the detection of the transit of the 1.27 Jupiter radii ($R_\text{J}$) planet, HD 209458 b, a planet which had been discovered by the RV method.
The first planet discovery using the transit method was OGLE-TR-56 b \parencite{Udalski_2002,Konacki_2003}.

\subsection{Space revolution}

From those initial detections, planet observations started to ramp up, with RV surveys led by newly developed spectrographs, with some notable examples being 
\textit{CORALIE} \parencite{Queloz_2000}, 
\textit{HIRES} \parencite[\textit{High Resolution Echelle Spectrometer};][]{Vogt_1994} and 
\textit{HARPS} \parencite[\textit{High Accuracy Radial velocity Planet Searcher};][]{Mayor_2003}.
This had the important consequence of leading to the discovery of planets in the Neptune \parencite{Santos_2004b,Butler_2004,McArthur_2004} and super-Earth \parencite{Mayor_2011} mass regimes.

This was accompanied by transit surveys, with multiple ground-based facilities sprouting, e.g. 
\textit{OGLE} \parencite[\textit{Optical Gravitational Lensing Experiment};][]{Udalski_2003}, 
\textit{HAT} \parencite[\textit{Hungarian Automated Telescope};][]{Bakos_2004}, 
\textit{TrES} \parencite[\textit{Trans-Atlantic Exoplanet Survey}; see][and references therein]{Alonso_2004} and 
\textit{WASP} \parencite[\textit{Wide Angle Search for Planets};][]{Pollacco_2006}.

However, atmospheric variability proved problematic for ground-based photometric observations, especially for lower brightness targets, and so innovation led to space, with \textit{CoRoT} \parencite[\textit{Convection Rotation and planetary Transits};][]{Auvergne_2009} leading the way and \textit{Kepler} \parencite{Borucki_2009} revolutionizing the field, with thousands of new exoplanet candidates being detected in the years following its launch \parencite{Batalha_2013,Morton_2016}.

\begin{figure}[!tpb]
    \centering
    \includegraphics[height=0.5\textheight, width=\textwidth, keepaspectratio]{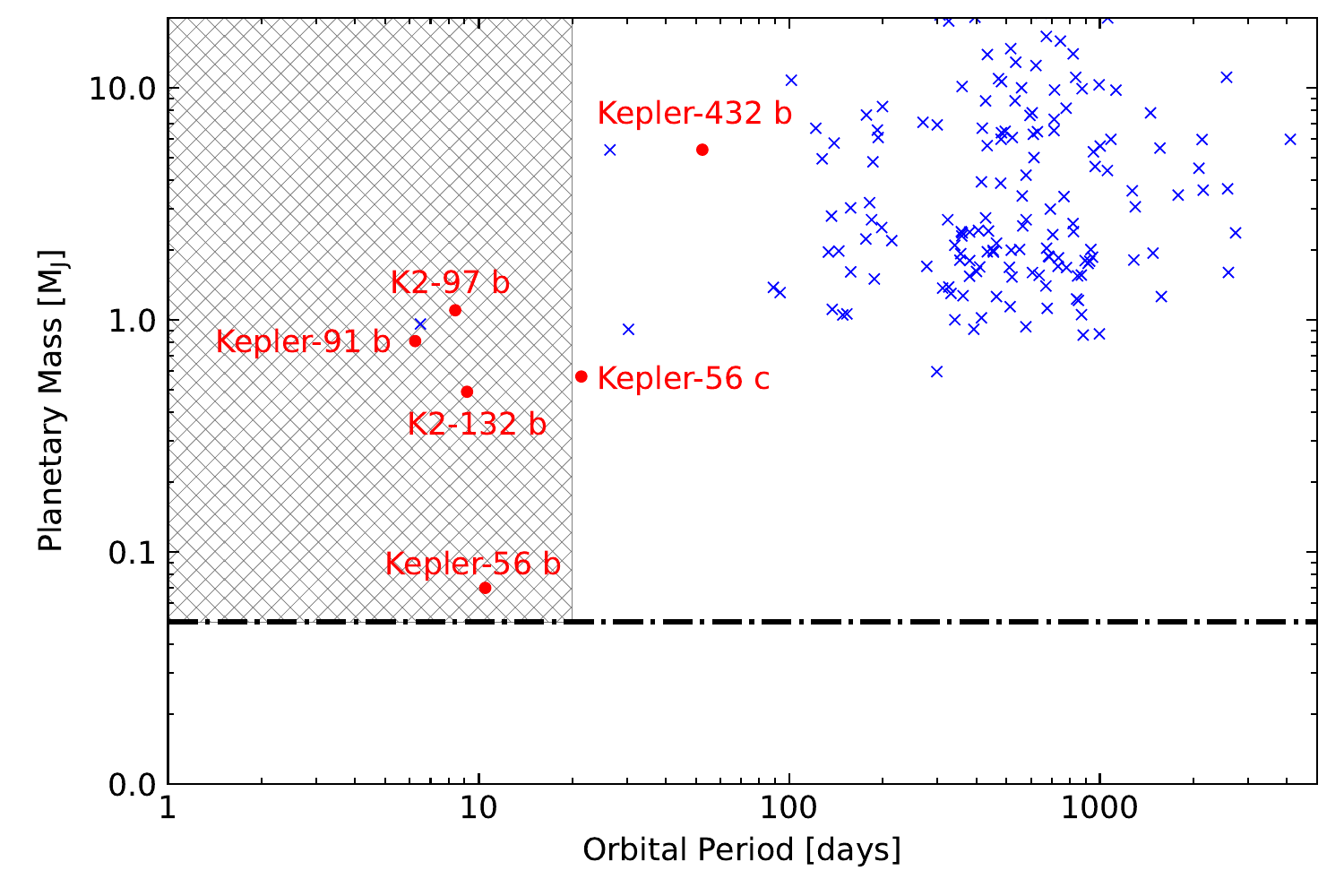}
    \caption[Mass-period diagram of known exoplanets orbiting red-giant branch stars]
    {Mass-period diagram of known exoplanets orbiting red-giant branch stars. 
    Planets detected by the transit method are depicted as red circles and those detected in RV surveys as blue crosses (in which case the masses are lower limits). 
    The dashed-dotted line marks the mass of Neptune. 
    The shaded area approximately corresponds to the parameter space probed by \textit{TESS} considering one sector (i.e. 27.4 days) of data ($M > M_\text{Neptune}$; $P \leq 20$d). 
    Inspired by figure 11 of \textcite{Campante_2016a}.}
    \label{fig:known_evolved_hosts}
\end{figure}

Most planets found with the \textit{Kepler} mission orbited dwarf stars, which was expected, given the sample of stars observed by the satellite \parencite{Batalha_2010}.
This choice was premeditated in an effort to expedite the discovery of a habitable Earth-like planet orbiting a Sun-like star.
Unfortunately, this left transiting planets orbiting evolved stars largely unexplored, with most systems being detected by RV surveys instead \parencite{Johnson_2007b,Johnson_2010b,Reffert_2015}, and only a handful from transits confirmed at the time \parencite[][]{Huber_2013a,Lillo-Box_2014,Barclay_2015,Quinn_2015} and an additional one confirmed years later \parencite{Chontos_2019}.

Following the failure of one of its reaction wheels, the \textit{Kepler} spacecraft was repurposed into the \textit{K2} mission \parencite{Howell_2014}.
Unlike its predecessor, where the satellite's 10$\times$10-degree field-of-view camera observed a single patch of the sky continuously for 4 years, \textit{K2} would observe a different field every $\sim$80 days along the ecliptic plane.
Due to the larger area of the sky covered throughout the mission and to a community-driven target selection, \textit{K2} greatly expanded on the number of evolved stars around which to look for planets.
This led to the discovery of additional planets transiting evolved hosts \parencite{Grunblatt_2016,Grunblatt_2017,Grunblatt_2019} as well as to the first ensemble studies of transiting exoplanets orbiting giant stars \parencite{Grunblatt_2018,Grunblatt_2019}.

Figure~\ref{fig:known_evolved_hosts} illustrates the state of the art after the \textit{K2} mission, highlighting all known transiting planets orbiting red-giant stars.

\subsection{\textit{TESS} - An all-sky survey}

\begin{figure}[!tpb]
    \centering
    \includegraphics[height=0.5\textheight, width=\textwidth, keepaspectratio]{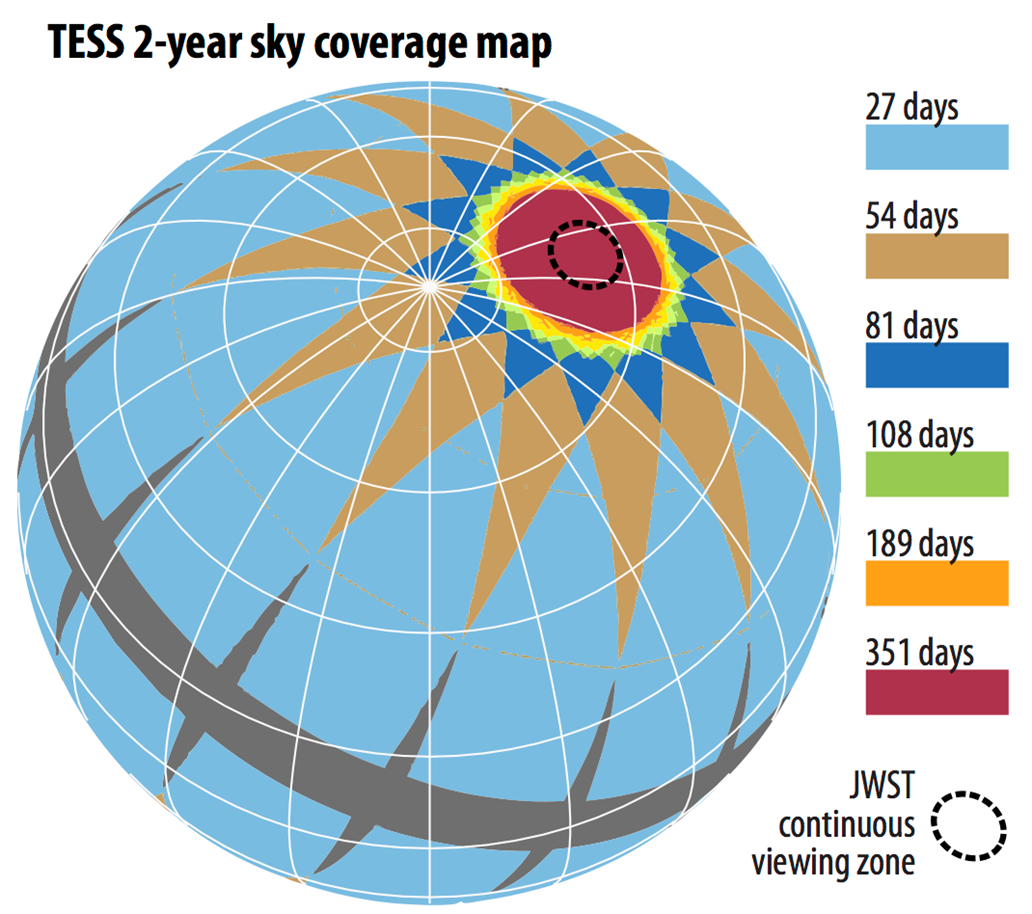}
    \caption[Pre-launch schematic showing the sky coverage of the four cameras from the \textit{TESS} satellite on the celestial sphere during the initial two years of the mission.]
    {Pre-launch schematic showing the sky coverage of the four cameras from the \textit{TESS} satellite on the celestial sphere during the initial two years of the mission. Each ecliptic hemisphere is observed by 13, 27.4-day sectors. Color-coded regions highlight different lengths of observations by \textit{TESS}, due to the areas covered by \textit{TESS} cameras overlapping near the poles during different sectors. The gray region, concentrated mostly near the ecliptic, is the region not observed by \textit{TESS}. The black dashed line around the ecliptic pole shows the expected region where the \textit{James Webb Space Telescope} (\textit{JWST}) will have continuous viewing capabilities. Figure from the \href{https://tess.mit.edu/}{\textit{TESS} MIT website}.}
    \label{fig:tess_coverage}
\end{figure}

Launched in April of 2018, NASA's \textit{TESS} mission \parencite[\textit{Transiting Exoplanet Survey Satellite};][]{Ricker_2015} is one of the latest photometric space missions to launch and the first to perform a (nearly) complete photometric survey of the sky with the intent of searching for exoplanets, during a two-year period (primary mission).

\textit{TESS}'s satellite is equipped with four identical cameras with a 24$\times$24-degree field of view, each with four charge-coupled devices (CCD).
Aligned vertically, the cameras cover the sky in 26 segments, named sectors, with 13 in the southern hemisphere, observed during the first year of the primary mission and the remaining 13 in the northern hemisphere, observed the following year.
Each sector is observed for a total of 27.4 days, but due to some overlapping of the sectors on each hemisphere, some areas of the sky, referred to as continuous viewing zones (CVZ), were observed for up to 351 days.
Figure~\ref{fig:tess_coverage} illustrates the sky coverage of the 4 cameras from \textit{TESS} for the 26 sectors covering the two hemispheres, including the multiple overlapping regions, color-coded according to the expected days of observations.

For every sector, the spacecraft observed and downlinked about 20,000 targets in two-minute cadence postage stamps, as well as full-frame images (FFIs) of the four cameras binned to 30-min cadence.
Additionally, \textit{TESS}'s extended mission has been approved and is already in progress\footnote{\href{https://heasarc.gsfc.nasa.gov/docs/tess/the-tess-extended-mission.html}{\textit{TESS} Extended Mission}}, with an increased observational cadence of 10-min for the FFIs (as well as adding a 20-second cadence observation mode for about 1000 targets per sector).
This ``second round'' from \textit{TESS} will not only (for the most part) double the number of available sectors of data of observed targets, which improves the confidence of previous planet detections and allows for new ones \parencite{Kane_2021}, but the shorter FFI cadence will increase the sample of targets for which asteroseismology can be applied, improving stellar characterization as well.

From these existing data, up to a few hundred new planets orbiting evolved stars are expected to be discovered \parencite{Campante_2016a,Barclay_2018}, an estimate that is highly dependent on the assumed occurrence rate.

%% file: chapters/introduction/synergies.tex
\section{Synergy with asteroseismology}
\label{sec:synergy_asteroseismology}

Since most exoplanets are detected through indirect methods, their fundamental properties (such as mass and radius) are only known in relation to their host's properties.
As such, a precise characterization of the planet is directly dependent on a precise characterization of its host.
In this section, I briefly introduce the field of asteroseismology, which has had a considerable impact on exoplanetary sciences in recent years.

\subsection{Introduction}
\label{sub:astero_intro}

Asteroseismology is the science that studies the internal structure of stars by interpreting their intrinsic global oscillations. 
The observed frequencies carry information on the internal structure and dynamics of the star, and may in turn be used to place constraints on the fundamental stellar properties.

There are two main types of oscillating stars. 
Classical oscillators have oscillations which are coherently excited and intrinsically unstable.
One famous example of this class of oscillators are the Cepheid variable stars, which have a very precise correlation between their pulsation period and luminosity.
The other type are solar-like oscillators, which have intrinsically stable oscillations and are stochastically driven by near-surface convection \parencite{Aerts_2010}.
Their name stems from the fact that the Sun exhibits this type of oscillations.

\begin{figure}[!tpb]
    \includegraphics[height=0.5\textheight, width=0.48\textwidth, keepaspectratio]{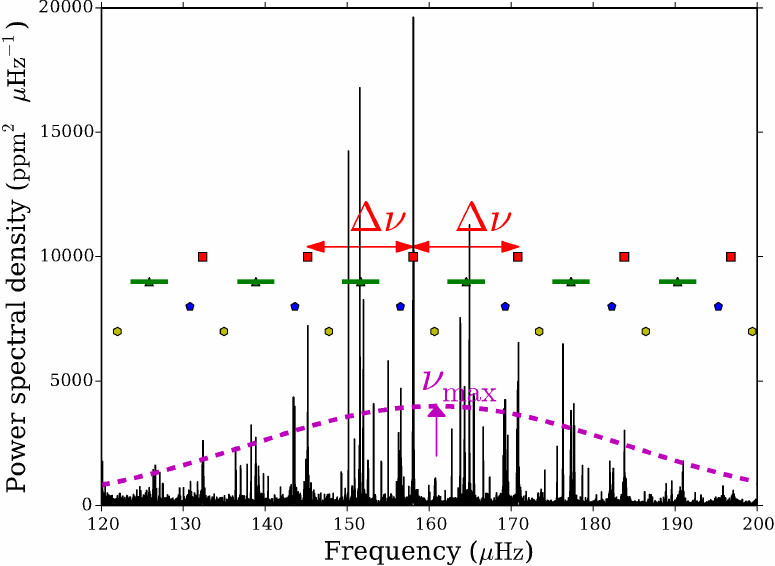}
    \includegraphics[height=0.5\textheight, width=0.48\textwidth, keepaspectratio]{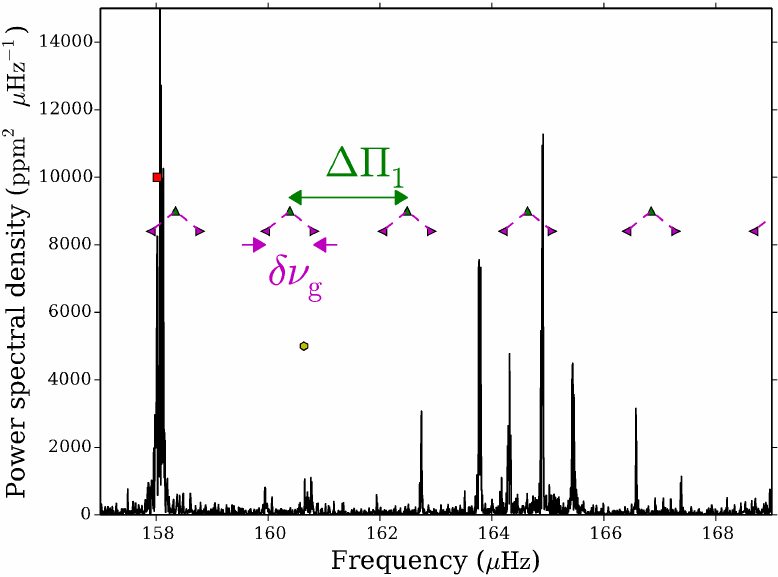}
    \caption[Oscillation spectrum of a \textit{Kepler} red-giant star]
    {Oscillation spectrum of a \textit{Kepler} red-giant star.
    On the left, an overview of all oscillations modes identified is shown. 
    The radial modes (red squares; $l = 0$) are equally spaced in frequency, called the large frequency separation, $\Delta \nu$.
    The green triangles with extended bars, blue pentagons and yellow hexagons represent the dipole modes ($l = 1$), quadruple modes ($l=2$) and octuple modes ($l=3$), respectively.
    The spectrum is centered on the frequency of the peak of the oscillations envelope, called the frequency of maximum oscillation, $\nu_\text{max}$.
    On the right, a zoom in of the left plot is shown.
    The green up triangles shown are the g-mode dipole frequencies, observed due to the mixed nature of the dipole modes.
    These mixed modes are equally spaced in period, called the period spacing, $\Delta \Pi_1$.
    The magenta left and right triangles show the rotational splittings of the dipole modes, along with the core rotational splitting parameter, $\delta \nu_\text{g}$.
    Both plots are from \textcite{Davies_2016a}.}
    \label{fig:oscillation_spectrum}
\end{figure}

Solar-like oscillators extend from low-mass main-sequence stars to evolved red giants and possibly beyond \parencite{Christensen-Dalsgaard_2001,Cunha_2020}.
Regarding the physical nature of the oscillations in these stars, they can either have the nature of standing acoustic waves, which are referred to as p-modes or pressure waves, or have the nature of internal gravity waves commonly referred to as g-modes or gravity waves. 
With the exception of radial p-modes (characterized by an angular degree $l = 0$), which travel through the entirety of the stellar interior, oscillations can either be trapped in the outer layers of the stars, where the restoring force is the pressure gradient (p-modes) or in the inner radiative region of the stars, where the restoring force is buoyancy (g-modes) \parencite[see ][]{Aerts_2010}.
As g-modes are trapped in the central regions of the star, only p-modes can be detected in main-sequence stars (and only with low angular degree, $l \leq 4$).
For evolved stars, modes of mixed character can also exist and be observed \parencite{Beck_2011}.
These are called mixed modes and behave like g-modes in the stars' inner region and like p-modes in the envelope, thus carrying information about the stellar core \parencite{Bedding_2011,Mosser_2011}.
Figure~\ref{fig:oscillation_spectrum} shows an oscillation spectrum from a \textit{Kepler} red-giant star, highlighting the oscillation modes visible.

\subsection{Asteroseismic stellar characterization}

Focusing solely on evolved solar-like oscillators, three global asteroseismic quantities can be extracted from the oscillation modes, which are of great interest due to their relation to stellar properties.

The quasi-regular frequency spacing between two adjacent radial modes of oscillation is known as the \textit{large frequency separation}, $\Delta\nu$, and is sensitive to the radius of the star and proportional to the square root of its mean density, $\bar{\rho}$ \parencite{Ulrich_1986}, through
\begin{equation}
    \Delta\nu \propto \sqrt{\bar{\rho}} \propto \sqrt{\frac{M}{R^3}},
    \label{eq:delta_nu_relation}
\end{equation}
where $M$ and $R$ are the stellar mass and radius, respectively.

The frequency of the peak of the oscillations envelope, known as the frequency of maximum power, $\nu_\text{max}$, varies due to both the surface gravity, $g$, and effective temperature, $T_\text{eff}$, of the star \parencite{Kjeldsen_1995a},
\begin{equation}
    \nu_\text{max} \propto g T_\text{eff}^{-1/2}.
    \label{eq:nu_max_relation}
\end{equation}
Rearranging the relations of these two quantities, both the mass and the radius of the star can be estimated, provided the stellar effective temperature is also known, using the following \textit{scaling relations} \parencite[see][]{Chaplin_2013},
\begin{align}
    \frac{R}{\text{R}_\odot} &\simeq \left( \frac{\nu_{\textnormal{max}}}{\nu_{\textnormal{max},\odot}} \right) \left( \frac{\langle \Delta\nu \rangle}{\langle \Delta\nu \rangle_\odot} \right)^{-2} \left( \frac{T_{\textnormal{eff}}}{T_{\textnormal{eff},\odot}} \right)^{0.5} , \\
    \frac{M}{\text{M}_\odot} &\simeq \left( \frac{\nu_{\textnormal{max}}}{\nu_{\textnormal{max},\odot}} \right)^3 \left( \frac{\langle \Delta\nu \rangle}{\langle \Delta\nu \rangle_\odot} \right)^{-4} \left( \frac{T_{\textnormal{eff}}}{T_{\textnormal{eff},\odot}} \right)^{1.5} ,
    \label{eq:scaling_relations}
\end{align}
where the denominator values are the solar reference values.
Derived stellar properties from these scaling relations have average uncertainties of $\sim 5\%$ in radius and $\sim 10\%$ in mass for both main sequence \parencite{SilvaAguirre_2012,Huber_2012,Huber_2013b,Huber_2017} and giant stars \parencite{Gaulme_2016,Huber_2017,Yu_2018}.

Finally, there is the period spacing, $\Delta \Pi_1$.
Period spacing is the regular spacing in period expected between two adjacent gravity-dominated modes.
As mixed modes have both g-mode and p-mode character, they exhibit a periodic spacing in period, which is larger the more the mixed mode is dominated by its gravity mode. 
Period spacing was first observed in a red-giant star by \textcite{Beck_2011}.
This quantity carries information from the inner regions of the star and can be used to infer stellar ages, together with $\Delta \nu$ and $\nu_\text{max}$, helping also to differentiate between different stellar evolutionary stages \parencite{Mosser_2012c,Stello_2013,Mosser_2014,Vrard_2016}.

As an alternative to global asteroseismic quantities, slower but more sophisticated methods to determine stellar properties from asteroseismic data include grid-based modeling \parencite{Stello_2009,Kallinger_2010a,Huber_2013b,SilvaAguirre_2015}, where additional asteroseismic information, such as individual modes of oscillation, can be used to infer the stellar structure.
Examples of such implementations include \textit{AIMS} \parencite[\textit{Asteroseismic Inference on a Massive Scale};][]{Rendle_2019}, \textit{BASTA} \parencite[\textit{BAyesian STellar Algorithm};][]{SilvaAguirre_2015} and \textit{PARAM} \parencite{Rodrigues_2017}.
With these methods, typical uncertainty values for \textit{Kepler} main sequence stars are of $\sim 1-3\%$ in radius, $\sim 5\%$ in mass and $\sim 10-15\%$ in age \parencite{Huber_2013b,SilvaAguirre_2015,SilvaAguirre_2017}.

Additionally, asteroseismic data can also be used to infer stellar inclinations \parencite{Huber_2013a,Campante_2016b}, and orbital properties, such as exoplanet eccentricities \parencite{Dawson_2012,Sliski_2014,VanEylen_2019}.

\subsection{\textit{TESS} asteroseismic potential}
\label{sub:tess_astero_potential}

\begin{figure}[!tpb]
    \includegraphics[height=0.5\textheight, width=\textwidth, keepaspectratio]{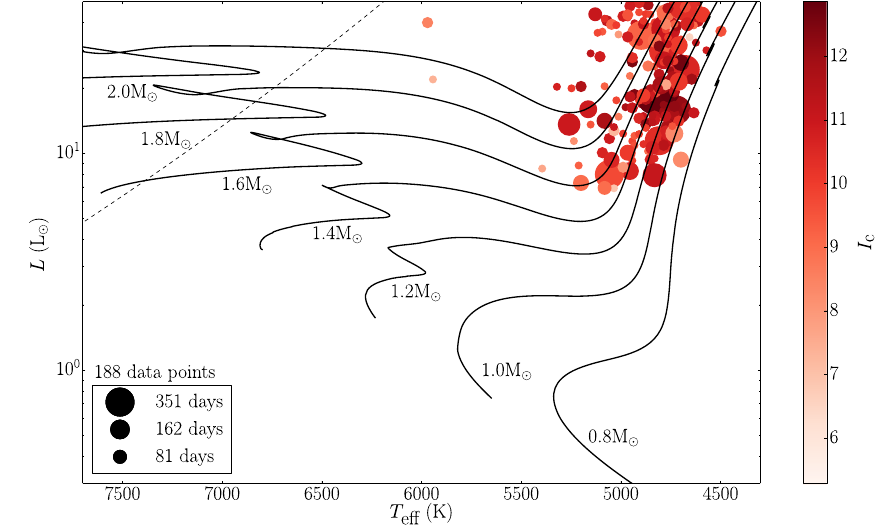}
    \caption[Predicted yield of \textit{TESS} evolved hosts with detectable solar-like oscillations]
    {Predicted yield of \textit{TESS} evolved hosts with detectable solar-like oscillations. 
    Data points are color-coded according to apparent magnitude (with the Johnson$-$Cousins $I_C$ magnitude used as a proxy for the \textit{TESS} magnitude) and their size is proportional to the observing length. 
    Solar-calibrated evolutionary tracks spanning the mass range $0.8-2.0$ $\text{M}_\odot$ are shown as continuous lines. 
    The slanted dashed line represents the red edge of the $\delta$ Scuti instability strip. 
    Most stars are located at the low-luminosity end of the red-giant branch, as per the focus of the work. 
    Figure from \textcite{Campante_2016a}.}
    \label{fig:asteroseismic_potential_tess}
\end{figure}
Initial results of asteroseismic studies of \textit{TESS} red-giant targets have already been published \parencite{SilvaAguirre_2020b,Mackereth_2021}.
\textcite{SilvaAguirre_2020b} analyzed a sample of 25 bright stars ($V$ magnitude < 7) with one or two sectors of \textit{TESS} data, quoting uncertainties for stellar radii, masses and ages of 6\%, 14\% and 50\%, respectively.
The same publication observes that these uncertainties improve significantly with the inclusion of parallax data from \textit{Gaia} DR2 \parencite{GaiaCollaboration_2018}, namely, to 3\%, 6\% and 20\%, respectively, close to values obtained for \textit{Kepler} targets.

\textcite{Mackereth_2021} focused on targets in the southern CVZ (more sectors of data), but considered fainter stars on average (\textit{Gaia} $G$ magnitude < 11), evaluating the probability of asteroseismic detections in \textit{TESS} data of red-giant stars.
They find an average detection yield of $\sim$36\% for all targets in their sample, which increases to $\sim$50\% when considering only targets with one year of data.

For asteroseismic exoplanet hosts, \textcite{Campante_2016a} conducted a study to predict the yield of evolved host stars in \textit{TESS} FFIs with detectable solar-like oscillations, pointing up to 200 low-luminosity red-giant branch (LLRGB) hosts, as illustrated in Figure~\ref{fig:asteroseismic_potential_tess}.
However, the authors note that this estimate was based on a synthetic stellar population of the Galaxy with adopted occurrence rates from \textcite{Fressin_2013} (for $T_\text{eff}$ > 4000 K) which do not account for evolutionary effects in the planetary systems as the hosts evolve off the main sequence.
At the same time, \textit{first-light} papers describing the asteroseismic characterization of evolved planetary hosts using \textit{TESS} data have already been published \parencite{Huber_2019b,Campante_2019,Nielsen_2019,Ball_2020,Jiang_2020}.
Provided that the oscillations envelope of giant stars is visible in \textit{TESS} FFI 30-min cadence data, asteroseismology of many evolved \textit{TESS} targets should be possible, with estimates of $\sim 3-5 \times 10^5$ red giants with detectable oscillations across the whole sky \parencite[even if only through global asteroseismic quantities;][]{SilvaAguirre_2020b,Mackereth_2021}.
\textit{TESS} thus presents a great opportunity for the detailed characterization of evolved hosts and their transiting exoplanets.

%% file: chapters/introduction/scientific_rationale.tex
\section{Giant planets orbiting giant stars}
\label{sec:giant_planets}

\begin{figure}[!tpb]
    \includegraphics[width=\textwidth, height=\textheight, keepaspectratio]{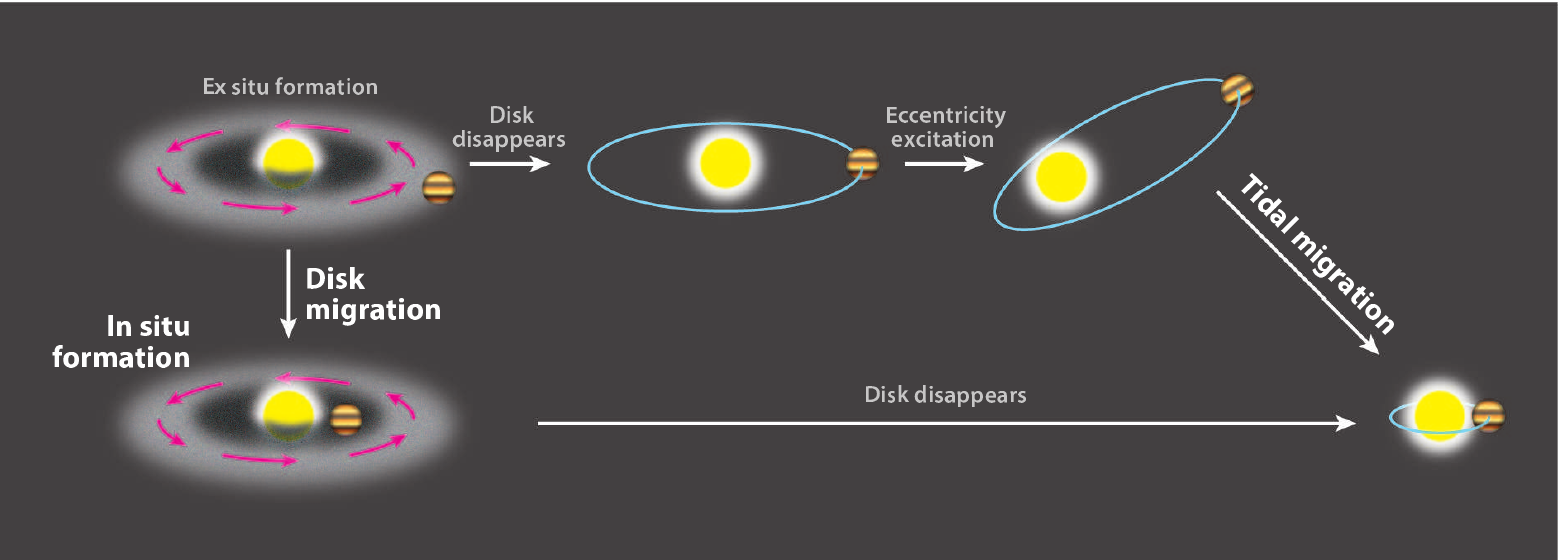}
    \caption[Schematic illustrating the three existing theories for the origin of hot Jupiters]
    {Schematic illustrating the three existing theories for the origin of hot Jupiters. Figure from \textcite{Dawson_2018}.}
    \label{fig:hot_jupiter_origin}
\end{figure}

The formation of gas-giant planets is believed to happen either through core accretion \parencite{Pollack_1996}, whereby gas from the protoplanetary disk accretes around a previously formed rocky/ice core, or gravitational instability \parencite{Boss_1997}, whereby a planet is formed because of a direct gravitational instability in the protoplanetary disk, in the same way that stars form from interstellar clouds.

Also, whilst long-period giant planets, beyond the ice-line, are believed to form in-situ, the situation is not so clear for hot Jupiters.
Three different hypotheses have been proposed in the literature to explain their origin, with varying degrees of agreement with the two theories of formation mentioned above.
The three are, in-situ formation, whereby the hot Jupiter is thought to form close to the star \parencite{Batygin_2016,Boley_2016}, disk migration, whereby the planet forms further away but migrates throughout the protoplanetary disk during stellar formation \parencite{Lin_1996,Ida_2008}, and high-eccentricity migration, whereby the planet forms away from the star, and is then excited to a highly eccentric orbit which is eventually circularized close to the star through tidal dissipation (several mechanisms have been proposed to explain the excitation of cold Jupiters' eccentricities; see \textcite{Dawson_2018}, and references therein).
Figure~\ref{fig:hot_jupiter_origin} illustrates these three hypotheses.

Population studies of giant planets with respect to their own structural properties, orbital properties or their hosts' properties can shed light on which of these theories may be at play.
Throughout this section, I explore the current state of the art in connection to these studies in order to motivate the goal of this thesis in finding more of these systems.
I also highlight the opportunities that open up with the exploitation of \textit{TESS}'s population of evolved planet hosts.

\subsection{Occurrence rate}
\label{sub:occurrence_rate}

For main-sequence host stars, the occurrence rate of giant planets, determined from RV studies, is of around 10\% \parencite{Marcy_2005,Cumming_2008,Mayor_2011}.
For transit surveys, \textcite{Fressin_2013} measured 5\%, though only for orbits shorter than $\sim$400 days, which is not comparable with the periods of up to $\sim$10 years considered in the above value from RVs.
As for smaller planets, estimates for the occurrence rate of planets with any mass lower or equal to Neptune's are of the order of 50\% \parencite{Mayor_2011,Petigura_2018a}.
Restricting it only to Earth-like planets, \textcite{Petigura_2013} estimate an occurrence rate of $\sim$26 \%.

With respect to hot Jupiters and still looking at main-sequence stars, RV surveys estimate a rate of about $\sim 1-1.5 \%$ \parencite{Marcy_2005,Cumming_2008,Mayor_2011,Wright_2012}. 
This value is at odds with the one found for \textit{Kepler} transiting planets, which is of about $0.4-0.5 \%$ \parencite{Howard_2012,Fressin_2013,Petigura_2018a}, a result that has been observed as well for \textit{TESS} \parencite{Zhou_2019}.
Hypotheses for these discrepancies have been attributed to different metallicities between samples \parencite{Howard_2012,Wright_2012} or to stellar multiplicity \parencite{Wang_2015b}.

Going beyond the main sequence and on to evolved hosts, some RV surveys found occurrence rates of giant planets similar to those of main-sequence stars, i.e. of about 10\% \parencite{Johnson_2007a,Wittenmyer_2020a}.
Other surveys have found discrepant values, such as the occurrence rate of 26\% found by \textcite{Bowler_2010}, a result that was also obtained by \textcite{Wittenmyer_2020a}, when the authors consider the entire Pan-Pacific Planet Search (PPPS) sample in their studies.
However, a common characteristic of all studies concerning evolved hosts is that of the larger masses of sample stars, with most targets being identified as evolved A stars.
Given the generally accepted correlation between stellar mass and giant-planet occurrence (see Section~\ref{sub:mass_correlation}), additional studies are necessary to understand the source of these discrepancies and determine the exact occurrence rate around evolved hosts. 

What is common with all RV surveys of evolved stars is the dearth of giant planets in short orbits, which points to differences in the hot-Jupiter populations around main-sequence and evolved stars \parencite{Johnson_2010a,Jones_2016}.
In the meantime, transit surveys have led to the discovery of a handful of systems around evolved stars \parencite{Huber_2013a,Lillo-Box_2014,Barclay_2015,Quinn_2015} and to an occurrence rate study for close-in planets orbiting \textit{K2} giant stars \parencite{Grunblatt_2019}, where the rate of $\sim 0.5$\%, matched the values found for main-sequence stars.
However, the latter study had only three confirmed planets in its sample, highlighting the need to increase the sample of known close-in planets orbiting evolved hosts, an opportunity presented by \textit{TESS} given its vast number of observed evolved targets.




\subsection{Metallicity correlation}

There is an established correlation between the general occurrence rate of giant planets and stellar metallicity for main-sequence stars \parencite{Gonzalez_1997,Santos_2001,Santos_2004a,Fischer_2005,Udry_2007,Sousa_2011a}.
This correlation establishes core accretion as the most likely formation mechanism for these planets \parencite{Boss_2001,Mordasini_2012}.
Moving on to subgiant stars, the trend also seems to be observed \parencite{Johnson_2010a,Jofre_2015b,Ghezzi_2018}.
However, for red-giant hosts, there has been no clear consensus, with some studies finding the same general correlation \parencite{Hekker_2007,Johnson_2010a,Reffert_2015,Jones_2016} and others finding none \parencite{Takeda_2008,Mortier_2013b,Maldonado_2013}.

\textcite{Pasquini_2007} try to explain this potential lack of correlation between metallicity and giant-planet occurrence around evolved stars by suggesting that the disk instability might be the preferential formation mechanism at work for intermediate-mass stars, removing the expected dependency on metallicity.
On the other hand, simulations by \textcite{Mordasini_2012}, based on the core accretion model, showed that stellar mass could also play a role in giant-planet formation, compensating for the lower metallicity of these planet-hosting evolved systems, when compared to main-sequence hosts.

As of yet, no clear consensus seems to have been reached, though \textit{TESS} population studies with asteroseismic hosts might provide some additional help, as asteroseismology can provide robust estimates of the surface gravity, $\log g$, lifting the degeneracy in spectroscopic estimations of the effective temperature, $T_\text{eff}$ and metallicity, [Fe/H].
Additionally, precise and (notionally) accurate asteroseismic masses might also help understand whether this correlation is present for all stellar masses \parencite{Santos_2012}.

\subsection{Mass correlation}
\label{sub:mass_correlation}

For stellar masses, the literature is clearer, with most results pointing to a correlation between the host's mass and the probability of having a giant planet, in agreement with core accretion theories of formation.
Giant stars are particularly useful for these studies, as they encompass a range of masses not probed with the sample of known main-sequence hosts.

\textcite{Johnson_2010a} found that for their sample, composed of stars ranging from M dwarfs with masses as low as 0.2 $\rm M_\odot$ up to intermediate-mass subgiants with masses of up to 1.9 $\rm M_\odot$, there was a trend between the stellar mass and the occurrence rate of giant planets.
However, disagreements with respect to the determination of the stellar masses of the evolved hosts and possible selection biases in the stellar sample used, called into question this mass correlation \parencite{Lloyd_2011,Lloyd_2013,Schlaufman_2013}, for masses larger than that of the Sun.
More recent works have weighed in on the masses of the stars in the sample using asteroseismology, thus improving the mass estimation \parencite{North_2017,Stello_2017a,Campante_2017}.

More recently, \textcite{Reffert_2015} extended the sample of \textcite{Johnson_2010a} up to 5 M$_\odot$, and observed that the giant-planet occurrence rate increased with stellar mass from 1.0 $\rm M_\odot$ to 1.9 $\rm M_\odot$ but then dropped rapidly for masses above 2.5 $\rm M_\odot$, with no planets found for stars with masses above 2.7 $\rm M_\odot$. 
Subsequent studies obtained similar results \parencite{Jones_2016,Ghezzi_2018}, with \textcite{Ghezzi_2018}, in particular, specifically addressing the issue of stellar mass determination and confirming the validity of the correlation found by \textcite{Johnson_2010a}.

\subsection{Eccentricities}

Looking at main-sequence stars, the population of hot Jupiters has mostly circular orbits, expected due to tidal interactions with its host \parencite{Hut_1981}.
As for warm Jupiters orbiting main-sequence hosts, the existence of a moderately eccentric population points to high-eccentricity migration being at least responsible for the origin of some of these planets \parencite{Dawson_2018}.

For evolved hosts, the planet population is expected to be different compared to that around main-sequence stars, due to dynamical interactions driven by stellar evolution \parencite{Veras_2016}.
Simulations have shown that the increased stellar radii can increase the angular momentum exchange between the planet and its host star, causing stellar tides to start to dominate over planetary tides and resulting in a transient population of moderately eccentric close-in giant planets \parencite{Villaver_2014}, non-existent around main-sequence stars.
This population has been observed already \parencite{Grunblatt_2018}, albeit with a small sample of planets, a case where \textit{TESS} once again can help.





\subsection{Radius inflation}
\label{sub:inflation}

\begin{figure}[!tp]
    \includegraphics[width=\textwidth, height=\textheight, keepaspectratio]{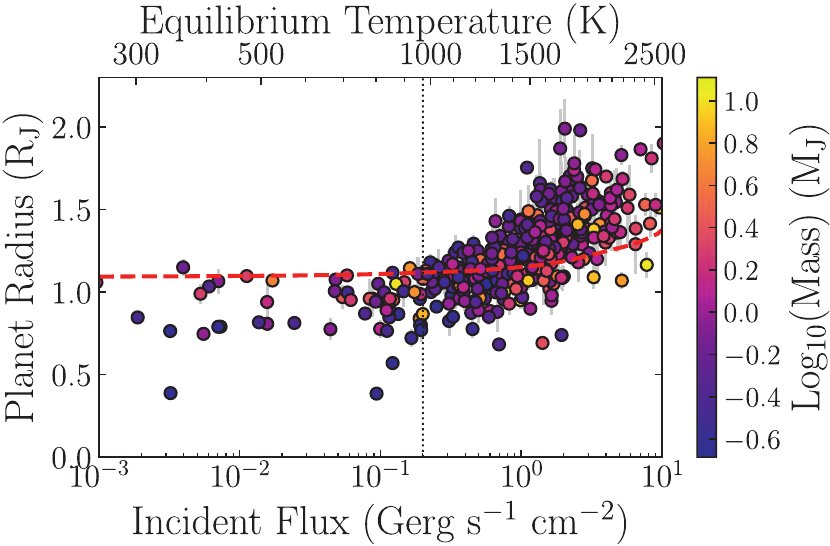}
    \caption[Radius of known giant exoplanets shown as a function of incident stellar flux]
    {Radius of known giant exoplanets shown as a function of incident stellar flux. 
    Masses from the planets are color-coded, going from 0.1 to 13 $M_\text{J}$. 
    The red dashed line shows the evolutionary model for a Jupiter-mass planet with no additional inflation effects considered, at 4.5 Gyr. 
    The vertical dashed line shows the flux cutoff below which no radius inflation is detected. 
    Figure from \textcite{Fortney_2021}}
    \label{fig:radius_inflation}
\end{figure}

A puzzling question since the discovery of the very first hot Jupiters has been their radii, which was found to be larger than what was postulated from evolutionary models \parencite{Bodenheimer_2001,Guillot_2002}.
This increased planetary radius has been shown to be directly correlated with the incident flux from the planet's host star \parencite{Laughlin_2011,Weiss_2013,Lopez_2016,Thorngren_2018}, as shown in Figure~\ref{fig:radius_inflation}.
Numerous theories have been presented to explain these inflated radii, which can be grouped into two different sets.

The first set proposes that the radii of these gas-giant planets, which are larger during formation and contract as the planet cools off afterwards, are kept inflated due to delayed cooling of the planet's interior.
Existing mechanisms proposed to cause this delay include enhanced atmospheric opacity \parencite{Burrows_2007} and reduction of the heat transport in planetary interiors \parencite{Chabrier_2007}.
A common counter-argument to both this mechanism and to delayed cooling of planetary interiors in general, is the difficulty in explaining the observed correlation of radius inflation with stellar irradiation.

The second set of theories suggests that the inflation is due to internal heating mechanisms that deposit heat into the planetary interior, either offsetting the cooling from formation or causing even a re-inflation of the radius.
Within this set, multiple hypotheses have been proposed, which can be separated into two distinct groups.
The first group postulates that the interior heating is directly related to tidal dissipation, either through high-eccentricity tidal migration \parencite{Bodenheimer_2001} or through thermal tides \parencite{Arras_2010,Socrates_2013}.
The second group posits that heat from stellar irradiation is directly deposited deep in the planet's interior, with different mechanisms suggested, namely, atmospheric circulation \parencite{Guillot_2002}, large scale atmospheric vertical mixing \parencite{Youdin_2010,Tremblin_2017} and ohmic dissipation \parencite{Batygin_2010,Ginzburg_2016}.

Whilst no single mechanism has managed to explain all existing observations, with some recent results even suggesting that more than one mechanism might be necessary to explain the available observations \parencite{Sarkis_2021}, there are some observational setups that can help filter out existing hypotheses.
Specifically, \textcite{Lopez_2016} argued that post-main-sequence hosts would be ideal candidates to evaluate the potential re-inflation of hot Jupiters, which would point to mechanisms that deposit stellar irradiation in the planet's interior being at play.
Evolved systems are older than their main-sequence counterparts, and have had longer time for their close-in giant planets to cool down from formation.
Additionally, stellar irradiation increases significantly as the host reaches the subgiant and giant branch.
Characterization of close-in giant planets orbiting evolved hosts should then shed light on whether re-inflation is at work or not in these planets.
Despite the general lack of known close-in transiting planets orbiting evolved hosts, initial studies have been done with the few that have been observed \parencite{Grunblatt_2016,Grunblatt_2017}, with results pointing to re-inflation being the most likely explanation for their inflated radii. 
Nonetheless, these studies were carried out based only on two planets, highlighting the need for additional systems to be observed and characterized, so that these theories can be re-evaluated with larger samples.

A more detailed and extensive description of the field of giant-planet inflation, along with multiple additional references can be found in \textcite{Spiegel_2013,Baraffe_2014,Dawson_2018}.

%% file: chapters/introduction/objectives.tex
\section{Objectives and structure}
\label{sec:objectives}

The overall goal of this thesis is to expand on our understanding of the processes that govern the formation and evolution of planetary systems, focusing on the study of giant planets orbiting giant stars.
This in turn can be pursued by trying to answer the many open-questions introduced in Section~\ref{sec:giant_planets}.

In particular, I wanted to focus on close-in giant planets (hot Jupiters) orbiting red-giant stars.
The origin behind planetary radius inflation and its efficiency are still not clearly understood, and the precise characterization of inflated giant planets orbiting evolved stars can provide some unique insights \parencite{Lopez_2016,Grunblatt_2017}.
At the same time, as few of these evolved systems with short orbital period are known to date, the discovery of additional systems could lead to better constraints on their occurrence rate \parencite{Grunblatt_2019}.

To tackle these questions, I needed to increase the existing sample of known close-in giant exoplanets orbiting red-giant hosts.
Moreover, given the impact of stellar signals in transiting light curves of evolved stars, a topic which we explore in Section~\ref{sec:stellar_signals}, I also needed to develop methods to improve the characterization of these systems.

The \textit{TESS} mission provided a clear opportunity for such an endeavour.
\textit{TESS} FFIs observed (almost) the entire sky for a minimum of $\sim$27.4 days (one sector), a length of observations which is sufficient when looking for hot Jupiters.
Additionally, \textit{TESS} asteroseismology is possible, even for targets with one sector of data (see Section~\ref{sub:tess_astero_potential}), which can improve the characterization of the host star, indispensable to determining accurate planetary properties.

To that end, I defined three major objectives.
The first objective was to develop and implement a method to characterize a planetary transit together with other stellar signals, all directly in the time domain.
The method should include physically motivated components to capture the most significant and expected stellar signals in the photometric light curves.

The second objective was to develop a pipeline, using third-party open-source software whenever possible, for the extraction and correction of light curves from the \textit{TESS} FFIs, and the subsequent search and validation of planetary transits.

Finally, I wanted to perform a systematic search in the \textit{TESS} FFIs for close-in transiting exoplanets orbiting red-giant stars.
This would entail the statistical classification of false-positive signals, as well as the identification of a list of potential physical transit/eclipse signals.
Additional validation of the physical signals would then further distinguish planetary transits from other (non-planetary) astrophysical signals.
All physical signals identified as planetary would then be ranked and classified as candidate planetary transits.

Any opportunities to obtain RV follow-up observations of candidate planets should also be pursued for confirmation.
All highly ranked planet candidates, whether confirmed or not, should then be modeled using the characterization tool developed.

This thesis is structured as follows.
Chapter~\ref{cha:transits} introduces the signals present in transiting light curves, starting with planetary transits and models to describe them, and also highlighting the accompanying stellar signals, often filtered out or ignored in planetary characterization.
Chapter~\ref{cha:characterization} then follows with the description and testing of the method developed for planetary transit characterization, which models both transits and stellar signals simultaneously in the time domain.
Chapter~\ref{cha:search_pipeline} details the development of the pipeline for the light curve extraction and transit search in the \textit{TESS} FFIs, describing each of the pipeline's multiple components.
Chapter~\ref{cha:planet_search} then presents the methodology and results of the \textit{TESS} FFI search, introducing the target list selected from the \textit{TESS} southern hemisphere and then discussing the data processing and exploitation done that led to the discovery of three new planet candidates.
Finally, in Chapter~\ref{cha:conclusions} I summarize the major results of this thesis and discuss future opportunities in the field of giants orbiting giants.


%% file: chapters/transits/main.tex
\chapter{Transit light curves}
\label{cha:transits}

Given the focus of the thesis on photometric data from \textit{TESS}, it is essential to understand some of the signals that are expected to be observed in a photometric light curve, in particular of red-giant stars.
In that respect, the first section of this chapter highlights the relevant physics at play in planetary transits, concluding with a description of the photometric signal observed for a transiting exoplanet.

Besides the signal from the planetary transit, the presence of a companion, in particular a massive one, can give rise to other deterministic out-of-transit variations in the light curve of the star.
In the context of characterizing the light curve of Kepler-91 b, a close-in giant planet orbiting a red-giant star, \textcite{Lillo-Box_2014} mention three main causes for such variations when considering closely packed planetary systems.
Namely, light from the planet (emitted or reflected), ellipsoidal variations (or tidal distortions) induced by the planet on the star and Doppler beaming due to the reflex motion of the star induced by the presence of a massive companion.
In my case, I do not attempt to characterize these variations, so the section details only the description of a planetary transit signal.

The second section discusses the remaining signals present in the light curve, all of stellar origin, which are stochastic in nature and often filtered out or ignored in planetary studies.
These play a leading role in the improvements to the characterization of transiting light curves discussed in this thesis.

\input{chapters/transits/transits}
\input{chapters/transits/stellar_signals}

%% file: chapters/transits/transits.tex
\section{Planetary transits}
\label{sec:transits}


In this section, I describe the observed signal that a transiting planet produces during an orbit around its host.
I start by introducing the equations necessary to describe the elliptical motion of a body around the center of mass of a two-body system.
Interpreting this motion from the perspective of a planet relative to its host star, I then obtain the expression that represents a planet's orbit in three dimensions.
Considering then the case for a transiting planet, I present the equations that outline its motion as it crosses the stellar disk.
Finally, I arrive at the signal that a transiting planet would cause on the observed light curve of its host star.
A complete derivation of planetary orbits, as well as additional insights, can be found in the work of \textcite{Murray_2010}.
For planetary transits, the work of \textcite{Winn_2010a} provides additional material.

\subsection{Orbits}
\label{sub:orbits}

Under Newton's law of universal gravitation, an orbiting system with two bodies orbits the center of mass of the system, called the barycenter.
For a star-planet system, this barycenter is often close to the center of the star, given its disproportionately larger mass in comparison to the planet.
Considering now the motion of a planet relative to its host, the orbit follows a closed ellipse with the star at one of the foci, and the distance $r$ from the star to any point on the orbit defined by its angular distance, $\theta$, is given by
\begin{equation}
    r = \frac{a(1 - e^2)}{1 + e \cos(\theta - \varpi)} ,
    \label{eq:kepler_first_law_theta}
\end{equation}
where $e$ is the orbit's eccentricity, constrained to values of $0 \le e < 1$ and $a$ is its \textit{semi-major axis}, related to the \textit{semi-minor axis}, $b$, through
\begin{equation}
    b^2 = a^2(1 - e^2) .
\end{equation}
The angle $\theta$, called the \textit{true longitude}, equates to the minimum and maximum of Equation~(\ref{eq:kepler_first_law_theta}) at $\theta = \varpi$ and $\theta = \varpi + \pi$, respectively. These points in the ellipse are referred to as the \textit{periapsis} and \textit{apoapsis}, respectively.
$\varpi$ is called the \textit{longitude of periapsis} as it represents the angular location of the periapsis of the ellipse with respect to a reference direction.
For the special case of $e = 0$, the expression simplifies to that of a circular orbit with $r = a$ and where $\varpi$ is undefined since all points in the orbit are now equidistant from the focus.
Equation~(\ref{eq:kepler_first_law_theta}) can also be defined in reference to the point of periapsis by introducing the \textit{true anomaly} angle, $f = \theta - \varpi$, with which we can rewrite Equation~(\ref{eq:kepler_first_law_theta}) so that the argument of the equation is the angular location with respect to the periapsis,
\begin{equation}
    r = \frac{a(1 - e^2)}{1 + e  \cos f} .
    \label{eq:kepler_first_law}
\end{equation}
\begin{figure}[!ptb]
    \centering
    \includegraphics[width=0.8\textwidth]{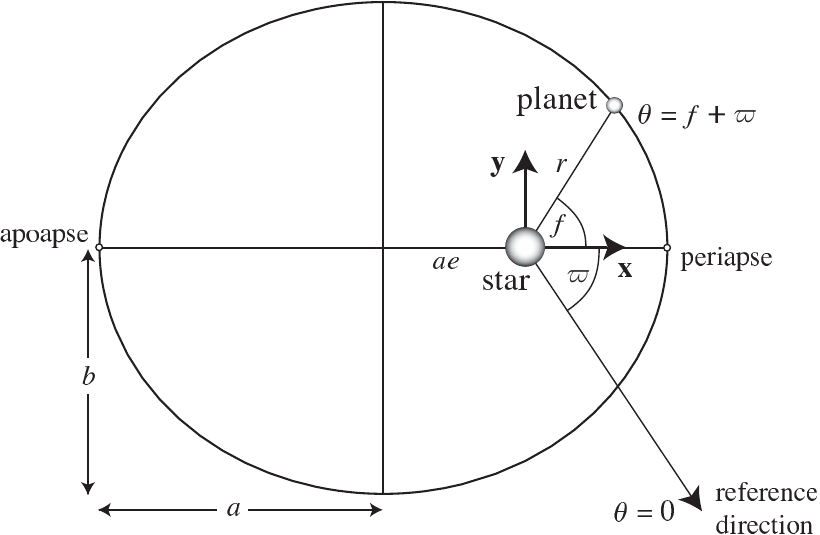}
    \caption[Schematic of an elliptical orbit in two dimensions, with the star at one of the foci]
    {Schematic of an elliptical orbit in two dimensions, with the star at one of the foci. 
    The orbit has semi-major axis $a$, semi-minor axis $b$, eccentricity $e$ and longitude of periapsis $\varpi$.
    Figure from \textcite{Murray_2010}.}
    \label{fig:elliptical_orbit_2d}
\end{figure}
Figure~\ref{fig:elliptical_orbit_2d} illustrates the geometry of such an elliptical orbit.

Notice that, despite describing an elliptical orbit, Equation~(\ref{eq:kepler_first_law}) does not depend on time, $t$, but on an angular quantity, the true anomaly $f$.
Additional expressions are required to express the position of the planet with respect to time.
To do so, observe that, although throughout the orbit the planet does not move at a constant angular rate, it always covers an angular distance of $2\pi$ over its orbital period, $P$, and so we define the \textit{mean motion} as
\begin{equation}
    n = \frac{2\pi}{P} ,
    \label{eq:mean_motion}
\end{equation}
from which we then define the \textit{mean anomaly}, $M$, at time $t$, as
\begin{equation}
    M = n(t - t_\text{per}) ,
    \label{eq:mean_anomaly}
\end{equation}
where $t_\text{per}$ is the \textit{time of periapsis passage}.
Finally, we introduce the \textit{eccentric anomaly}, $E$, with which we can rewrite Equation~(\ref{eq:kepler_first_law}) as
\begin{equation}
    r = a(1 - e\cos E) ,
    \label{eq:ecc_anomaly}
\end{equation}
and which can be related to $M$ through Kepler's equation,
\begin{equation}
    M = E - e \sin E .
    \label{eq:kepler_equation}
\end{equation}

The location of the orbiting planet in the elliptical plane can now be fully described, given the parameters $a$, $e$, $P$, and $t_\text{per}$.
For any time, $t$, we:
\begin{itemize}[noitemsep, topsep=1ex]
    \item Find $M$ using Equations~(\ref{eq:mean_anomaly} and \ref{eq:mean_motion}),
    \item Find $E$ by solving Kepler's equation (Equation~\ref{eq:kepler_equation}),
    \item Find $r$ using Equation~(\ref{eq:ecc_anomaly}),
    \item Find $f$ using Equation~(\ref{eq:kepler_first_law}).
\end{itemize}

\begin{figure}[!ptb]
    \centering
    \includegraphics[width=0.8\textwidth]{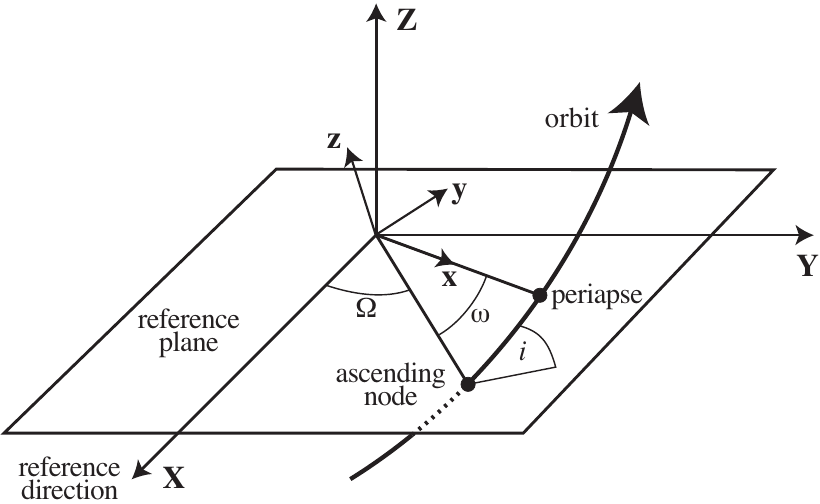}
    \caption[Schematic of an elliptical orbit in three dimensions]
    {Schematic of an elliptical orbit in three dimensions. 
    Extending the orbit's two-dimensional representation to three dimensions requires three additional parameters: the orbital inclination $i$, the longitude of the ascending node $\Omega$ and the argument of periapsis $\omega$. 
    Figure from \textcite{Murray_2010}.}
    \label{fig:elliptical_orbit_3d}
\end{figure}

To now add a third dimension to our representation of the orbit, we first define a reference plane parallel to the cartesian $x$ and $y$-axis, as illustrated in Figure~\ref{fig:elliptical_orbit_3d}.
This reference plane is oriented so as to match the observer's plane of the sky and perpendicular to the observer's line of sight, which we position in the positive direction of the $z$-axis.
In this scheme, the \textit{line of nodes} is the line that intersects both the reference plane and the orbital plane.
Of the two points of intersection between the line of nodes and the orbit-line, we call the one where the orbit crosses the reference plane from below to above the \textit{ascending node}.

From this configuration we can now define three angular quantities necessary to describe the orbit in three dimensions.
First we define the \textit{orbital inclination}, $i$, which measures the angle from the the reference plane to the orbital plane.
For $i = 0^\circ$, the observer sees the orbit face-on with the orbital plane perpendicular to the line of sight, whilst for $i = 90^\circ$, the orbit is edge-on, with the orbital plane parallel to the observer's line of sight.
Then we define the \textit{longitude of the ascending node}, $\Omega$, as the angle measured from the reference line (the $x$-axis) to the ascending node.
Finally, we define the \textit{argument of periapsis}, $\omega$, as the angle measured from the ascending node to the periapsis of the orbit, measured in the orbital plane and following the orbit's direction of motion.
For circular orbits this angle is undefined, similarly to $\varpi$, since there is no periapsis.
Using these three angular quantities, we now map our previous orbital plane positions, described by $r$ and $f$, to the three cartesian coordinate axes $x$, $y$ and $z$, through the expressions
\begin{align}
    x &= r \left( \cos\Omega \cos(\omega + f) - \sin\Omega \sin(\omega + f) \cos i \right) , \label{eq:general_orbit_cartesian_x}\\
    y &= r \left( \sin\Omega \cos(\omega + f) + \cos\Omega \sin(\omega + f) \cos i \right) , \label{eq:general_orbit_cartesian_y}\\
    z &= r \sin(\omega + f) \sin i . \label{eq:general_orbit_cartesian_z}
\end{align}

Adding it all together, we get a set of all orbital parameters required to describe the position of a planet in an elliptical orbit in three dimensions: the semi-major axis $a$, the eccentricity $e$, the orbital period $P$, the time of periapsis passage $t_\text{per}$, the inclination $i$, the longitude of the ascending node $\Omega$, and the argument of periapsis $\omega$.

\subsection{Transits}
\label{sub:transits}

An \textit{eclipse} is the obscuration of a celestial body by another. When the bodies have vastly different sizes, i.e., for the case of a planet and its host star, then the passage of the smaller body in front of the larger one is called a \textit{transit} and the passage of the smaller body behind the larger one is called an \textit{occultation}.

\begin{figure}[!ptb]
    \centering
    \includegraphics[width=0.7\textwidth, height=\textheight, keepaspectratio]{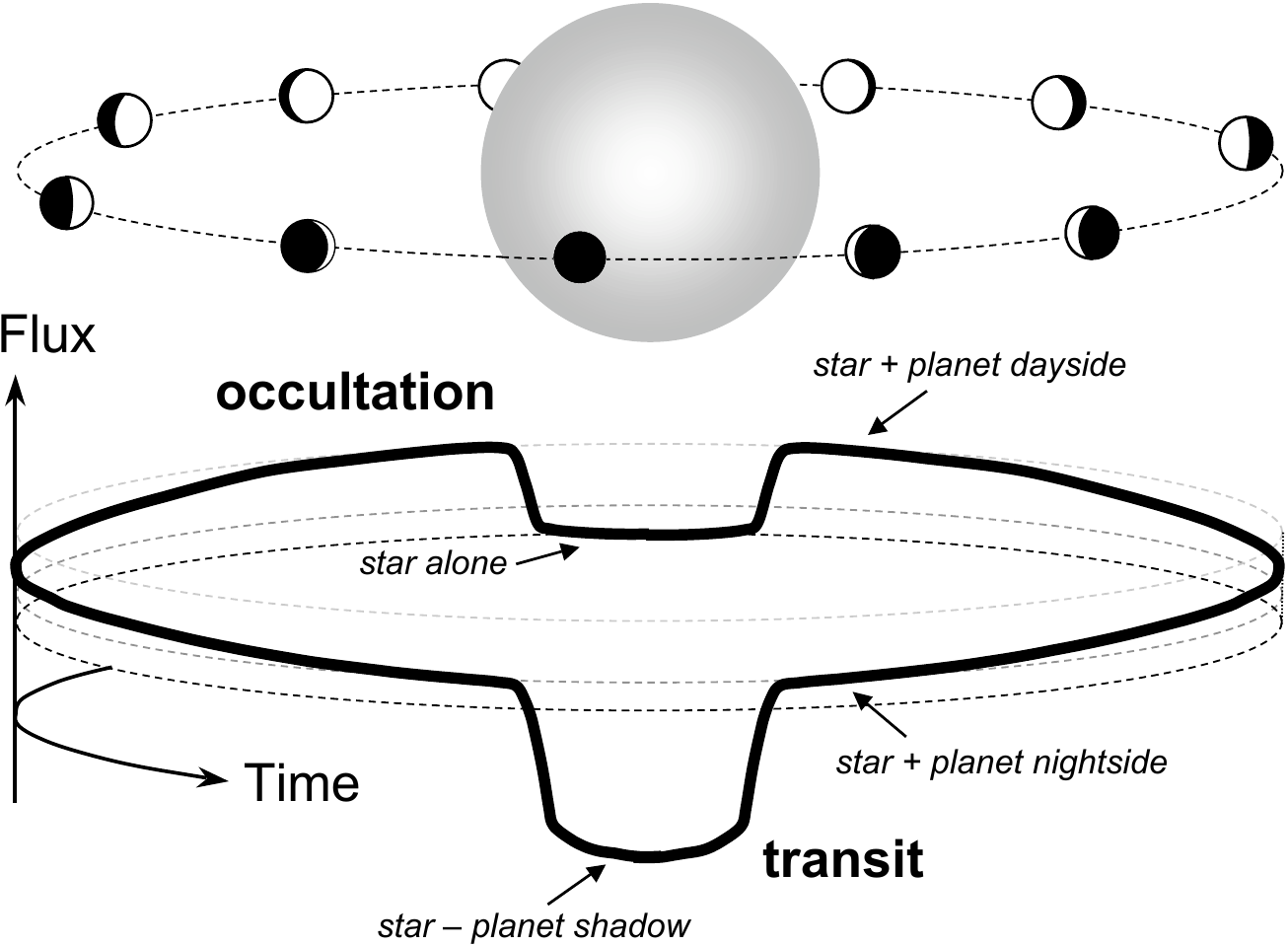}
    \caption[Schematic of transits and occultations]
    {Schematic of transits and occultations. 
    The flux observed contains both contributions from the star as well as reflected light from the planet. 
    The reduction in flux is greater during a transit as the planet covers a portion of the stellar surface. 
    In the occultation, the star covers the contribution from the planet's reflected light, causing another, less deep, reduction in flux observed. 
    Figure from \textcite{Winn_2010a}.}
    \label{fig:transits_occulations}
\end{figure}

Since we are interested in planetary transits specifically, we consider a planet of mass $M_\text{p}$ and radius $R_\text{p}$ orbiting a star of mass $M_\star$ and radius $R_\star$.
Figure~\ref{fig:transits_occulations} illustrates this scenario and the flux observed throughout time due to the variations from the concealments.

\begin{figure}[!ptb]
    \centering
    \includegraphics[width=0.7\textwidth, height=\textheight, keepaspectratio]{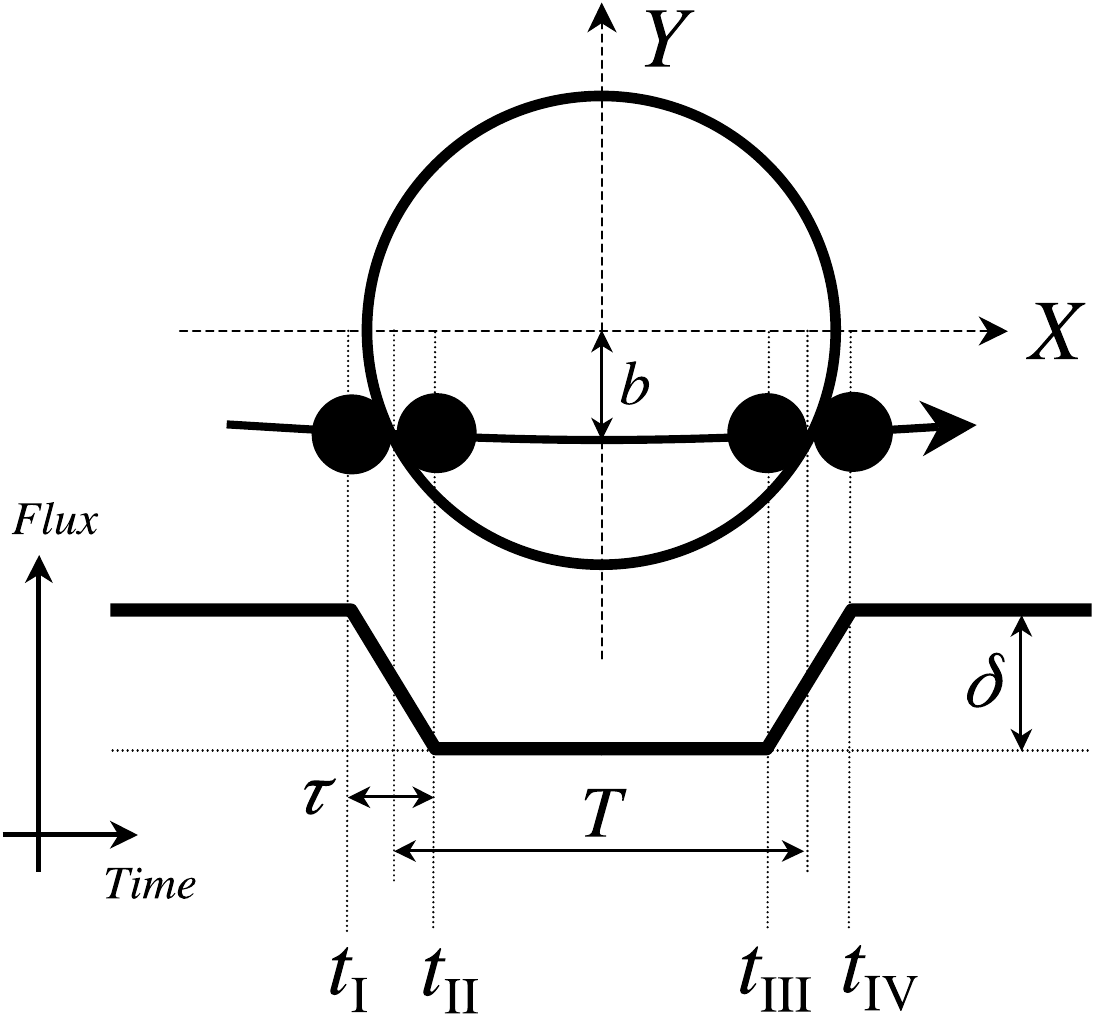}
    \caption[Illustration of a transit viewed (almost) edge-on, with the planet orbit parallel to the $x$-axis, going from negative to positive values during a transit]
    {Illustration of a transit viewed (almost) edge-on, with the planet orbit parallel to the $x$-axis, going from negative to positive values during a transit. 
    The impact parameter $b$ shows the distance between the center of the planet and the center of the star, which is aligned with the center of the coordinate system.
    The duration of the transit is $\tau_\text{tra} = t_\text{IV} - t_\text{I}$, whilst the duration of ingress and egress are $\tau_\text{in} = t_\text{II} - t_\text{I}$ and $\tau_\text{eg} = t_\text{IV} - t_\text{III}$, respectively. 
    Figure from \textcite{Winn_2010a}.}
    \label{fig:transit_edge_on_view}
\end{figure}

To derive an equation to describe the flux variations, we first reorient the reference plane so that the line of nodes lines up with the $x$-axis, with the descending node in positive $x$, giving $\Omega=180^\circ$ (see Figure~\ref{fig:elliptical_orbit_3d}).
This simplifies Equations~(\ref{eq:general_orbit_cartesian_x}$-$\ref{eq:general_orbit_cartesian_z}) to
\begin{align}
    x &= -r \cos(\omega + f) , \label{eq:transit_cartesian_x} \\
    y &= -r \sin(\omega + f) \cos i , \label{eq:transit_cartesian_y} \\
    z &= r \sin(\omega + f) \sin i . \label{eq:transit_cartesian_z}
\end{align}
Figure~\ref{fig:transit_edge_on_view} illustrates the geometry of the orbit in this new orientation.
In this configuration, the sky-projected distance between the centers of the star and the planet is $d = \sqrt{x^2 + y^2}$. 
Using Equations~(\ref{eq:transit_cartesian_x}) and (\ref{eq:transit_cartesian_y}), 
\begin{equation}
    d = \frac{a(1 - e^2)}{1 + e  \cos f} \sqrt{1 - \sin^2 (\omega + f) \sin^2 i} .
    \label{eq:r_sky}
\end{equation}
From Equation~(\ref{eq:r_sky}) and Figure~\ref{fig:transit_edge_on_view}, it is clear that both eclipses happen when $d$ is at its minimum.
For an estimate of this minimum we follow the approximations from \textcite{Winn_2010a} where the times of eclipses, defined where $x = 0$, and shown in terms of the true anomaly, $f$, are
\begin{equation}
    f_\text{tra} = \frac{\pi}{2} - \omega, \qquad f_\text{occ} = -\frac{\pi}{2} - \omega ,
    \label{eq:true_anomaly_transit}
\end{equation}
where ``tra'' refers to the transits and ``occ'' to the occultations.
This approximation is valid for most orbits, failing only for very eccentric orbits and close orbits with grazing eclipses (when the bodies never cross completely in front of the other one).

The \textit{impact parameter}, $b$ (not to be confused with the ellipse's semi-minor axis), is the sky-projected distance between the bodies when $x = 0$, usually represented in units of stellar radius:
\begin{align}
    b_\text{tra} = \frac{a \cos i}{R_\star} \left( \frac{1 - e^2}{1 + e \sin \omega} \right) , \label{eq:transit_impact_param} \\
    b_\text{occ} = \frac{a \cos i}{R_\star} \left( \frac{1 - e^2}{1 - e \sin \omega} \right) . \label{eq:occultation_impact_param}
\end{align}

Figure~\ref{fig:transit_edge_on_view} also shows four relevant times during a transit ($t_\text{I-IV}$). The duration of the transit, $\tau_\text{tra}$ lasts from $t_\text{I}$ to $t_\text{IV}$. From $t_\text{I}$ to $t_\text{II}$ is the \textit{ingress} of the transit, starting from the first contact between the disks of the planet and the star and ending at the time when the planet disk is completely within the stellar disk.
On the opposite side, we have the \textit{egress} of the transit, from $t_\text{III}$ to $t_\text{IV}$, starting when the planetary disk touches the opposite stellar limb and ending when the disks stop overlapping completely.

Re-introducing Equation~(\ref{eq:struve_transit}),
\begin{equation}
    \Delta F \equiv \frac{F}{F_0} \propto \left( \frac{R_P}{R_\star} \right)^2 ,
\end{equation}
we have that the normalized reduction in flux during a transit is proportional to the ratio between the radii of the planet and star.
This expression works under the assumption that both the planet and the star are spheres, producing sky-projected circular disks, which we assume also for the model.

To determine the reduction in flux at any point during the orbit of the planet, we define the sky-projected distance between the centers of the star and the planet, scaled by the stellar radius, $z \equiv d / R_\star$, and the ratio of the radii, which we define as $p \equiv R_\text{p} / R_\star$.
Assuming a uniform stellar flux, $F_0$, and assuming the planet is a dark sphere which does not reflect light, then the normalized transit flux observed in the light curve during the planet's orbit depends only on the fraction of the stellar disk covered by the planet's disk at any point in the orbit, given by a function $\lambda$ which depends on $z$ and $p$ (see Equation~1 of \textcite{Mandel_2002} for the complete expression), 
\begin{equation}
    \frac{F}{F_0} = 1 - \lambda(z, p).
    \label{eq:obscured_flux}
\end{equation}

Leaving the assumption of uniform stellar flux and considering now the effects of stellar limb darkening, we have that the stellar brightness has a higher central observed value which falls radially to the stellar edge (the limb).
This dimming depends on the angle between the stellar surface normal and the observer, $\theta$, usually parametrized through its cosine, $\mu = \cos \theta$, which essentially represents a radial coordinate in the stellar disk.
Multiple limb-darkening laws exist to describe the falloff in brightness.
Notable mentions are the historical linear law \parencite{Milne_1921,Grygar_1972,Claret_1990}, the quadratic law \parencite{Claret_1990,Mandel_2002}, the square-root law \parencite{Diaz-Cordoves_1992}, the non-linear law \parencite{Claret_2000} and the "power-2" law \parencite{Morello_2017}.
For our model we adopt a quadratic limb-darkening law, widely used in the literature
\begin{eqnarray}
    \frac{I(\mu)}{I_0} = 1 - u_1(1 - \mu) u_2(1 - \mu)^2 ,
\end{eqnarray}
where $I(\mu)$ is the stellar surface brightness at $\mu$, $I_0$ is the surface brightness at the center of the stellar disk and $u_1$ and $u_2$ are the quadratic limb-darkening coefficients, which follow $u_1 + u_2 < 1$.
Using this limb-darkening model, the determination of the normalized stellar flux during the planet's orbit is now more complex and, during the transit, requires the integration of the surface brightness $I(\mu)$ over the area of the stellar disk obscured by the planet.
The complete analytical expression for the resulting light curve is far too lengthy and we refer to section~4 of \textcite{Mandel_2002} for details.
Here we denote it simply as a function of its four parameters: $z$, $p$, $u_1$ and $u_2$, 
\begin{equation}
    \frac{F}{F_0} = f \left( z, p, u_1, u_2 \right) .
    \label{eq:transit_quadratic_ld}
\end{equation}



With a chosen limb-darkening model for the stellar flux, we now have all the necessary ingredients to describe the light curve of a transiting planet.
From Equation~(\ref{eq:transit_quadratic_ld}), we have 4 parameters that need to be defined.
However, the normalized distance between centers, $z$ is calculated with $d$ from Equation~(\ref{eq:r_sky}), which is determined by the orbit of the planet, and consequently depends on all its parameters.
All in all, 9 parameters have to be defined to fully characterize the planet's light curve:
\begin{itemize}[noitemsep, topsep=1ex]
    \item $a/R_\star$, the semi-major axis in stellar radius units (so that $z$ is determined instead of $d$ using Equation~(\ref{eq:r_sky}))
    \item $e$, the eccentricity
    \item $P$, the orbital period
    \item $t_\text{per}$, the time of periapsis passage or, alternatively, $t_0$, the time of inferior conjunction (or transit epoch)
    \item $i$, the orbital inclination (though for transiting systems this is almost always $\sim 90^\circ$)
    \item $\omega$, the argument of periapsis
    \item $p$, the ratio between the planetary and stellar radii
    \item $u_1$ and $u_2$, the coefficients for a quadratic limb-darkening model
\end{itemize}
Notice how $\Omega$ is no longer a parameter, as we define our frame of reference for transits so that $\Omega=180^\circ$, as mentioned earlier.

For any time, $t$, we then calculate the transiting planet light curve flux by:
\begin{itemize}[noitemsep, topsep=1ex]
    \item Finding $M$ using Equations~(\ref{eq:mean_anomaly} and \ref{eq:mean_motion}),
    \item Finding $E$ by solving Kepler's equation (Equation~\ref{eq:kepler_equation}),
    \item Finding $r$ using Equation~(\ref{eq:ecc_anomaly}),
    \item Finding $f$ using Equation~(\ref{eq:kepler_first_law}),
    \item Finding $z$ using Equations~(\ref{eq:transit_cartesian_x}, \ref{eq:transit_cartesian_y} and \ref{eq:r_sky}), with $a/R_\star$ instead of $a$,
    \item Finding $\frac{F}{F_0}$ using Equation~(\ref{eq:transit_quadratic_ld}).
\end{itemize}


%% file: chapters/transits/stellar_signals.tex
\section{Stellar signals}
\label{sec:stellar_signals}

A transit signal is not the only signal present in photometric transit time series.
The surface of the star cannot be resolved and so, any variation in the surface's integrated flux is also going to be present in the flux measurements. 
These surface flux variations are caused by different physical phenomena in a star and their timescales and amplitudes vary not only between different spectral types but also between different stages of stellar evolution.

In the context of photometric time series observations of solar-like stars, signals corresponding to physical processes in stars have been identified in the power spectrum of solar-like stars.
The most significant ones are the different scales of granulation \parencite{Harvey_1985,Aigrain_2004,Michel_2008,Mathur_2011}, solar-like oscillations \parencite{Kjeldsen_1995b,Kallinger_2010d,Bedding_2010}, signals from active regions \parencite{Harvey_1985,Aigrain_2004,Corsaro_2015a} and possibly faculae \parencite{Karoff_2012,Karoff_2013}.
These are stochastic processes, with no deterministic models to describe them and so, their characterization has to be carried out through their statistical properties, i.e., their characteristic timescales and amplitudes, which has historically been done by modeling their contributions in the power spectrum of the observed light curve.

\begin{figure}[!ptb]
    \includegraphics[width=\textwidth, height=\textheight, keepaspectratio]{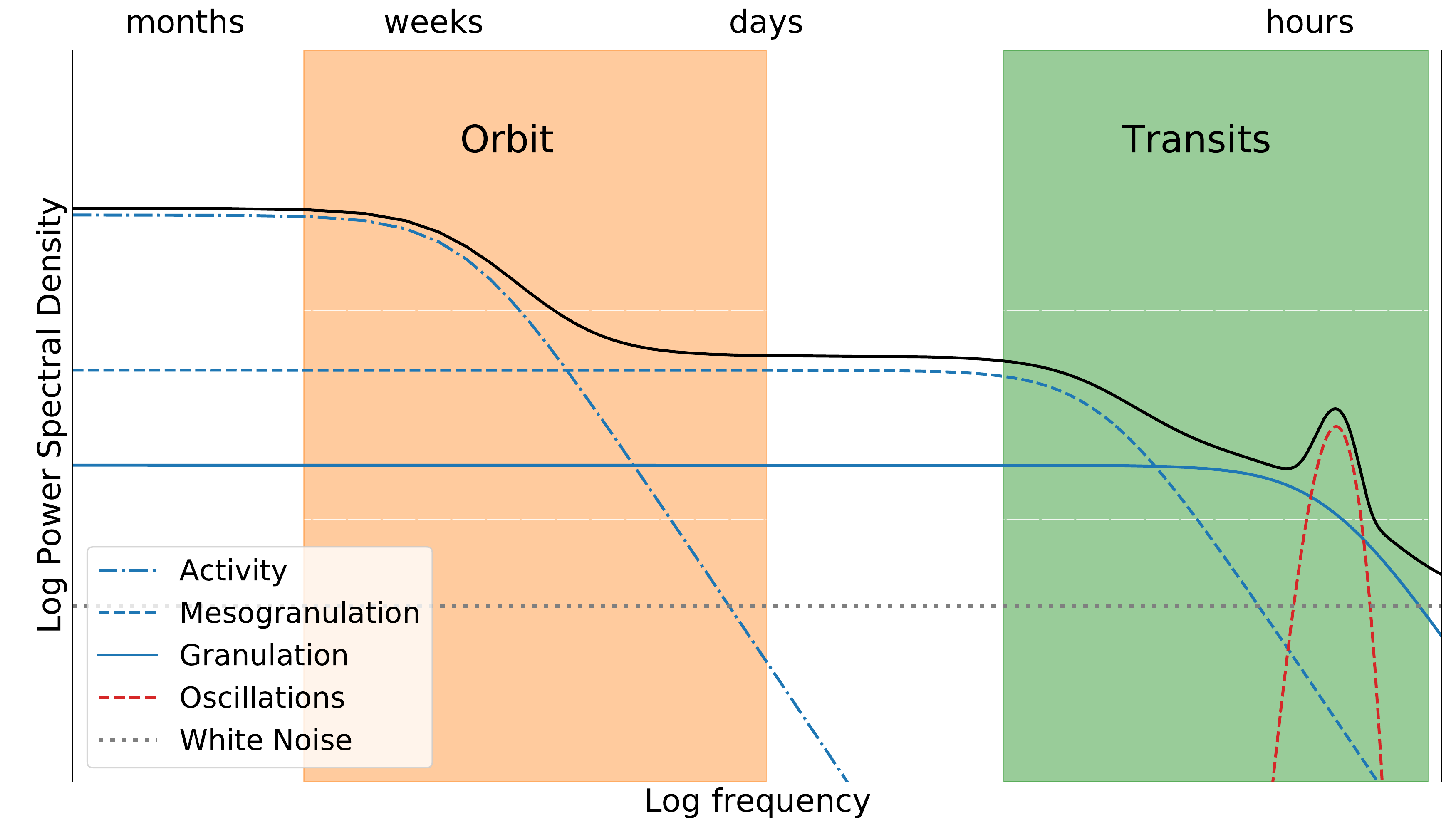}
    \caption[Illustrative example of the power spectrum of a low-luminosity red-giant branch star]
    {Illustrative example of the power spectrum of a low-luminosity red-giant branch star.
    Colored lines highlight different stellar signals present in the data. 
    The blue lines show the signals of the granulation, mesogranulation and active regions, according to line style, as per the legend.
    The red dashed line shows the oscillations envelope signal.
    The gray dotted line shows the white noise signal expected for a \textit{Kepler} star.
    The black solid line denotes the sum of all previous components and is representative of the effective signal seen when computing the power spectral density of the star's light curve. 
    Shaded regions highlight the characteristic timescales associated with the planet's orbit and transit durations (i.e. relevant to this thesis), in orange and green respectively.
    Notice how the mesogranulation, granulation and oscillations all overlap with the timescales of the transit durations, which, coupled with the high amplitudes of these signals, makes it difficult to filter them out when characterizing a transit.}
    \label{fig:psd_llrgb}
\end{figure}
Figure~\ref{fig:psd_llrgb} shows a representative power spectrum of a LLRGB star, highlighting the most important processes visible in the power spectrum of such a star and illustrating their typical timescales and amplitudes.
The typical timescales for transit durations and planet orbits are also added for a comparison with the stellar signals. 
Unlike their main-sequence counterparts, evolved hosts also have oscillations on the timescales of the transit durations of close-in Jupiter-sized planets, along with the mesogranulation and granulation \parencite{Mathur_2011}. 
This makes the stellar signals harder to filter out from the light curve, due to the risk of removing some of the transit signal \parencite{Carter_2009}.

It is important to note that these power spectrum models are empirical models guided by theoretical predictions and statistical observations of each phenomenon, and found to fit the data adequately, with models evolving and becoming more sophisticated over time.
Below, I describe the models for both the granulation and the solar-like oscillations.

\subsection{Granulation}

Granulation is the irregular cellular pattern at the surface of a star caused by the turbulent convection, where the granules evolve on a wide range of timescales and amplitudes \parencite{Harvey_1985,Aigrain_2004,Mathur_2011}. 
Due to this wide range, granulation is usually modeled as multiple processes, with \textcite{Harvey_1985} identifying three different signals in the Sun, which he named granulation (sharing the same name as the overall process), mesogranulation and supergranulation, all of which were modeled in the power spectrum as a Lorentzian function:
\begin{equation}
    P(\nu) = \frac{4 a^2 \tau}{1 + (2 \pi \nu \tau)^2} ,
    \label{eq:granulation_psd_harvey}
\end{equation}
where the power of the process $P$ at frequency $\nu$ depends on the rms brightness fluctuation, $a$, and the characteristic timescale, $\tau$, of the process. 
Additionally, \textcite{Harvey_1985} also found the same model to be accurate at modeling the active regions of the star (related with magnetic activity), characterized by longer timescales than the granulation.

Later works found that a different exponent in the denominator of Equation~(\ref{eq:granulation_psd_harvey}) was more appropriate in modeling the flux variations caused by the granulation and the stellar activity both in the Sun \parencite{Michel_2009} and in evolved stars \parencite{Kallinger_2010d}. 
In my work, I follow an expression commonly cited in the literature and adopted by both \textcite{Kallinger_2014} and \textcite{Corsaro_2015a},
\begin{equation}
    P(\nu) = \frac{2 \sqrt{2}}{\pi} \frac{a^2 / b}{ \left(\nu / b \right)^4 + 1} ,
    \label{eq:granulation_psd_kallinger}
\end{equation}
where the power of the process again depends on the characteristic amplitude, $a$, and a characteristic frequency, $b$, defined as $b = (2 \pi \tau)^{-1}$. 
Notice the new exponent in the denominator, which is now 4.

In the context of evolved stars specifically, \textcite{Kallinger_2014} explored the number of background models required to best characterize the power spectrum of long-cadence \textit{Kepler} time series, concluding that three components with the same functional form as Equation~(\ref{eq:granulation_psd_kallinger}) are necessary, constraining the properties of granulation, mesogranulation and active regions, respectively. 
The authors also mention that one single granulation component, instead of two, is sufficient in cases where the instrumental white noise dominates the background or is not negligible.
These signals are depicted in Figure~\ref{fig:psd_llrgb} by the three blue lines with different styles (see legend).

\subsection{Solar-like oscillations}

As for solar-like oscillations (introduced in Section~\ref{sub:astero_intro}), if the oscillation modes can be identified individually, they are usually modeled as Lorentzian functions \parencite{Kallinger_2010d}.
For the purpose of describing the background signal coming from a star, or when there are no enough data to resolve the oscillation modes individually, the power of the entire envelope of solar-like oscillations is commonly approximated with a Gaussian function following
\begin{equation}
    P(\nu) = P_\text{g} \exp \left( \frac{-(\nu - \nu_\text{max})^2}{2 \sigma^2} \right) ,
    \label{eq:envelope_gaussian_psd}
\end{equation}
which is centered on the frequency of maximum power, $\nu_{\text{max}}$, and has a maximum power of $P_\text{g}$ with a width of $\sigma$. 
This frequency of maximum power, whilst not providing the same amount of information as the individual modes, still conveys global asteroseismic information on the star, closely related to the star's physical properties \parencite{Kjeldsen_1995a}. 
This signal is depicted in Figure~\ref{fig:psd_llrgb} by the red dashed line.

\subsection{Impact on transit photometry}

Unfortunately, all these statistical models that describe their respective physical processes cannot be easily translated into the time domain, so as to separate their contribution from the transiting signal and allow the planet characterization to not be hindered by the stellar signals present. 
For main-sequence stars, since the timescales of most of these signals do not overlap with the duration of transits of close-in giant planets, the issue can be mitigated by binning the time series observations such that the stellar signals are filtered out. 
For evolved stars the situation is less favorable, as the timescales of granulation and oscillations are of the same order of magnitude as the transit durations of those planets, which rules out averaging over the observations. 

The solution is then to find a method of characterizing these same power spectrum models in the time series, that is, to model non-parametric processes characterized by their statistical properties, directly in time domain. 
One such method is Gaussian process regression, as we will see in the next chapter.

%% file: chapters/characterization/main.tex
\chapter{Transiting system characterization}
\label{cha:characterization}

\input{chapters/characterization/gps}
\input{chapters/characterization/modelling}
\input{chapters/characterization/applications}

%% file: chapters/characterization/gps.tex
\section{Gaussian processes}
\label{sec:gps}

\textcite{Rasmussen_2006}, the often cited source on Gaussian processes, define them as a ``collection of random variables, any finite number of which have a joint Gaussian distribution''. 
To better understand this definition, we start by looking at the univariate Gaussian distribution of a random variable $x$,
\begin{equation}
    x \sim \mathcal{N} \left( \mu, \sigma^2 \right) ,
    \label{eq:univariate_gaussian_notation}
\end{equation}
with mean $\mu$ and variance $\sigma^2$, and probability density given by
\begin{equation}
    f(x) = \frac{1}{\sigma \sqrt{2 \pi}} \exp \left[ -\frac{1}{2} \left( \frac{x - \mu}{\sigma} \right) \right] .
    \label{eq:univariate_gaussian_expression}
\end{equation}

By increasing the number of variables to an arbitrary number $n$, i.e., $\bm{x} = (x_1, \ldots, x_n)$ is now our $n$-dimensional input vector of random variables, and admitting correlations between them, we get a multivariate joint Gaussian distribution
\begin{equation}
    \bm{x} \sim \mathcal{N} \left( \bm{\mu}, \Sigma \right) ,
    \label{eq:multivariate_gaussian_notation}
\end{equation}
with $n$-dimensional mean vector $\bm{\mu}$ and $n \times n$ covariance matrix $\Sigma$, where each matrix entry expresses the covariance between pairs of variables $(x_i, x_j)$. 
The probability density of this distribution is then given by
\begin{equation}
    f(\bm{x}) = \frac{1}{\sqrt{(2 \pi)^n \vert \Sigma \vert}} \exp \left[ -\frac{1}{2} (\bm{x} - \bm{\mu})^T \Sigma^{-1} (\bm{x} - \bm{\mu}) \right] .
    \label{eq:multivariate_gaussian_expression}
\end{equation}

If we now imagine expanding our input space $\bm{x}$ to infinity, then our Gaussian distribution would have to be defined for each possible random variable $x$, a mean value $\mu$ and a covariance value between variable $x$ and every other variable $x'$. We do this by defining a mean function $m(x)$ and a covariance function $k(x, x')$, and defining a Gaussian process (GP) as
\begin{equation}
    f(\bm{x}) \sim \mathcal{GP} \left( m(\bm{x}), k(\bm{x}, \bm{x}') \right) ,
\end{equation}

which can be interpreted as a distribution over functions of $\bm{x}$, constrained on the properties imposed by its mean and covariance functions. 
To determine the covariance for any point $x$ would require us to determine its covariance with an infinite number of other points $x'$, which is not feasible. 
However, since any finite collection of variables from this infinite distribution is a joint Gaussian distribution itself and its properties are consistent with all infinitely remaining variables in the Gaussian process, whether they are taken into consideration or not, Gaussian processes can be used to predict the distribution of functions for any finite set of variables in our input space $\bm{x}$, by considering the joint Gaussian distribution of only the set of variables we are interested in and ignoring all others. 
Additionally, GPs allow us to further constrain this distribution of functions by a given set of observations $\mathcal{D}={(x_i, y_i)}$, with the resulting distribution of functions being conditioned by these observed variables.

\begin{figure}[!ptb]
    \centering
    \includegraphics[width=0.56\textwidth]{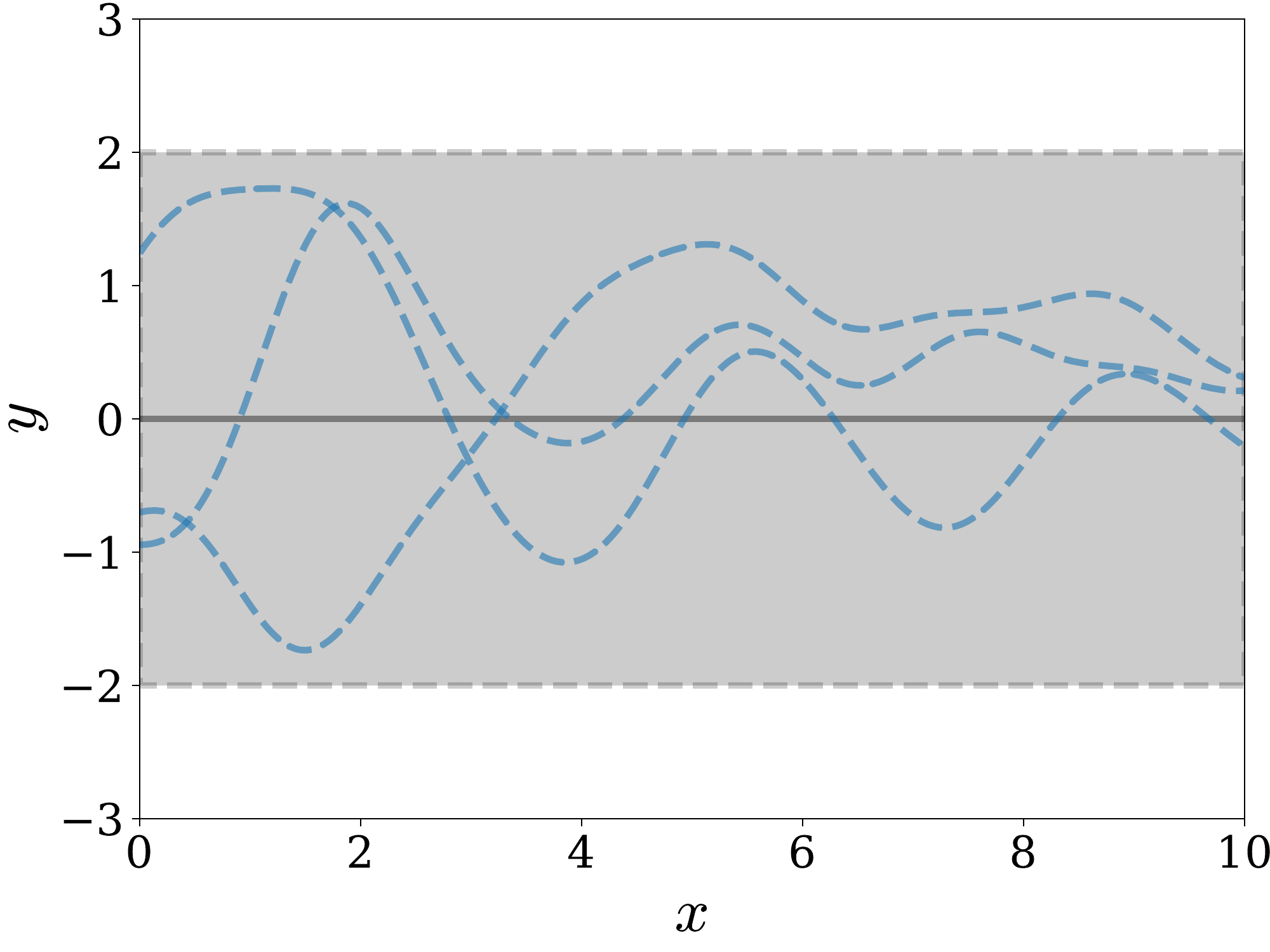}
    \includegraphics[width=0.43\textwidth]{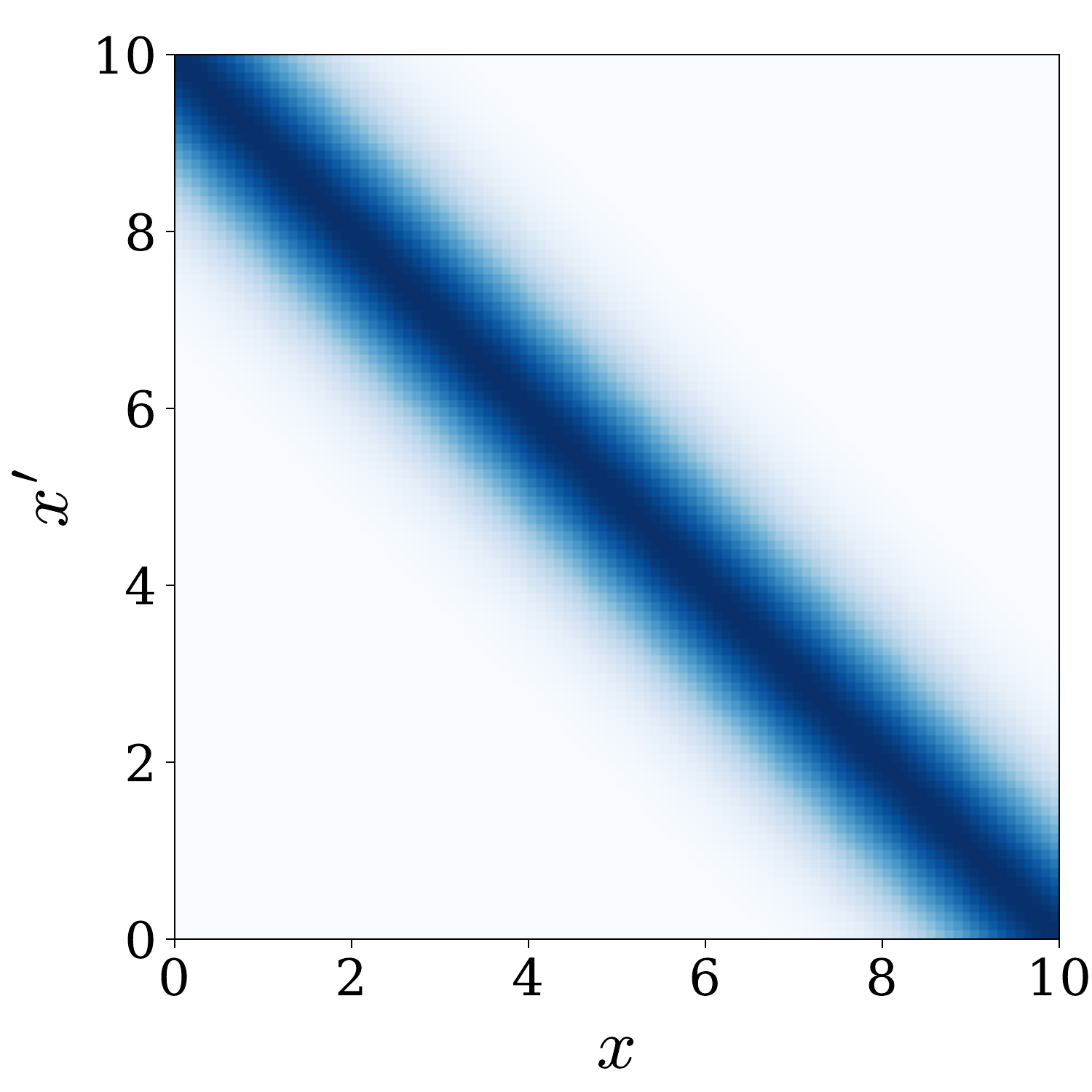}
    \includegraphics[width=0.56\textwidth]{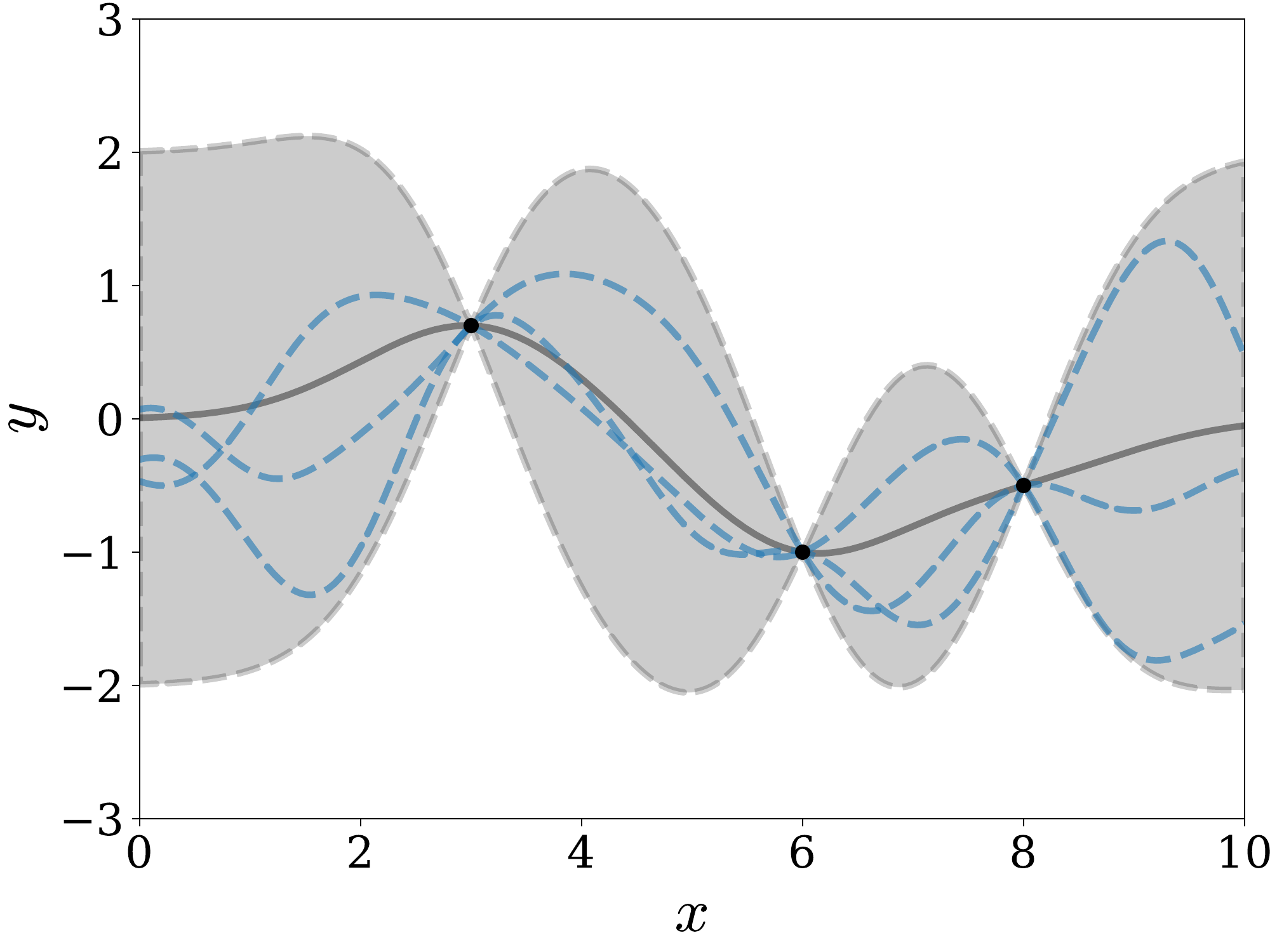}
    \includegraphics[width=0.43\textwidth]{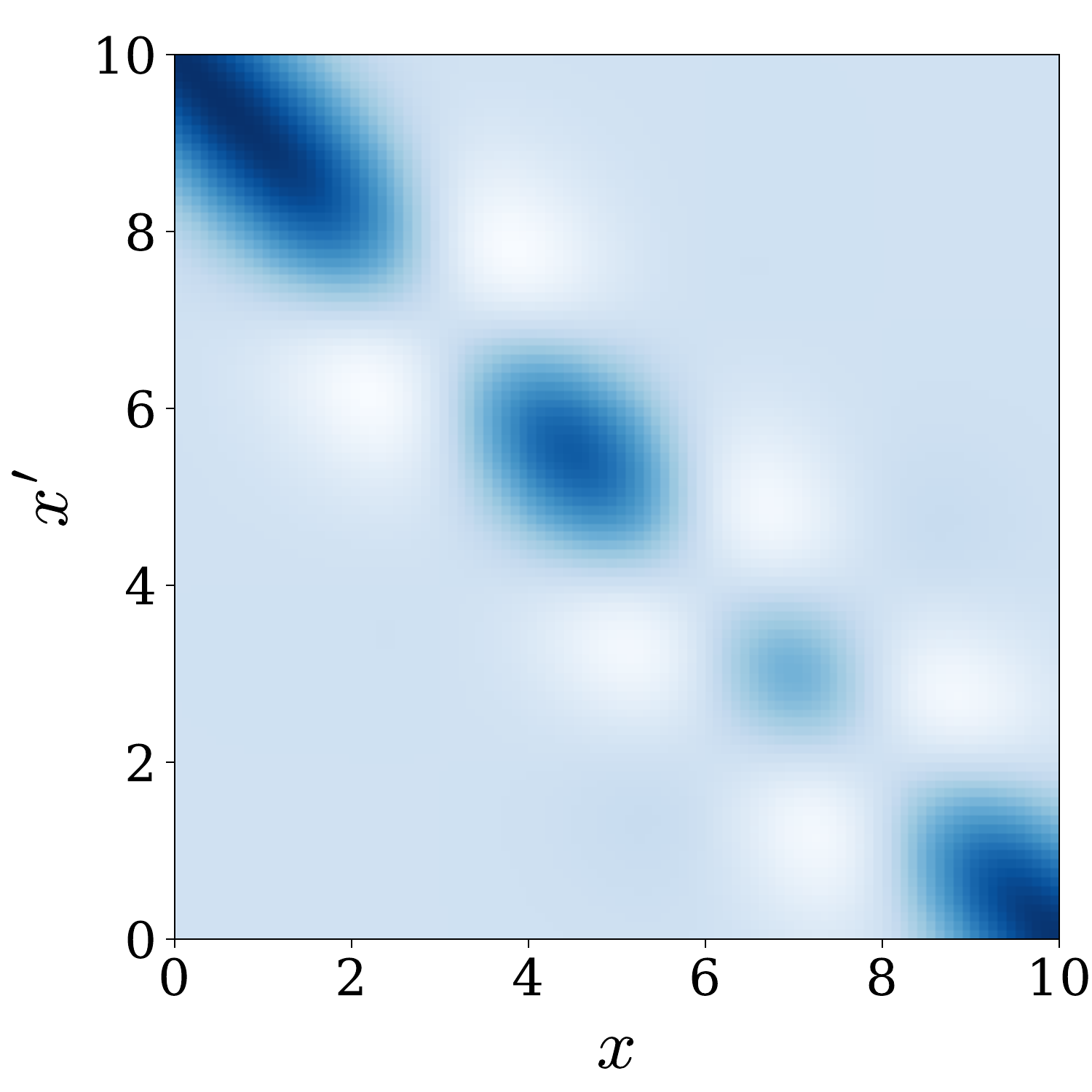}
    \caption[Two Gaussian process distributions and respective covariance matrices for 100 points in $x$ between 0 and 10]
    {Two Gaussian process distributions and respective covariance matrices for 100 points in $x$ between 0 and 10.
    On the left, the black solid line and gray shaded region denote the mean $\mu$ and 2-$\sigma$ of the distributions, respectively. 
    Overlapped are also 3 samples drawn from that same distribution.
    On the right, the covariance matrices generated by the GP's covariance function are shown for all possible pairs of points, highlighting the correlations between pairs (where bluer color corresponds to higher correlation). 
    Above, the plots show a GP prior distribution with no constraints imposed from observed data and where the samples follow only the properties imposed by the mean and covariance functions of the GP. 
    Below, three observations are now introduced, constraining the possible correlations between points as seen in the covariance matrix, and consequently constraining the GP distribution and the samples drawn from it.}
    \label{fig:gp_prior_posterior_example}
\end{figure}

To illustrate this, Figure~\ref{fig:gp_prior_posterior_example} shows two GP distributions with mean zero and a squared exponential covariance (described in Section~\ref{sub:kernels}) for a set of equidistant points in $x$ between 0 and 10.
On the left, the mean and two sigma interval of the joint Gaussian distribution is shown for all points of $x$ along with 3 samples drawn from that same distribution.
On the right, a visual representation of the covariance matrix is shown for that same Gaussian distribution, with darker blue regions highlighting larger correlations between pairs of points.
In the top plot, the distribution follows only the properties of the mean and covariance functions, whilst on the bottom, additional constraints are introduced in the form of observed data, which conditions the GP distribution, and its mean and covariance functions, to only consider models that include those data points.

Interpreting the distributions in Figure~\ref{fig:gp_prior_posterior_example} in a Bayesian context, we would say that the GP distribution on the top defines a prior distribution over functions that follow the properties defined by its mean and covariance functions, such as smoothness and periodicity, without any additional constraints. 
By adding observed data, this distribution is then conditioned to those observations, resulting in the new posterior distribution over functions on the bottom, where only the functions that include the observed data points are considered.

\subsection{Kernels}
\label{sub:kernels}

Defining the properties of a Gaussian process means defining its mean and covariance functions. 
Here we focus exclusively in exploring covariance functions, often referred to as kernels, which control properties of functions such as its smoothness. 

Kernels are functions following the form $k(x, x')$ and the only requirement for a function to be considered a covariance function is that their resulting covariance matrices are positive semidefinite. 
Additionally, kernels themselves also depend on one or more parameters.
These define general properties of the kernel itself, such as its characteristic length-scale, amplitude or period, and modifying them can further alter the characteristics of the covariance between variables.
These kernel parameters are often referred to as hyper-parameters and we represent their vector by the subscript $\alpha$.

\begin{figure}[!ptb]
    \centering
    \begin{subfigure}{0.49\textwidth}
        \caption{Diagonal kernel (Eq.~\ref{eq:diagonal_kernel})}
        \includegraphics[width=\textwidth]{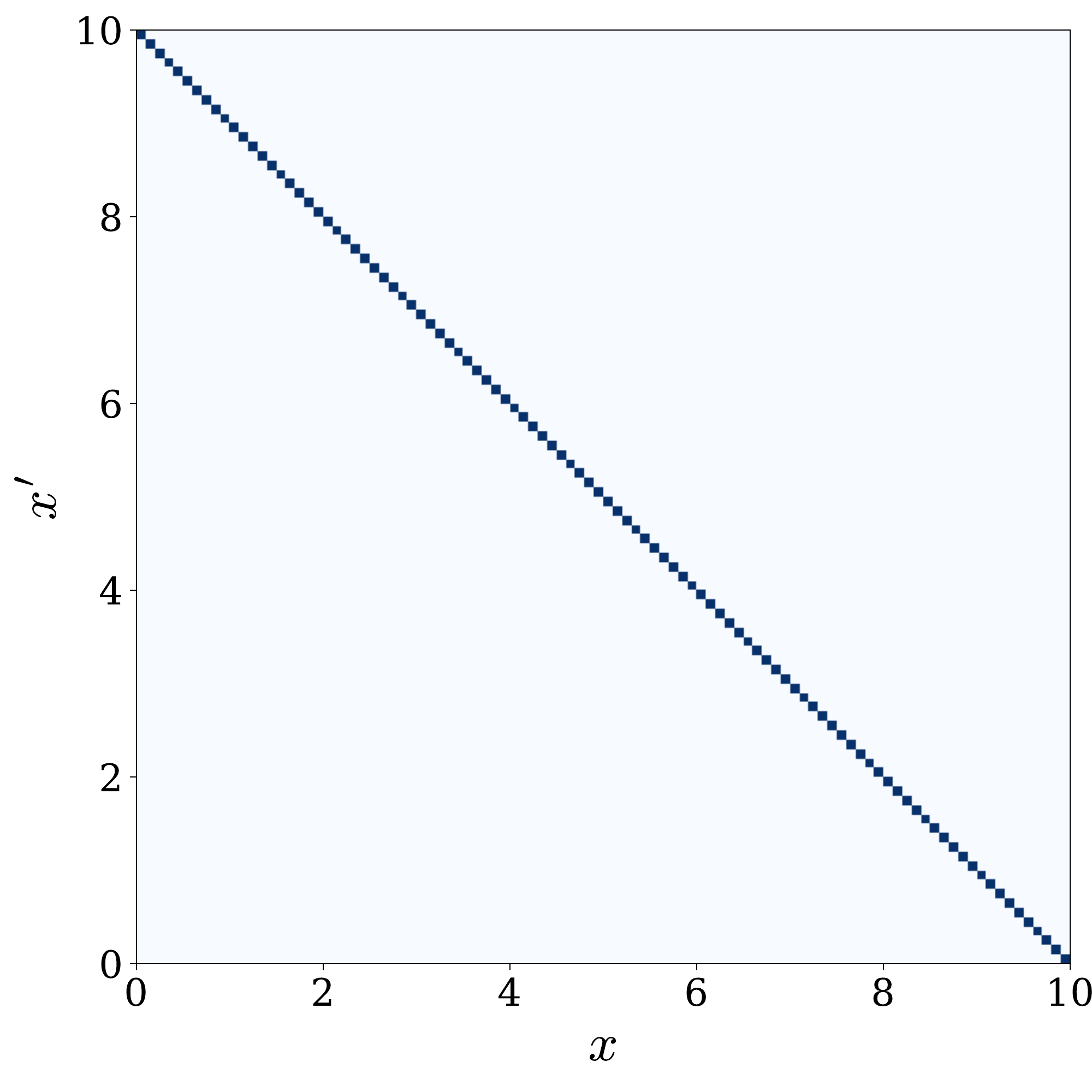}
        \label{fig:diagonal_cov}
    \end{subfigure}
    \begin{subfigure}{0.49\textwidth}
        \caption{Squared Exponential kernel (Eq.~\ref{eq:se_kernel})}
        \includegraphics[width=\textwidth]{characterization/gps/se_kernel_cov.pdf}
        \label{fig:se_cov}
    \end{subfigure}
    \begin{subfigure}{0.49\textwidth}
        \caption{Periodic kernel (Eq.~\ref{eq:periodic_kernel})}
        \includegraphics[width=\textwidth]{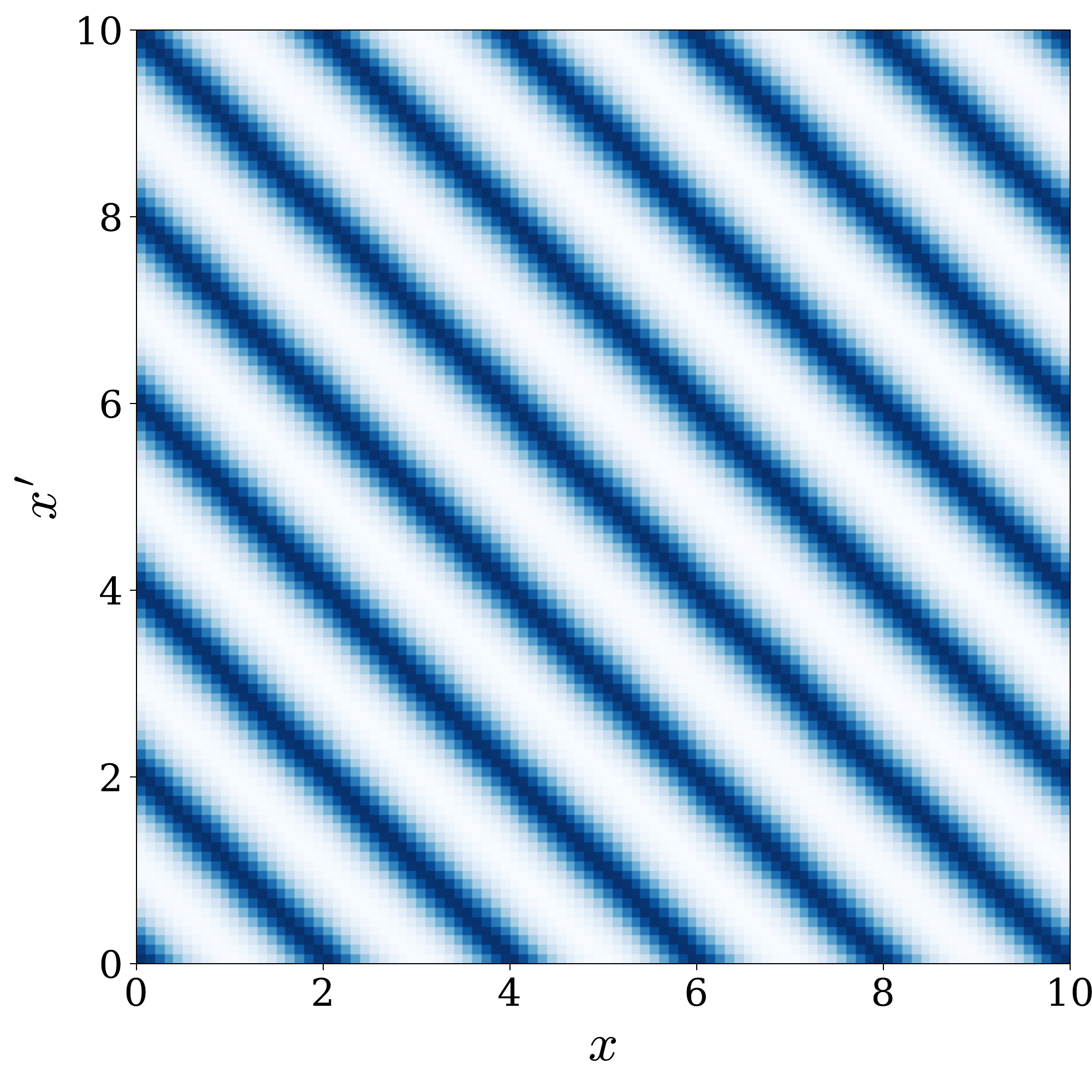}
        \label{fig:periodic_cov}
    \end{subfigure}
    \caption[Covariance matrices for the kernel examples listed in Section~\ref{sub:kernels}, for a set of equidistant variables in $x$ between 0 and 10]
    {Covariance matrices for the kernel examples listed in Section~\ref{sub:kernels}, for a set of equidistant variables in $x$ between 0 and 10. 
    Correlation between variables increases from white (lowest) to dark blue (highest).}
    \label{fig:kernel_cov_matrices}
\end{figure}

Some common kernels in the general Gaussian process literature include: 

\begin{itemize}
    \item diagonal kernel
\end{itemize}
\begin{equation}
    k_\alpha(x, x') = \sigma^2 \delta_{xx'} ,
    \label{eq:diagonal_kernel}
\end{equation}
with $\alpha = \{\sigma^2\}$ and where the Kronecker delta $\delta_{xx'}$ evaluates to unity when $x = x'$ and is null for all other combination pairs. 
This kernel attributes the value $\sigma^2$, associated with the variance, to all pairs of points in the covariance matrix diagonal, and 0 everywhere else. 
Since this kernel does not consider any covariance between points, it is representative of white noise.
Its covariance matrix is shown in Figure~\ref{fig:diagonal_cov}.

\begin{itemize}
    \item squared exponential kernel (SE kernel)
\end{itemize}
\begin{equation}
    k_\alpha(x, x') = a \exp \left( - \frac{\vert x - x' \vert^2}{2 l^2} \right) ,
    \label{eq:se_kernel}
\end{equation}
with $\alpha = \{a, l\}$. 
This kernel establishes a correlation between variables that is stronger the closer they are in input space $x$, scaled by a characteristic amplitude $a$ and that decays exponentially for farther away variables, following a characteristic length-scale $l$.
Its covariance matrix is shown in Figure~\ref{fig:se_cov}.

\begin{itemize}
    \item periodic kernel
\end{itemize}
\begin{equation}
    k_\alpha(x, x') = a \exp \left[ - \frac{2}{l^2} \sin^2 \left( \pi \frac{\vert x - x' \vert}{p} \right) \right] ,
    \label{eq:periodic_kernel}
\end{equation}
with $\alpha = \{a, l, p\}$.
This kernel defines a periodic correlation stronger between variables distanced by multiples of period $p$, scaled by a characteristic amplitude $a$ and that decays following a characteristic length-scale $l$.
Its covariance matrix is shown in Figure~\ref{fig:periodic_cov}.



\begin{figure}[!ptb]
    \centering
    \begin{subfigure}{0.49\textwidth}
        \caption{$a = 1.0, l = 1.0$}
        \includegraphics[width=\textwidth]{characterization/gps/gp_se_prior.pdf}
    \end{subfigure} \\
    \begin{subfigure}{0.49\textwidth}
        \caption{$a = 1.0, l = 5.0$}
        \includegraphics[width=\textwidth]{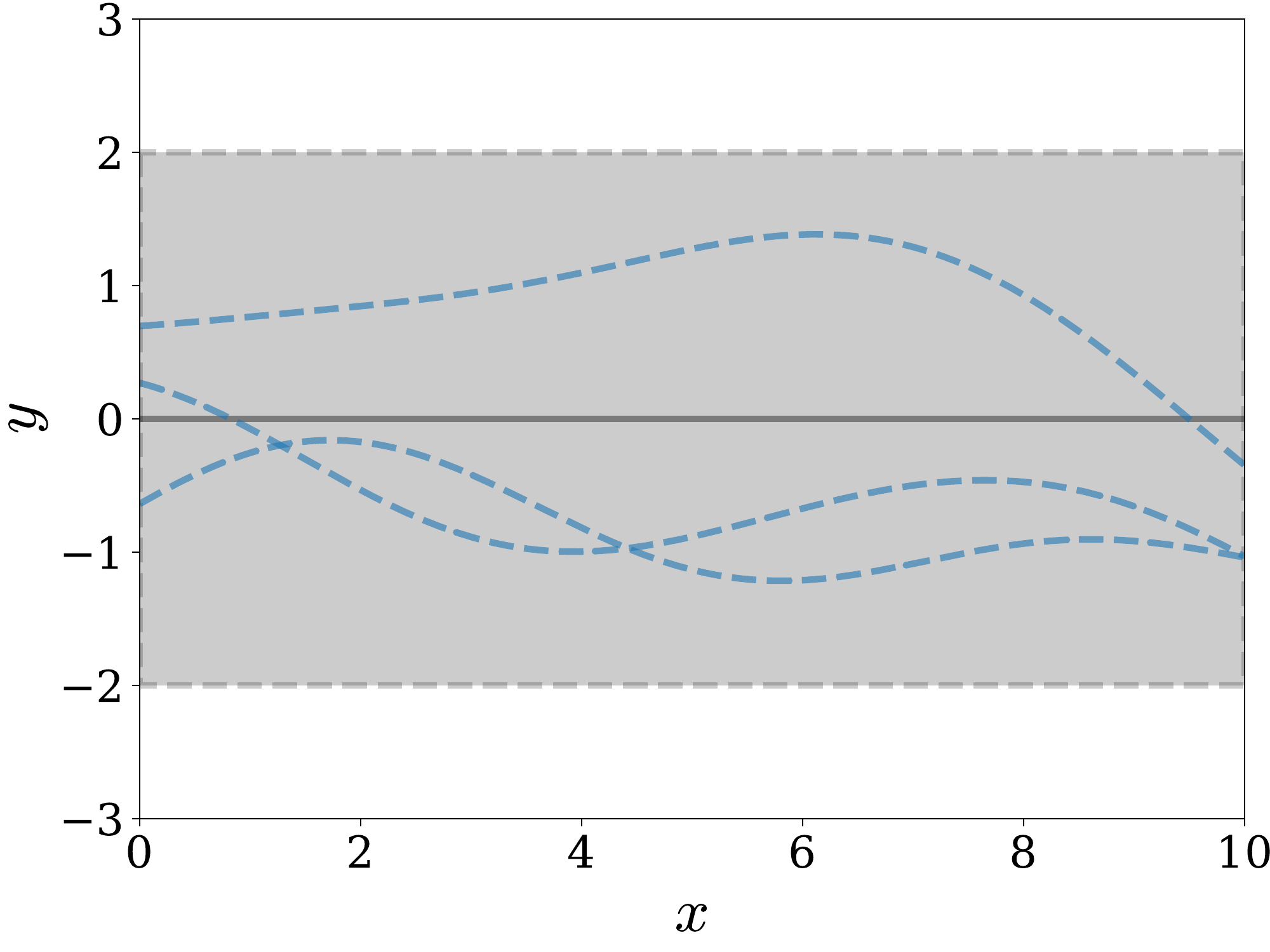}
    \end{subfigure}
    \begin{subfigure}{0.49\textwidth}
        \caption{$a = 1.0, l = 0.2$}
        \includegraphics[width=\textwidth]{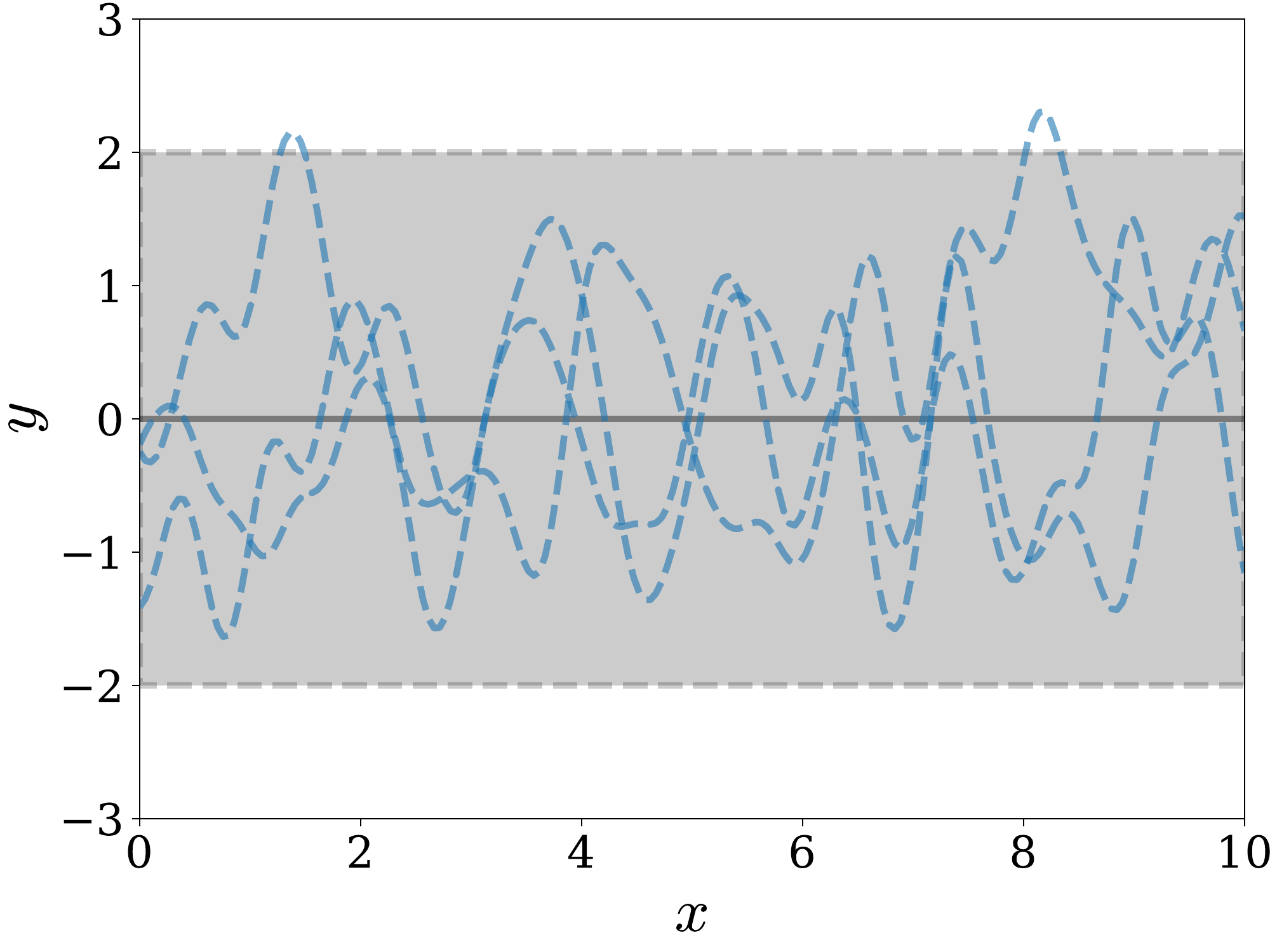}
    \end{subfigure}
    \caption[Impact of different characteristic length-scale hyper-parameters $l$ on the properties of functions sampled from a GP distribution with a squared exponential kernel]
    {Impact of different characteristic length-scale hyper-parameters $l$ on the properties of functions sampled from a GP distribution with a squared exponential kernel.
    The black solid line and gray shaded region denote the mean $\mu$ and 2-$\sigma$ region of the distribution, respectively.
    The values of the parameters used for the kernel are shown in each of the subcaptions.
    Notice how, despite the GP distribution being seemingly the same, the 3 samples, shown as blue dashed lines, follow the varying length-scale.}
    \label{fig:gp_hyperparameters_example}
\end{figure}

To demonstrate the impact that modifying the kernel hyper-parameters can have on the properties of the functions represented in the GP distribution, Figure~\ref{fig:gp_hyperparameters_example} shows the effect of varying the characteristic length-scale $l$ of the squared exponential kernel (Eq.~\ref{eq:se_kernel}) on the GP prior distribution from the top-left plot of Figure~\ref{fig:gp_prior_posterior_example} and its samples.

Finally, since both sums and products of kernels also define valid kernels, all these examples and any other covariance functions that fulfill the positive semidefinite property can be combined to create more elaborate models, broadening the possibilities considerably.

\subsection{Estimating hyper-parameters}
\label{sub:gp_parameter_optimization}

Earlier in this section we introduced the applicability of GPs to the prediction of unknown data given the definition of a mean and covariance functions and their respective parameters as well as a set of $n$ observations $\mathcal{D}={(x_i, y_i)}$. However, for our use case, which is the characterization of stellar signals, we are more interested in estimating the mean and kernel parameters themselves, as those are the unknown quantities that describe the variations in flux in the time series.
To do so, we interpret now the problem as estimating what is the probability for a set of mean function parameters $\bm{\theta}$ and kernel hyper-parameters $\bm{\alpha}$ given the set of observations $\mathcal{D}$. In Bayesian notation this would be
\begin{equation}
    p(\bm{\alpha}, \bm{\theta} \vert \bm{y}, \bm{x}) = \frac{p(\bm{y} \vert \bm{x}, \bm{\alpha}, \bm{\theta}) p(\bm{\alpha}, \bm{\theta})}{p(\bm{y} \vert \bm{x})} ,
    \label{eq:hyperparameter_posterior}
\end{equation}
where the posterior probability $p(\bm{\alpha}, \bm{\theta} \vert \bm{y}, \bm{x})$ for a set of parameters is proportional to the likelihood $p(\bm{y} \vert \bm{x}, \bm{\alpha}, \bm{\theta})$ that those parameters adequately explain the dataset multiplied by the prior probability for the parameters, $p(\bm{\alpha}, \bm{\theta})$, which contains our prior knowledge for each of the parameters without taking into account any data from the dataset. The denominator $p(\bm{y} \vert \bm{x})$ is the marginal likelihood or model evidence, and represents the cumulative likelihood for all combinations of values of the parameters $\bm{\alpha}$ and $\bm{\theta}$ considered.

The expression to evaluate the likelihood function is just the probability density function for the $n$-dimensional joint Gaussian distribution, given by Equation~(\ref{eq:multivariate_gaussian_expression}). This likelihood is often presented in the literature in its log-form, which is more easily calculated computationally:
\begin{equation}
    \log p(\bm{y} \vert \bm{x}, \bm{\alpha}, \bm{\theta}) = - \frac{1}{2} (\bm{y} - \bm{\mu})^T \Sigma^{-1} (\bm{y} - \bm{\mu}) - \frac{1}{2} \log \left( \vert \Sigma \vert \right) - \frac{n}{2} \log \left( 2 \pi \right) ,
    \label{eq:hyperparameter_likelihood}
\end{equation}
where $\bm{\mu}$ is the mean vector and $\Sigma$ is the covariance matrix. The mean vector is obtained from the GP mean function evaluated at points $\bm{x}$ and depends on the parameters $\bm{\theta}$. The covariance matrix is obtained from the kernel function evaluated for each pair of points $(\bm{x}, \bm{x'})$ and depends on the set of hyper-parameters $\bm{\alpha}$. $\vert \Sigma \vert$ is the covariance matrix's determinant.

To estimate the probability distributions of each of the parameters in the model, we need to evaluate the posterior probability of Equation~(\ref{eq:hyperparameter_posterior}) and marginalize over each parameter. 
Practically, calculating the posterior is often not possible analytically and numerically calculating the evidence (denominator) is usually too expensive computationally. 
An alternative approximation is to maximize the numerator, i.e. maximize the likelihood function (Eq.~\ref{eq:hyperparameter_likelihood}) weighted by the parameter priors. 
This approach is known as type II maximum likelihood \parencite[ML-II;][]{Berger_1985,Rasmussen_2006} and is the one used in my work.

%% file: chapters/characterization/modelling.tex
\section{Modeling stellar signals in the time domain}
\label{sec:modelling}


The application of GPs to the characterization of stellar signals has already seen some developments in recent years.
Both \textcite{Dawson_2014} and \textcite{Barclay_2015} used an SE kernel (Equation~\ref{eq:se_kernel}) to try to capture out-of-transit variations in their respective models of photometric time series, with \textcite{Barclay_2015} even getting a definite confirmation on the planetary nature of Kepler-91 b \parencite{Esteves_2013,Sliski_2014,Lillo-Box_2014}.
More recently, \textcite{Grunblatt_2017} extended these GP models to add some more physical motivation to the selection of kernels.
The same has also been seen for RV time series, with examples such as \textcite{Brewer_2009} and \textcite{Faria_2016}.

A common disadvantage of GP regression is the computational cost of evaluating the inverse covariance matrix $\Sigma^{-1}$ of the likelihood function (Equation~\ref{eq:hyperparameter_likelihood}), which has time complexity $\mathcal{O}(n^3)$, meaning the computation cost scales with the cube of the number of variables $n$ in the set chosen from the input space $\bm{x}$.
Although computers are very fast machines nowadays, the likelihood function has to be evaluated for every sample of the posterior and when considering \textit{TESS} light curves, which are the focus of this work, the number of flux observations ranges from $\sim 10^{3}$ to $\sim 10^{4}$ data points, making the calculation required for each sample prohibitively long.

In this work, I first identified and evaluated a GP regression framework that accelerated massively the computational time of the likelihood estimation, by imposing some limitations on the available kernel functions.
Then, I took the idea of choosing a physically motivated model a step further, by finding, within the limitations imposed in this framework, individual kernels that could best capture the stellar signals discussed in Section~\ref{sec:stellar_signals}, in the time domain.
Finally, I implemented these kernels in an open-source framework such that they share the same parameters as the stellar signals themselves, making it more intuitive to reason about them and their prior values.

%
\subsection{Speeding it up with \texttt{celerite}}
\label{sub:celerite}

To solve the issue of computational complexity, my method is built on top of \texttt{celerite} \parencite{Foreman-Mackey_2017}, an open-source \texttt{Python} package that implements a Gaussian process framework with some imposed restrictions in order to improve computational times. 
The first limitation is that only one-dimensional datasets are allowed, meaning the values of the input space $\bm{x}$ have to be scalars, although this limitation has been expanded to two dimensions, under certain conditions \parencite[see ][]{Gordon_2020}. 
This is not an issue for this use case, since the models I am trying to define are describing non-parametric functions in time-series data, where the input space is time $\bm{t}$. 
The second limitation imposed by the method is that the allowed kernels in the \texttt{celerite} framework have to follow a specific functional form that is stationary, meaning the covariance between two variables $t_i$ and $t_j$ depends only on the distance between variables,
\begin{equation}
    k(t_i, t_j) = k(\tau_{ij}) \qquad \textnormal{with} \qquad \tau_{ij} = \vert t_i - t_j \vert .
    \label{eq:stationary_kernel}
\end{equation}

However, stationarity implies that covariance functions have the form of autocorrelation functions, whose Fourier transform gives its power spectral density.
In this way, this limitation ends up providing a benefit, since I am looking for the equivalent time-domain functions of the frequency-domain stellar-signal equations introduced in Section~\ref{sec:stellar_signals}.

To arrive at the functional form imposed by the method in \textcite{Foreman-Mackey_2017}, I follow their notation and procedure, and start with the work of \textcite{Rybicki_1995} which defined a Gaussian process kernel of the form
\begin{equation}
    k_\alpha(\tau_{ij}) = \sigma_i^2 \delta_{ij} + a \exp (-c \tau_{ij}) ,
    \label{eq:rybicki_kernel}
\end{equation}
where $\sigma_i^2$ is the measurement uncertainty for data observed at $t_i$, $\delta_{ij}$ is the Kronecker delta, and $\alpha = \{a, c \}$ are the kernel's hyper-parameters. 
With this restricted form, the covariance matrix is tridiagonal and so inverting the matrix has a reduced time complexity of $\mathcal{O}(n)$. 
Equation~(\ref{eq:rybicki_kernel}) can be generalized to form an arbitrary mixture of exponentials,
\begin{equation}
    k_\alpha(\tau_{ij}) = \sigma_i^2 \delta_{ij} + \sum_{m=1}^M a_m \exp (-c_m \tau_{ij}) ,
    \label{eq:mixture_kernel}
\end{equation}
whilst increasing the time complexity to $\mathcal{O}(nM^2)$. \textcite{Foreman-Mackey_2017} took the idea further by introducing complex parameters $a_m \rightarrow a_m \pm i \ b_m$ and $c_m \rightarrow c_m \pm i \ d_m$ and rewriting the equation as a sum of sines and cosines, resulting in an expression for a mixture of quasi-periodic oscillators,
\begin{equation}
    k(\tau_{ij}) = \sigma_i^2 \delta_{ij} + \sum^M_{m=1} \left[ a_m \exp (-c_m \tau_{ij}) \cos (d_m \tau_{ij}) + b_m \exp (-c_m \tau_{ij}) \sin (d_m \tau_{ij}) \right] ,
	\label{eq:sum_sines_kernel}
\end{equation}
where the hyper-parameters are now $\alpha = \{a_m, b_m, c_m, d_m\}$. The PSD of this kernel, obtained by computing its Fourier transform, is then
\begin{equation}
    P(\omega) = \sum^M_{m=1} \sqrt{\frac{2}{\pi}} \frac{(a_m c_m + b_m d_m)(c_m^2 + d_m^2) + (a_m c_m - b_m d_m) \omega^2}{\omega^4 + 2(c_m^2 - d_m^2)\omega^2 + (c_m^2 + d_m^2)^2} ,
    \label{eq:sum_sines_psd}
\end{equation}
where $\omega = 2 \pi \nu$ is the angular frequency.

To attribute a clearer physical meaning to the kernel in Equation~(\ref{eq:sum_sines_kernel}), \textcite{Foreman-Mackey_2017} established a relationship between its PSD (Equation~\ref{eq:sum_sines_psd}) and the PSD of a stochastically driven damped harmonic oscillator, given by
\begin{equation}
    P(\omega) = \sqrt{\frac{2}{\pi}} \frac{S_0 \omega_0^4}{(\omega^2 - \omega_0^2)^2 + \omega_0^2 \omega^2 / Q^2} ,
    \label{eq:harmonic_oscillator_psd}
\end{equation}
where $\omega_0$ is the frequency of the undamped oscillator, $Q$ is the oscillator’s quality factor, and $S_0$ is an amplitude measure proportional to the power of the spectrum at $\omega = \omega_0$, 
\begin{equation}
    P(\omega_0) = \frac{2}{\pi} S_0 Q^2 .
    \label{eq:harmonic_oscillator_psd_power_at_max}
\end{equation}

By matching the PSDs from Equations~(\ref{eq:sum_sines_psd}) and (\ref{eq:harmonic_oscillator_psd}), the hyper-parameters $\alpha = \{a_m, b_m, c_m, d_m\}$ can be equated to the set of parameters $\{S_0, \omega_0, Q\}$ from the harmonic oscillator and so the kernel of Equation~(\ref{eq:sum_sines_kernel}) can be rewritten as
\begin{equation}
    \begin{array}{ll}
        k(\tau_{ij}; S_0, Q, \omega_0) = S_0 \omega_0 Q \exp \left( -\frac{\omega_0 \tau_{ij}}{2Q} \right) \\[10pt]
        \hspace{6.78em} \times \left\{
            \renewcommand{\arraystretch}{1.2}
            \setlength{\arraycolsep}{0pt}%
            \begin{array}{ll}
                \cosh (\eta \omega_0 \tau_{ij}) + \frac{1}{2 \eta Q} \sinh (\eta \omega_0 \tau_{ij}), &\quad 0 < Q < 1/2 \\
                2(1 + \omega_0 \tau_{ij}), &\quad Q = 1/2 \\
                \cos (\eta \omega_0 \tau_{ij}) + \frac{1}{2 \eta Q} \sin (\eta \omega_0 \tau_{ij}), &\quad 1/2 < Q       
            \end{array}
        \right. ,
    \end{array}
    \label{eq:celerite_sho_kernel}
\end{equation}
where $\eta = \vert 1- (4Q^2)^{-1}\vert^{1/2}$.

%
\subsection{Kernels for the granulation and oscillations envelope}
\label{sub:celerite_kernels}

Specific limits of the kernel in Equation~(\ref{eq:celerite_sho_kernel}) are of physical interest and I explore them to define kernels to characterize in the time domain the stellar signals discussed in Section~\ref{sec:stellar_signals}. 

First, by considering the limit of $Q = 1/\sqrt{2}$, the kernel simplifies to 
\begin{equation}
	k(\tau_{ij}) = S_0 \omega_0 \exp \left( -\frac{1}{\sqrt{2}} \omega_0 \tau_{ij} \right) \text{cos} \left( \frac{\omega_0 \tau_{ij}}{\sqrt{2}} - \frac{\pi}{4} \right) ,
	\label{eq:celerite_granulation_kernel}
\end{equation}
with the corresponding PSD becoming
\begin{equation}
	P(\omega) = \sqrt{\frac{2}{\pi}} \frac{S_0}{ \left( \omega / \omega_0 \right)^4 + 1} .
	\label{eq:celerite_granulation_psd}
\end{equation}

\begin{figure}[!ptb]
	\centering
	\includegraphics[width=\textwidth, height=\textheight, keepaspectratio]{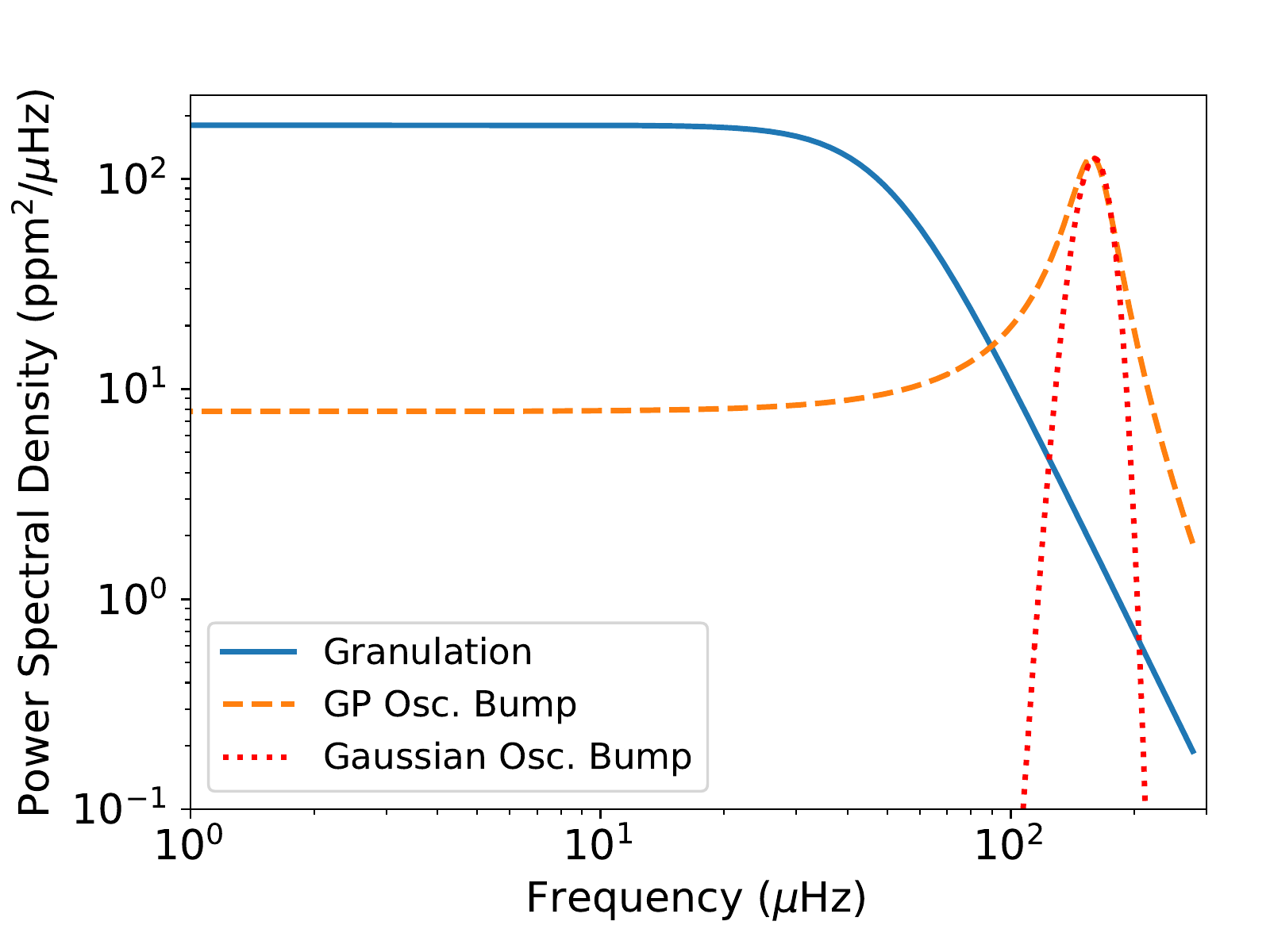}
	\caption[Illustration and comparison of PSD profiles for different stellar signals]
    {Illustration and comparison of PSD profiles for different stellar signals.
    Solid (blue) curve depicts the PSD of a granulation profile (Equation~\ref{eq:celerite_granulation_psd}).
    Dashed (orange) and dotted (red) curves are the PSDs of the functions used to capture the signal from the oscillations envelope in the GP (Equation~\ref{eq:sum_sines_psd}) and power spectrum analysis (Equation~\ref{eq:envelope_gaussian_psd}), respectively.}
	\label{fig:comparison_kernel_psd}
\end{figure}

The PSD in Equation~(\ref{eq:celerite_granulation_psd}) has the same functional form as the PSD from Equation~(\ref{eq:granulation_psd_kallinger}) used to model granulation. 
The parameters in these equations can then be related by ensuring that both equations are equally normalized, the procedure for which I show in Appendix~\ref{app:celerite_normalization}.
The following relation between parameters is obtained:
\begin{equation}
    \begin{split}
        a_\text{gran} =& \ \sqrt{\sqrt{2} S_{0,\text{gran}} \omega_{0,\text{gran}}} ,\\[5pt]
        b_\text{gran} =& \ \frac{\omega_{0,\text{gran}}}{2\pi} .
    \end{split}
\end{equation}
The blue solid line in Figure~\ref{fig:comparison_kernel_psd} illustrates the shape of this PSD.

Another limit of interest for the kernel in Equation~(\ref{eq:celerite_sho_kernel}) is given for values of $Q > 1$. 
In this case, the shape of its PSD function near the peak frequency $\omega_0$ approaches a Lorentzian. 
This limit, although an approximation, was chosen in this work as the model to capture the signal from the oscillations envelope, characterized in the frequency domain analysis by Equation~(\ref{eq:envelope_gaussian_psd}). 
Figure~\ref{fig:comparison_kernel_psd} shows both models, highlighting the differences between them, such as the low-frequency tail of power and the less pronounced decay of power at higher frequencies of the PSD of the GP kernel model.
Although no clear relation can be drawn between the width of the Gaussian, $\sigma$, in Equation~(\ref{eq:envelope_gaussian_psd}) and the harmonic oscillator's parameters $\{S_0, \omega_0, Q\}$, the remaining two parameters from the Gaussian can be expressed as
\begin{equation}
    \begin{split}
        P_g =& \ 4 S_{0,\text{env}} Q_\text{env}^2 ,\\[5pt]
        \nu_\text{max} =& \ \frac{\omega_{0,\text{env}}}{2\pi} .
    \end{split}
\end{equation}


\subsection{Putting it all together - \textit{gptransits}}
\label{sub:gptransits}

I have implemented both models described in the previous section in an open-source \texttt{Python} package named \texttt{gptransits}.
The package is hosted on my Github repository (\url{https://github.com/Fill4/gptransits}), where technical details and instructions on installation and usage are discussed.
This implementation, as mentioned earlier, builds on top of \texttt{celerite} and provides three kernel components to characterize stellar signals: granulation, oscillations envelope and white noise.
Since any combination (sums or additions) of kernels is still a valid kernel, it is possible to design various stellar signal models, with a straightforward example being the inclusion of two granulation components, capturing signals at different characteristic frequencies.

The package also includes a parametric planetary transit model.
Specifically, I include the quadratic limb-darkened parametric model from Equation~(\ref{eq:transit_quadratic_ld}), by using its implementation in the \texttt{Python} package \texttt{batman} \parencite{Kreidberg_2015}, which has seen extensive adoption in the literature.


To sample the parameter posterior distributions (Equation~\ref{eq:hyperparameter_posterior}), I use the \texttt{emcee}\footnote{\url{https://github.com/dfm/emcee}} \texttt{Python} package \parencite{Foreman-Mackey_2013}. 
This package provides an implementation of Goodman $\&$ Weare’s Affine Invariant Markov chain Monte Carlo (MCMC) ensemble sampler \parencite{Goodman_2010} and can be used to perform Bayesian parameter estimation given a prior distribution for each of the parameters in the model.

This tool was tested extensively in the characterization of stellar signals both with and without the inclusion of planetary transits, the methodology and results of which are described in Section~\ref{sec:applications}.

%% file: chapters/characterization/applications.tex
\section{Testing the method}
\label{sec:applications}


To evaluate the applicability of the framework to the characterization of stellar signals, I performed two applications of the method to different sets of light curves without planets.
For these tests I define two slightly different models of stellar signals.

The first application is based on simulated \textit{TESS} light curves, where the accuracy of the extracted signal parameters is evaluated in comparison with the values used to generate the data.
The second one is based on subsets of \textit{Kepler} light curves, with the goal of gauging the accuracy and precision obtained for the signal parameters in comparison with a standard characterization in the frequency domain.

Finally, a transit component is added to the models and a transit signal is injected in the simulated \textit{TESS} light curves, with the goal of evaluating the accuracy and improvement in the recovery of the planetary properties.

The results presented in this section comprise the scope of the publication I led introducing the method developed for simultaneous time-domain characterization of planetary transits and stellar signals \parencite{Pereira_2019}.

\subsection{Stellar signal models}
\label{sub:test_models}

As mentioned before, two models were considered for the tests used to evaluate the framework.

Model 1 contains a single granulation kernel (Equation~\ref{eq:celerite_granulation_kernel}) to capture the mesogranulation, an oscillations envelope kernel (Equation~\ref{eq:celerite_sho_kernel}), where the limit $1.2 < Q < 18$ was imposed empirically, to follow as closely as possible the Gaussian envelope found in power-spectrum fitting, and a white noise kernel (Equation~\ref{eq:diagonal_kernel}), modeling extra variance at each point in time, not related to each point's measurement uncertainty, which is added in quadrature to the diagonal of the covariance matrix.
The resulting GP kernel is
\begin{equation}
	\begin{split}
		&k(\tau_{ij}) = \sigma_i^2 \delta_{ij} + \\
		&+ \ S_{0,gran} \omega_{0,gran} \exp \left( -\frac{1}{\sqrt{2}} \omega_{0,gran} \tau_{ij} \right) \text{cos} \left( \frac{\omega_{0,gran} \tau_{ij}}{\sqrt{2}} - \frac{\pi}{4} \right) + \\
		&+ \ S_{0,env} \omega_{0,env} Q \exp \left( -\frac{\omega_{0,env} \tau_{ij}}{2Q} \right) \left[ \cos (\eta \omega_{0,env} \tau_{ij}) + \frac{1}{2 \eta Q} \sin (\eta \omega_{0,env} \tau_{ij}) \right] ,
	\end{split}
	\label{eq:model_1_kernel}
\end{equation}
where, again, $\eta = \vert 1- (4Q^2)^{-1}\vert^{1/2}$. 
Model 2 considers the addition of an extra granulation kernel at higher frequencies, which is imposed by the priors (see $b_\text{gran,2}$ in Table~\ref{tab:priors_model2}).

Looking at the methodology of \textcite{Kallinger_2014}, both the granulation and mesogranulation components should be adopted when estimating the stellar background signal for red-giant stars, owing to granulation being relevant at different timescales. 
The reason for also considering Model 1 in our tests is due to the shorter temporal coverage of available \textit{TESS} data (most targets are observed for a duration of 27.4 days), as well as the higher white noise level expected, when compared with the \textit{Kepler} stars on which the aforementioned study was performed.
For that same reason, no model was considered with a third component, which is sometimes added in the literature to characterize long-term variations \parencite[on the order of 1 cycle per 10 days;][]{Corsaro_2015a}.


Only uniform priors were considered for our models' parameters, represented by the symbol $\mathcal{U}$ and a lower and upper limit.
We also defined considerably larger prior intervals than usual to evaluate the capacity of the method to converge even with reduced prior knowledge.
The direct correspondence to frequency-domain models ensures that power-spectrum fitting analysis of the same data can inform and restrict these priors in future applications.

Tables~\ref{tab:priors_model1} and \ref{tab:priors_model2} present the uniform prior bounds chosen for the prior distributions of all parameters in Models 1 and 2, respectively.
\begin{table}[!ptb]
	\centering
	\setlength{\tabcolsep}{12pt}
	\begin{tabular}{l c}
		\hline
		Parameter                   & Prior \\
		\hline
		$a_\text{gran,1}$ [ppm]     & $\mathcal{U}$(10, 400)	\\
		$b_\text{gran,1}$ [$\mu$Hz] & $\mathcal{U}$(10, 200)	\\
		$P_\text{g}$ [ppm]          & $\mathcal{U}$(10, 1800)	\\
		$Q_\text{env}$              & $\mathcal{U}$(1.2, 18)	\\
		$\nu_\text{max}$ [$\mu$Hz]  & $\mathcal{U}$(80, 220)	\\
		White Noise [ppm]           & $\mathcal{U}$(0, 400)		\\
		\hline
	\end{tabular}
	\caption[Lower and upper bounds chosen for the uniform distributions used as priors for the parameters in Model 1]
	{Lower and upper bounds chosen for the uniform distributions used as priors for the parameters in Model 1.}
	\label{tab:priors_model1}
\end{table}
\begin{table}[!ptb]
	\centering
	\setlength{\tabcolsep}{12pt}
	\begin{tabular}{l c}
		\hline
		Parameter                   & Prior \\
		\hline
		$a_\text{gran,1}$ [ppm]     & $\mathcal{U}$(10, 400)	\\
		$b_\text{gran,1}$ [$\mu$Hz] & $\mathcal{U}$(10, 70)		\\
		$a_\text{gran,2}$ [ppm]     & $\mathcal{U}$(10, 400)	\\
		$b_\text{gran,2}$ [$\mu$Hz] & $\mathcal{U}$(80, 300)	\\
		$P_\text{g}$ [ppm]          & $\mathcal{U}$(10, 1800)	\\
		$Q_\text{env}$              & $\mathcal{U}$(1.2, 18)	\\
		$\nu_\text{max}$ [$\mu$Hz]  & $\mathcal{U}$(80, 220)	\\
		White Noise [ppm]           & $\mathcal{U}$(0, 400)		\\
		\hline
	\end{tabular}
	\caption[Lower and upper bounds chosen for the uniform distributions used as priors for the parameters in Model 2]
	{Lower and upper bounds chosen for the uniform distributions used as priors for the parameters in Model 2.}
	\label{tab:priors_model2}
\end{table}

\subsection{Parameter estimation and convergence metrics}
\label{sub:convergence}

To explore the parameter space and sample the posterior probability, we initialized multiple MCMC chains, each of which produced 25,000 posterior samples. 
The number of chains varied for each of the two models used as it was defined as 4 times the number of parameters in the model being tested.

To evaluate the convergence of the chains, a range of tests were calculated based on the samples obtained. 
The Geweke statistic \parencite{Geweke_1992} was applied to all the chains to estimate the burn-in period of each chain. 
Chains that did not pass this test were removed from the determination of the final parameters. 
In the next step, chains with low posterior probabilities were discarded, specifically, chains with an average posterior probability of their samples lower than the 10th percentile of the posterior probability distribution of all samples.
This was done to avoid considering any chains that converged to local minima.

For the remaining chains, we determined the univariate and multivariate Gelman-Rubin diagnostics \parencite{Gelman_1998}, which evaluate convergence of Markov chains by comparing the between-chains and within-chain variance. 
Whilst the univariate approach analyzes each parameter of the model independently, the multivariate version of this diagnostic takes into account covariances between the parameters, so it is a more demanding diagnostic for convergence. 
While these tests are not strictly valid when applied to these chains, since \texttt{emcee} produces correlated chains, the diagnostics still provide a numerical evaluation of convergence of the results.
When determining both of these diagnostics, up to 50\% of the samples from each chain (starting from the initial samples) were allowed to be discarded in order to improve the convergence of the remaining samples. 
A maximum value of 1.1 was defined as the threshold enforced for the Gelman-Rubin diagnostic.
The tests applied showed that, after selection of the best samples and chains, even in the rare case where only 25\% of all the initial samples where considered, the number of samples was large enough to achieve accurate estimations on all parameters.


With the final samples selected, both the median and mode were calculated for each of the model's parameters along with two highest posterior density (HPD) intervals. 
The first HPD interval calculated was a 68.3\% HPD, which was used as a measure of the lower and upper uncertainty on the value of each parameter (so that meaningful uncertainties could also be determined for non-Gaussian sample distributions). 
The second was a 95\% HPD interval, which was compared with the width of the uniform prior defined for each of the model's parameters in order to determine whether the converged samples were exploring a small subset of the prior space or if the entire prior had similar posterior probability, in which case the specific parameter was flagged as not being well constrained.

\subsection{\textit{TESS} artificial light curves}
\label{sub:test_tess_lcs}

The first application of our framework was to \textit{TESS}-like artificial time series, where the objective was to determine the accuracy of the recovered parameters that characterize the stellar signals present in the simulated light curves. 
For this test, we considered a set of 20 LLRGB stars (with effective temperature $4800 < T_\text{eff} < 5500\:\text{K}$, frequency of maximum oscillation amplitude $105 < \nu_\text{max} < 185\:\mu\text{Hz}$, and apparent magnitude $V < 11$). 

Generation of the artificial light curves was performed originally in the frequency domain by using scaling relations, after which an inverse Fourier transform was applied, taking into consideration the 30-min cadence of \textit{TESS} FFIs. 
We used a photometric noise model for \textit{TESS} \parencite{Sullivan_2015,Campante_2016a} in order to predict the rms noise for a given exposure time, and included a systematic noise term of $20\:\text{ppm}\,\text{hr}^\text{1/2}$. 
To model the granulation PSD, a scaled version (to predict \textit{TESS} granulation amplitudes) of model F of \textcite[][see Equation~\ref{eq:granulation_psd_kallinger}]{Kallinger_2014} was adopted, which contains two granulation components, the granulation and mesogranulation. 
No aliased granulation power was considered. 
Individual radial, (mixed) dipole and quadrupole modes were also modeled \parencite{Kuszlewicz_2019}.

For each of the 20 simulated stars, 10 independent 27.4-day time series were generated. 
The values and corresponding uncertainties computed for each of the model parameters took into account the analysis of at least 5 and up to all 10 of these independent time series (chosen according to their performance in the evaluated convergence statistics, as detailed in Section~\ref{sub:convergence}), so as to reduce the possibility of systematic errors. 
Specifically, each parameter was estimated as the median of the values obtained for the chosen independent time series. 
Its uncertainty was defined as the sum in quadrature of the uncertainty associated with the median value and the standard deviation of the values determined for the remaining chosen time series. 
Additionally, if a parameter from any of the chosen runs was identified as being not constrained, the final parameter calculated from the results of all runs was flagged.

\subsubsection{Results}

\begin{figure}[!pt]
	\centering
	\includegraphics[width=\textwidth, height=\textheight, keepaspectratio]{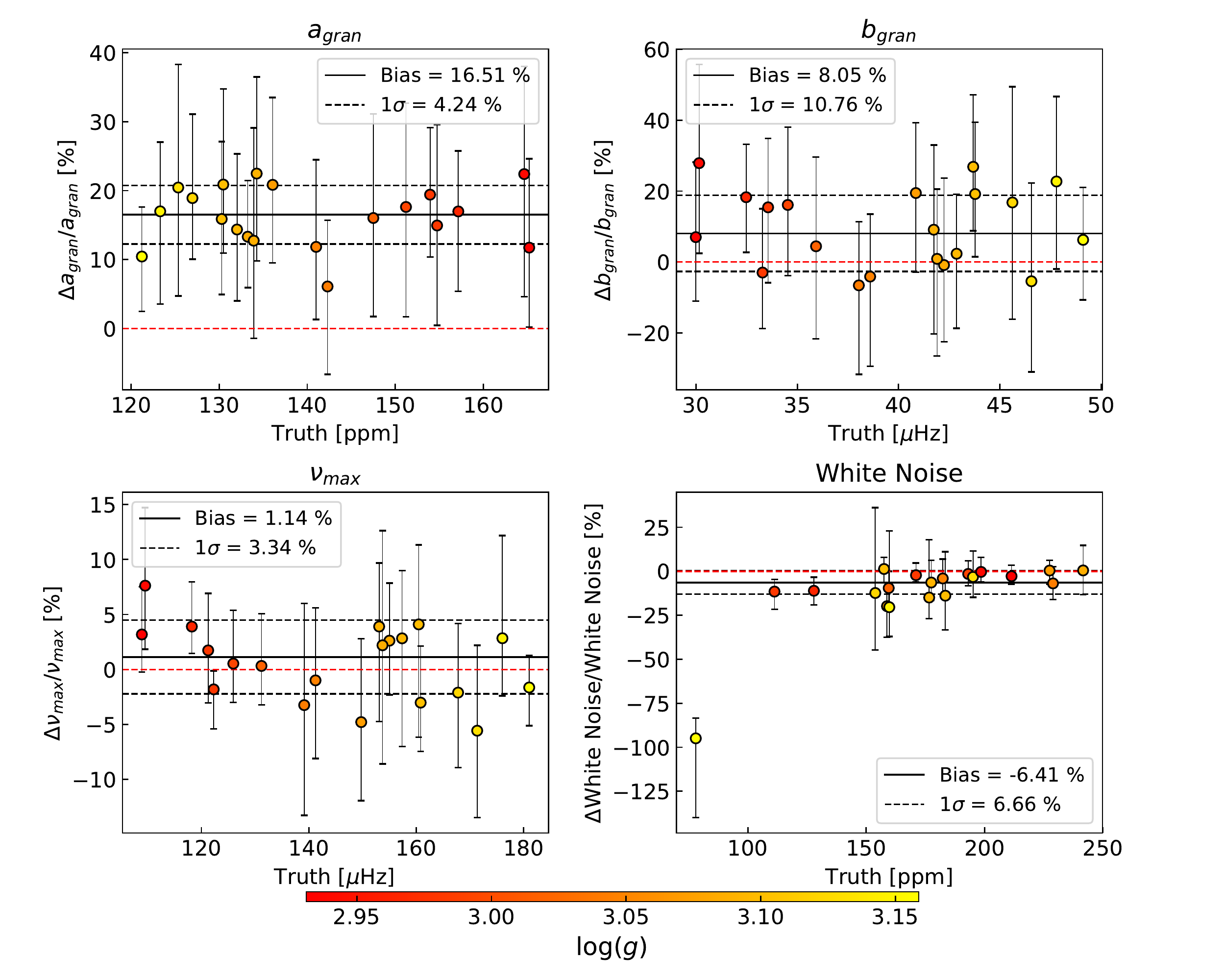}
	\caption[Comparison of the parameters in the fit of Model 1 to the \textit{TESS}-like artificial data with the input used to generate those data]
	{Comparison of the parameters in the fit of Model 1 to the \textit{TESS}-like artificial data with the input used to generate those data. 
	Data points represent the relative deviation with respect to the input value, with error bars corresponding to the uncertainties returned by the Bayesian sampling. 
	Black solid and dashed lines represent the median and standard deviation of the data points, respectively, with their numerical values shown in the inset. 
	The red dashed line denotes a null offset. 
	Data points are color-coded according to a star's surface gravity, $\log g$.}
	\label{fig:comparison_tess_model1}
\end{figure}

\begin{figure}[!pt]
	\centering
	\includegraphics[width=\textwidth, height=\textheight, keepaspectratio]{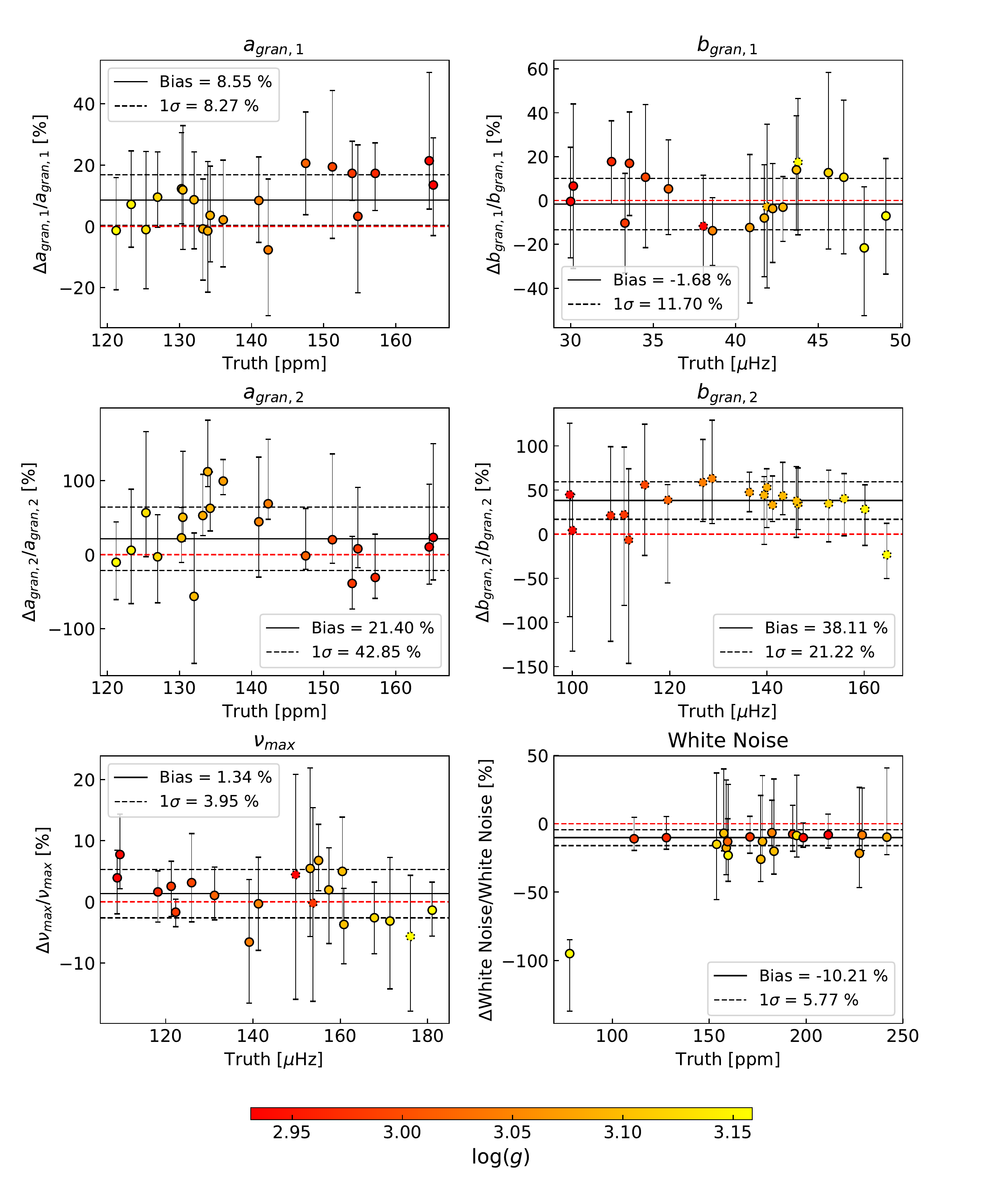}
	\caption[Comparison of the parameters in the fit of Model 2 to the \textit{TESS}-like artificial data with the input used to generate those data]
	{Comparison of the parameters in the fit of Model 2 to the \textit{TESS}-like artificial data with the input used to generate those data. 
	Data points represent the relative deviation with respect to the input value, with error bars corresponding to the uncertainties returned by the Bayesian sampling. 
	Parameters that have been flagged as not being constrained (see Section~\ref{sub:convergence}) have dotted edges. 
	Black solid and dashed lines represent the median and standard deviation of the data points, respectively, with their numerical values shown in the inset. 
	The red dashed line denotes a null offset. 
	Data points are color-coded according to a star's surface gravity, $\log g$.}
	\label{fig:comparison_tess_model2}
\end{figure}

We applied our method twice to all data sets, once for each of the models, i.e., Models 1 and 2. 
The parameters derived from the GP regression were then compared to the input parameters used when generating the light curves. 
Figures \ref{fig:comparison_tess_model1} and \ref{fig:comparison_tess_model2} show such a comparison for the parameters of Models 1 and 2, respectively. 

The parameters shown in Figure~\ref{fig:comparison_tess_model1} for Model 1 are the amplitude and characteristic frequency of the mesogranulation component in the model, $a_\text{gran,1}$ and $b_\text{gran,1}$, respectively, the frequency of maximum oscillation, $\nu_\text{max}$, and the white noise. 
Figure \ref{fig:comparison_tess_model2} depicts the same parameters with the addition of the amplitude and characteristic frequency of the granulation component, $a_\text{gran,2}$ and $b_\text{gran,2}$, respectively. 
Both the oscillator's quality factor, $Q$, and the power at $\nu\!=\!\nu_\text{max}$, $P_\text{g}$, are not shown, as the artificial light curves result from simulating individual oscillation modes and hence there are no \textit{bona fide} input values to compare with.

In what follows, a given parameter is considered to have been accurately determined if the null offset (red dashed line) is within the $1\sigma$ interval (black dashed lines) associated with the median of the data points (or bias; black solid line).
Looking at the results obtained when applying Model 1 (Figure~\ref{fig:comparison_tess_model1}), the amplitude of the mesogranulation component, $a_\text{gran,1}$, is not correctly retrieved, showing a bias of 16.51\% (4.24\% scatter) relative to the input values in the mesogranulation component in the data, whilst its characteristic frequency, $b_\text{gran,1}$, is correctly recovered with a bias of 8.05\% (10.76\% scatter). 
This is not too surprising, since the model being considered is incomplete: the mesogranulation amplitude is being overestimated in an attempt to capture the power in the two granulation components present in the data. 
It should be noted that the low-frequency tail of the oscillation envelope profile (see Section~\ref{sub:celerite_kernels} and Figure~\ref{fig:comparison_kernel_psd}) contributes to somewhat attenuating this offset. 
Nevertheless, the estimation of $\nu_\text{max}$ is robust, with this parameter being accurately (1.14\% bias) and precisely (3.34\% scatter) recovered. 
Finally, the white noise level is recovered to within 7\% of the input value. 
The slight, overall underestimation of the white noise level is to be expected because of the non-negligible contribution of the oscillation envelope profile at high frequencies (see Figure~\ref{fig:comparison_kernel_psd}). 
Concerning the outlying artificial star, it has the highest value of $\nu_\text{max}$ amongst the stars in the sample. 
Upon inspection of its PSD, it became clear that the proximity of the oscillations to the Nyquist frequency ($\nu_\text{Nyq} \sim 283$ $\mu$Hz for the 30-min cadence of the simulated light curves) prevents the white noise level from being robustly determined. 
For this reason, this star was not considered when determining the bias and scatter values shown for the white noise comparison.

Concerning Model 2 (Figure~\ref{fig:comparison_tess_model2}), the introduction of a second granulation component in the model leads to an improvement in the fit to the mesogranulation signal, with the amplitude and characteristic frequency within 8.55\% (8.27\% scatter) and $-$1.68\% (11.70\% scatter) of the input values, respectively. 
Results are, however, noticeably less robust for the added granulation component, with the correct amplitude of the granulation being within uncertainties only due to the high scatter (21.40\% bias and 42.85\% scatter) and the characteristic frequency not being constrained at all (38.11\% bias and 21.22\% scatter). 
Finally, the estimation of $\nu_\text{max}$ continues to be robust (1.34\% bias and 3.95\% scatter) whilst the correct white noise level could not be recovered ($-$10.61\% bias and 5.77\% scatter).

All in all, the introduction of the second granulation component does improve the fit to the mesogranulation but its characteristic frequency cannot be constrained for any star and both the characteristic frequency of the mesogranulation and $\nu_\text{max}$ show unconstrained results for some of the stars. 
The white noise is also not constrained when adding the extra component to the model. 
Considering the high white noise levels expected for \textit{TESS} data, as well as the short duration of typical \textit{TESS} time series, Model 1 with only the mesogranulation should be better suited to accurately find the stellar signals of RGB stars observed by \textit{TESS}.

\begin{figure}[!tbp]
	\centering
	\includegraphics[width=\textwidth, height=\textheight, keepaspectratio]{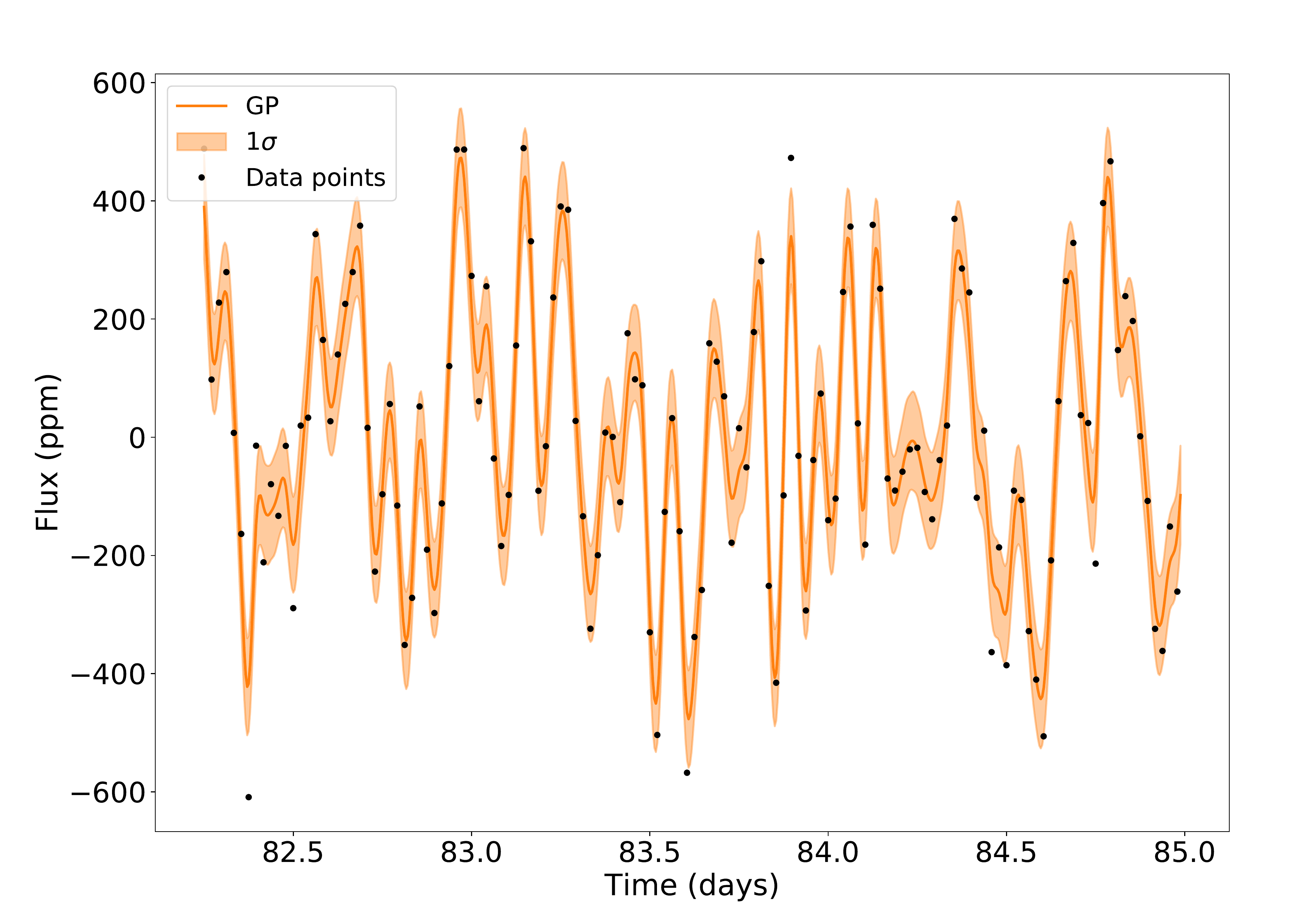}
	\caption[Predictive model (Model 1) output by the GP regression (mean and $1 \sigma$ interval) when applied to one of the artificial \textit{TESS}-like time series]
	{Predictive model (Model 1) output by the GP regression (mean and $1 \sigma$ interval) when applied to one of the artificial \textit{TESS}-like time series. 
	The plot is zoomed in on the first $\sim 3$ days of simulated data to improve visualization.}
	\label{fig:tess_gp_fit}
\end{figure}

\begin{figure}[!tbp]
	\centering
	\includegraphics[width=\textwidth, height=\textheight, keepaspectratio]{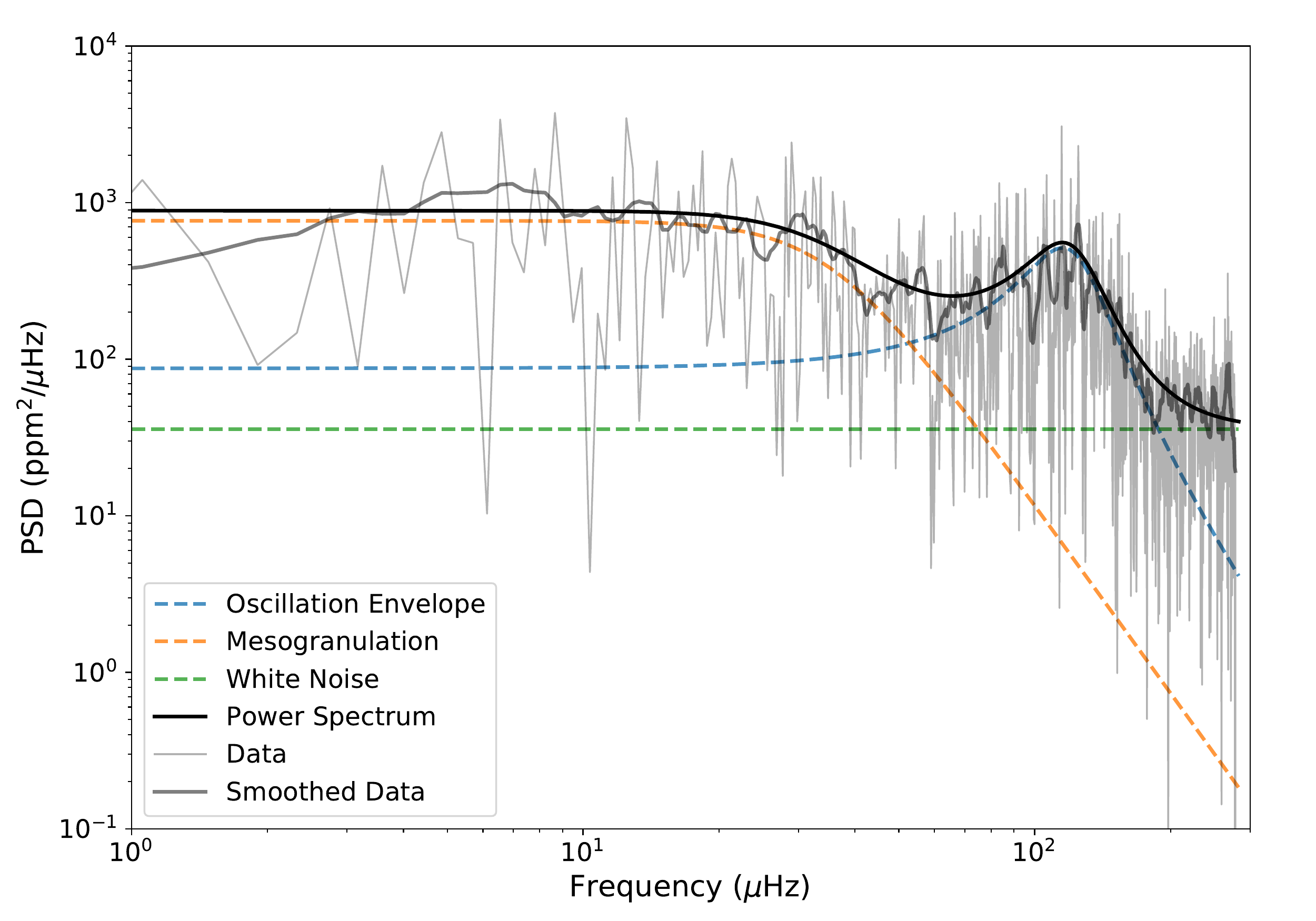}
	\caption[Power spectral density of the same (full) light curve depicted in Figure~\ref{fig:tess_gp_fit}]
	{Power spectral density of the same (full) light curve depicted in Figure~\ref{fig:tess_gp_fit}. 
	The PSD of the light curve is shown in light gray, with a slightly smoothed version overlapped in dark gray. 
	The PSD of the GP regression output (Model 1) is shown as a solid black curve, with individual components identified by different line styles and colors (see legend).}
	\label{fig:tess_psd_fit}
\end{figure}

As an illustrative example of the performance of our GP regression method in the characterization of the stellar signal in the time domain, Figure~\ref{fig:tess_gp_fit} shows a blowup of the fit performed to one of the artificial time series. 
Figure \ref{fig:tess_psd_fit} shows the PSD of that same GP regression output compared to the PSD of the light curve. 
Both figures show the results obtained when fitting Model 1 to the data.

\subsection{\textit{Kepler} light curves}
\label{sub:test_kepler}

For the application to \textit{Kepler} light curves, the objective was now to compare the characterization of the stellar signals in the time domain achieved by our GP regression method to a typical approach based on the modeling of the PSD of the time series in the frequency domain.
The sample of stars considered for the test was the same sample of 19 \textit{Kepler} LLRGB stars considered in \textcite{Corsaro_2015a}. 
In order to mimic the typical amount of data expected from \textit{TESS} and to account for systematics that might be present in the light curve, 10 non-overlapping subsets of 27.4 days of observations were extracted from the full \textit{Kepler} light curve of each star.

Once again, the values and corresponding uncertainties computed for each of the model parameters take into account the analysis of at least 5 and up to all 10 of these independent subsets, as was the case for the previous application. 


To compare against the GP regression in the time domain, a standard fit to the power spectrum was done using models with similar components to the ones included in our GP kernel (see Section~\ref{sub:test_models}).
The similarity concerns only the adopted model for the oscillations envelope, which in the PSD fitting method is characterized by a Gaussian envelope (Equation~\ref{eq:envelope_gaussian_psd}), since the remaining components have equivalent models in both the time and frequency-domain.
The PSD analysis was performed using the \texttt{\textsc{Diamonds}} code\footnote{\url{https://github.com/EnricoCorsaro/DIAMONDS}} \parencite{Corsaro_2014}, which fits the power spectrum and determines the model parameters within a Bayesian framework and has had peer-reviewed applications in the literature \parencite{Corsaro_2015a,Corsaro_2017}.


\subsubsection{Results}

\begin{figure}[!pt]
	\centering
	\includegraphics[width=\textwidth, height=\textheight, keepaspectratio]{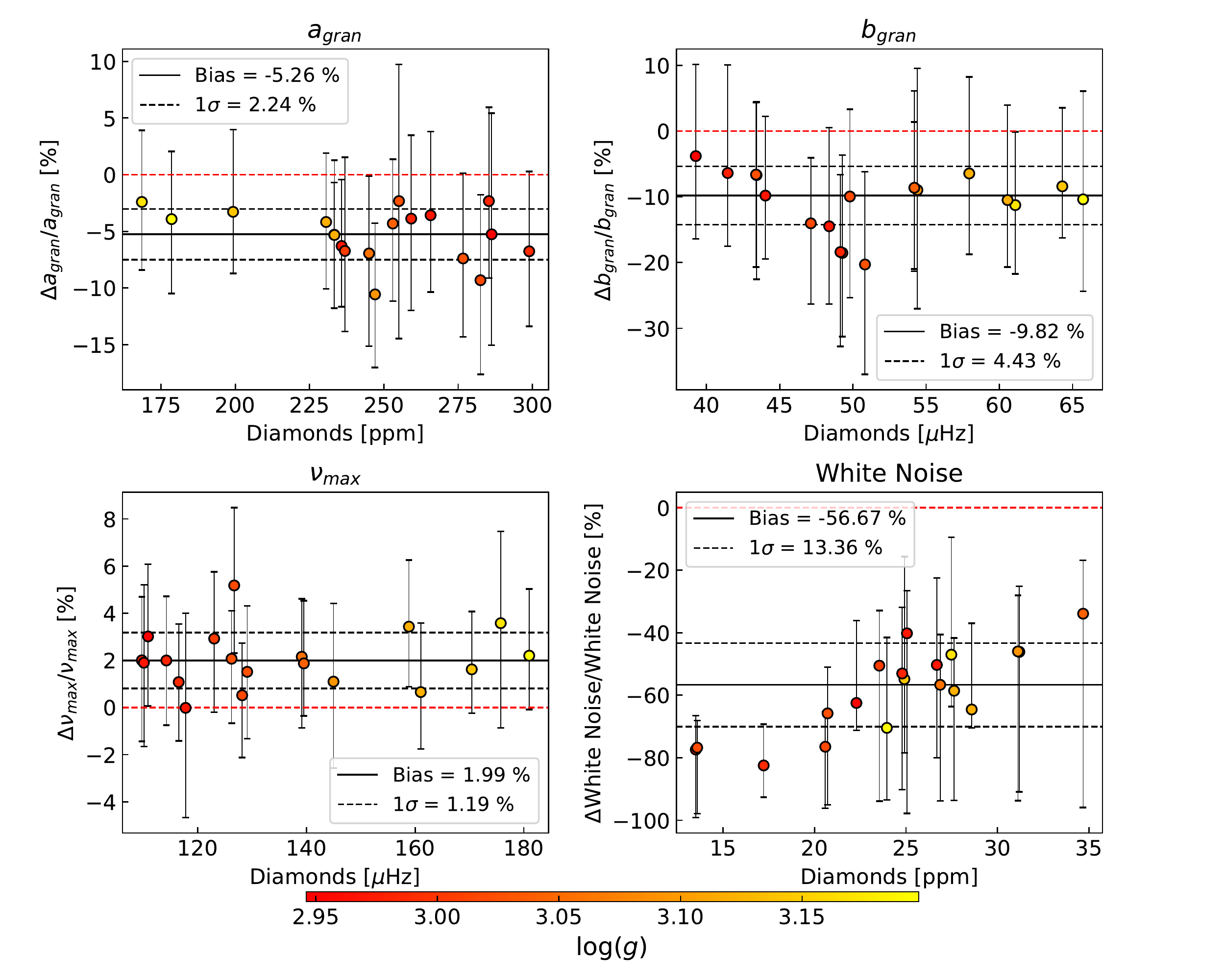}
	\caption[Comparison of the parameters in the fit of Model 1 to the \textit{Kepler} time series data both by means of a GP regression and power-spectrum fitting]
	{Comparison of the parameters in the fit of Model 1 to the \textit{Kepler} time series data both by means of a GP regression and power-spectrum fitting. 
	Data points represent the relative deviation with respect to the value determined using the PSD-fitting procedure, with error bars corresponding to the sum in quadrature of the uncertainties of both methods. 
	Black solid and dashed lines represent the median and standard deviation of the data points, respectively, with their numerical values shown in the inset. 
	The red dashed line denotes a null offset. 
	Data points are color-coded according to a star's surface gravity, $\log g$.}
	\label{fig:comparison_diamonds_model1}
\end{figure}

\begin{figure}[!pt]
	\centering
	\includegraphics[width=\textwidth, height=\textheight, keepaspectratio]{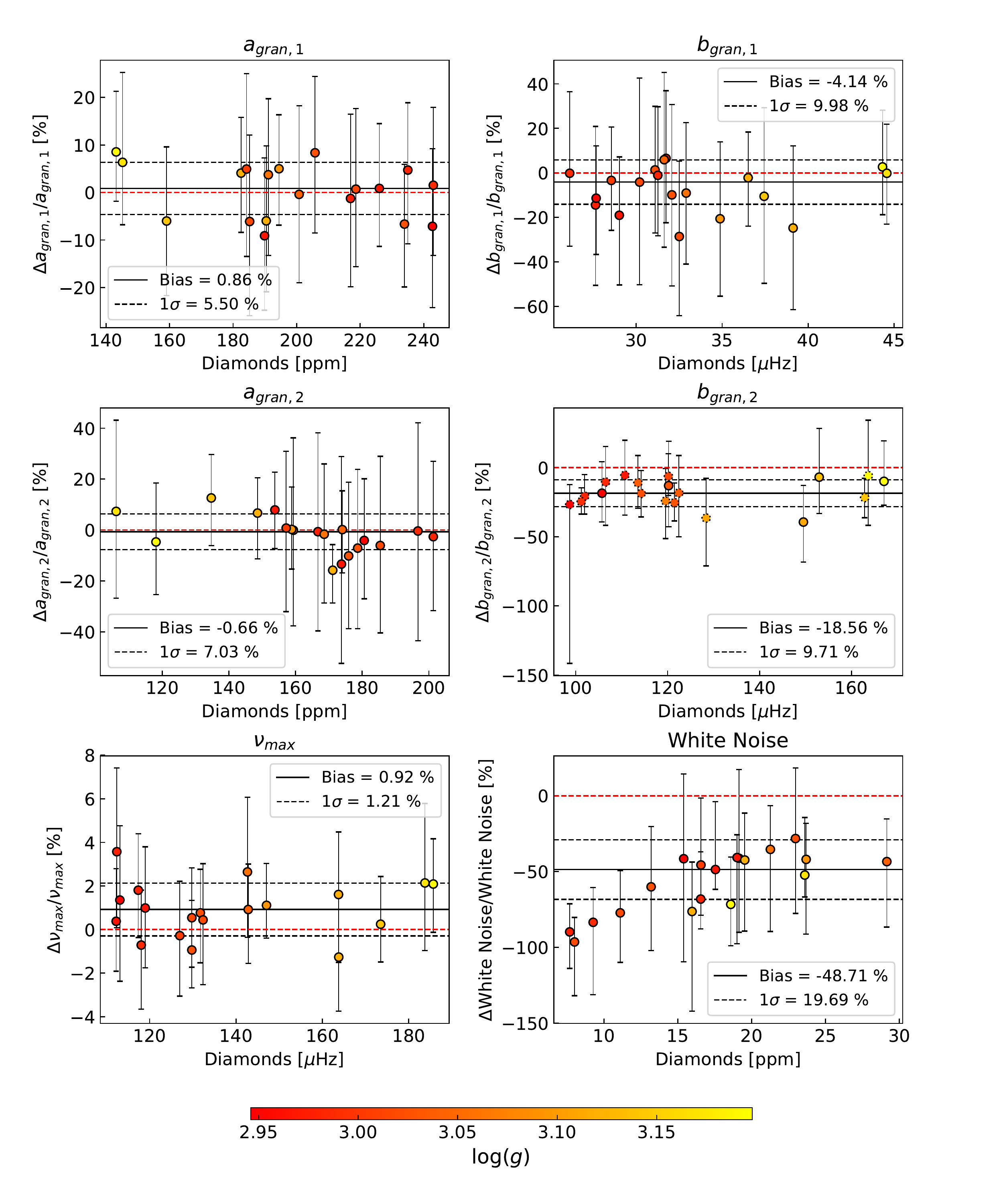}
	\caption[Comparison of the parameters in the fit of Model 2 to the \textit{Kepler} time series data both by means of a GP regression and power-spectrum fitting]
	{Comparison of the parameters in the fit of Model 2 to the \textit{Kepler} time series data both by means of a GP regression and power-spectrum fitting. 
	Data points represent the relative deviation with respect to the value determined using the PSD-fitting procedure, with error bars corresponding to the sum in quadrature of the uncertainties of both methods. 
	Parameters that have been flagged as not being constrained (see Section~\ref{sub:convergence}) have dotted edges. 
	Black solid and dashed lines represent the median and standard deviation of the data points, respectively, with their numerical values shown in the inset. 
	The red dashed line denotes a null offset. 
	Data points are color-coded according to a star's surface gravity, $\log g$.}
	\label{fig:comparison_diamonds_model2}
\end{figure}

We applied both our method and the PSD fitting method twice to all data sets, once for each of the models, i.e., Models 1 and 2.
A comparison between parameters obtained with the time domain GP and the power spectrum fit is shown in Figures~\ref{fig:comparison_diamonds_model1} and \ref{fig:comparison_diamonds_model2} for Models 1 and 2, respectively. 
The parameters depicted are the same as in Figures~\ref{fig:comparison_tess_model1} and \ref{fig:comparison_tess_model2}.

%

Contrary to the application to simulated light curves, the method is now tested on real \textit{Kepler} data for a sample of well-studied LLRGB stars. 
By doing this, the parameters derived through GP regression can be compared to the equivalent parameters obtained when performing a standard fit to the power spectrum.
This test thus allows for a comparison with the more traditional methodology used in asteroseismic studies of stellar light curves.

Looking at the results obtained when considering Model 1 (Figure~\ref{fig:comparison_diamonds_model1}), the parameters describing the mesogranulation component in the model, $a_\text{gran,1}$ and $b_\text{gran,1}$, are underestimated, with relative biases of about $-$5\% and $-$10\%, respectively. 
This underestimation is expected considering the low-frequency tail present in the oscillation envelope profile (see Figure~\ref{fig:comparison_kernel_psd}). 
To confirm that these offsets between parameters were only due to the differences in the models considered (specifically, the component that captures the signal from the oscillations), a second fit to the power spectrum was performed, where the components chosen followed the exact power spectrum equations of the kernels used in the GP model, including the approximated one.
Results from this test confirmed that, when the models are an exact match, both the GP regression and the frequency domain fit recover the same results.

Concerning $\nu_\text{max}$, a small relative bias of 1.99\% (1.19\% scatter) is found between methods. 
Finally, results for the white noise level show a relative bias of about $-$56\% (13.36\% scatter) between the two methods. 
However, it should be noted that white noise levels for the \textit{Kepler} stars in the sample are relatively low (\textit{Kepler}'s effective collecting area is larger than that of the individual \textit{TESS} cameras by a factor of $\sim\!10^2$), which, coupled with differences in the models adopted between methods (see Figure~\ref{fig:comparison_kernel_psd}), results in large relative differences. 
Inspection of the absolute value of this same bias reveals differences no greater than 15 ppm and similar between stars, which results in the trend seen in the relative offsets, where stars with higher white noise (as determined by \texttt{\textsc{Diamonds}}) have a lower relative offset.

With respect to Model 2 (Figure~\ref{fig:comparison_diamonds_model2}), excellent agreement is seen between the parameters describing the mesogranulation component, $a_\text{gran,1}$ and $b_\text{gran,1}$, with relative differences of 0.86\% and $-$4.14\%, respectively. 
The low-frequency tail of the oscillation envelope profile does not appear to be affecting this component. 
It does, however, impact the parameters describing the higher frequency granulation component. 
Whilst its amplitude, $a_\text{gran,2}$, is accurately recovered with a bias of $-$0.66\% (7.03\% scatter), the characteristic frequency is systematically shifted, showing a bias of $-$18.56\% (9.71\% scatter), with more than half the stars having parameters that have not been well constrained (dotted edges). 
Regarding $\nu_\text{max}$, just like with Model 1, a small relative bias of 0.92\% (1.21\% scatter) is found between methods. 
For the white noise level, results show a less pronounced discrepancy compared to the one seen for Model 1, with the relative bias between methods now being of $-$48.71\% (16.89\% scatter) and the absolute differences never exceeding 6 ppm.

\begin{figure}[!tbp]
	\centering
	\includegraphics[width=\textwidth, height=\textheight, keepaspectratio]{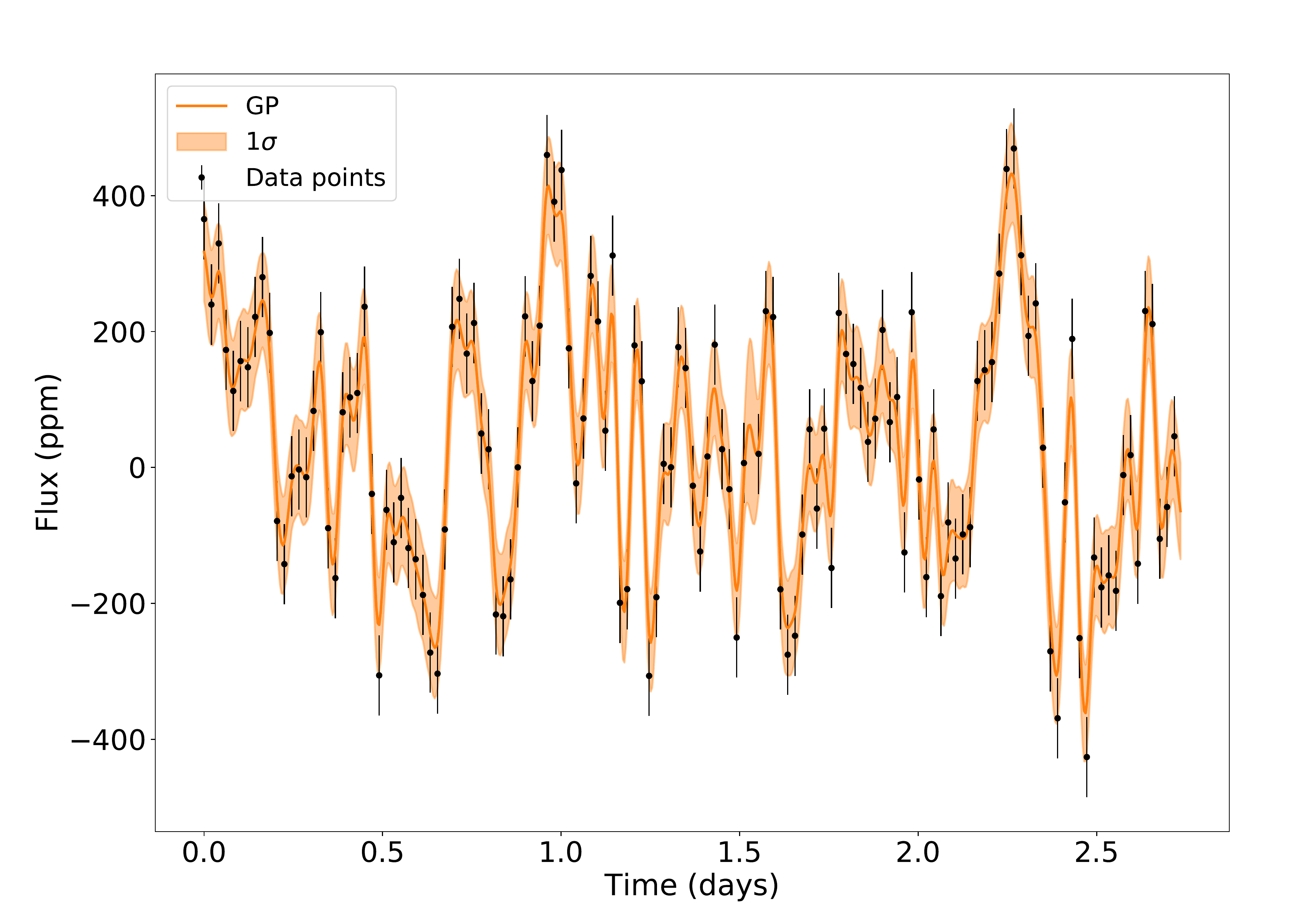}
	\caption[Predictive model (Model 2) output by the GP regression (mean and $1\sigma$ interval) when applied to one of the \textit{Kepler} LLRGB stars in the sample]
	{Predictive model (Model 2) output by the GP regression (mean and $1\sigma$ interval) when applied to one of the \textit{Kepler} LLRGB stars in the sample. 
	The plot is zoomed in on the first $\sim3$ days of observations to improve visualization.}
	\label{fig:kepler_gp_fit}
\end{figure}
\begin{figure}[!tbp]
	\centering
	\includegraphics[width=\textwidth, height=\textheight, keepaspectratio]{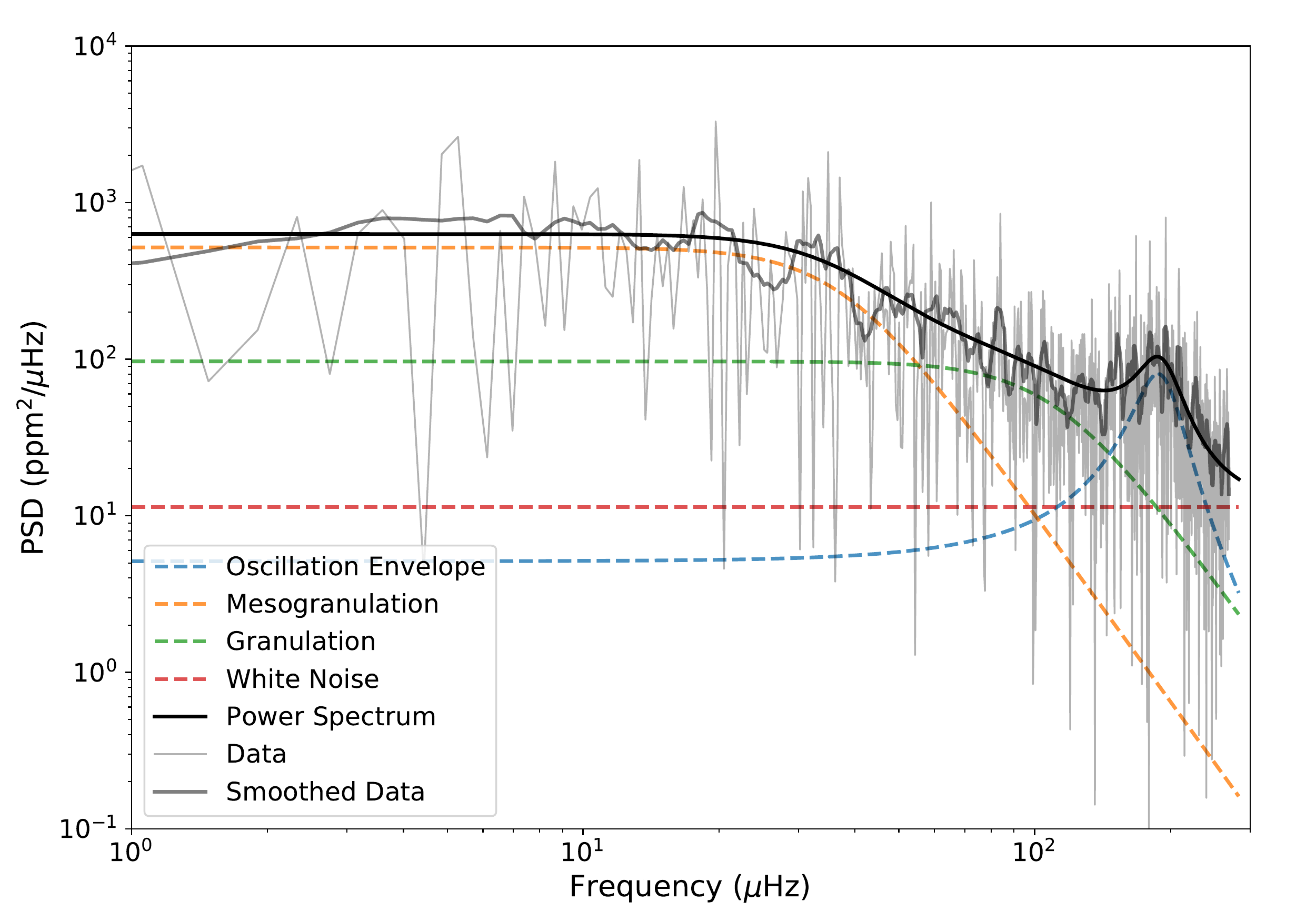}
	\caption[Power spectral density of the same (full) light curve depicted in Figure~\ref{fig:kepler_gp_fit}]
	{Power spectral density of the same (full) light curve depicted in Figure~\ref{fig:kepler_gp_fit}. 
	The PSD of the light curve is shown in light gray, with a slightly smoothed version overlapped in dark gray. 
	The PSD of the GP regression output (Model 2) is shown as a solid black curve, with individual components identified by different line styles and colors (see legend).}
	\label{fig:kepler_psd_fit}
\end{figure}

To illustrate the result from the GP regression, a blowup of the output of the GP regression when applied to one of the \textit{Kepler} LLRGB stars in the sample is shown in Figure~\ref{fig:kepler_gp_fit}. 
Figure \ref{fig:kepler_psd_fit} shows the PSD of that same GP regression result compared to the PSD of the light curve. 
Both figures show the results obtained when fitting Model 2 to the data.

\subsubsection{Uncertainties in $\nu_\text{max}$}

We now take a closer look at the uncertainties determined in the estimation of $\nu_\text{max}$ by both methods. 
Taking into account results from asteroseismology of red giants from the first four months of \textit{Kepler} data, \textcite{Kallinger_2010a} defined the following relation to determine $\sigma_{\nu_\text{max}}$, a lower limit for the uncertainty in $\nu_\text{max}$:
\begin{equation}
	\sigma_{\nu_\text{max}} = \nu_\text{res} \left( 1 + \frac{4}{(HBR/\sigma_\text{g})^{2/3}} \right),
	\label{eq:sigma_numax_kallinger}
\end{equation}
where \textit{HBR} is the height-to-background ratio, defined as the ratio between the power of the oscillations bump, $P_\text{g}$, and the background signal at $\nu\!=\!\nu_\text{max}$, $B_{\nu_\text{max}}$. 
$\sigma_\text{g}$ is the width of the oscillations bump and $\nu_\text{res}$ is the frequency resolution, which is the inverse of the dataset length.

Estimates of $P_\text{g}$, $B_{\nu_\text{max}}$ and $\sigma_\text{g}$ can be obtained from the scaling relations found in \textcite{Mosser_2012b},
\begin{alignat}{2}
	& P_\text{g} = 2.03 \times 10^7 \times \nu_\text{max}^{-2.38}, \\[5pt]
	& B_{\nu_\text{max}} = 6.37 \times 10^6 \times \nu_\text{max}^{-2.41},
\end{alignat}
and in \textcite{Campante_2016a},
\begin{alignat}{2}
	& \text{FWHM}_\text{g} = \frac{\nu_\text{max}}{2}, \\[5pt]
	& \sigma_\text{g} = \frac{\text{FWHM}_\text{g}}{2 \sqrt{2 \ \text{ln}(2)}},
\end{alignat}
where $\text{FWHM}_\text{g}$ is the full width at half maximum of the oscillations bump.

Using the values of $\nu_\text{max}$ obtained by the GP method, Equation~\eqref{eq:sigma_numax_kallinger} was used to determine an estimate of $\sigma_{\nu_\text{max}}$, which was then compared to the uncertainties in the determination of $\nu_\text{max}$ calculated by both the GP method, $\sigma_\text{GP}$, and the frequency-domain method using \texttt{\textsc{Diamonds}}, $\sigma_\texttt{\textsc{Diamonds}}$.
This comparison was only done for the results considering Model 1, with the average absolute and relative values obtained for these quantities, considering all 19 stars in the sample, being:
\begin{itemize}
	\item $\left \langle \sigma_{\nu_\text{max}} \right \rangle \!=\! 7.15 \ \mu$Hz, $(5.20\%)$
	\item $\left \langle \sigma_\text{GP} \right \rangle \!=\! 3.43 \ \mu$Hz, $(2.46\%)$
	\item $\left \langle \sigma_\texttt{\textsc{Diamonds}} \right \rangle \!=\! 1.91 \ \mu$Hz, $(1.44\%)$
\end{itemize}
The values above would suggest that both methods applied are underestimating by a factor of 2 the uncertainties in $\nu_\text{max}$. 
However, since the relation from Equation~(\ref{eq:sigma_numax_kallinger}) was determined based on the uncertainties estimated by \textcite{Kallinger_2010a}, following the methodology described in \textcite{Gruberbauer_2009}, the difference between the expected uncertainty, $\sigma_{\nu_\text{max}}$, and the calculated uncertainties, $\sigma_\text{GP}$ and $\sigma_\textsc{Diamonds}$, might just be related to differences in the methods adopted.
In any case, the uncertainty determination in the GP method, described in Section~\ref{sub:convergence}, estimated similar, albeit slightly higher uncertainties, than the ones obtained with \texttt{\textsc{Diamonds}}, providing some confidence in their conservative values (see Section~4.5 in \textcite{Corsaro_2014} for a description of uncertainty estimation with \texttt{\textsc{Diamonds}}).

\subsection{Adding transits}
\label{sub:test_transits}

Having confirmed the applicability of our method to the characterization of stellar signals in the time domain, our next test was to determine the impact of including the stellar model in the characterization of the planetary transit.

For our transit model, we considered the quadratic limb-darkened transit model from \textcite{Mandel_2002}, described in Section~\ref{sub:transits}, as is implemented in our method (see Section~\ref{sub:gptransits}).
Since the transit model is parametric, we can directly set it as the mean of our Gaussian process model for the stellar signals (which assumed a mean of zero when only stellar signals were characterized). 
Considering our objective is to model \textit{TESS} light curves, we select Model 1 as our model for the stellar signals (Equation~\ref{eq:model_1_kernel}), following the conclusions from Section~\ref{sub:test_tess_lcs}. 
This new combined model should be capable of capturing both the stellar and the transit signals simultaneously when modeling the light curve of a star. 
For comparison, we defined a simpler transit model including only the same parametric transit component and an additional Gaussian uncorrelated white noise component. 
This simpler model followed a standard configuration found often in planetary transit characterization, so as to evaluate the potential improvements in the determined transit parameters when including the stellar signals simultaneously with the transit signal.

For the test, simulated transits of close-in giant planets (orbital periods from 0.5 to 27.4 days and radii from $\sim$0.4 to 2 $R_\text{J}$) were injected into the sample of simulated light curves from Section~\ref{sub:test_tess_lcs}, as described in section~3.1 of \textcite{Campante_2018}. 
The detection of the injected transits was evaluated using the Box-fitting Least Squares method \parencite[BLS;][]{Kovacs_2002}, with all detections with a signal detection efficiency (SDE) above 5.98 (a threshold determined empirically for 1-sector light curves) most likely not being statistical false positives (see Section~\ref{sec:transit_search} for an extensive description of this methodology for transit searches). 
For all likely detections, the light curves were characterized using both models and the recovered transit parameters compared with the true values used for the injected transits.

\subsubsection{Results}

\begin{figure}[!ptb]
	\centering
	\includegraphics[width=\textwidth, height=\textheight, keepaspectratio]{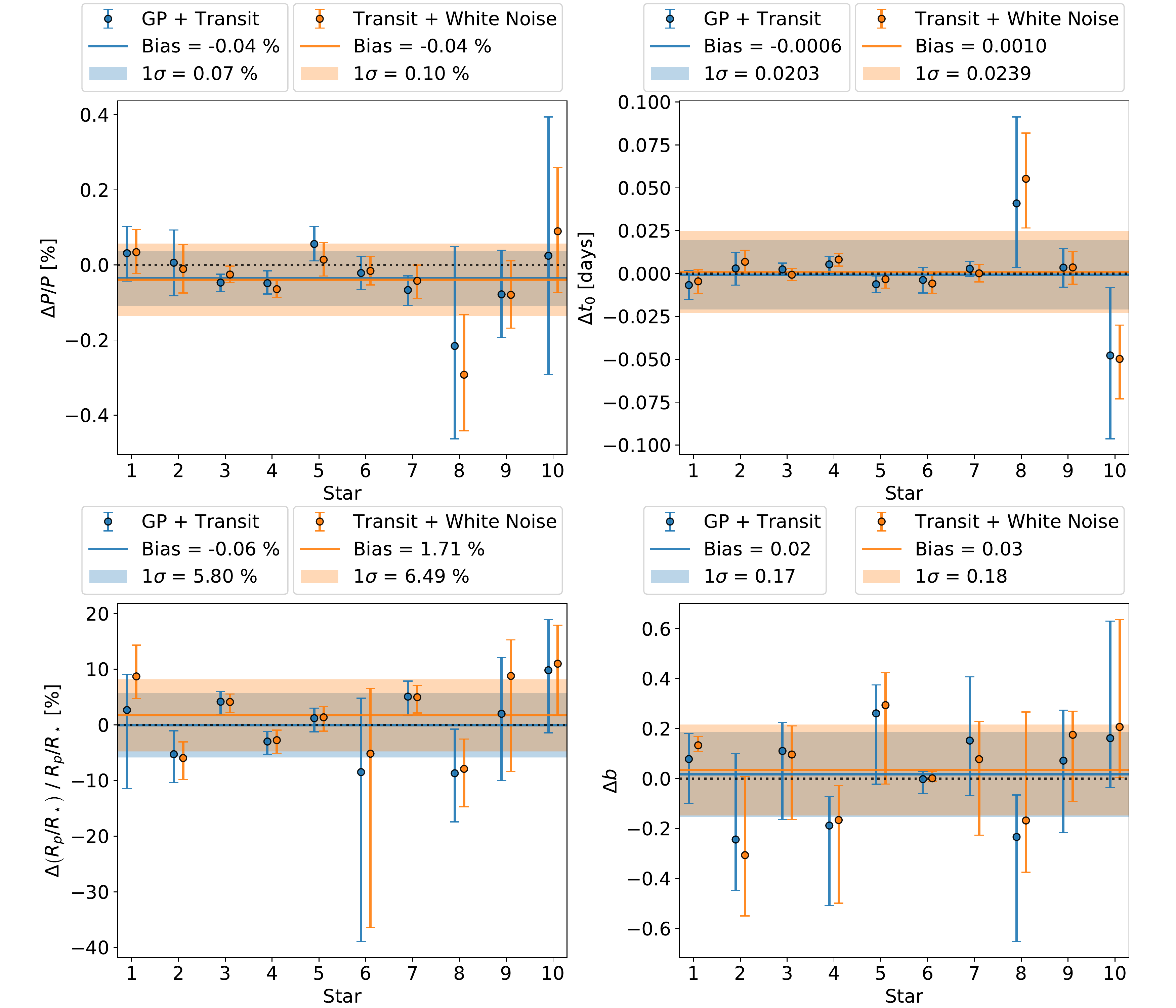}
	\caption[Comparison of the transit parameters recovered by both transit models considered with the input parameters used in the simulated transits injected in the \textit{TESS}-like artificial light curves]
	{Comparison of the transit parameters recovered by both transit models considered with the input parameters used in the simulated transits injected in the \textit{TESS}-like artificial light curves.
	The blue points correspond to our complete light curve model, with transit and stellar signals characterized simultaneously, whilst the orange points correspond to the simpler transit model, including only the transit model and gaussian noise.
	Data points represent the relative deviation with respect to the input value, with error bars corresponding to the uncertainties returned by the Bayesian sampling.
	Solid colored lines and shaded regions represent the median and standard deviation of the data points for each of the methods, respectively, with their values shown in the legend above.
	The black dotted line denotes the null offset. 
	Stars in the x-axis are ordered by decreasing SDE (ranging from 6.47 to 4.72).}
	\label{fig:transit_parameter_comparison}
\end{figure}

Figure~\ref{fig:transit_parameter_comparison} shows a comparison of the values for the transit period, $P$, transit epoch, $t_0$, ratio of radii, $R_\text{p}/R_\star$ and impact parameter, $b$, recovered using both our model with GPs (in blue) and with a simpler transit model (in orange), with the input values used to simulate the transits injected in the \textit{TESS} artifical light curves from Section~\ref{sub:test_tess_lcs}.
Only the impact parameter is shown in the place of the semi-major axis over stellar radius, $a/R_\star$, eccentricity, $e$, inclination, $i$ and argument of periapsis, $\omega$, since all these parameters are degenerate when only photometric data is available.
The quadratic limb-darkening coefficients are also omitted given the interest in exploring the recovery of the transit parameters and not stellar properties.

Following the same methodology as before, we have that, for both the period and epoch, both models recover essentially the same results, with equally small, negligible biases. 
For the period, the relative bias is -0.04\% for both models, whilst for the epoch our model obtains a bias of -6$\times$10$^{-4}$ (10$^{-3}$ for the simpler model).
Our GP model does improve slightly on the scatter, with the comparison in terms of periods having values of 0.07\% and 0.10\% for our model and the simpler model, respectively, and the epoch comparison having values of 0.0203 and 0.0239 for our model and the simpler one, respectively.
As for the impact parameter, the value is less relevant, since it represents a variety of other parameters jointly.
Nevertheless, our model obtains a bias of 0.02 (0.17 scatter) compared to a bias of 0.03 (0.18 scatter) for the simpler model.
Finally, we have the comparison of the recovered ratio of radii between models.
Here, our model not only recovers a noticeably more accurate value, with a bias of -0.06\%, compared to a bias of 1.71\% for the simpler one, but the scatter is also reduced, with our model obtaining 5.80\% compared to the simpler model's 6.49\%.
This preliminary result suggests that our new combined model is capable of recovering both more precise and accurate planetary radii, assuming a similar measurement of the stellar radius.

\begin{figure}[!ptb]
    \centering
    \includegraphics[width=0.91\textwidth, height=\textheight, keepaspectratio]{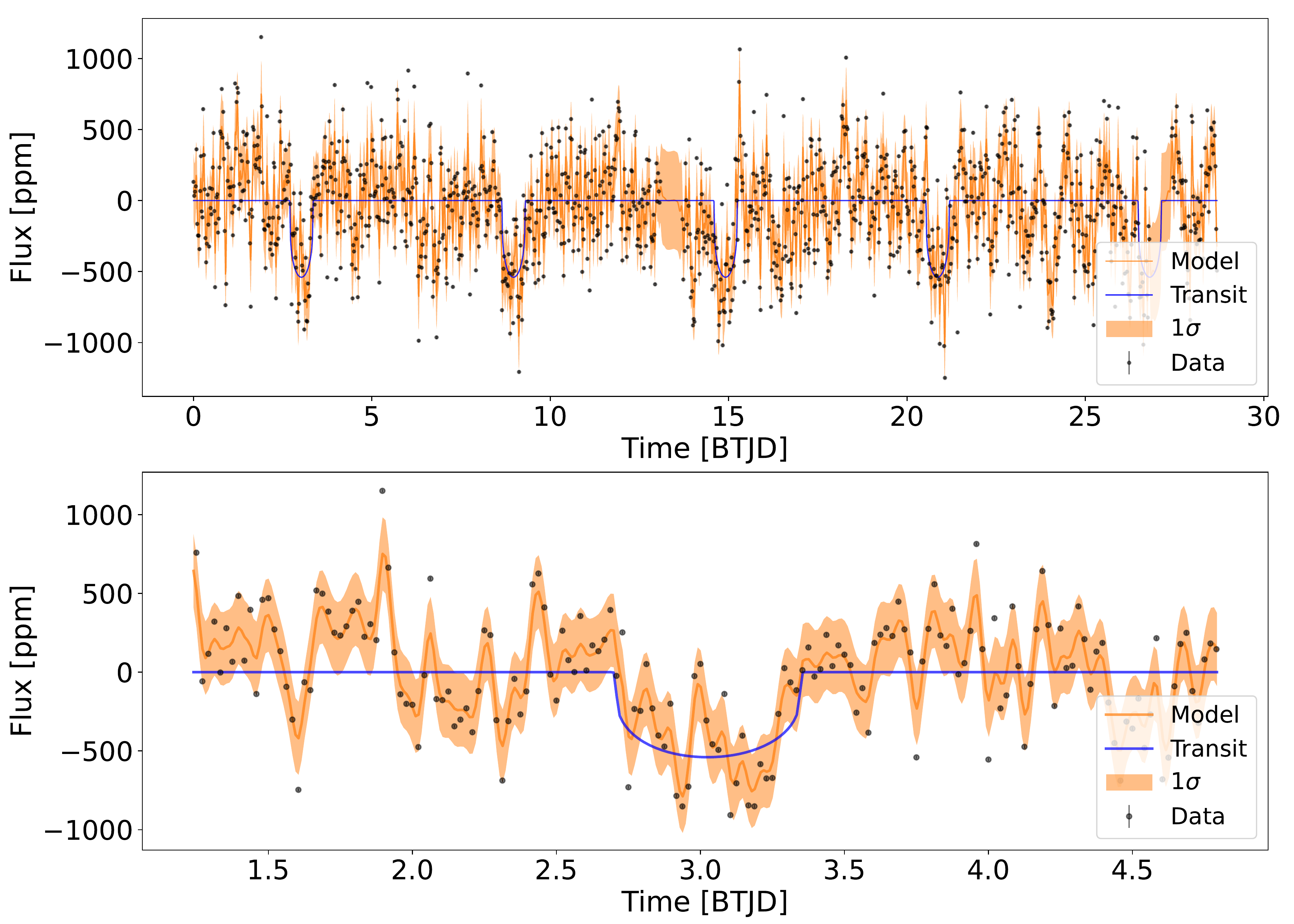}
    \includegraphics[width=0.91\textwidth, height=\textheight, keepaspectratio]{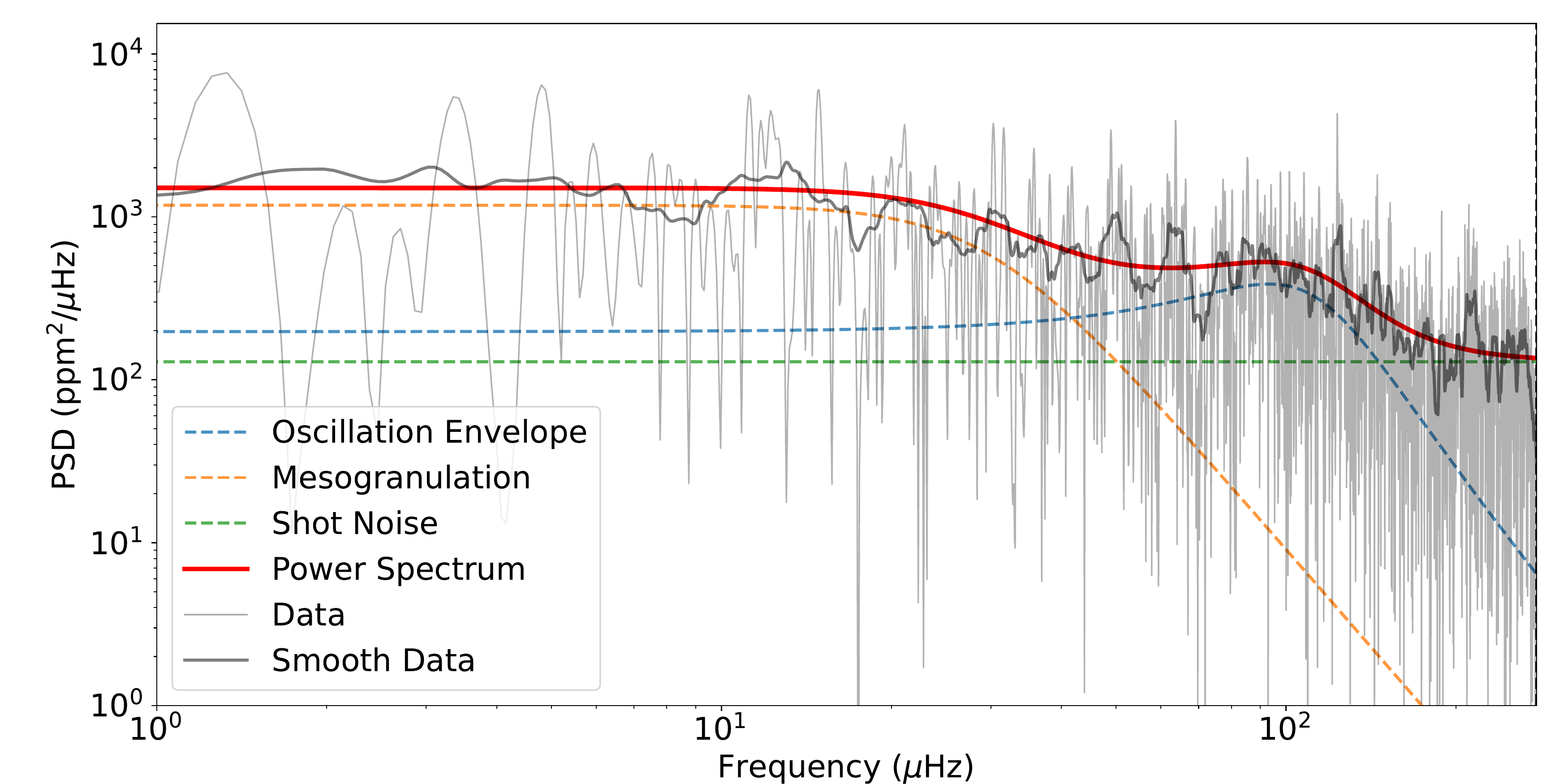}
    \caption[Light curve fit and corresponding power spectrum from the characterization of a \textit{TESS} simulated light curve with an injected transit.]
    {\textit{Top}: Fit of a \textit{TESS} simulated light curve with an injected transit.
    Black points and error bars correspond to the light curve data.
    The GP predictive model is shown in orange, with the solid central line denoting the median of the distribution and the shaded area the 1-$\sigma$ interval.
    The transit component of the model is depicted by the blue solid line.
    The panel below shows a zoom in of the first transit of the light curve.
	\textit{Bottom}: Power spectrum of the light curve in the panels above after removing the transit signal.
    The PSD of the light curve is shown in light gray, with a slightly smoothed version overlapped in dark gray.
    The PSD of the sum of the GP components in the model is shown as a solid red curve, with individual components identified by different colors (see legend).}
    \label{fig:simulated_lc_fit}
\end{figure}

Figure~\ref{fig:simulated_lc_fit} shows an example of the fit achieved by our model to one of the simulated light curves with an injected transit.
In the upper and middle panels, the complete model is shown, in orange, as well as the transit component separately (in blue), over the light curve data.
The middle panel shows a zoom in of the first transit event.
The bottom panel shows the PSD of the GP model, and its individual components, overplotted on the power spectrum of the same light curve, with the contributions from the transit removed.

%% file: chapters/pipeline/main.tex
\chapter{\textit{TESS} FFI transit search pipeline}
\label{cha:search_pipeline}

\textit{TESS} delivers two different data products from its observations of the sky, produced from the 2-second full-frame images (FFI) obtained by each camera.
For a select number of targets, pixel cutouts from around the target's position are extracted from each image, which are then stacked and summed into 2-minute cadence postage stamps.
For all other targets, their data have to be extracted from the FFIs, which are stacked and summed to 30-min cadence.

Of all \textit{TESS} data, only postage stamps have their data processed and delivered to the community directly by NASA's Science Payload Operations Center (SPOC) pipeline \parencite{Jenkins_2016}, which provides corrected light curves as well as other intermediate data products.
In this work, I wanted to observe evolved stars, the majority of which were not included in the sample of fast-cadence targets from \textit{TESS}, given the mission's main objective of finding planets orbiting nearby bright dwarf stars.
For the former stars, their data has to be processed from the 30-min cadence FFIs, which requires the creation of a data processing pipeline to extract and correct light curves from the FFIs.

To that end, in this section I describe my own pipeline, composed mostly of existing open-source software, adapted according to my use case and implemented to work well together and to flow data automatically between each other. 
I also implement some additional data processing methods when necessary.

The pipeline is divided in four components:
\begin{itemize}[noitemsep, topsep=1ex]
    \item Aperture photometry (Section~\ref{sec:photometry})
    \item Systematics correction (Section~\ref{sec:corrections})
    \item Transit search (Section~\ref{sec:transit_search})
    \item Transit validation (Section~\ref{sec:transit_validation})
\end{itemize}

\input{chapters/pipeline/photometry}
\input{chapters/pipeline/corrections}
\input{chapters/pipeline/transit_search}
\input{chapters/pipeline/validation}

%% file: chapters/pipeline/photometry.tex
\section{Aperture photometry}
\label{sec:photometry}

For the purpose of light curve extraction and correction from the \textit{TESS} FFIs, I used the open-source \texttt{Python} package \texttt{eleanor}\footnote{\href{https://github.com/afeinstein20/eleanor}{https://github.com/afeinstein20/eleanor}}, described in \textcite{Feinstein_2019}.

\texttt{eleanor} is a tool that was originally designed specifically to extract target pixel files from \textit{TESS} FFIs and produce systematics-corrected light curves for any star observed by \textit{TESS}.
By providing a target's TIC ID or coordinates (right ascension and declination), the tool first identifies whether the target is visible by any of \textit{TESS}'s cameras during the multiple sectors of observations. 
For each sector where the target is observed, the tool determines the pixel position of the target on the FFIs.
Then, it extracts a target pixel file (TPF) from the FFIs, i.e. a collection of pixel cutouts centered on the calculated position of the target.
One cutout is produced for each exposure, and its size can be defined by the user (in pixels, with a default of 13 $\times$ 13).

From the TPF, the tool then extracts a light curve using aperture photometry.
Aperture photometry first requires the selection of an aperture mask, essentially a mask to define the pixels, and their weights, from which the stellar flux will be measured.
Here, \texttt{eleanor} does not choose a specific aperture (though this can be overridden by the user) and instead measures the photometry using a library of different apertures (see figure~3 of \textcite{Feinstein_2019}). 
The final aperture is only chosen at the end of the light curve extraction process by evaluating a metric, called the Combined Differential Photometric Precision (CDPP), which was defined for the \textit{Kepler} mission and attempts to quantify the ease with which terrestrial transit signatures can be detected \parencite{Christiansen_2012}.

For each trial aperture, \texttt{eleanor} extracts a light curve by summing up the flux contributions of the star from the aperture mask pixels and then estimates and removes the background signal from the remaining pixels (not included in the aperture).
This background-corrected light curve, referred to as the ``raw'' light curve by \texttt{eleanor}, is then corrected for systematics by removing linear models fitted to the \textit{x} and \textit{y} pixel positions and measured background, with time, which, in essence, removes signals correlated with any of these quantities.
It is the light curve resulting from this linear-model correction, referred to as the ``corrected'' light curve by \texttt{eleanor}, that is used to determine the CDPP, so that the aperture and consequent light curve that maximize the metric's value can be selected.

\texttt{eleanor} also provides an additional method to further correct the light curve.
This method does principal component analysis (PCA) correction by utilizing the community cotrending basis vectors (CBVs), calculated and shared by the SPOC pipeline \parencite{Jenkins_2016}.
These CBVs are essentially light curves generated by extracting the most common systematics trends across all stars of each camera and during each sector of \textit{TESS}. 
\texttt{eleanor}'s PCA correction method subtracts the three most impactful CBVs for a given target from the light curve, a similar procedure to the methodology the SPOC pipeline uses for short-cadence \textit{TESS} targets.
The resulting light curve from this additional correction method is returned to the user alongside the already mentioned ones, and is referred to as the ``pca'' light curve.

\begin{figure}[!ptb]
    \centering
    \includegraphics[width=\textwidth, height=\textheight, keepaspectratio]{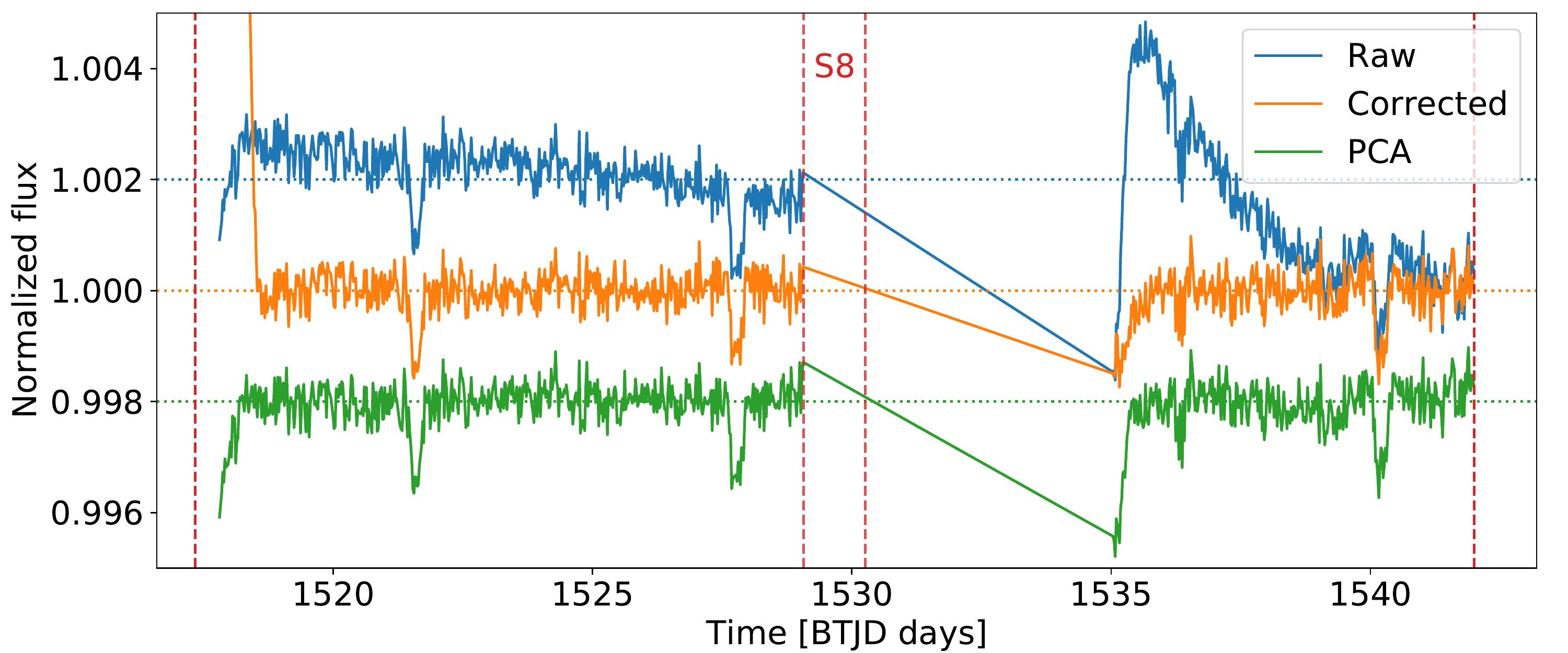}
    \caption[Example of the three light curves returned by \texttt{eleanor}, extracted from the \textit{TESS} FFIs.]
    {Example of the three light curves returned by \texttt{eleanor}, extracted from the \textit{TESS} FFIs. 
    The ``raw'' light curve contains the extracted flux from the target pixel files with only background correction. 
    The ``corrected'' and ``pca'' light curves include additional corrections, with the former including some linear-model corrections to remove long-term trends and the latter including corrections using the community cotrending basis vectors (CBVs) provided by NASA (see text for details).
    The vertical red dashed lines denote the start and end of the two \textit{TESS} orbits for the target's sector (in this case Sector 8, as shown in red text between the two central lines).
    The target is TIC 348835438, a red-giant star with a $V$-band magnitude of 10.9.}
    \label{fig:eleanor_light_curves}
\end{figure}

Figure~\ref{fig:eleanor_light_curves} depicts an example of the three light curves returned by \texttt{eleanor}, highlighting the removal of long-term trends from the ``raw'' light curve in both the ``corrected'' and the ``pca'' light curves.

%% file: chapters/pipeline/corrections.tex
\section{Systematics correction}
\label{sec:corrections}

Light curves returned by \texttt{eleanor}, albeit classified as corrected, are still often not adequate for transit search. 
In this section, I describe the data processing and additional corrections carried out to prepare the light curves.
First, \texttt{eleanor} returns one light curve per sector where the target is observed, so an initial step is to normalize these individual sector light curves so that all sectors have the same median level of the flux.

\begin{figure}[!ptb]
    \centering
    \includegraphics[width=\textwidth, height=\textheight, keepaspectratio]{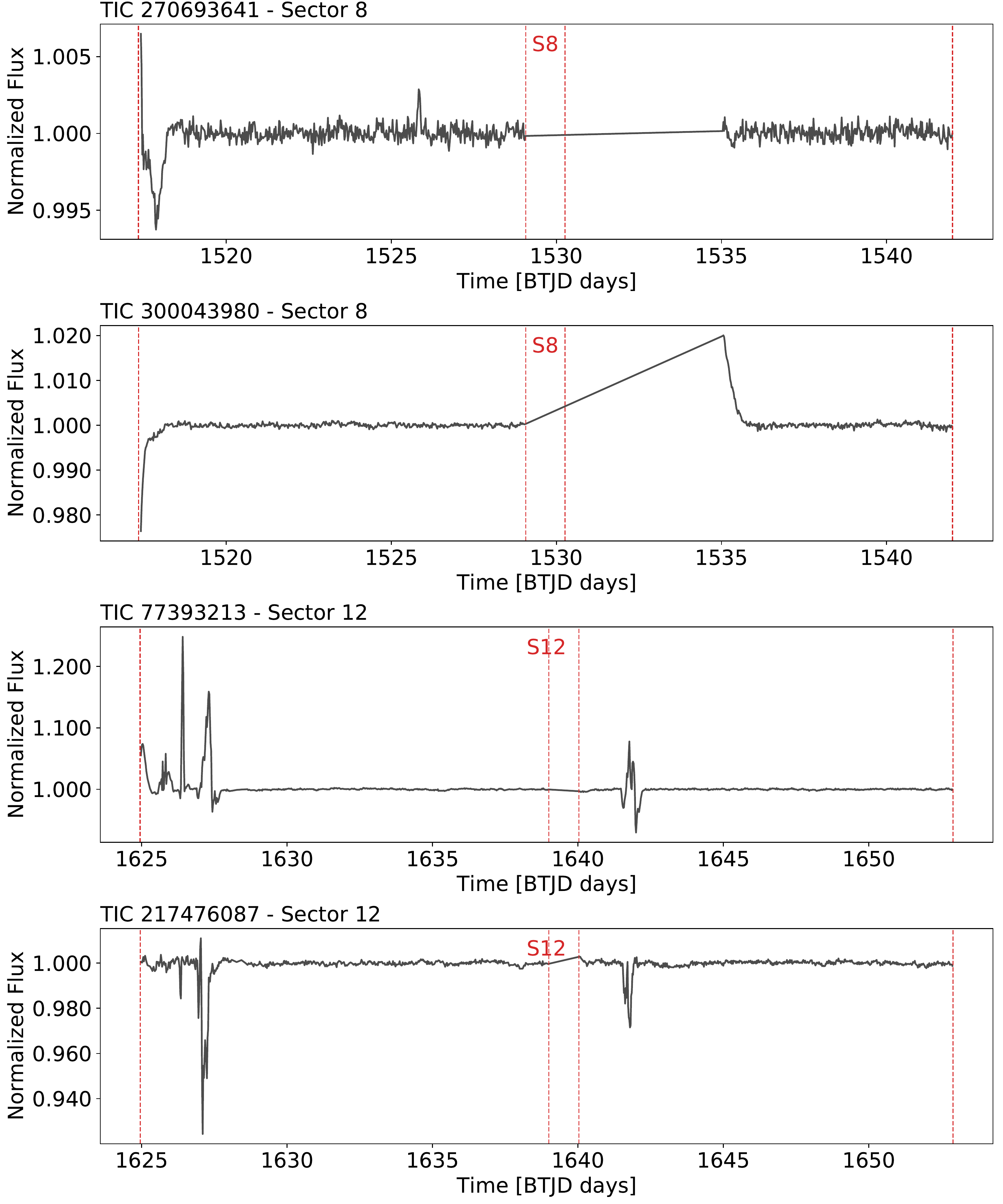}
    \caption[\textit{TESS} light curves highlighting the noisy patterns still present after extraction and correction using \texttt{eleanor}]
    {\textit{TESS} light curves highlighting the noisy patterns still present after extraction and correction using \texttt{eleanor}.
    The TIC ID of each target and the sector from which the light curve was extracted are shown in the top left of each panel.
    The vertical red dashed lines denote the start and end of the two \textit{TESS} orbits for each target's sector.
    Note how light curves from the same sector exhibit noise patterns at similar time ranges, especially near the start and end of the \textit{TESS} orbits.}
    \label{fig:light_curves_noise_sector}
\end{figure}
Inspection of these normalized light curves revealed that not only are there outlier flux measurements of non-physical nature, but multiple light curves also show additional noise patterns with long duration, which also appeared to be of non-physical nature. 
By inspecting multiple targets in the same sector, I concluded that this noise seemed to be related to each individual sector, with similar noise patterns showing up at similar observation times for targets observed by different cameras but during the same sector, for both corrected light curves produced by \texttt{eleanor}.
Figure~\ref{fig:light_curves_noise_sector} shows some examples of the more extreme cases of noise patterns found for certain sectors, in particular two targets in Sector 8 and two in Sector 12.
These systematics are not only undesirable due to their non-physical signal causing issues for any study of the stellar signals (e.g., asteroseismology), but I also found that their high amplitude and typical timescales lead to non-physical transit-like signals being detected in lieu of potentially real, physical ones.

One observation that can be drawn from the examples shown in Figure~\ref{fig:light_curves_noise_sector} and that, for the most part, was consistent for all examples explored, is that the noisy patterns tend to show up near the start and end of each of \textit{TESS}'s orbits, in particular at the start.
To try and understand the reason behind these unexpected changes in flux, I examined the NASA Data Release Notes\footnote{\href{https://archive.stsci.edu/tess/tess_drn.html}{https://archive.stsci.edu/tess/tess\_drn.html}} for each sector.
From these documents, I verified that strong scattered light was observed at the start of each orbit, as well as close to times when the satellite exhibits anomalous behavior, which was consistent with the presence of most noisy patterns in the light curves.
Also according to the documents, the time-stamps when excessive scattered light was observed were not considered for the systematics correction applied by NASA to short-cadence targets (i.e. in the determination of the CBVs).
Since the CBVs do not contain information on the scattered light, its presence cannot be accounted for by the corrections, and so the extracted light curves exhibit the aforementioned noisy patterns.
As such, I had to devise an alternative way of dealing with these signals.

\begin{table}[!ptb]
    \centering
    \setlength{\tabcolsep}{10pt}
    \begin{tabular}{cccc}
        \textbf{Sector} & \multicolumn{2}{c}{\textbf{Orbital ranges [BTJD]}} & \textbf{Additional [BTJD]} \\
        \midrule
        \midrule
        \multirow{2}{*}[-3pt]{Sector 1} & 1325.29--1325.59 & 1338.23--1338.53 & \multirow{2}{*}[-3pt]{1347.0--1349.51} \\ 
        \cmidrule{2-3} \\[-14pt] 
        & 1339.64--1339.94 & 1352.89--1353.19 &  \\ 
        \midrule
        \multirow{2}{*}[-3pt]{Sector 2} & 1354.09--1354.39 & 1366.86--1367.16 &  \\ 
        \cmidrule{2-3} \\[-14pt] 
        & 1368.58--1368.88 & 1381.22--1381.52 &  \\ 
        \midrule
        \multirow{2}{*}[-3pt]{Sector 3} & 1382.71--1386.2 & 1395.15--1395.45 &  \\ 
        \cmidrule{2-3} \\[-14pt] 
        & 1396.65--1396.95 & 1404.99--1409.38 &  \\ 
        \midrule
        \multirow{2}{*}[-3pt]{Sector 4} & 1410.89--1411.56 & 1423.22--1423.52 &  \\ 
        \cmidrule{2-3} \\[-14pt] 
        & 1424.54--1424.84 & 1436.56--1436.86 &  \\ 
        \midrule
        \multirow{2}{*}[-3pt]{Sector 5} & 1437.98--1438.28 & 1449.91--1450.21 &  \\ 
        \cmidrule{2-3} \\[-14pt] 
        & 1451.54--1451.84 & 1463.44--1464.31 &  \\ 
        \midrule
        \multirow{2}{*}[-3pt]{Sector 6} & 1465.2--1465.5 & 1476.73--1477.03 &  \\ 
        \cmidrule{2-3} \\[-14pt] 
        & 1478.09--1478.39 & 1489.76--1490.06 &  \\ 
        \midrule
        \multirow{2}{*}[-3pt]{Sector 7} & 1491.6--1492.1 & 1502.75--1503.05 &  \\ 
        \cmidrule{2-3} \\[-14pt] 
        & 1504.69--1505.19 & 1515.8--1516.1 &  \\ 
        \midrule
        \multirow{2}{*}[-3pt]{Sector 8} & 1517.33--1518.33 & 1528.78--1529.08 & \multirow{2}{*}[-3pt]{1531.74--1536.0} \\ 
        \cmidrule{2-3} \\[-14pt] 
        & 1530.25--1530.85 & 1541.71--1542.01 &  \\ 
        \midrule
        \multirow{2}{*}[-3pt]{Sector 9} & 1543.21--1544.71 & 1555.55--1555.85 &  \\ 
        \cmidrule{2-3} \\[-14pt] 
        & 1556.71--1557.71 & 1568.29--1568.79 &  \\ 
        \midrule
        \multirow{2}{*}[-3pt]{Sector 10} & 1569.42--1571.37 & 1581.5--1581.8 &  \\ 
        \cmidrule{2-3} \\[-14pt] 
        & 1582.45--1585.22 & 1595.39--1595.69 &  \\ 
        \midrule
        \multirow{2}{*}[-3pt]{Sector 11} & 1596.76--1600.44 & 1609.41--1609.71 &  \\ 
        \cmidrule{2-3} \\[-14pt] 
        & 1610.77--1614.7 & 1623.6--1623.9 &  \\ 
        \midrule
        \multirow{2}{*}[-3pt]{Sector 12} & 1624.94--1628.45 & 1638.71--1639.01 &  \\ 
        \cmidrule{2-3} \\[-14pt] 
        & 1640.02--1642.63 & 1652.6--1652.9 &  \\ 
        \midrule
        \multirow{2}{*}[-3pt]{Sector 13} & 1653.9--1657.6 & 1667.4--1667.7 &  \\ 
        \cmidrule{2-3} \\[-14pt] 
        & 1668.61--1668.91 & 1682.07--1682.37 &  \\
        \bottomrule
    \end{tabular}
    \caption[Ranges of time stamps removed from light curve data for each of the first 13 sectors of \textit{TESS} data]
    {Ranges of time stamps removed from light curve data for each of the first 13 sectors of \textit{TESS} data.
    For each sector, an interval at both the start and end of each of the two orbits was defined, as well as any additional intervals that were deemed necessary given the noisy patterns observed in the representative light curves selected for each sector. 
    Some intervals are especially large due to anomalous behavior from the \textit{TESS} satellite, which caused longer time ranges to be discarded directly by the \textit{TESS} mission.}
    \label{tab:sector_cuts}
\end{table}
\begin{figure}[!ptb]
    \centering
    \includegraphics[width=\textwidth, height=\textheight, keepaspectratio]{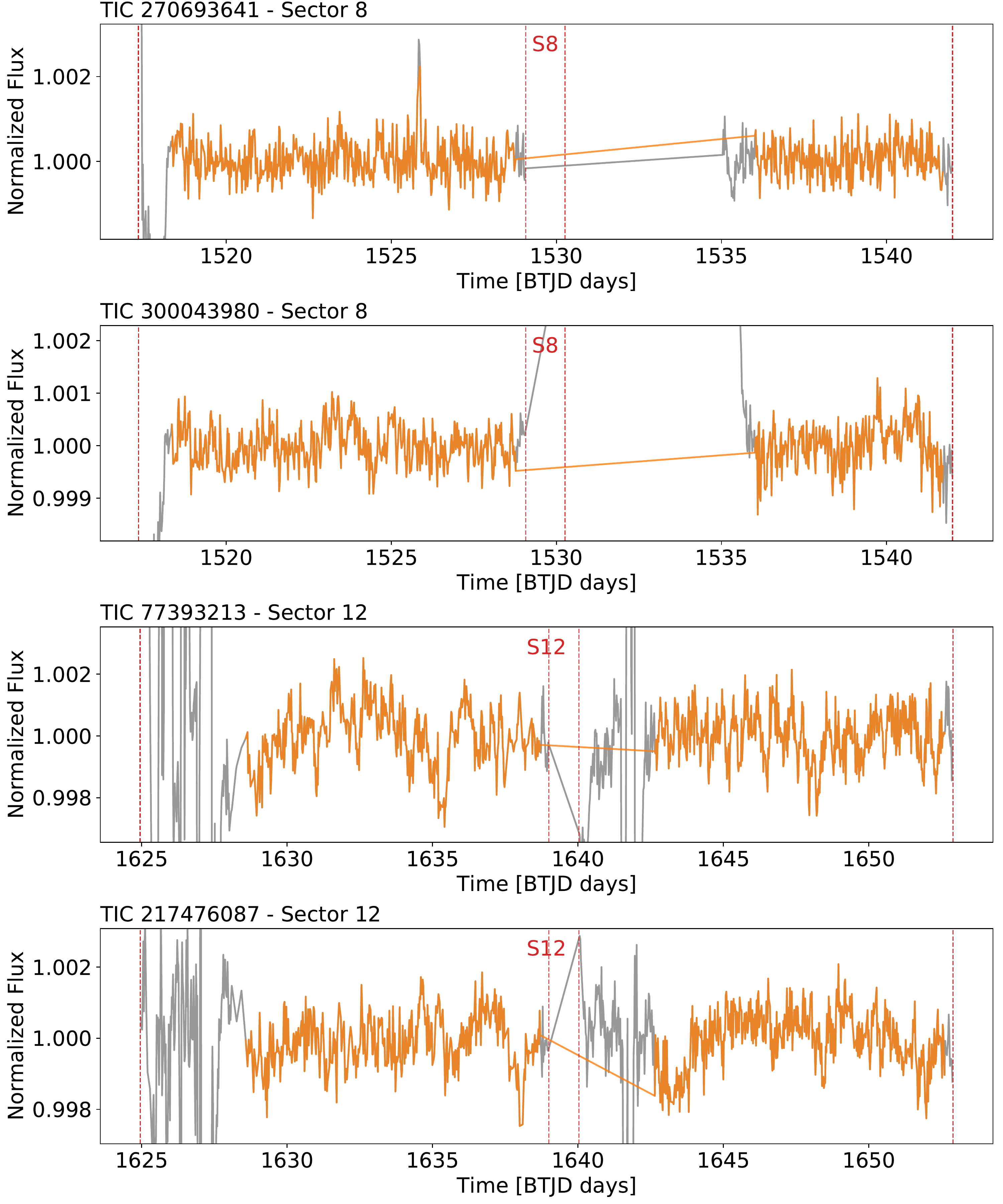}
    \caption[Improvement in the light curves from the removal of the time-stamp ranges defined in Table~\ref{tab:sector_cuts}]
    {Improvement in the light curves from the removal of the time-stamp ranges defined in Table~\ref{tab:sector_cuts}.
    The orange sections correspond to the portions of the original (gray) light curve that were kept.
    The vertical red dashed lines denote the start and end of the two \textit{TESS} orbits for each target's sector.
    The four light curves are the same as in Figure~\ref{fig:light_curves_noise_sector}.
    Note how the removed time-stamp ranges concentrate mostly on the noisy patterns present in each light curve.
    The TIC ID of each target and the sector from which the light curve was extracted are shown in the top left of each panel.}
    \label{fig:light_curves_cuts}
\end{figure}
For that, I decided to heuristically determine a set of time stamps per sector to exclude from all light curves. 
Specifically, from my sample of LLRGB stars (see Section~\ref{sec:target_selection}), exclusively composed of targets from \textit{TESS}'s southern hemisphere, I selected, for each sector, 5 stars observed by each camera, for a total of 20 stars per sector.
I then visually inspected all selected targets and identified ranges of time stamps that had to be excluded, per sector, so that the noisy patterns present in all their light curves would be removed.
The objective was to minimize the number of light curves that would have extreme systematics present, whilst at the same time maximizing their temporal coverage (since this procedure essentially reduces the length of observations by removing data at the edges of the \textit{TESS} orbits).
The defined, per-sector, time-stamp ranges are shown in Table~\ref{tab:sector_cuts}, whilst Figure~\ref{fig:light_curves_cuts} illustrates the results of removing the flagged time stamps from the four noisy light curves shown in Figure~\ref{fig:light_curves_noise_sector}.
It is important to note that, since at the time of the determination of these cuts, only the southern hemisphere of \textit{TESS} observations had been completed (i.e., the first year of observations), I have not yet determined a list of ranges for the northern sectors, which will undoubtedly be necessary to account for identical problematic signals in those targets.

Besides the time-stamp exclusions, due to some spurious and often isolated flux measurement outliers showing very extreme values, I also performed sigma clipping at 6-$\sigma$, which I found to be conservative enough to not remove any points from potential transits.



%% file: chapters/pipeline/transit_search.tex
\section{Transit search}
\label{sec:transit_search}

\begin{figure}[!ptb]
    \centering
    \includegraphics[width=\textwidth, height=0.4\textheight, keepaspectratio]{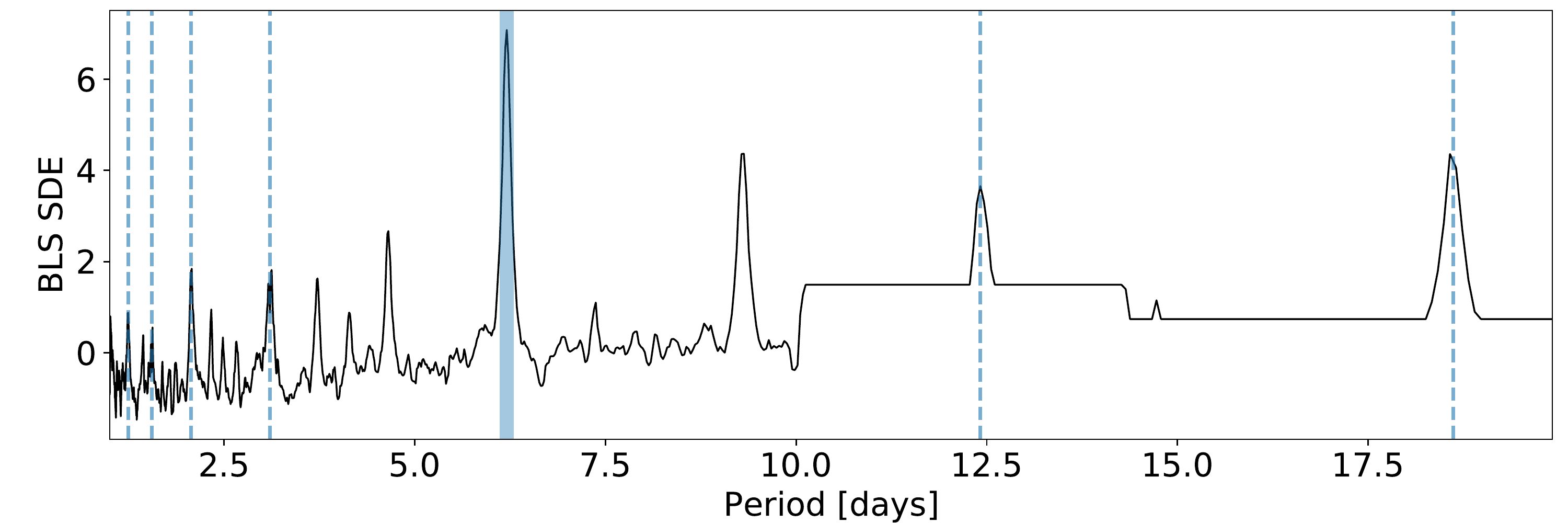}
    \includegraphics[width=\textwidth, height=0.4\textheight, keepaspectratio]{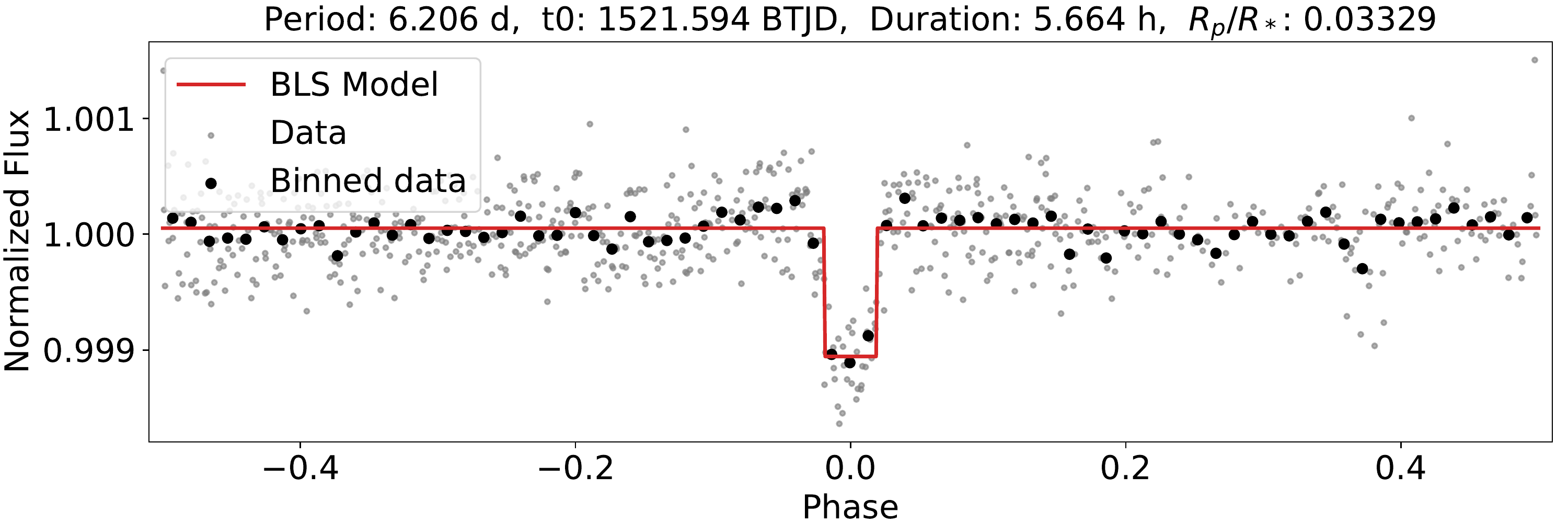}
    \caption[Example of a periodogram and corresponding phase-folded transit obtained with the Box-fitting Least Squares (BLS) algorithm]
    {Example of a periodogram and corresponding phase-folded light curve and transit, obtained with the Box-fitting Least Squares (BLS) algorithm.
    \textit{Top}: BLS periodogram obtained for a sample \textit{TESS} target, which shows for every trial period (x-axis), the corresponding signal detection efficiency (SDE) metric (y-axis), which measures the goodness of fit of a transit model, with that period, to the data.
    The blue solid line highlights the period with the highest SDE, which is chosen as the most significant transit for this light curve.
    The blue dashed lines are the harmonics of the period corresponding to the highest peak.
    \textit{Bottom}: Phase-folded light curve of the same sample target, plotted with a BLS transit model.
    The gray and black points are the phase-folded, original and binned (10-point bins) flux measurements, respectively.
    The period used for the phase-folding is taken from the blue solid line in the periodogram above.
    The model of that same transit is shown by the box-shaped solid red line.}
    \label{fig:bls_light_curve}
\end{figure}

Searching for transits in light curves has been commonly carried out in the literature using an implementation of the Box-fitting Least Squares (BLS) algorithm \parencite{Kovacs_2002}.
In this work, I used the implementation from the open-source \texttt{Python} package \texttt{astropy}\footnote{\href{https://github.com/astropy/astropy}{https://github.com/astropy/astropy}} \parencite{AstropyCollaboration_2013,AstropyCollaboration_2018}.
This method characterizes a transit signal as a periodic signal of period $P$ that can take only two possible values, $H$ and $L$. 
During the transit, the signal has constant value $L$ and its duration is defined as a fraction of the period, $qP$, where $q$ is a small value ($\sim 0.01-0.05$).
Essentially, the transit is modeled as a periodic box-shaped signal, as illustrated in the bottom panel of Figure~\ref{fig:bls_light_curve}.

By considering a grid of potential periods and durations for a transit, the method finds, for each trial period, the transit duration, epoch and depth that best characterize the light curve by performing a least-squares fit to the data.
For each trial period, the BLS method then returns those best fit parameters as well as a goodness-of-fit value, referred to as Signal Residue (SR) in \textcite{Kovacs_2002}.
From this list of SR values, the authors define a Signal Detection Efficiency (SDE) metric, as
\begin{equation}
    {\rm SDE} = \frac{{\rm SR} - \mu_{\rm SR}}{\sigma_{\rm SR}},
    \label{eq:sde_bls}
\end{equation}
where SR is the goodness-of-fit value for each trial period, with $\mu_\text{SR}$ and $\sigma_\text{SR}$ corresponding to its average and standard deviation when considering all periods, respectively.
Computing then the SDE for each trial period and plotting these values with respect to the period results in a spectrum of all trial periods, commonly referred to as the BLS periodogram and illustrated in the top panel of Figure~\ref{fig:bls_light_curve}.
This periodogram, with all its corresponding SDE metrics, is then inspected to ascertain whether a potential transit signal is present or not in the light curve.

\begin{figure}[!ptb]
    \centering
    \includegraphics[width=\textwidth, height=0.4\textheight, keepaspectratio]{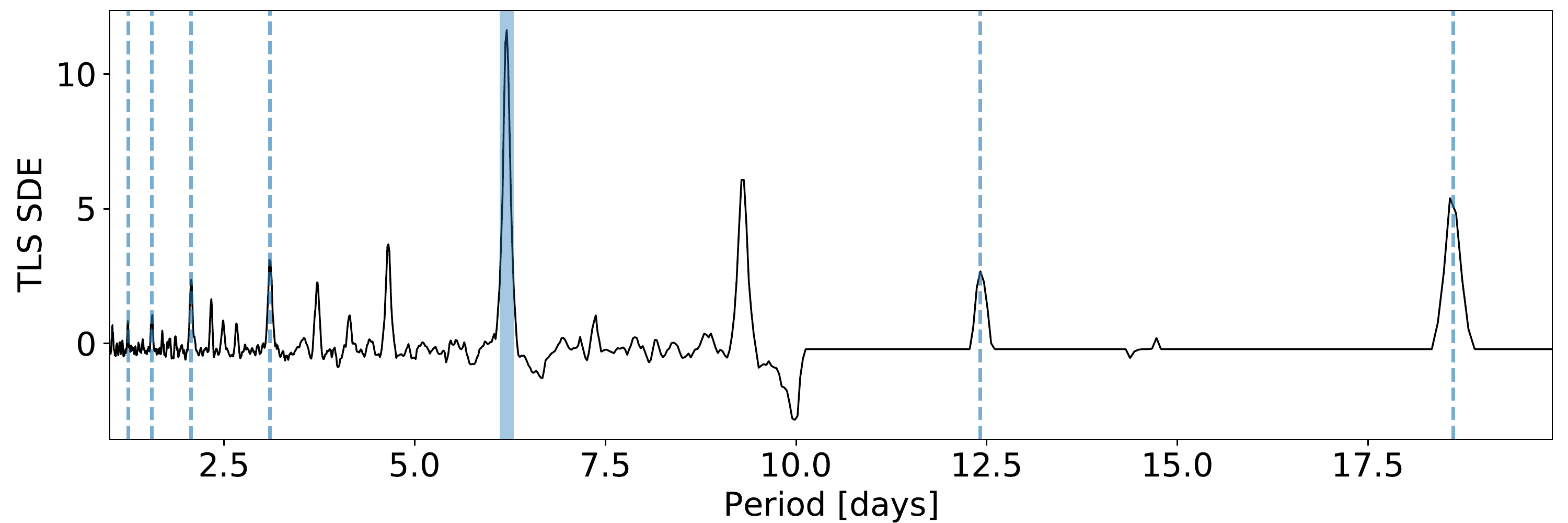}
    \includegraphics[width=\textwidth, height=0.4\textheight, keepaspectratio]{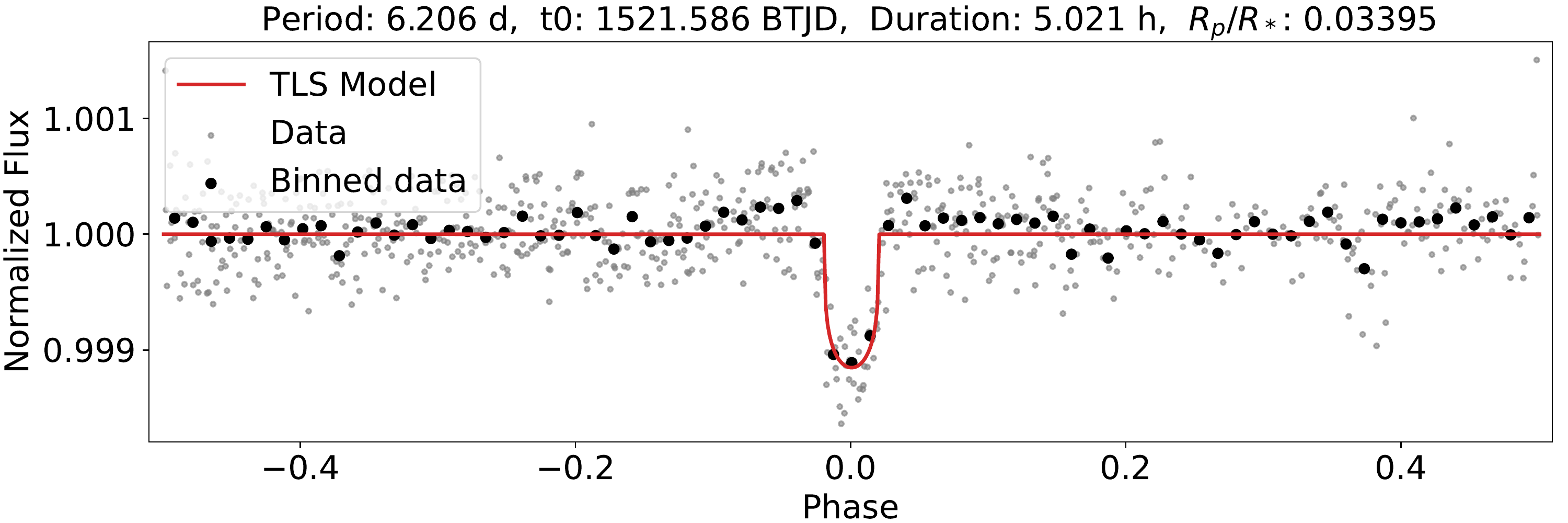}
    \caption[Example of a periodogram and corresponding phase-folded light curve and transit, obtained with the Transit Least Squares (TLS) algorithm]
    {Example of a periodogram and corresponding phase-folded light curve and transit, obtained with the Transit Least Squares (TLS) algorithm.
    \textit{Top}: TLS periodogram obtained for a sample \textit{TESS} target, which shows for every trial period (x-axis), the corresponding signal detection efficiency (SDE) metric (y-axis), which measures the goodness of fit of a transit model, with that period, to the data.
    The blue solid line highlights the period with the highest SDE, which is chosen as the most significant transit for this light curve.
    The blue dashed lines are the harmonics of the period corresponding to the highest peak.
    \textit{Bottom}: Phase-folded light curve of the same sample target, plotted with a TLS transit model.
    The gray and black points are the phase-folded, original and binned (10-point bins) flux measurements, respectively.
    The period used for the phase-folding is taken from the blue solid line in the periodogram above.
    The model of that same transit is shown by the transit-like solid red line, in contrast to the box-shaped BLS model shown in Figure~\ref{fig:bls_light_curve}.}
    \label{fig:tls_light_curve}
\end{figure}
Additionally, I also used the Transit Least Squares (TLS) method \parencite{Hippke_2019}, implemented in the open-source \texttt{transitleastsquares}\footnote{\href{https://github.com/hippke/tls}{https://github.com/hippke/tls}} \texttt{Python} package.
This method follows closely the BLS method, introducing some changes to attempt to improve the detection of planetary transits.
The most significant one is the introduction of limb-darkening in the transit model considered (see Section~\ref{sec:transits} for more details), to more closely match the expected physical signal of a transit, as is illustrated in the bottom panel of Figure~\ref{fig:tls_light_curve}.
Obviously, this added complexity in the transit model introduces additional computational costs.
To offset these costs, the authors of TLS found ways to optimize the selection of the grid of transit periods and durations, including an informed non-linear spacing.
Similarly to BLS, the TLS method also returns a slightly modified SDE metric for each trial period (see equation~3 of \textcite{Hippke_2019}), resulting in the TLS periodogram, illustrated in the top panel of Figure~\ref{fig:tls_light_curve}.
Additionally, the TLS method also computes the significance (in units of standard deviations) of the difference between the depths of the odd and even eclipses, a useful value to identify if potential transits are instead produced by eclipsing binaries (see Section~\ref{sec:transit_validation}).

The reason for still including the BLS method, given the existence of the new TLS implementation, is to complement the TLS search, given that BLS has stood the test of time despite its less sophisticated model.
This way, potential planet transit detections with TLS can be confirmed also with BLS, serving as an extra measure of confidence.

\begin{sidewaysfigure}[!ptb]
    \centering
    \includegraphics[width=0.9\textwidth, keepaspectratio]{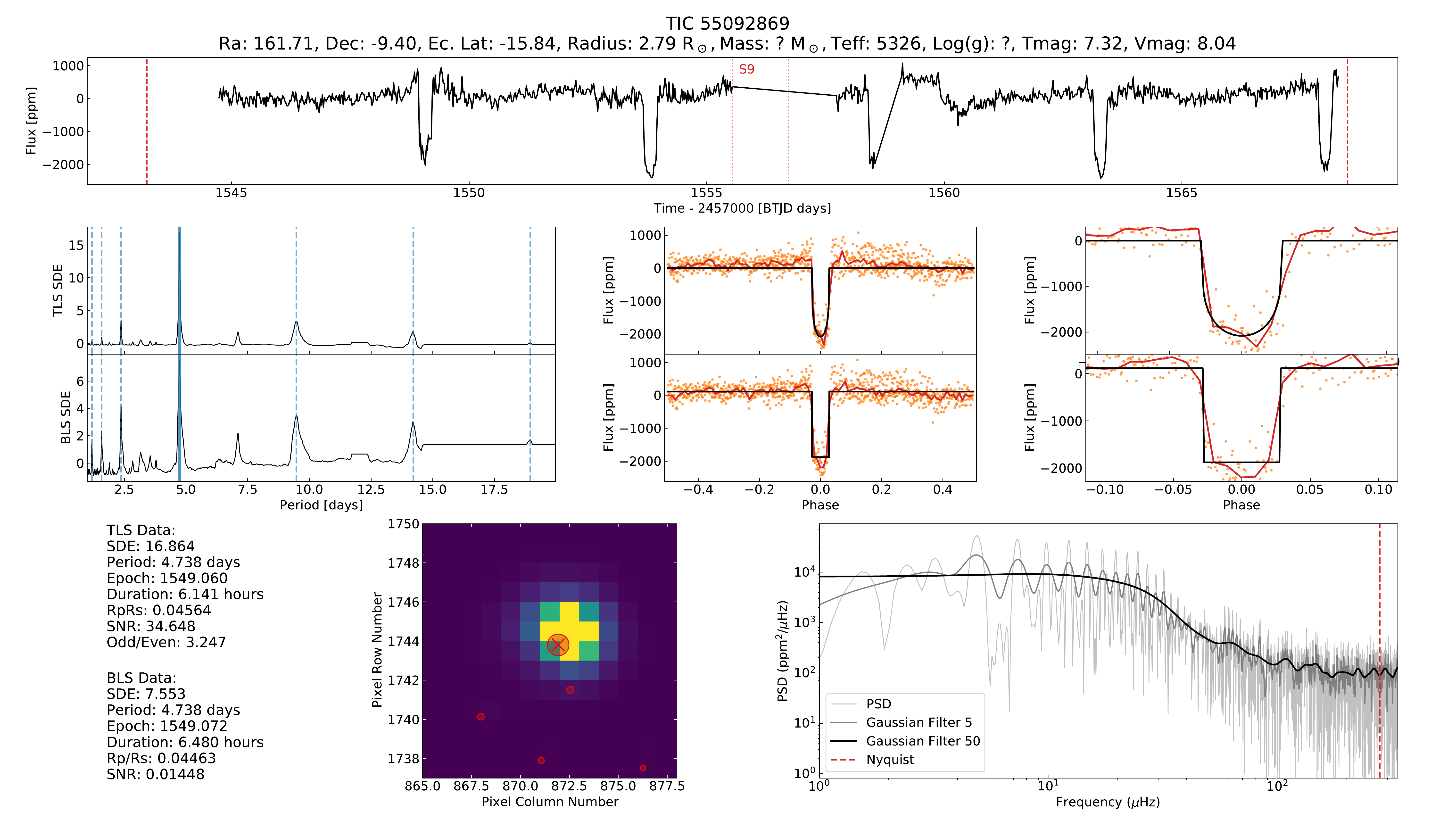}
    \caption[Summary plot of target TIC 55092869, known as KELT-11, showing the transit of its known orbiting planet, KELT-11 b]
    {Summary plot of target TIC 55092869, known as KELT-11, showing the transit of its known orbiting planet, KELT-11 b.
    \textit{Top:} Final light curve for the target, after the extraction from the FFIs and initial corrections by \texttt{eleanor} and including the additional corrections described in Section~\ref{sec:corrections}.
    Above the plot is also shown some information and properties from the target, taken from the \textit{TESS} Input Catalog (when available and otherwise marked with a question mark).
    \textit{Middle:} From left to right, the periodogram, phase-folded transit and zoom in obtained from the search using TLS (above) and BLS (below), similar to the plots depicted in Figures~\ref{fig:bls_light_curve} and \ref{fig:tls_light_curve}.
    \textit{Bottom:} On the left, information related to the most significant transit found by BLS and TLS, shown by the blue solid line in the periodograms above.
    To its right, a sample of a target pixel file, extracted from the \textit{TESS} FFIs.
    Red circles denote all stars in the field of the observed target that are brighter than \textit{Gaia} $G$ magnitude 16, with the size scaled to that same magnitude (the circle with a cross in its center denotes the target).
    Finally, a periodogram of the final light curve, useful for visual inspection of the presence of oscillations.}
    \label{fig:summary_plot_example}
\end{sidewaysfigure}

The entire process of the pipeline is automated until this stage, at which point, for any target for which the pipeline successfully completed light curve extraction, correction and transit search, a summary figure is created that contains the most relevant information and plots concerning the target.
Figure~\ref{fig:summary_plot_example} depicts the summary plot for TIC 55092869, known as KELT-11, a known host of a $\sim 1.35 \ R_\text{J}$ planet on a $\sim 4.74$-day orbit \parencite{Pepper_2017,Beatty_2017}.

%% file: chapters/pipeline/validation.tex
\section{Transit validation}
\label{sec:transit_validation}

As seen in the previous section, a transit search on a light curve produces a periodogram, i.e. a spectrum of periods for which a transit model is fitted to the light curve data and a resulting metric, the SDE, is calculated.
Stars can have more than one transiting planet, in which case different peaks in the periodogram could correspond to different planetary signals.
However, as discussed in Section~\ref{sub:occurrence_rate}, hot Jupiters are very rare, either when orbiting main-sequence or evolved stars. 
They are also usually found in single planet systems \parencite{Latham_2011}.
In addition, there are added challenges concerning the detection of planetary transits in red giants (discussed in Section~\ref{sec:stellar_signals}).
For all these reasons, I focused my efforts on looking for only a single planet around each star, with a period from 1 to 20 days.
In that respect, the highest peak in a periodogram is taken to be the representative transit for that target, and it is the one that undergoes validation to ascertain its nature.

In this pipeline, I introduced a two-step approach to the validation of transits. 
The first step is a statistical procedure that identifies statistical false-positive signals up to a chosen confidence level, so as to set apart the ones that have a higher probability of being of physical origin.
The second approach is an automated method applied to those signals that have been identified as astrophysical in nature.
Its objective is to attribute a probability for the signal to be either of planetary origin, or some other astrophysical scenario.

\subsection{Step 1 -- Identifying statistical false-positives}
\label{sub:validation_statistical}

To identify signals that have a high probability of being statistical false positives, I start by defining a threshold for the SDE metric and classifying all signals whose measured SDE is below this threshold as false-positives.
A simple exercise to estimate this SDE threshold is to perform a search on simulated light curves without transits and studying the distribution of SDEs obtained.
\textcite{Campante_2018} did just that for \textit{TESS} simulated light curves of LLRGB stars using the BLS algorithm, obtaining a statistical false-positive rate of 1\% when considering only signals with an SDE above 6.34.
The authors also determined thresholds for either 1 or 2-sector light curves, again for a false-positive rate of 1\%, with the corresponding SDE values being 5.98 and 6.39, respectively.

For my search, given the usage of two different search methods (i.e. BLS and TLS), I adopted a slightly different approach.
Specifically, the SDE thresholds used for the classification of signals were instead determined by examining the SDE distributions obtained for all targets in the search.
Section~\ref{sec:sample_analysis} describes all the data exploration steps and transit validation applied to my \textit{TESS} southern search.

Additionally, other metrics could also be used to further differentiate targets with statistical physical signals. 
An example is the Combined Differential Photometric Precision (CDPP) (mentioned in Section~\ref{sec:photometry}), which attempts to quantify the ease with which terrestrial transit signatures could be detected in \textit{Kepler} light curves \parencite{Christiansen_2012}.
However, I did not implement any additional metrics in the pipeline besides the SDE.

\subsection{Step 2 -- Identifying astrophysical false-positives}
\label{sub:validation_astrophysical}

For the identification of astrophysical false positives I applied \texttt{VESPA} \parencite{Morton_2012,Morton_2015c}, which is a computationally expensive tool that performs automatic astrophysical false-positive evaluation of transit signals.
\texttt{VESPA} works under the assumption that the candidate transit is of physical nature.
The tool makes use of provided properties of both the transit and the stellar host to determine whether the transit signal is expected to originate from a planetary system scenario or from another astrophysical configuration that can also cause transit-like signals, which it considers as being a false positive.
In particular, \texttt{VESPA} evaluates three other astrophysical system configurations: eclipsing binaries (EB), background eclipsing binaries (BEB) and hierarchical triple stellar systems (HEB).

As a first step, \texttt{VESPA} uses the transit host's celestial position, together with the Galactic simulation code \texttt{TRILEGAL} \parencite{Girardi_2005}, to estimate the prior probability of each considered system configuration existing in the Galactic region given by the target's coordinates.
Then, using provided stellar parameters for the host, such as apparent stellar magnitudes (in particular, \textit{TESS} magnitude is required for \textit{TESS} targets) and effective temperature, the tool simulates populations of each of the considered physical scenarios, including planetary systems, such that the observational properties of the simulated systems match the ones provided.
For each system in the simulated populations, \texttt{VESPA} then determines the likelihood that such system could produce a transit light curve with the same properties as the observed one.
The information from the priors and likelihoods is then combined into a posterior probability as well as a false-positive probability (FPP), which states the probability of this transit being caused by a non-planetary system.

The FPP is estimated as
\begin{equation}
    \text{FPP} = 1 - P_\text{p},
\end{equation}
where the planetary scenario probability, $P_\text{p}$, is estimated by
\begin{equation}
    P_\text{p} = \frac{\mathcal{L}_\text{p} \pi_\text{p}}{\mathcal{L}_\text{p} \pi_\text{p} + \mathcal{L}_\text{FP} \pi_\text{FP}},
    \label{eq:vespa_planet_posterior}
\end{equation}
with $\mathcal{L}_i$ and $\pi_i$ representing the likelihoods and priors, respectively, and the subscripts \textbf{p} and \textbf{FP} corresponding to the planetary scenario and the remaining false-positive scenarios, respectively.
\texttt{VESPA} considers the limit of FPP < 0.01 to be an adequate one to determine a planetary nature for the transit.

\begin{figure}[!ptb]
    \centering
    \fbox{\includegraphics[width=\textwidth, height=\textheight, keepaspectratio]{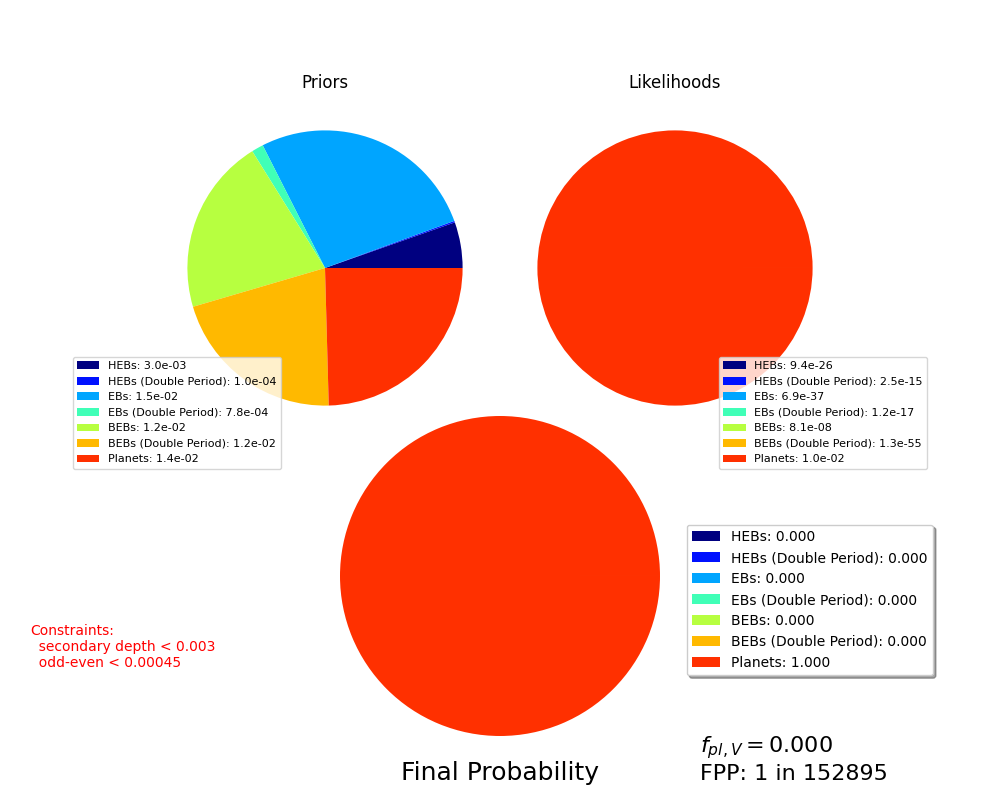}}
    \caption[False-positive probability summary plot from \texttt{VESPA}, for the transit of TIC 55092869.01 (KELT-11 b)]
    {False-positive probability (FPP) summary plot from \texttt{VESPA}, for the transit of TIC 55092869.01 (KELT-11 b).
    The upper left pie chart shows the prior probability for each considered astrophysical scenario, taking into consideration the Galactic location of the target (given by its coordinates).
    The upper right pie chart then shows the likelihood of each scenario being responsible for the observed transit, evaluated from the transit properties and phase-folded shape.
    Finally, the bottom pie chart shows the posterior probability for each of the possible astrophysical scenarios, determined using both the likelihoods and priors.
    In this case, only the planetary scenario has any posterior probability of explaining the transit.
    The red text on the lower left corner denotes the mismatch between the odd and even transits and the depth of the secondary eclipse, both important metrics where high values usually indicate astrophysical false-positive scenarios.
    On the lower right, $f_\text{pl,V}$ is the ``specific occurrence rate'' necessary for validation, as defined in section~3.4 of \textcite{Morton_2012}.
    Below it, the false-positive probability, $\text{FPP}$, denotes the probability that the transit is caused by a non-planetary astrophysical system.}
    \label{fig:vespa_summary_example}
\end{figure}

As output, \texttt{VESPA} condenses most of the information from its analysis into a single summary figure, an example of which is shown in Figure~\ref{fig:vespa_summary_example}.
The figure shows the results obtained for the light curve of TIC 55092869 (KELT-11).
Worth noting in the figure is the bottom pie chart, which shows the posterior probabilities for each of the considered physical scenarios.
In this case, the tool correctly identified the transit as that of a planetary candidate.
Note also the FPP value shown in the lower right corner, of $\text{FPP} < \; \sim 1 \text{ in } 10^{5}$, which gives a very confident indication that this transit was due to a planetary companion.

%% file: chapters/search/main.tex
\chapter{\textit{TESS} southern hemisphere search}
\label{cha:planet_search}

With the pipeline to explore \textit{TESS} FFI data and the method to characterize planetary transits around evolved stars ready, my next step was to carry out a systematic search for giant planets orbiting giant stars.
And for that, I first needed to choose which stars to look at.

I start this section by describing the steps taken for target selection, introducing all limits imposed in order to select a sample of bright LLRGB stars in the southern hemisphere of \textit{TESS}.
Then, I explore the results obtained from applying the search pipeline (Chapter~\ref{cha:search_pipeline}) to the stellar sample and detail all the steps taken to identify and rank planet candidates.
Afterwards, I introduce all planet candidates found and discuss the results from modeling their light curves using my tool for transit characterization (Chapter~\ref{cha:characterization}).
Finally, I weigh in on the results of this search in the context of my objectives for this work.

An important note concerning my efforts to search for giant planets in the \textit{TESS} data is that it has been conducted in the context of a collaboration.
This collaboration has not only allowed for additional confirmation on planet candidates found, ensuring they are identified using independent pipelines, but has also resulted in the pursuit of additional planet candidates, not included in my southern sample, which I also introduce in this chapter.


\input{chapters/search/target_selection}
\input{chapters/search/analysis}
\input{chapters/search/candidates}
\input{chapters/search/summary}

%% file: chapters/search/target_selection.tex
\section{Target selection}
\label{sec:target_selection}

For the target list, targets were specifically limited to low-luminosity red-giant branch (LLRGB) stars.
The reasons for this choice are twofold.
First, LLRGBs are evolved enough, when compared to subgiants, that the characteristic frequencies of their oscillations, or equivalently, their $\nu_\text{max}$, is lower than the Nyquist frequency of long-cadence \textit{TESS} observations, making it possible to detect oscillations, provided there are sufficient data.
Second, the stars are small enough when compared to high-luminosity red-giant branch and red-clump stars, that a transiting Jupiter-sized planet should be detectable in their light curves.
As an example, a Jupiter-sized planet transiting an 8 $\rm R_\odot$ star produces a 150 ppm decrease in flux, on the order of the white noise level observed in my sample of \textit{TESS} light curves, as demonstrated by the $\sigma$ parameter obtained in the light curve fit of TIC 348835438 (see Table~\ref{tab:tic348_params}), a target close to the sample's imposed limit on \textit{TESS} magnitude of 10 (see Table~\ref{tab:stellar_properties}).


\begin{figure}[!hpt]
    \centering
    \includegraphics[width=\textwidth, height=\textheight, keepaspectratio]{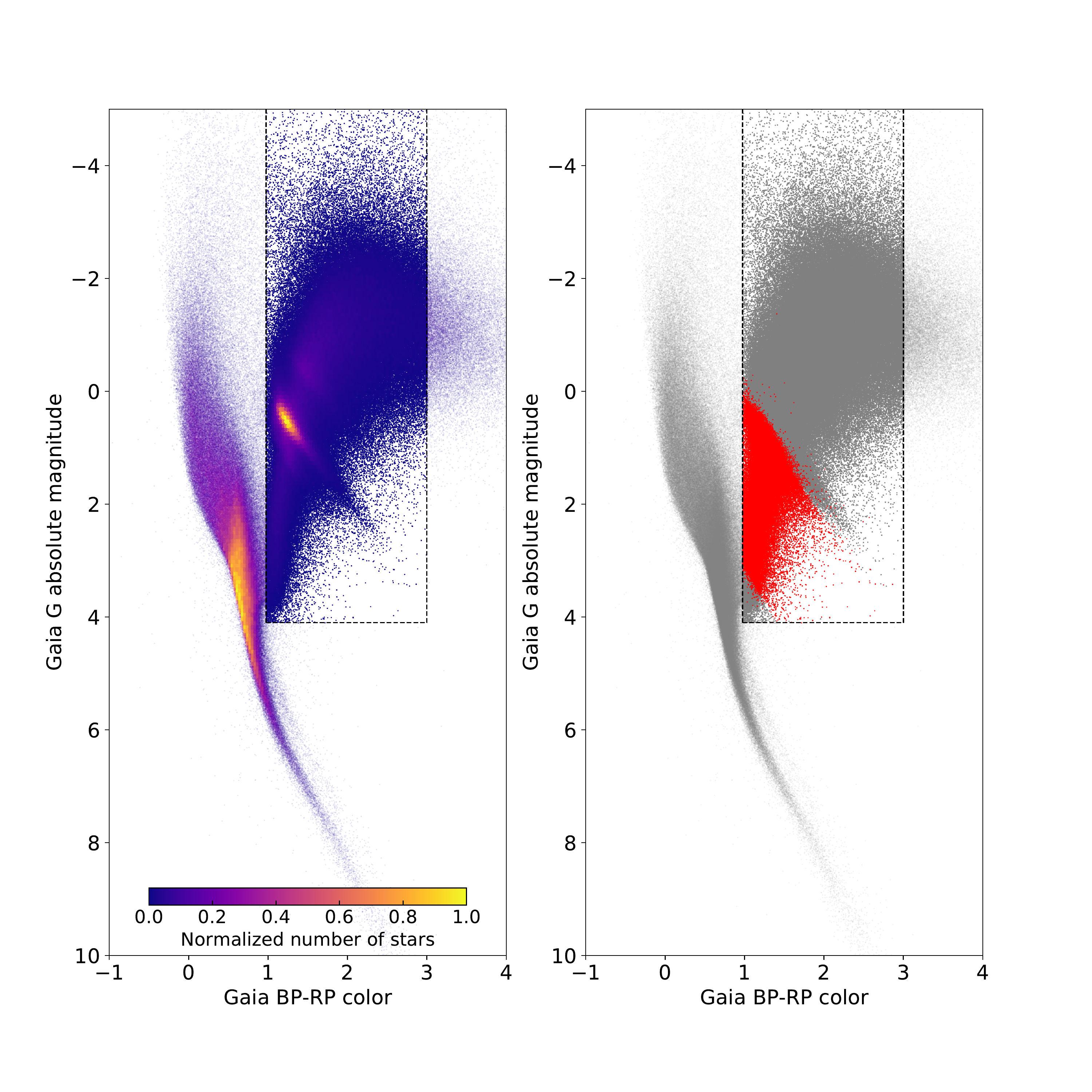}
    \caption[HR Diagram showing the target selection of all giant stars (on the left) and LLRGB stars (on the right, in red) from the \textit{TESS} Input Catalog]
    {HR Diagram showing the target selection of all giant stars (on the left) and LLRGB stars (on the right, in red) from the \textit{TESS} Input Catalog.
    The selection criteria concerning the left panel were based on cuts in both color and magnitude, described in Equation~(\ref{eq:mag_cuts}).
    The selection concerning the right panel was defined by considering only stars with a radius between 2.5 R$_\odot$ and 10 R$_\odot$, according to the value available in the TIC.}
    \label{fig:hr_target_selection}
\end{figure}
Stellar properties for the targets were taken from the \textit{TESS} Input Catalog \parencite[TIC;][]{Stassun_2019}, with the selection criteria being comprised of a set of empirical, yet physically motivated limits imposed on these properties.
The first set of limits defined had the intent of singling out the giant stars from the entire sample of the TIC.
These limits were chosen by visual inspection of an HR diagram, such as the one in Figure~\ref{fig:hr_target_selection}, in order to remove main-sequence and subgiant stars from the sample.
Specifically, the limits were imposed using the magnitudes of the \textit{Gaia} blue and red photometer bands, $G_\text{BP}$ and $G_\text{RP}$, respectively, as well as the \textit{Gaia} $G$-band absolute magnitude, $G_\text{abs}$, following
\begin{equation}
    \begin{split}
        0.975 < {G}_\text{BP} - {G}_\text{RP} &< 3.0 , \\
        {G}_\text{abs} &< 4.1 .
    \end{split}
    \label{eq:mag_cuts}
\end{equation}
The \textit{Gaia} absolute magnitude is not directly available in the TIC and was estimated following 
\begin{equation}
    {G}_\text{abs} = {G} + 5 (\log_{10}(\pi) + 1) ,
\end{equation}
where $G$ is the \textit{Gaia} $G$-band magnitude and $\pi$ is the \textit{Gaia} parallax.
The left plot of Figure~\ref{fig:hr_target_selection} illustrates this first set of limits, represented by the dashed-line delimited area in the HR diagram, inside which are the giant stars.

The second selection criterion defined was intended to distinguish the LLRGB targets from the rest of the giant sample.
To do so, the estimated stellar radius measurements present in the TIC were used.
These radii were determined using the Stefan--Boltzmann relation, with distances calculated using \textit{Gaia} parallaxes, and effective temperatures obtained either from spectroscopy (when available) or derredened colors (see section 2.3.5 of \textcite{Stassun_2019} for more details).
When comparing their radii to asteroseismic radii from \textcite{Huber_2017}, whose sample includes stars ranging from dwarfs to red giants, \textcite{Stassun_2019} find a mean difference of 0.64\% (7.03\% scatter), for all stars in common between both works.

Initial radius limits for the selection of LLRGB stars were set between 3 $\rm R_\odot$ and 8 $\rm R_\odot$.
Then, assuming a conservative estimate of the systematics in the TIC radii of over 15\%, the limits were expanded to radii between 2.5 $\rm R_\odot$ and 10 $\rm R_\odot$.
This ensured that the sample would include all desired targets, with the drawback of adding a residual amount of targets of less interest for my objectives, i.e., subgiants and red clump stars.
The right panel of Figure~\ref{fig:hr_target_selection} highlights the location of this LLRGB sample in an HR diagram, in red.

Additionally, in order to improve the chances of detecting solar-like oscillations whilst keeping the number of stars to a minimum, the sample was limited to brighter targets, by defining a ceiling in \textit{TESS} magnitude of 10 \parencite{Mackereth_2021}.
Stars with \textit{TESS} magnitude brighter than 6.8 were also removed (which represented $\sim$2\% of the targets), since their brightness is expected to saturate \textit{TESS}'s CCDs \parencite{Sullivan_2015} and \texttt{eleanor} does not have any implemented method to specifically handle such cases \parencite{Feinstein_2019}.

An important note on the target list is that, although all selected targets are present in the TIC, not all of them have been observed by \textit{TESS}, since the satellite's sky coverage during the nominal mission has a $\sim$12$^{\circ}$ gap in latitude, centered on the ecliptic plane, as well as small gaps between the various cameras and CCDs within the cameras.

To conclude, after all selection limits were considered, the sample contained about 100,000 stars, approximately evenly divided between the northern and southern \textit{TESS} hemispheres.
Out of this all-sky sample, the search was restricted to the southern targets, as at the time this work was carried out, the second year of \textit{TESS} observations was still underway.
This southern sample has around 50,000 LLRGB stars, with 40,772 observed by \textit{TESS}.

%% file: chapters/search/analysis.tex
\section{Sample analysis}
\label{sec:sample_analysis}

For all targets in the target list, I ran my pipeline (Chapter~\ref{cha:search_pipeline}) up to the transit search and saved all results for data analysis and transit validation.

As was discussed in Section~\ref{sec:transit_validation}, transit validation was conceptually separated into two steps.
Initially, I was looking to try and identify whether the detected signals were physical in nature, or merely the result of noise.
Then, for signals of likely physical nature, I wanted to identify whether they were caused by a planet or some other astrophysical false-positive scenario.

\begin{figure}[!ptb]
    \centering
    \includegraphics[width=\textwidth, height=\textheight, keepaspectratio]{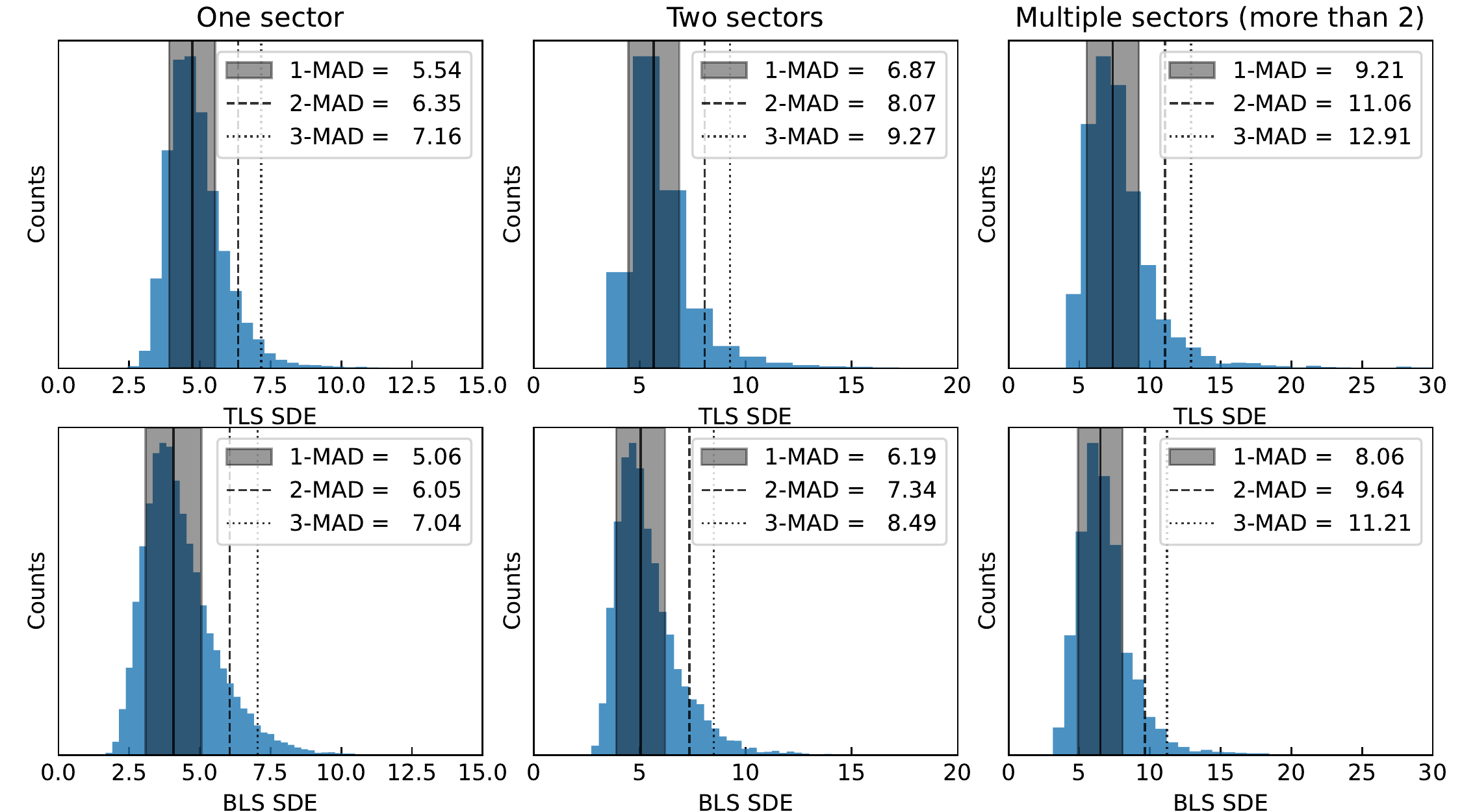}
    \caption[Distribution of the signal detection efficiency (SDE) metric obtained for all light curves in my \textit{TESS} southern search]
    {Distribution of the signal detection efficiency (SDE) metric obtained for all light curves in my \textit{TESS} southern search.
    Each plot shows the SDE for both the TLS (top) and the BLS (bottom) methods.
    The SDE values are also grouped according to the number of sectors of data available for the targets, with groups defined for 1-sector, 2-sector and multi-sector light curves, going from the left to right.}
    \label{fig:sde_hist}
\end{figure}
For the identification of physical signals, the initial step was to classify statistical false-positives, following the methodology described in Section~\ref{sec:transit_validation}.
As the SDE thresholds are sensitive to the amount of data available for each target, I opted to separate targets into different groups according to the available number of sectors.
Ultimately, three groups of targets were considered, thus leading to the separation of all targets into 1-sector, 2-sector or multi-sector ($\ge$3 sectors) targets.

The SDE distributions for targets in their respective groups are depicted in Figure~\ref{fig:sde_hist}, for both the TLS (top) and BLS (bottom) methods.
The figure also highlights the values of the median SDE plus one, two or three mean absolute deviations (MAD) for each of the distributions.
As shown in the figure, there is a shift towards higher values in the distributions' median SDE, as light curves with more sectors are considered, which was taken into consideration for the selection of three bins of targets (I explored additional binning configurations, but found them to be less significant and, given the relatively small number of multi-sector targets, opted for choosing three bins, as described above).


To be conservative, the SDE values selected as thresholds for the classification of statistical false positives were the median plus one MAD from each of the distributions.
This corresponded to two SDE thresholds per bin, one for each search method.
Based on these thresholds, all light curves where the signal SDE was above the thresholds for both search methods were shortlisted.
Applying this selection criterion reduced the number of stars to about 3000, from the initial number of 40,772 targets observed by \textit{TESS}.

\begin{figure}[!ptb]
    \centering
    \includegraphics[width=\textwidth, height=\textheight, keepaspectratio]{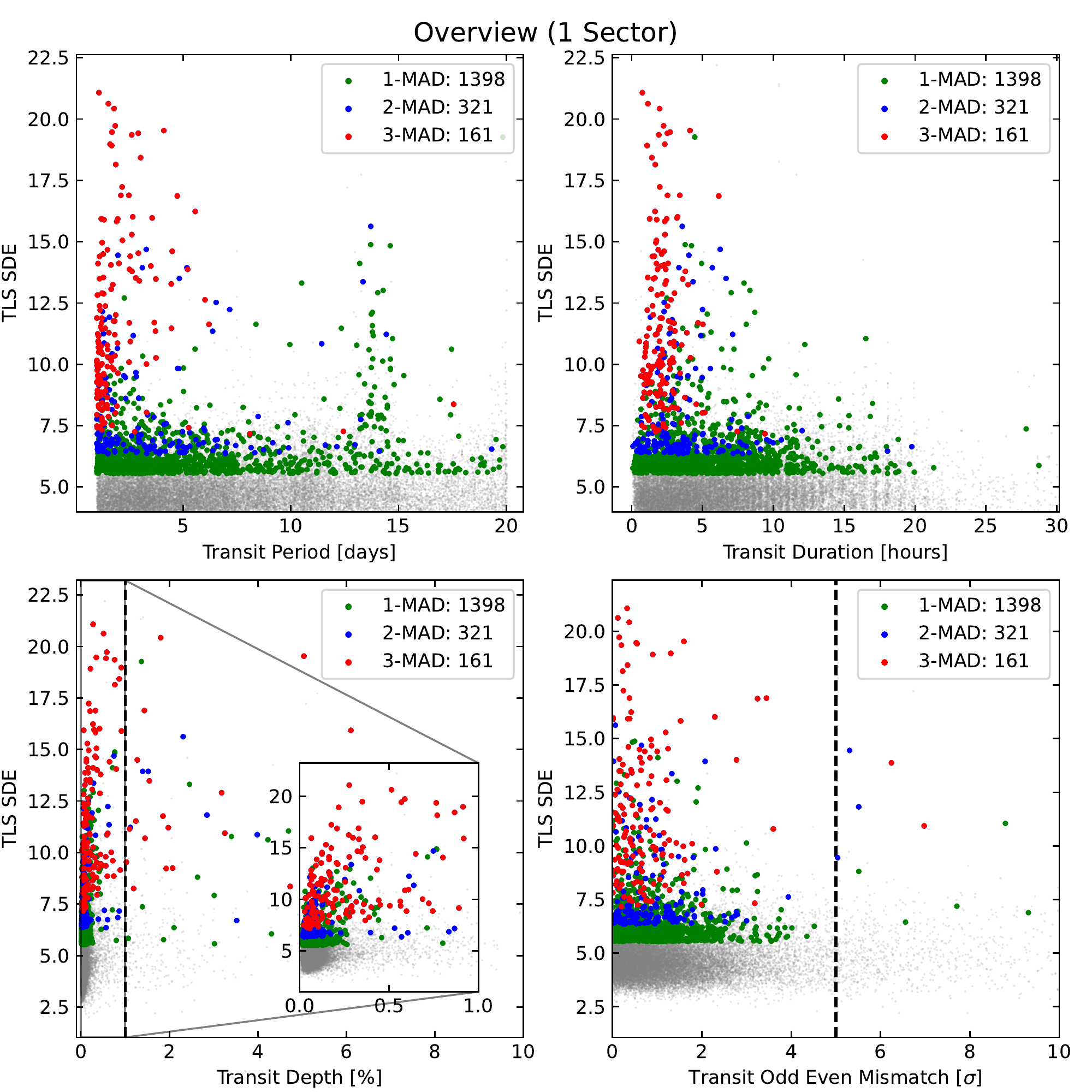}
    \caption[Overview of the signal properties for all targets in our \textit{TESS} southern search that have one sector of data]
    {Overview of the signal properties for all targets in our \textit{TESS} southern search that have one sector of data.
    The gray dots in the background represent all signals classified as statistical false positives, whilst the green, blue and red points are the signals that pass the median plus one, two and three MAD SDE thresholds.
    From top to bottom and from left to right, the panels depict the signal period, duration, depth and mismatch between odd and even eclipses against the TLS SDE, respectively.
    The dashed black line in the signal depth panel denotes the 1\% signal depth, which was defined as an upper limit above which a signal cannot be of planetary nature (upper limit in the inset).
    In the odd-even mismatch panel, the dashed black line denotes the $5$-$\sigma$ limit, also defined as an upper limit above which the signal cannot be considered of a planetary nature and is most likely originating from an eclipsing binary.}
    \label{fig:overview_one_sector}
\end{figure}
\begin{figure}[!ptb]
    \centering
    \includegraphics[width=\textwidth, height=\textheight, keepaspectratio]{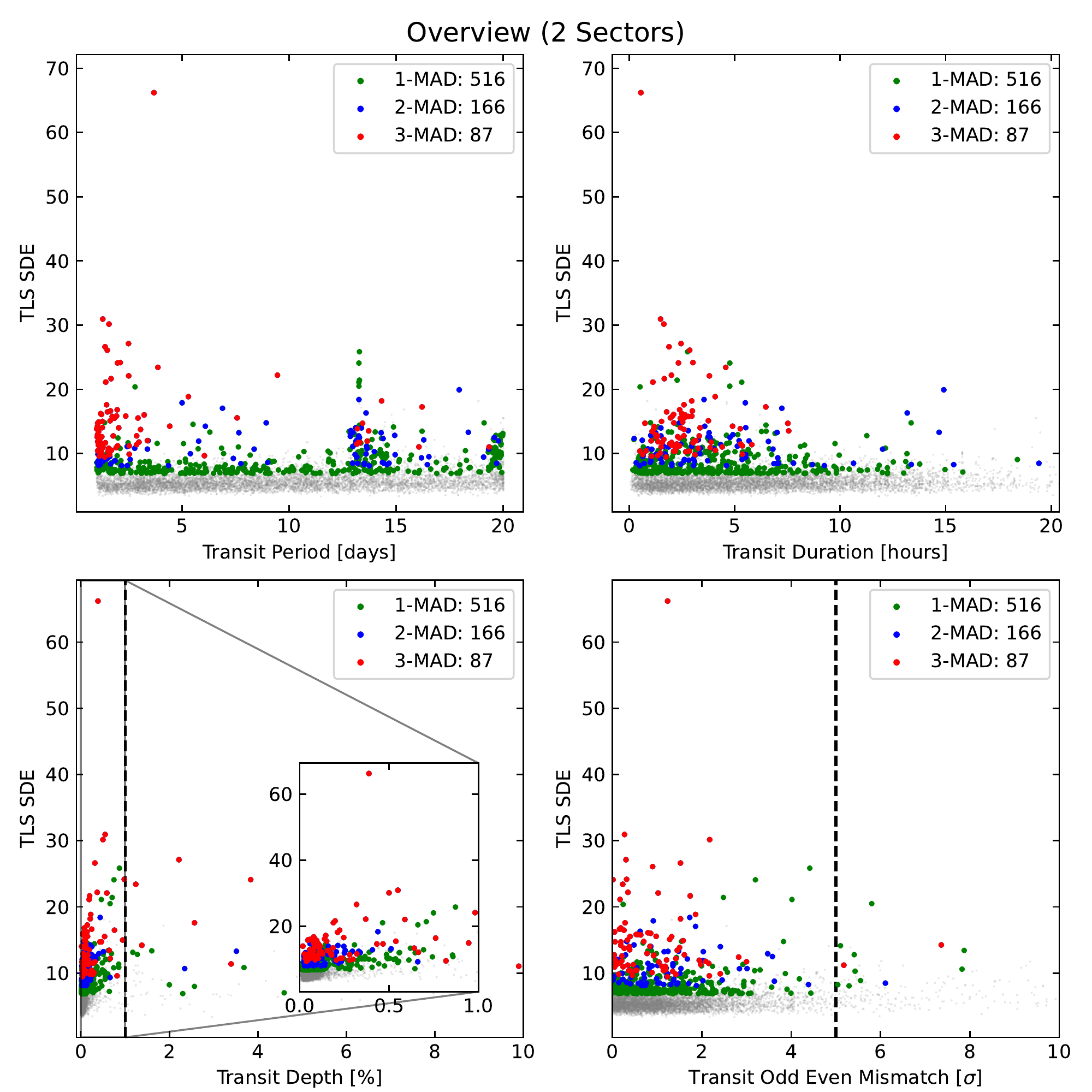}
    \caption[Overview of the signal properties for all targets in our \textit{TESS} southern search that have two sectors of data]
    {Overview of the signal properties for all targets in our \textit{TESS} southern search that have two sectors of data.
    The gray dots in the background represent all signals classified as statistical false positives, whilst the green, blue and red points are the signals that pass the median plus one, two and three MAD SDE thresholds.
    From top to bottom and from left to right, the panels depict the signal period, duration, depth and mismatch between odd and even eclipses against the TLS SDE, respectively.
    The dashed black line in the signal depth panel denotes the 1\% signal depth, which was defined as an upper limit above which a signal cannot be of planetary nature (upper limit in the inset).
    In the odd-even mismatch panel, the dashed black line denotes the $5$-$\sigma$ limit, also defined as an upper limit above which the signal cannot be considered of a planetary nature and is most likely originating from an eclipsing binary.}
    \label{fig:overview_two_sector}
\end{figure}
\begin{figure}[!ptb]
    \centering
    \includegraphics[width=\textwidth, height=\textheight, keepaspectratio]{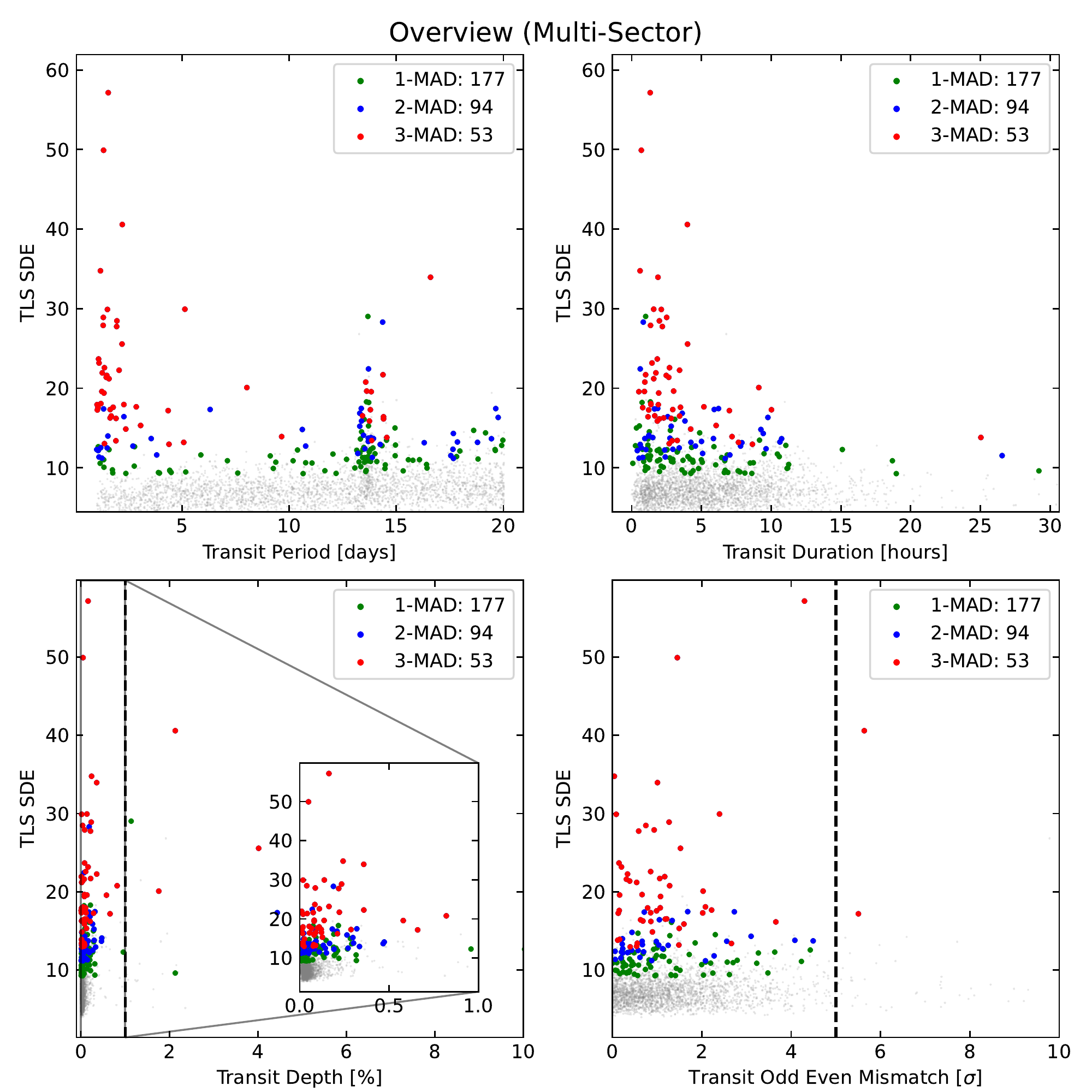}
    \caption[Overview of the signal properties for all targets in our \textit{TESS} southern search that have multiple ($\geq$3) sectors of data]
    {Overview of the signal properties for all targets in our \textit{TESS} southern search that have multiple ($\geq$3) sectors of data.
    The gray dots in the background represent all signals classified as statistical false positives, whilst the green, blue and red points are the signals that pass the median plus one, two and three MAD SDE thresholds.
    From top to bottom and from left to right, the panels depict the signal period, duration, depth and mismatch between odd and even eclipses against the TLS SDE, respectively.
    The dashed black line in the signal depth panel denotes the 1\% signal depth, which was defined as an upper limit above which a signal cannot be of planetary nature (upper limit in the inset).
    In the odd-even mismatch panel, the dashed black line denotes the $5$-$\sigma$ limit, also defined as an upper limit above which the signal cannot be considered of a planetary nature and is most likely originating from an eclipsing binary.}
    \label{fig:overview_multi_sector}
\end{figure}
Figures~\ref{fig:overview_one_sector}, \ref{fig:overview_two_sector} and \ref{fig:overview_multi_sector} highlight some of the signal properties of these 3000 transiting candidates, for 1-sector, 2-sector and multi-sector targets, respectively.
The figures show all targets classified as statistical false positives as gray points in the background, whilst in the foreground the targets that pass the median plus one, two and three MAD SDE thresholds are highlighted in green, blue and red, respectively.
From top to bottom and from left to right, the panels depict the signal period (in days), duration (in hours), fractional depth and mismatch between odd and even eclipses (in standard deviations) against the TLS SDE, respectively.
The dashed black line in the signal depth panel denotes the 1\% signal depth, which was defined as an upper limit above which a signal cannot be of planetary nature (given the properties defined for my sample of LLRGB targets, a depth of 1\% will, at best, correspond to a 2.5 $R_\text{J}$ orbiting body).
For the odd-even mismatch panel, the dashed black line denotes the $5$-$\sigma$ limit, also defined as an upper limit above which the signal cannot be considered of a planetary nature and its most likely originating from an eclipsing binary.

Of particular note in these figures is the unexpectedly large number of shortlisted targets with signal periods $\sim 12-15$ days, more pronounced for the multi-sector targets.
This is similar to the pattern initially observed in the search which led to the corrections introduced in Section~\ref{sec:corrections}, albeit now being considerably less noticeable.
The cause of this excess of signals is the uncorrected scattered light near the start and end of the \textit{TESS} orbits, an issue that the added corrections targeted.

This sample of $\sim$3000 targets constitutes a reproducible, magnitude-limited sample of candidate transit signals statistically validated as significant, pertaining to the entire \textit{TESS} southern hemisphere.
It can potentially be used as a starting point for occurrence rate studies (see Section~\ref{sub:future_work}), requiring additional physical validation of all signals, for example through a systematic application of the \texttt{VESPA} tool.
For this thesis, I decided to instead focus on the confirmation of the most promising planet candidates in the sample, so I adopted a more informal approach.
This approach aimed at identifying and ranking a smaller sample of potential planet candidates, for which RV follow-up could be procured.

For that, all targets in the sample of 3000 were visually inspected, an exercise done primarily by analyzing the targets' corresponding summary plot (for an example, see Figure~\ref{fig:summary_plot_example}), with particular attention given to the signal's shape and properties.
This was done in an attempt to distinguish all signals that could be categorized as physical in nature (either from a planet or another astrophysical system) and resulted in 254 targets classified as physical signals.

The signals of all targets in this physical sample were then evaluated with \texttt{VESPA} to identify astrophysical false positives (EBs, BEBs, HEBs or other unknown signals; see Section~\ref{sec:transit_validation}).
Additionally, the visual analysis was repeated, this time aimed at determining the physical nature of the signal, as an additional result for comparison with the verdict of \texttt{VESPA}.
Specifically, I searched for patterns of certain astrophysical scenarios that can be easily identified visually or from a signal's properties, which can either reinforce, if there is agreement, or call into question, if not, the conclusions reached by \texttt{VESPA}.
Some common visual patterns and properties I looked for in the candidate transits were:
\begin{itemize}[noitemsep, topsep=1ex]
    \item Transit with a V-shape, indicative of an eclipsing binary.
    \item Transit signals are only present in some of the \textit{TESS} sectors, indicative of a signal not originating from the target, possibly a foreground eclipsing binary.
    \item Transit depths larger than 1\%, indicative that the signal is (likely) not from a planet.
    \item Candidates with a significant secondary eclipse, indicative of an eclipsing binary.
    \item Candidates with a significant difference between odd and even transits ($\sigma > 5$), indicative of a secondary eclipse, and consequently an eclipsing binary.
\end{itemize}

\begin{figure}[!ptb]
    \centering
    \includegraphics[width=\textwidth, height=\textheight, keepaspectratio]{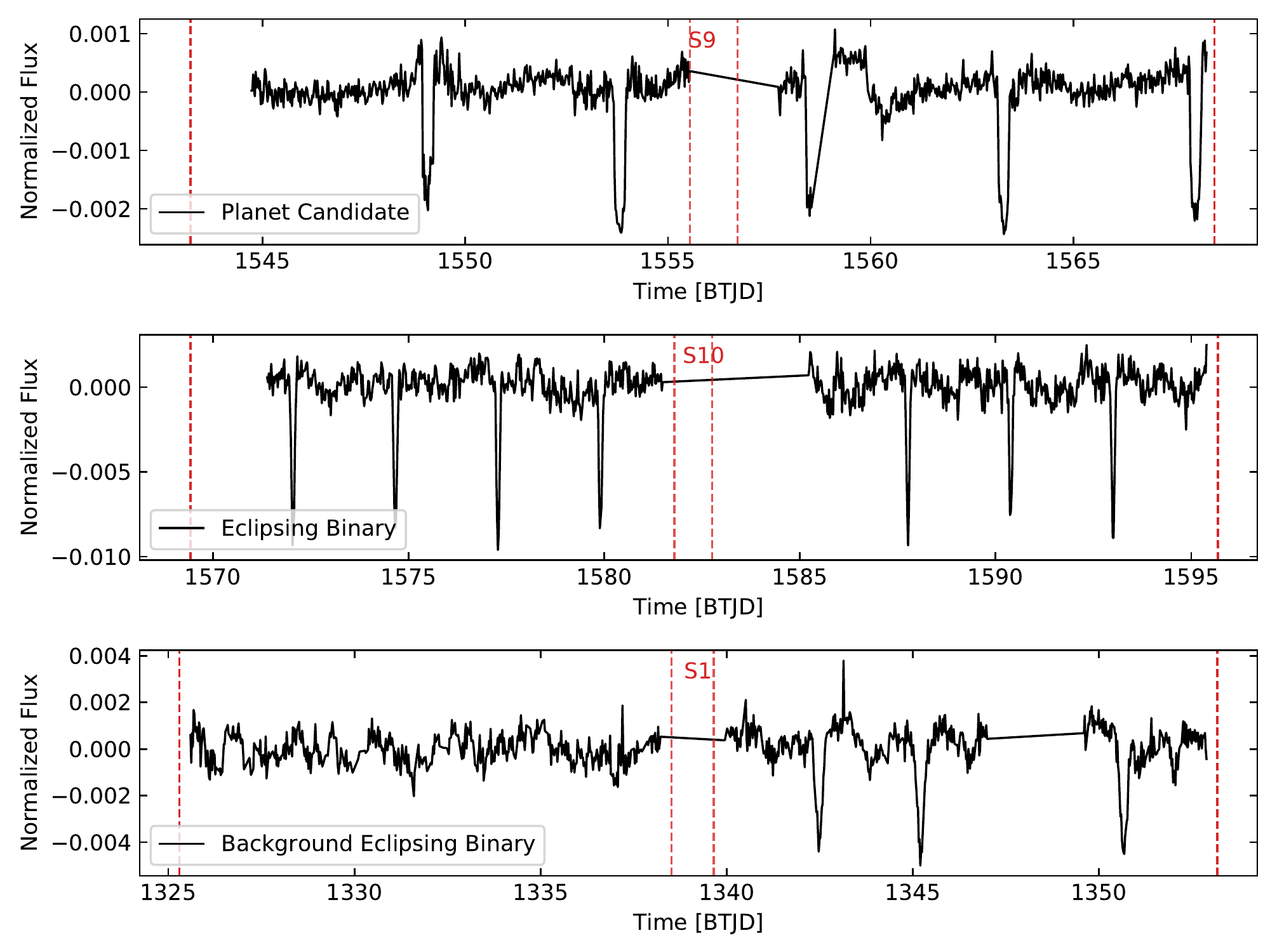}
    \caption[Three light curves demonstrating expected eclipse signals of different astrophysical scenarios]
    {Three light curves demonstrating expected eclipse signals of different astrophysical scenarios.
    The vertical red dashed lines denote the start and end of the two \textit{TESS} orbits for each target's sector.
    \textit{Top}: Light curve has signals with a round shape and relatively shallow depth, and was visually identified as a planet candidate.
    \textit{Middle}: Light curve is V-shaped and has deep signals, commonly associated with eclipsing binaries.
    \textit{Bottom}: Light curve with signals similar to those in the middle panel, but which are only present in the second orbit of the depicted sector.
    This signal was classified visually as originating from a background eclipsing binary after inspection of the target pixel files from which the light curve was extracted (see text for details).}
    \label{fig:astrophysical_examples_transits}
\end{figure}
Figure~\ref{fig:astrophysical_examples_transits} depicts examples of light curves examined during this search and illustrates some of those characteristic patterns for specific astrophysical scenarios, namely for a planet candidate, an eclipsing binary and a background eclipsing binary, from top to bottom, respectively.
The bottom panel of the figure, in particular, highlights an unexpected example of a light curve that required a more comprehensive analysis to determine its source, which we detail below.

Finally, taking into consideration the results of both \texttt{VESPA} and the visual inspection, transits from potential planet candidates were singled out and ranked.
All in all, I selected 4 targets as having a high-probability of hosting a planet, out of the 254 signals identified as being of physical origin.
This points to $\sim 1.6$\% of all physical signals identified being of planetary origin or, the other way around, $\sim 98.4$\% of the physical signals being astrophysical false positives.
This false-positive rate is in agreement with the value found by \textcite{Sliski_2014} for \textit{Kepler}'s giant stars with a single transiting object of $70\% \pm 30\%$.

\begin{figure}[!ptb]
    \centering
    \includegraphics[width=\textwidth, height=\textheight, keepaspectratio]{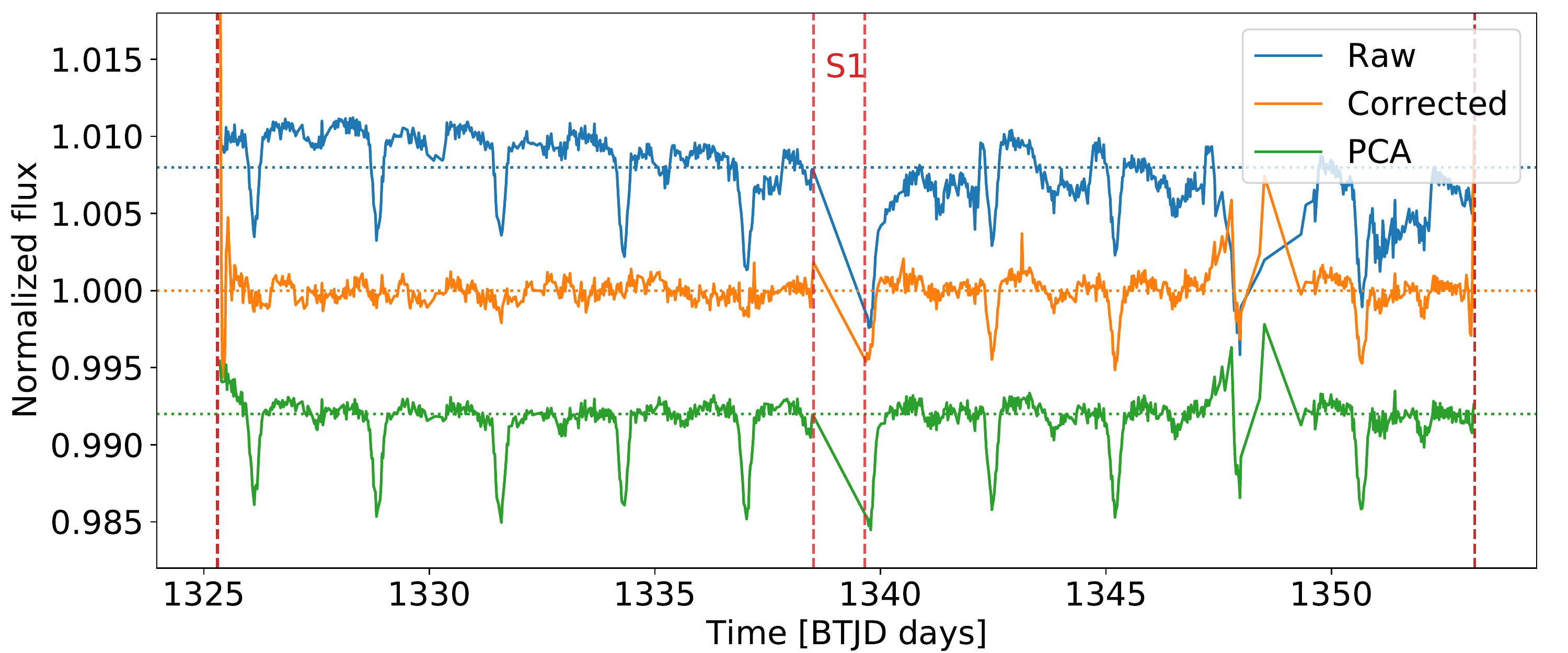}
    \caption[Raw, corrected and pca light curves produced by \texttt{eleanor} for the background eclipsing binary in the lower panel of Figure~\ref{fig:astrophysical_examples_transits}]
    {Raw, corrected and pca light curves produced by \texttt{eleanor} for the background eclipsing binary in the lower panel of Figure~\ref{fig:astrophysical_examples_transits}.
    The vertical red dashed lines denote the start and end of the two \textit{TESS} orbits for the target's sector.
    Note how the corrected light curve does not have a V-shaped transit signal during the first orbit, unlike the raw and pca ones.}
    \label{fig:beb_lcs}
\end{figure}

\begin{figure}[!ptb]
    \centering
    \includegraphics[width=0.6\textwidth, height=\textheight, keepaspectratio]{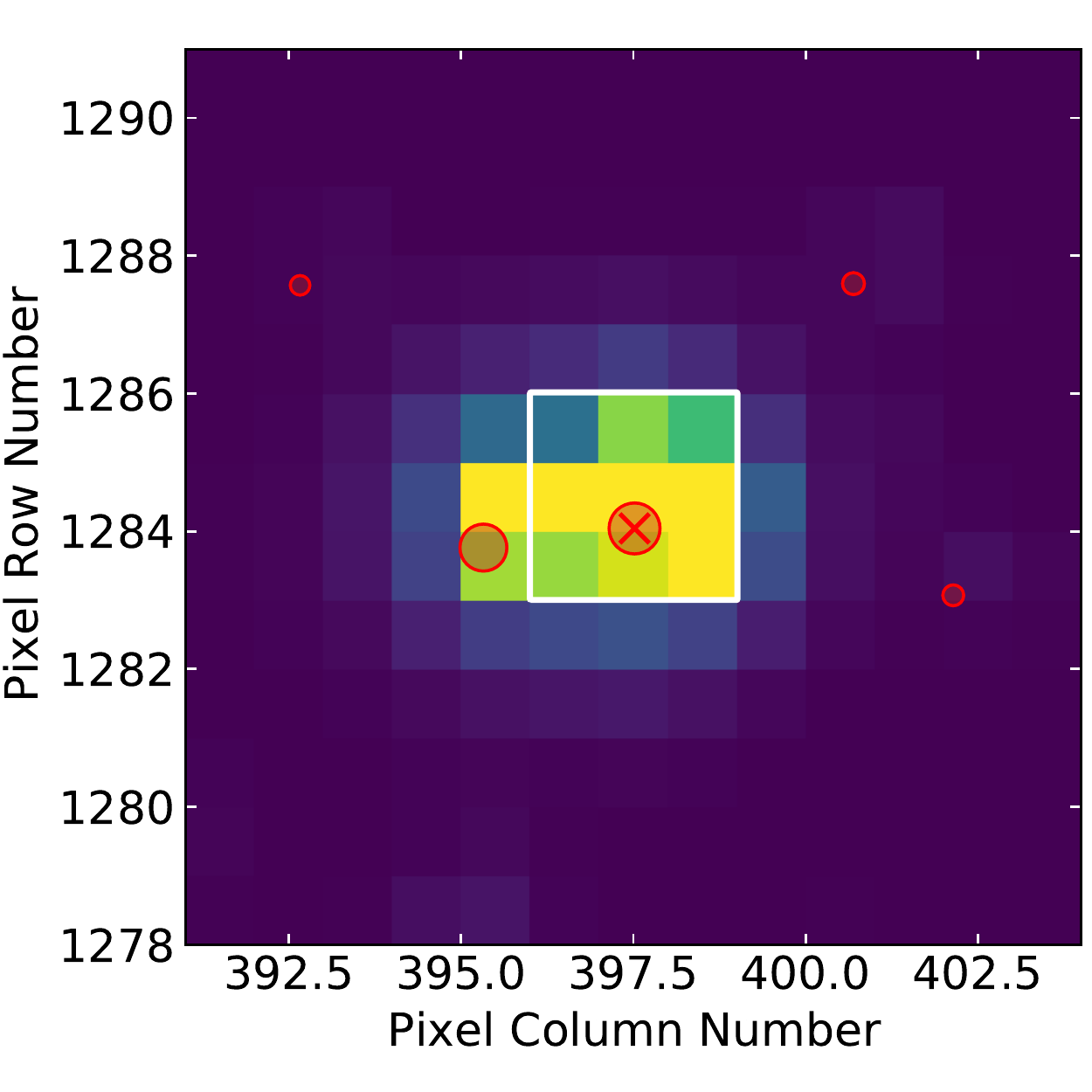}
    \includegraphics[width=0.9\textwidth, height=\textheight, keepaspectratio]{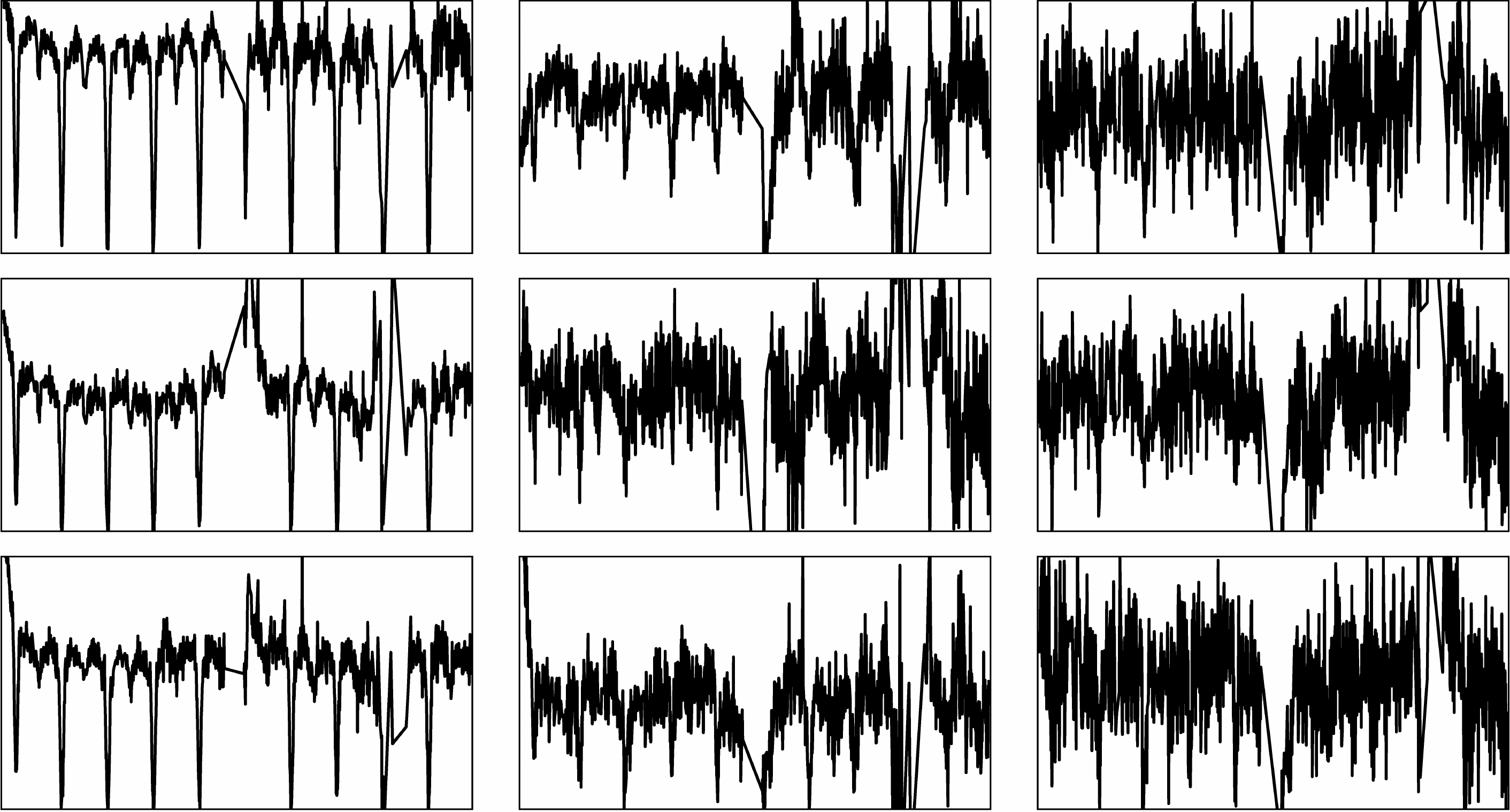}
    \caption[Pixel file inspection of the background eclipsing binary in the lower panel of Figure~\ref{fig:astrophysical_examples_transits}]
    {Pixel file inspection of the background eclipsing binary in the lower panel of Figure~\ref{fig:astrophysical_examples_transits}.
    \textit{Top}: Single frame from the target pixel file of the target.
    Red circles denote all stars in the field of the observed target that are brighter than \textit{Gaia} $G$ magnitude 16, with the size scaled to that same magnitude (the circle with a cross in its center denotes the target).
    The white solid line denotes the aperture selected by \texttt{eleanor} to extract the target's light curve.
    Note the bright star to the left of the main target, close to the bounds of the aperture.
    \textit{Bottom}: Light curves corresponding to the individual flux contributions of each of the nine pixels inside the target's selected aperture, throughout the entire sector.
    The presence of the signal from an eclipsing binary concentrated only on the pixels on the left. 
    Together with the presence of a star close to the bounds of the aperture on this same side suggests that the V-shaped signals are coming from this external star.}
    \label{fig:beb_pixel_analysis}
\end{figure}

Taking a closer look at the signal classified as a background eclipsing binary in the bottom panel of Figure~\ref{fig:astrophysical_examples_transits}, an inspection of the three light curves produced by \texttt{eleanor} for this target, shown in Figure~\ref{fig:beb_lcs}, revealed that only the ``corrected'' light curve did not show the V-shaped signal during the target's first orbit.
To understand why that was the case, Figure~\ref{fig:beb_pixel_analysis} shows two diagnostic panels enabling a closer look at the data in the pixel files.

On the top, a single frame from the pixel files is shown, with all stars in the field of the observed target denoted by red circles (and the target itself with an additional red cross in its center).
The white solid line denotes the aperture selected by \texttt{eleanor} to extract the target's light curve.
Of particular note is the bright star to the left of the main target, close to the bounds of the aperture.
Turning now to the bottom panel, nine light curves are shown, corresponding to the individual flux contributions of the nine pixels included in the selected aperture.
The conspicuous V-shaped signal is clearly visible in the light curves from the leftmost pixels, on the same side of the nearby bright star close to the aperture.
This pointed to the source of the eclipsing binary signal being from this close-by star.

The next step was to understand why this signal is only present in the second orbit of the ``corrected'' light curve, unlike the ``pca'' light curve. 
Both these light curves are produced from a different set of corrections applied to the ``raw'' light curve, as mentioned in Section~\ref{sec:photometry}, indicating that the absence of the eclipse-like signal during the first orbit of the ``corrected'' light curve was likely caused by these corrections.

The ``pca'' light curve has corrections applied based on the CBVs provided by NASA (see Section~\ref{sec:photometry}), which are calculated per \textit{TESS} camera, meaning the contribution of a single nearby star is not taken into account.
On the other hand, the ``corrected'' light curve includes background correction, calculated using all pixels from the target pixel file with the exception of the ones from the aperture.
In this case, it seems plausible that \texttt{eleanor} might have been able to correctly remove this V-shaped signal only for data from the first orbit, leading to this strange signal in the light curve.

%% file: chapters/search/candidates.tex
\section{Planet candidates}
\label{sec:planet_candidates}

From the search for transits in the \textit{TESS} southern sample, four planet candidates were identified, TIC 55092869.01, TIC 441462736.01, TIC 348835438.01 and TIC 204650483.01.
Of these, the former two are known hosts, and the latter two had not been yet identified as planet candidates.
In this section, I introduce each candidate and their host, show the results from the characterization of the system using the GP time-domain model and discuss some of its properties.
For one of these systems, for which RV follow-up data are available, I also do an additional joint characterization of its transit and radial-velocity signals, to obtain both the mass and radius of the planet, and consequently, its mean density.

Furthermore, I introduce an additional candidate, TIC 394918211.01, not part of the sample, which was identified by colleagues in our collaboration and for which RV follow-up observations have also been acquired. 
For that candidate I also characterized both the \textit{TESS} light curve individually and the transit and RVs simultaneously.

Table~\ref{tab:vespa_results} shows the results from the \texttt{VESPA} automated analysis for these five candidates.
The middle column denotes the astrophysical false-positive probability, that is, the probability that the transit is not of planetary nature but instead originates from another astrophysical scenario.
The rightmost column points to the figure associated with the target's analysis, in Appendix~\ref{app:vespa_plots}.
\begin{table}[!ptb]
    \centering
    \setlength{\tabcolsep}{14pt}
    \renewcommand{\arraystretch}{1.1}
    \begin{tabular}{lll}
        \toprule
        \toprule
        Target                      & FPP                       & Figure                                \\
        \midrule
        TIC 55092869.01             & 6.54 $\times$ 10$^{-6}$   & Figure~\ref{fig:vespa_summary_tic550} \\
        TIC 441462736.01            & 1.32 $\times$ 10$^{-2}$   & Figure~\ref{fig:vespa_summary_tic441} \\
        TIC 348835438.01            & 8.99 $\times$ 10$^{-1}$   & Figure~\ref{fig:vespa_summary_tic348} \\
        TIC 204650483.01            & 2.77 $\times$ 10$^{-1}$   & Figure~\ref{fig:vespa_summary_tic204} \\
        TIC 394918211.01            & 6.64 $\times$ 10$^{-3}$   & Figure~\ref{fig:vespa_summary_tic394} \\
        \bottomrule
    \end{tabular}
    \caption[\texttt{VESPA} false-positive probabilities inferred for each of the five planet candidates identified in this work's \textit{TESS} southern search for transits]
    {\texttt{VESPA} false-positive probabilities inferred for each of the five planet candidates identified in this work's \textit{TESS} southern search for transits.
    The middle column denotes the astrophysical false-positive probability, that is, the probability that the transit is not of planetary nature but due to another astrophysical scenario.
    The rightmost column points to the figure associated with the target's analysis, in Appendix~\ref{app:vespa_plots}.}
    \label{tab:vespa_results}
\end{table}

Considering a 1\% false-positive rate as the limit to consider a candidate as of planetary nature, three of the five candidates in this list would not be statistically classified as planets.
One of these three, TIC 441462736.01, is a known planet \parencite{Huber_2019b}.
The other two have also been tentatively confirmed as planets, as I will show below (Pereira et al. in prep., Grunblatt et al. in prep.).

These classifications highlight one important aspect of the application of \texttt{VESPA}, that is, whilst it is a very powerful tool for population studies \parencite{Morton_2016}, when used on a case-by-case basis, care must be taken.
This reinforces the decision to do a visual classification of the physical signals in addition to the analysis based on \texttt{VESPA}.
It was by considering the results of both these analyses, which is inevitably subjective due to the nature of a visual classification, that all five planet candidates were selected.
It bears mention that promising transit signals were discussed and cross-checked with other pipelines within the context of our collaboration, emphasizing the meticulousness of the process followed to select and rank planet candidates for RV follow-up.

Table~\ref{tab:stellar_properties} details the stellar properties of the five candidates' hosts.
For the most part, the values are taken from the \textit{TESS} Input Catalog \parencite{Stassun_2019}, with a dash meaning there are no available data (no stellar masses are present in the TIC for any of the five targets).
For the two known hosts, stellar parameters are taken directly from the discovery papers, as noted by the ``Source'' column in the table.
\begin{table}[!ptb]
    \centering
    \setlength{\tabcolsep}{12pt}
    \renewcommand{\arraystretch}{1.0}
    \begin{tabular}{lll}
        \toprule
        \toprule
        Property                                        & Value                     & Source                        \\
        \midrule
        \multicolumn{3}{c}{TIC 55092869 (KELT-11)}                                                                  \\
        \midrule
        \textit{TESS} magnitude                         & 7.32                      & TIC, \textcite{Stassun_2019}  \\ 
        \textit{V} magnitude                            & 8.04                      & TIC, \textcite{Stassun_2019}  \\
        Parallax, $\pi$ [mas]                           & 10.056 $\pm$ 0.0526       & TIC, \textcite{Stassun_2019}  \\
        Effective temperature, $T_\text{eff}$ [K]       & 5370 $\pm$ 50             & \textcite{Pepper_2017}        \\
        Stellar radius, $R_\star$ [$\rm R_\odot$]       & 2.72 $\pm$ 0.20           & \textcite{Pepper_2017}        \\
        Stellar mass, $M_\star$ [$\rm M_\odot$]         & 1.438 $\pm$ 0.06          & \textcite{Pepper_2017}        \\
        \midrule
        \multicolumn{3}{c}{TIC 441462736 (TOI 197)}                                                                 \\
        \midrule
        \textit{TESS} magnitude                         & 7.36                      & TIC, \textcite{Stassun_2019}  \\ 
        \textit{V} magnitude                            & 8.15                      & TIC, \textcite{Stassun_2019}  \\
        Parallax, $\pi$ [mas]                           & 10.518 $\pm$ 0.080        & \textcite{Huber_2019b}        \\
        Effective temperature, $T_\text{eff}$ [K]       & 5080 $\pm$ 90             & \textcite{Huber_2019b}        \\
        Stellar radius, $R_\star$ [$\rm R_\odot$]       & 2.943 $\pm$ 0.064         & \textcite{Huber_2019b}        \\
        Stellar mass, $M_\star$ [$\rm M_\odot$]         & 1.212 $\pm$ 0.074         & \textcite{Huber_2019b}        \\
        \midrule
        \multicolumn{3}{c}{TIC 348835438}                                                                           \\
        \midrule
        \textit{TESS} magnitude                         & 9.99                      & TIC, \textcite{Stassun_2019}  \\ 
        \textit{V} magnitude                            & 10.92                     & TIC, \textcite{Stassun_2019}  \\
        Parallax, $\pi$ [mas]                           & 2.513 $\pm$ 0.041         & TIC, \textcite{Stassun_2019}  \\
        Effective temperature, $T_\text{eff}$ [K]       & 4859 $\pm$ 122            & TIC, \textcite{Stassun_2019}  \\
        Stellar radius, $R_\star$ [$\rm R_\odot$]       & 3.98                      & TIC, \textcite{Stassun_2019}  \\
        Stellar mass, $M_\star$ [$\rm M_\odot$]         & ---                       & TIC, \textcite{Stassun_2019}  \\
        \midrule
        \multicolumn{3}{c}{TIC 204650483}                                                                           \\
        \midrule
        \textit{TESS} magnitude                         & 9.03                      & TIC, \textcite{Stassun_2019}  \\ 
        \textit{V} magnitude                            & 9.97                      & TIC, \textcite{Stassun_2019}  \\
        Parallax, $\pi$ [mas]                           & 4.602 $\pm$ 0.052         & TIC, \textcite{Stassun_2019}  \\
        Effective temperature, $T_\text{eff}$ [K]       & 4956 $\pm$ 122            & TIC, \textcite{Stassun_2019}  \\
        Stellar radius, $R_\star$ [$\rm R_\odot$]       & 3.38                      & TIC, \textcite{Stassun_2019}  \\
        Stellar mass, $M_\star$ [$\rm M_\odot$]         & ---                       & TIC, \textcite{Stassun_2019}  \\
        \midrule
        \multicolumn{3}{c}{TIC 394918211}                                                                           \\
        \midrule
        \textit{TESS} magnitude                     & 10.79                     & TIC, \textcite{Stassun_2019}  \\ 
        \textit{V} magnitude                        & 11.71                     & TIC, \textcite{Stassun_2019}  \\
        Parallax, $\pi$ [mas]                       & 2.165 $\pm$ 0.023         & TIC, \textcite{Stassun_2019}  \\
        Effective temperature, $T_\text{eff}$ [K]   & 4913 $\pm$ 122            & TIC, \textcite{Stassun_2019}  \\
        Stellar radius, $R_\star$ [$\rm R_\odot$]       & 3.41                      & TIC, \textcite{Stassun_2019}  \\
        Stellar mass, $M_\star$ [$\rm M_\odot$]         & ---                       & TIC, \textcite{Stassun_2019}  \\
        \bottomrule
    \end{tabular}
    \caption[Available stellar properties for the five stars with planet candidates identified in this work's \textit{TESS} southern search for transits]
    {Available stellar properties for the five stars with planet candidates identified in this work's \textit{TESS} southern search for transits.
    Most information is taken from the \textit{TESS} Input Catalog \parencite{Stassun_2019}, with stellar parameters of the two known hosts taken from their respective discovery papers.
    A dash means no available values in the TIC.}
    \label{tab:stellar_properties}
\end{table}


\subsection{TIC 55092869.01}
\label{sec:kelt11b}

\begin{table}[!ptb]
    \centering
    \setlength{\tabcolsep}{14pt}
    \renewcommand{\arraystretch}{1.1}
    \begin{tabular}{lllll}
        \toprule
        \toprule
        Parameter                            & Prior                                & Posterior        & 84\%             & 16\%            \\
        \midrule
        $a_\text{meso}$ [ppm]                & $\mathcal{U}$(10, 1000)              & 262.91           & +46.75           & -33.66          \\ 
        $b_\text{meso}$ [$\mu$Hz]            & $\mathcal{U}$(1, 280)                & 5.01             & +0.90            & -0.82           \\ 
        $\sigma$ [ppm]                       & $\mathcal{U}$(10, 1000)              & 158.72           & +3.75            & -3.84           \\ 
        $P$ [days]                           & $\mathcal{U}$(4.2, 5.2)              & 4.73522          & +0.00049         & -0.00051        \\ 
        $t_0$ [BTJD]                         & $\mathcal{U}$(1548.0, 1550.0)        & 1549.0796        & +0.0017          & -0.0019         \\ 
        $R_\text{p} / R_\star$               & $\mathcal{LU}$(0.001, 0.1)           & 0.0456           & +0.0009          & -0.0006         \\ 
        $a / R_\star$                        & $\mathcal{LU}$(1.5, 20)              & 4.46             & +0.88            & -0.90           \\ 
        $i$ [deg]                            & $\mathcal{U}$(60.0, 90.0)            & 85.65            & +2.92            & -4.99           \\ 
        $e$                                  & $\mathcal{U}$(0.0, 0.5)              & 0.26             & +0.17            & -0.17           \\ 
        $\omega$ [deg]                       & $\mathcal{U}$(0.0, 360.0)            & 123.66           & +107.68          & -68.48          \\ 
        $u_1$                                & $\mathcal{N}$(0.6, 0.1)              & 0.49             & +0.06            & -0.06           \\ 
        $u_2$                                & $\mathcal{N}$(0.09, 0.1)             & -0.03            & +0.09            & -0.09           \\ 
        \bottomrule
    \end{tabular}
    \caption[Prior and posterior distributions for all parameters in the GP + transit model obtained in the characterization of the \textit{TESS} light curve of TIC 55092869.01 (KELT-11 b)]
    {Prior and posterior distributions for all parameters in the GP + transit model obtained in the characterization of the \textit{TESS} light curve of TIC 55092869.01 (KELT-11 b).
    Figure~\ref{fig:tic550_param_hist} shows an alternative look at the posterior distributions of each parameter, through an histogram of all the posterior samples drawn during the fit.}
    \label{tab:tic550_params}
\end{table}
Planet candidate TIC 55092869.01 is a known giant planet, commonly known as KELT-11 b\footnote{\url{https://exofop.ipac.caltech.edu/tess/target.php?id=55092869}} and discovered by \textcite{Pepper_2017}.
The planet has a radius of $1.37_{-0.12}^{+0.15} \ R_\text{J}$ and a mass of $0.195_{-0.018}^{+0.019} \ M_\text{J}$, and orbits its host star on a period of $4.736529^{+0.000068}_{-0.000059}$ days.
The host star, KELT-11, is a subgiant with a radius of $2.72_{-0.17}^{+0.21}$ $\rm R_\odot$.

Table~\ref{tab:tic550_params} shows the prior and posterior distributions of each parameter of the GP + transit model fitted to the \textit{TESS} light curve of this target.
Priors follow uniform ($\mathcal{U}$), log-uniform ($\mathcal{LU}$) and normal ($\mathcal{N}$) distributions.
Figure~\ref{fig:tic550_param_hist} shows an alternative look at the posterior distributions of each parameter, through an histogram of all the posterior samples drawn during the fit.
Note that the estimates on the eccentricity $e$ and argument of periapsis $\omega$ are unconstrained since there is a degeneracy between these parameters when just fitting the light curve data, something also observed in all subsequent characterizations with the GP + transit model.

\begin{figure}[!ptb]
    \centering
    \includegraphics[width=0.91\textwidth, height=\textheight, keepaspectratio]{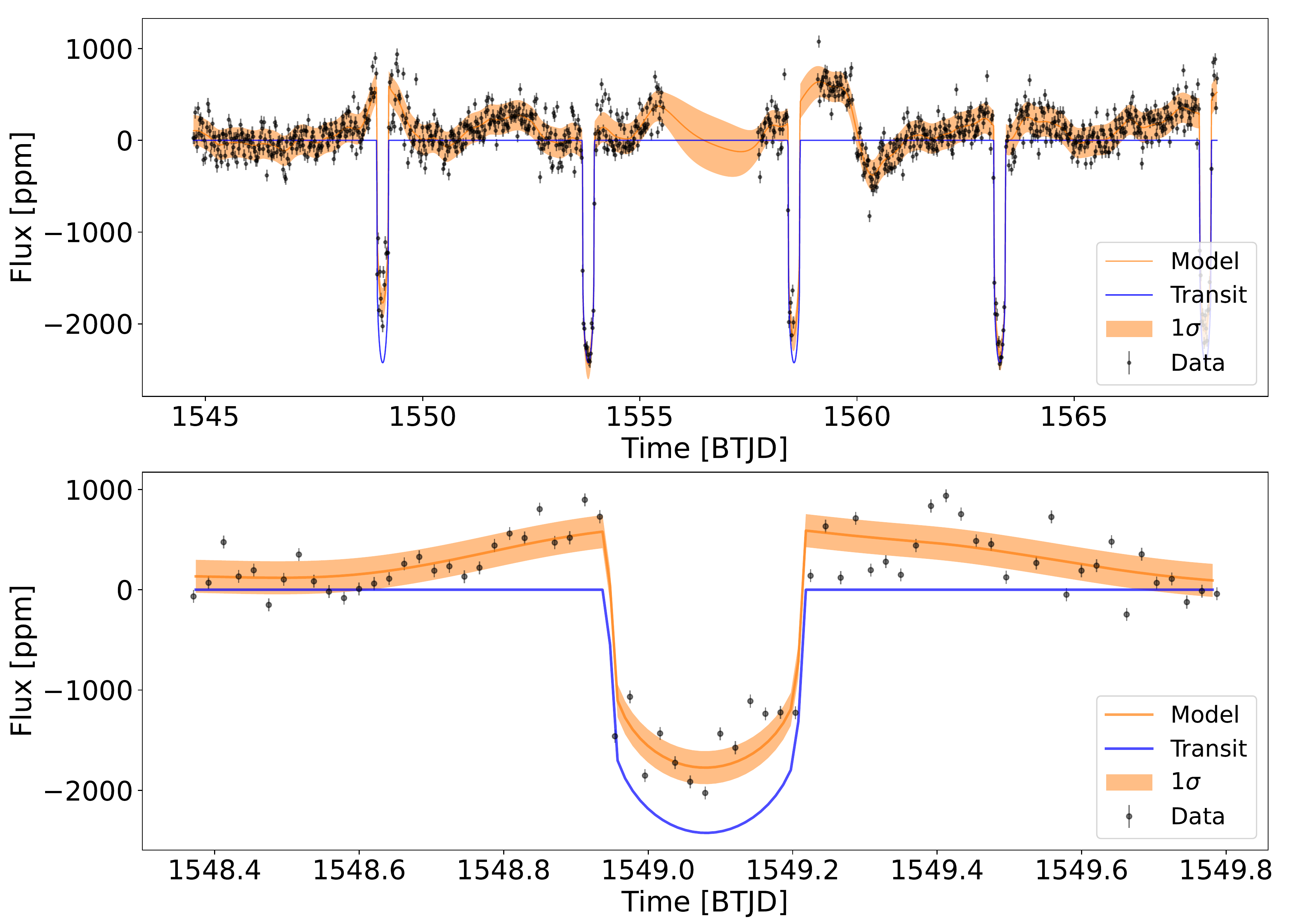}
    \includegraphics[width=0.91\textwidth, height=\textheight, keepaspectratio]{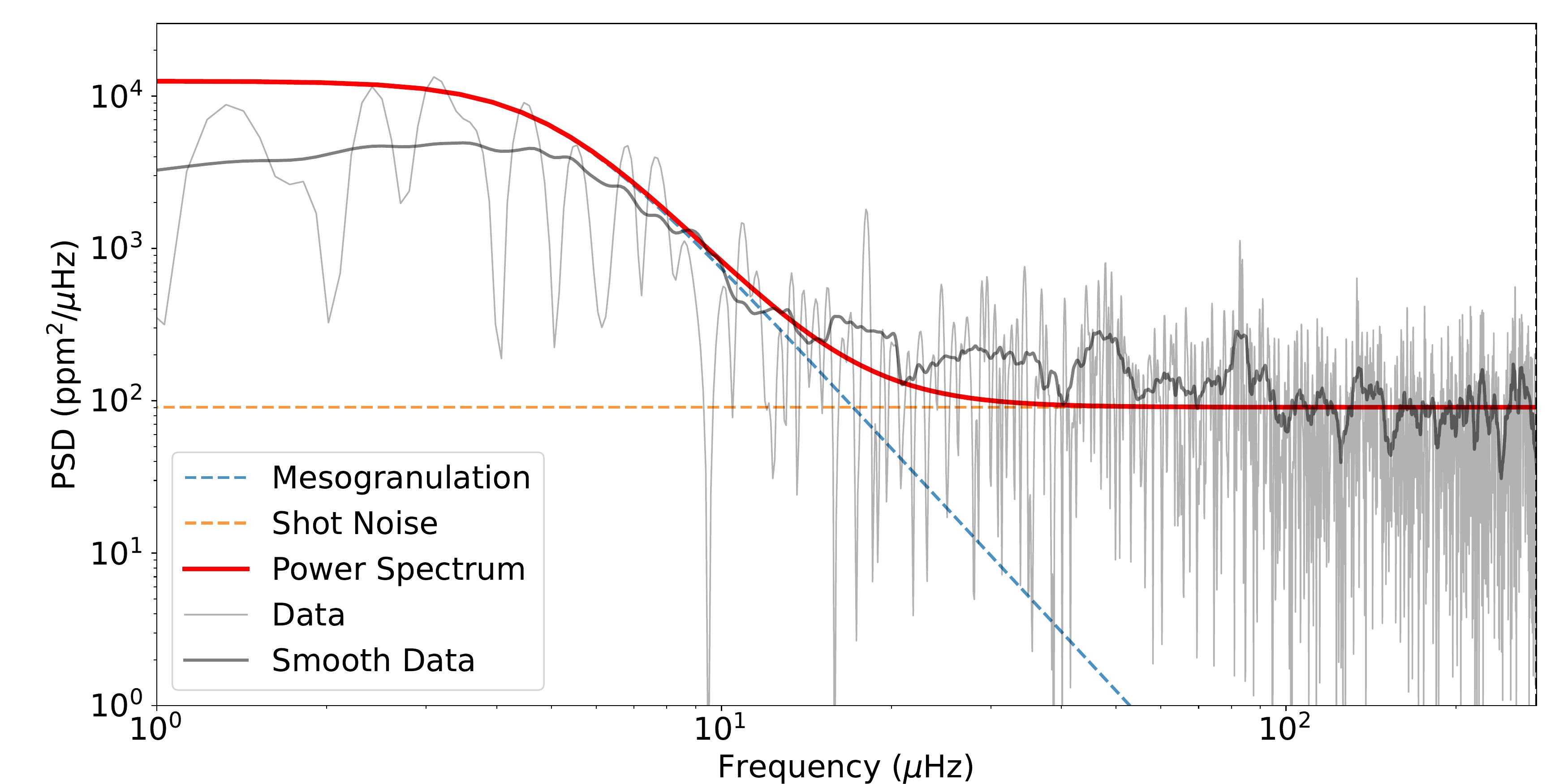}
    \caption[Light curve fit and corresponding power spectrum from the characterization of the \textit{TESS} light curve of planet candidate TIC 55092869.01 (KELT-11 b)]
    {\textit{Top}: Light curve fit of planet candidate TIC 55092869.01 (KELT-11 b).
    Black points and error bars correspond to the light curve data.
    The GP predictive model is shown in orange, with the solid central line denoting the median of the distribution and the shaded area the 1-$\sigma$ interval.
    The transit component of the model is depicted by the blue solid line.
    The panel below it has a zoom in of the first transit of the light curve.
    \textit{Bottom}: Power spectrum of the light curve in the panel above.
    The PSD of the light curve is shown in light gray, with a slightly smoothed version overlapped in dark gray.
    The PSD of the sum of the GP components in the model in the panels above is shown as a solid red curve, with individual components identified by different colors (see legend).}
    \label{fig:tic550_fit}
\end{figure}

Figure~\ref{fig:tic550_fit} shows the light curve with the complete model in orange, as well as the transit component separately (in blue) in the two top panels.
Below is the PSD of the GP model, and its individual components, overplotted on the power spectrum of the light curve, with the contributions from the transit removed.
The oscillations envelope component was not included in the model since the oscillations for this target are expected to be present at higher frequencies than the Nyquist frequency of \textit{TESS} 30-min cadence data ($\sim 283 \mu \rm Hz$).

Looking at the light curve in the upper panel of Figure~\ref{fig:tic550_fit}, one noticeable feature that can be observed are the out-of-transit variations.
These are not found in the light curve from \textcite{Pepper_2017}, although, given that their data originated from ground-based observations, the noise level might have been too large for such a detection.
The start of Chapter~\ref{cha:transits} describes common causes for these out-of-transit variations, mentioning, in particular, the work of \textcite{Lillo-Box_2014} on the close-in giant planet Kepler-91 b, which orbits a red-giant star. 
In my case, my model does not attempt to characterize these variations, and is limited to the transit signal and the GP with components for the mesogranulation and the white noise.
Nevertheless, a characterization of the \textit{TESS} transit of KELT-11 b accounting for this out-of-transit variability would be an interesting exercise for a comparison with published pre-\textit{TESS} analyses.

Comparing the derived planetary parameters to the ones from \textcite{Pepper_2017}, the periods are in agreement within $3$-$\sigma$.
As for the planetary radius, and considering the stellar radius from \textcite{Pepper_2017}, I obtain a value of $1.207^{+0.117}_{-0.091} \ R_\text{J}$, in agreement within $2$-$\sigma$ with the published one.
This difference in radius might be due to the larger scatter in the photometric data from \textcite{Pepper_2017}, when compared to the \textit{TESS} light curve (see figures~1 and 2 of \textcite{Pepper_2017}).


\subsection{TIC 441462736.01} 

\begin{table}[!ptb]
    \centering
    \setlength{\tabcolsep}{14pt}
    \renewcommand{\arraystretch}{1.1}
    \begin{tabular}{lllll}
        \toprule
        \toprule
        Parameter                            & Prior                                & Posterior        & 84\%             & 16\%            \\
        \midrule
        $a_\text{meso}$ [ppm]                & $\mathcal{U}$(10, 1000)              & 82.22            & +8.07            & -7.28           \\ 
        $b_\text{meso}$ [$\mu$Hz]            & $\mathcal{U}$(1, 280)                & 15.35            & +4.42            & -3.82           \\ 
        $\sigma$ [ppm]                       & $\mathcal{U}$(10, 1000)              & 115.42           & +3.24            & -3.01           \\ 
        $P$ [days]                           & $\mathcal{U}$(13.8, 14.8)            & 14.27346         & +0.00725         & -0.00695        \\ 
        $t_0$ [BTJD]                         & $\mathcal{U}$(1356.0, 1358.0)        & 1357.0156        & +0.0055          & -0.0053         \\ 
        $R_\text{p} / R_\star$               & $\mathcal{LU}$(0.001, 0.1)           & 0.0263           & +0.0021          & -0.0016         \\ 
        $a / R_\star$                        & $\mathcal{LU}$(1.5, 20)              & 8.90             & +2.99            & -2.26           \\ 
        $i$ [deg]                            & $\mathcal{U}$(60.0, 90.0)            & 86.18            & +2.78            & -4.57           \\ 
        $e$                                  & $\mathcal{U}$(0.0, 0.5)              & 0.27             & +0.17            & -0.19           \\ 
        $\omega$ [deg]                       & $\mathcal{U}$(0.0, 360.0)            & 123.03           & +129.54          & -80.99          \\ 
        $u_1$                                & $\mathcal{N}$(0.6, 0.1)              & 0.57             & +0.09            & -0.09           \\ 
        $u_2$                                & $\mathcal{N}$(0.09, 0.1)             & 0.07             & +0.10            & -0.10           \\ 
        \bottomrule
    \end{tabular}
    \caption[Prior and posterior distributions for all parameters in the GP + transit model obtained in the characterization of the \textit{TESS} light curve of TIC 441462736.01 (TOI 197.01; HD 221416 b)]
    {Prior and posterior distributions for all parameters in the GP + transit model obtained in the characterization of the \textit{TESS} light curve of TIC 441462736.01 (TOI 197.01; HD 221416 b).
    Figure~\ref{fig:tic441_param_hist} shows an alternative look at the posterior distributions of each parameter, through an histogram of all the posterior samples drawn during the fit.}
    \label{tab:tic441_params}
\end{table}
Planet candidate TIC 441462736.01 is also a known planet, discovered in short-cadence \textit{TESS} data by \textcite{Huber_2019b}, given the denomination of TOI 197.01\footnote{\url{https://exofop.ipac.caltech.edu/tess/target.php?id=441462736}} and also known as HD 221416 b.
According to the discovery paper, the planet has a radius of $0.836^{+0.031}_{-0.028} \ R_\text{J}$ and a mass of $0.190 \pm 0.018 \ M_\text{J}$, orbiting its host star on a period of $14.2767 \pm 0.0037$ days.
The host is a subgiant with a radius of $2.943 \pm 0.064$ $\rm R_\odot$ and a mass of $1.212 \pm 0.074$ $\rm M_\odot$.
    
\begin{figure}[!ptb]
    \centering
    \includegraphics[width=0.91\textwidth, height=\textheight, keepaspectratio]{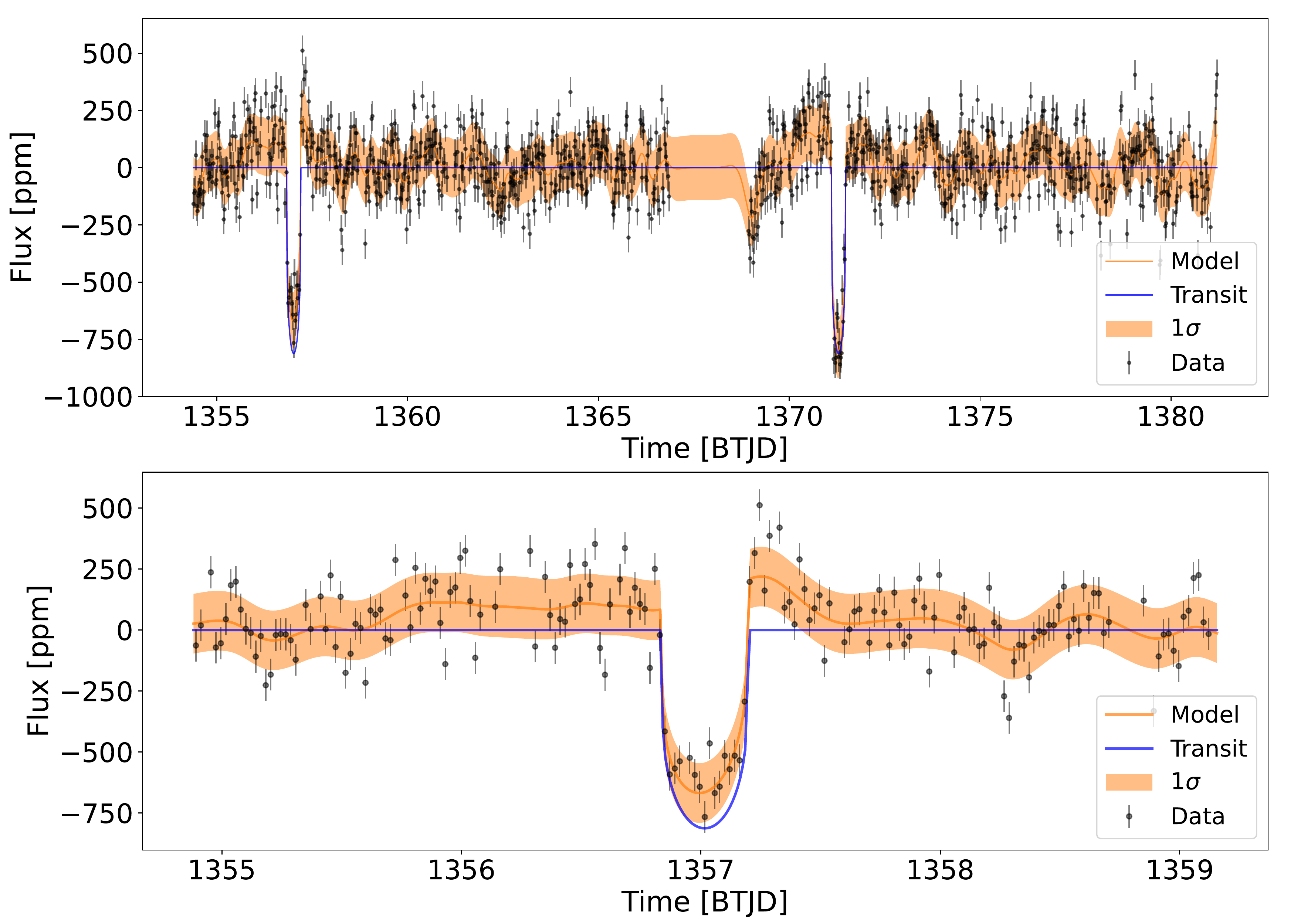}
    \includegraphics[width=0.91\textwidth, height=\textheight, keepaspectratio]{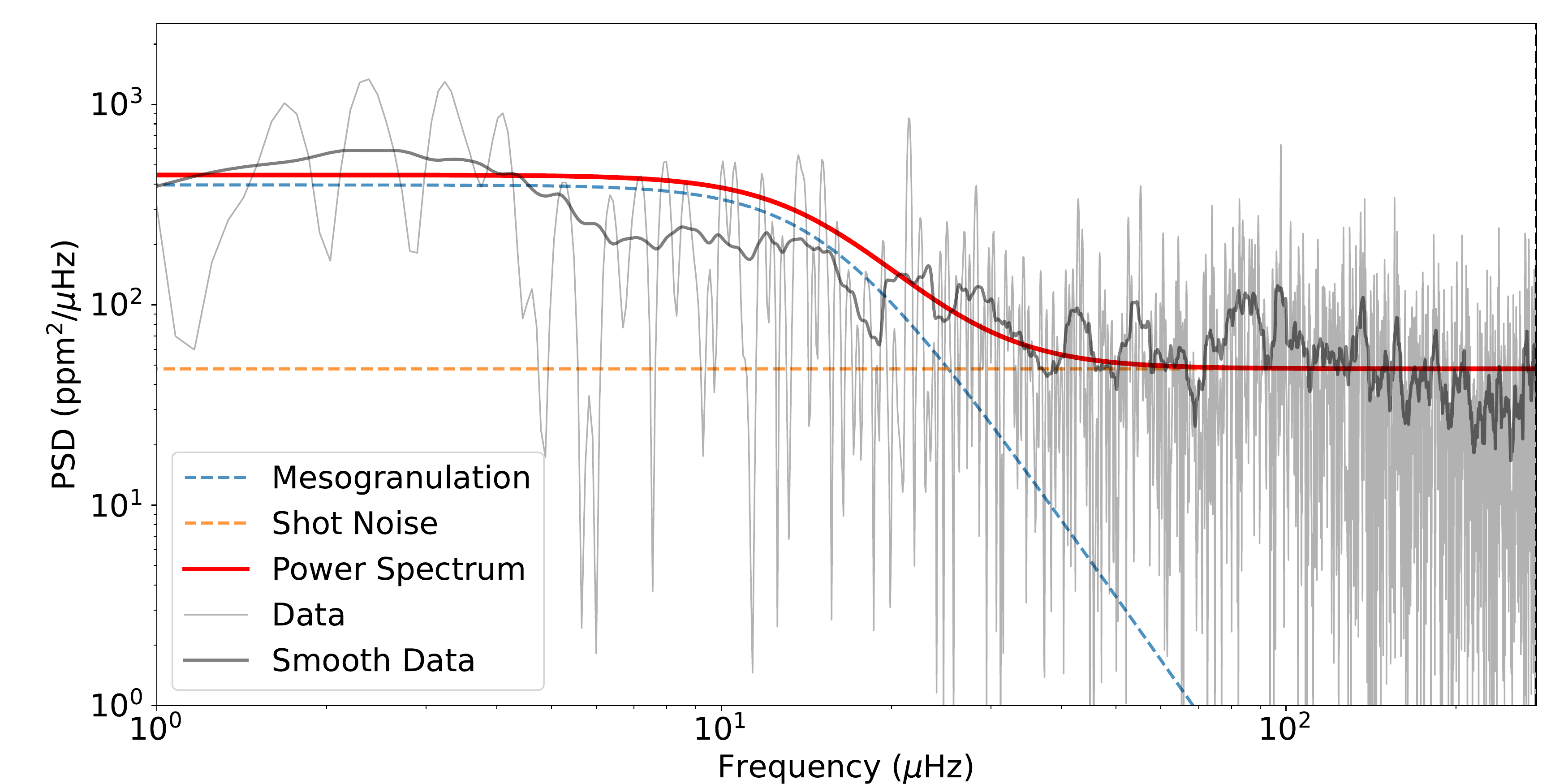}
    \caption[Light curve fit and corresponding power spectrum from the characterization of the \textit{TESS} light curve of planet candidate TIC 441462736.01 (TOI 197.01; HD 221416 b)]
    {\textit{Top}: Light curve fit of planet candidate TIC 441462736.01 (TOI 197.01; HD 221416 b).
    Black points and error bars correspond to the light curve data.
    The GP predictive model is shown in orange, with the solid central line denoting the median of the distribution and the shaded area the 1-$\sigma$ interval.
    The transit component of the model is depicted by the blue solid line.
    The panel below it has a zoom in of the first transit of the light curve.
    \textit{Bottom}: Power spectrum of the light curve in the panel above.
    The PSD of the light curve is shown in light gray, with a slightly smoothed version overlapped in dark gray.
    The PSD of the sum of the GP components in the model in the panels above is shown as a solid red curve, with individual components identified by different colors (see legend).}
    \label{fig:tic441_fit}
\end{figure}

Table~\ref{tab:tic441_params} shows the prior and posterior distributions of each parameter of the GP + transit model fitted to the \textit{TESS} light curve of this target.
Figure~\ref{fig:tic441_param_hist} shows an alternative look at the posterior distributions of each parameter, through an histogram of all the posterior samples drawn during the fit.
The oscillations envelope component was not included in the GP since the oscillations for this target are present at higher frequencies than the Nyquist frequency of \textit{TESS} 30-min cadence data ($\sim 283 \mu \rm Hz$), with \textcite{Huber_2019b} estimating $\nu_\text{max}=430 \pm 18$ $\mu$Hz in their study using \textit{TESS} short-cadence data.
Figure~\ref{fig:tic441_fit} shows the light curve with the complete model in orange, as well as the transit component separately (in blue) in the top two panels.
Below is the PSD of the GP model, and its individual components, overplotted on the power spectrum of the light curve, with the contributions from the transit removed.

Comparing the estimated planetary parameters to the ones from \textcite{Huber_2019b}, shows that the periods are in agreement within $1$-$\sigma$.
Since this planet was first discovered with \textit{TESS} short-cadence data, the transit epoch $t_0$ can also be compared between both works.
\textcite{Huber_2019b} cite a value of $t_0 = 1357.0149^{+0.0025}_{-0.0026}$ BTJD, again in agreement within $1$-$\sigma$.
As for the planetary radius, considering the stellar radius from \textcite{Huber_2019b}, I obtain a value of $0.753^{+0.051}_{-0.043} \ R_\text{J}$, in agreement with their (seismic) radius within $2$-$\sigma$.


\subsection{TIC 348835438.01}

\begin{table}[!ptb]
    \centering
    \setlength{\tabcolsep}{12pt}
    \renewcommand{\arraystretch}{1.1}
    \begin{tabular}{lllll}
        \toprule
        \toprule
        Parameter                            & Prior                                & Posterior        & 84\%             & 16\%            \\
        \midrule
        $a_\text{meso}$ [ppm]                & $\mathcal{U}$(10, 1000)              & 154.19           & +23.41           & -19.16          \\ 
        $b_\text{meso}$ [$\mu$Hz]            & $\mathcal{U}$(1, 280)                & 43.57            & +13.19           & -10.61          \\ 
        $\sigma$ [ppm]                       & $\mathcal{U}$(10, 1000)              & 266.73           & +9.41            & -9.96           \\ 
        $P$ [days]                           & $\mathcal{U}$(5.7, 6.7)              & 6.2022           & +0.0069          & -0.0076         \\ 
        $t_0$ [BTJD]                         & $\mathcal{U}$(1520.5, 1522.5)        & 1521.592         & +0.017           & -0.028          \\ 
        $R_\text{p} / R_\star$               & $\mathcal{LU}$(0.001, 0.1)           & 0.0321           & +0.0059          & -0.0028         \\ 
        $a / R_\star$                        & $\mathcal{LU}$(1.5, 20)              & 4.68             & +1.95            & -1.67           \\ 
        $i$ [deg]                            & $\mathcal{U}$(60.0, 90.0)            & 80.70            & +6.84            & -12.62          \\ 
        $e$                                  & $\mathcal{U}$(0.0, 0.5)              & 0.27             & +0.16            & -0.18           \\ 
        $\omega$ [deg]                       & $\mathcal{U}$(0.0, 360.0)            & 148.56           & +119.77          & -90.23          \\ 
        $u_1$                                & $\mathcal{N}$(0.6, 0.1)              & 0.62             & +0.10            & -0.11           \\ 
        $u_2$                                & $\mathcal{N}$(0.09, 0.1)             & 0.11             & +0.09            & -0.10           \\ 
        \bottomrule
    \end{tabular}
    \caption[Prior and posterior distributions for all parameters in the GP + transit model obtained in the characterization of the \textit{TESS} light curve of TIC 348835438.01]
    {Prior and posterior distributions for all parameters in the GP + transit model obtained in the characterization of the \textit{TESS} light curve of TIC 348835438.01.
    Figure~\ref{fig:tic348_param_hist} shows an alternative look at the posterior distributions of each parameter, through an histogram of all the posterior samples drawn during the fit.}
    \label{tab:tic348_params}
\end{table}
Planet candidate TIC 348835438.01 was identified by my pipeline, as well as independently identified by our collaborators, cementing our confidence in the transit's planetary nature.
Radial-velocity follow-up observations of the target have, in the meantime, been conducted and a confirmation of the planet is underway (Grunblatt et al., in prep.).

\begin{figure}[!ptb]
    \centering
    \includegraphics[width=0.91\textwidth, height=\textheight, keepaspectratio]{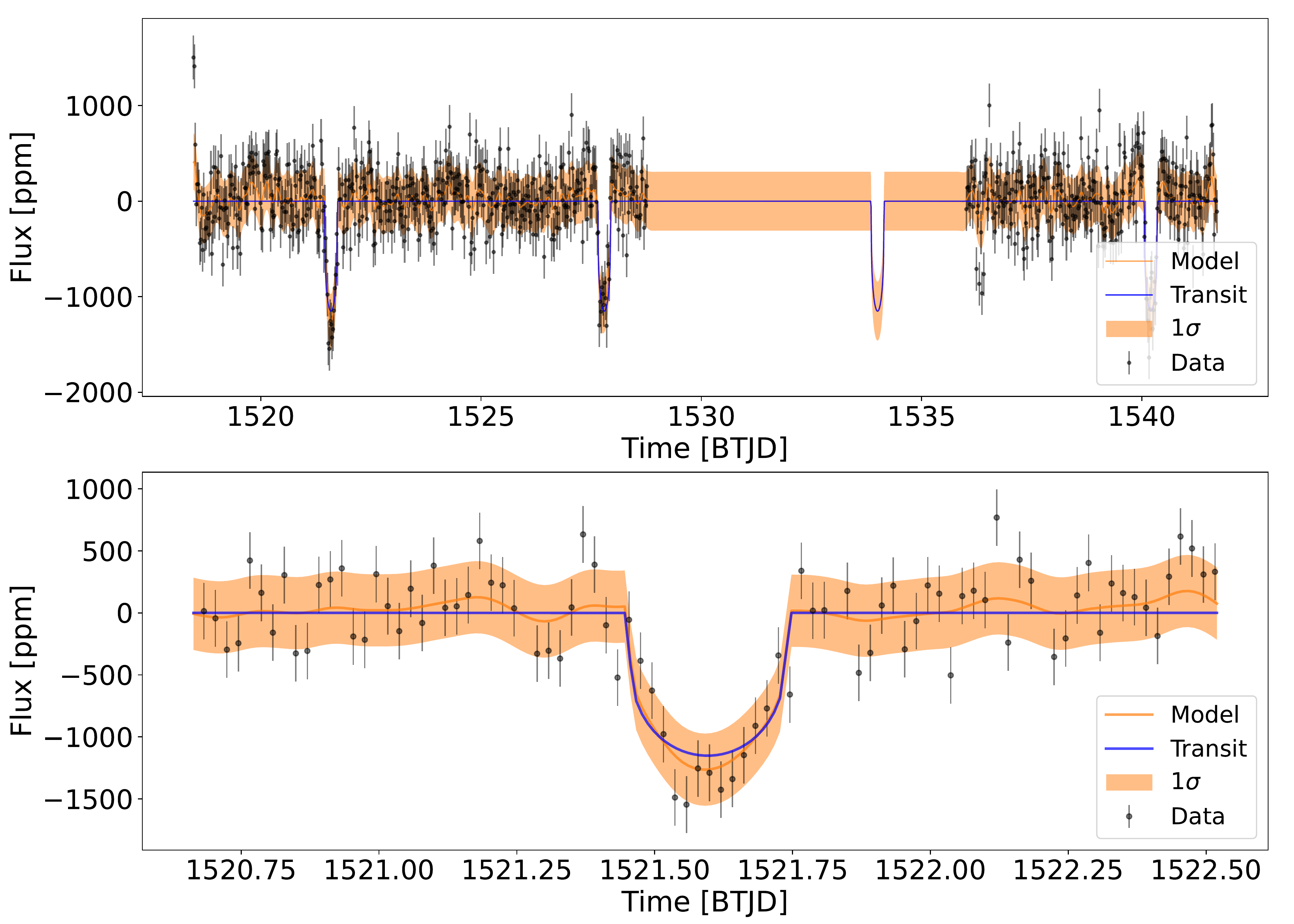}
    \includegraphics[width=0.91\textwidth, height=\textheight, keepaspectratio]{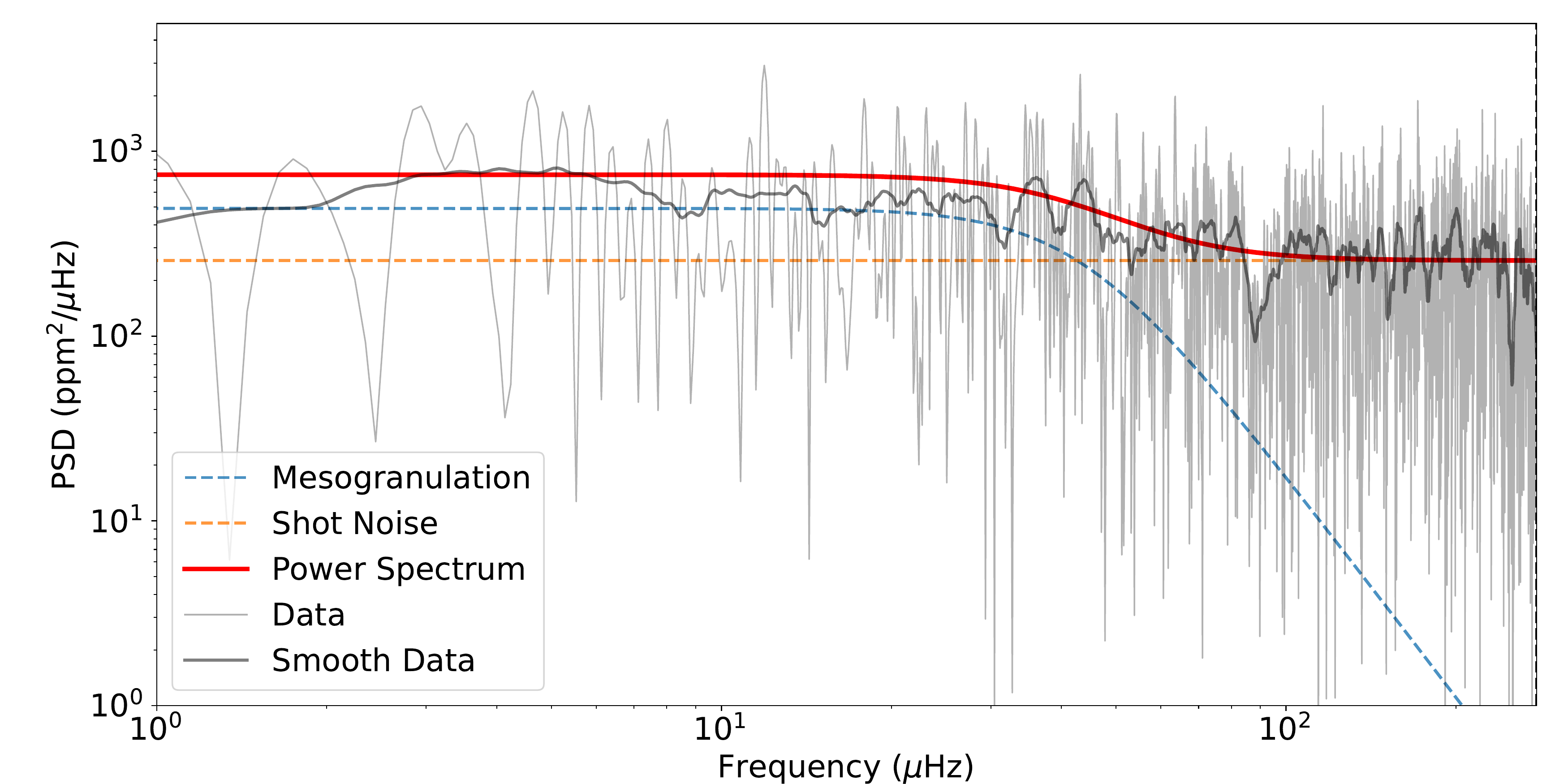}
    \caption[Light curve fit and corresponding power spectrum from the characterization of the \textit{TESS} light curve of planet candidate TIC 348835438.01]
    {\textit{Top}: Light curve fit of planet candidate TIC 348835438.01.
    Black points and error bars correspond to the light curve data.
    The GP predictive model is shown in orange, with the solid central line denoting the median of the distribution and the shaded area the 1-$\sigma$ interval.
    The transit component of the model is depicted by the blue solid line.
    The panel below it has a zoom in of the first transit of the light curve.
    \textit{Bottom}: Power spectrum of the light curve in the panel above.
    The PSD of the light curve is shown in light gray, with a slightly smoothed version overlapped in dark gray.
    The PSD of the sum of the GP components in the model in the panels above is shown as a solid red curve, with individual components identified by different colors (see legend).}
    \label{fig:tic348_fit}
\end{figure}

Table~\ref{tab:tic348_params} shows the prior and posterior distributions of the parameters of the GP + transit model fitted to the \textit{TESS} light curve of this target.
Figure~\ref{fig:tic348_param_hist} shows an alternative look at the posterior distributions of each parameter, through an histogram of all the posterior samples drawn during the fit.
Figure~\ref{fig:tic348_fit} shows the light curve with the complete model in orange, as well as the transit component separately (in blue) in the top two panels.
Below is the PSD of the GP model, and its individual components, overplotted on the power spectrum of the light curve, with the contributions from the transit removed.
Due to the high white noise level in the target's power spectrum, the oscillations envelope is not visible, despite being expected to reside below the Nyquist frequency of \textit{TESS} long-cadence data.
For this reason, the oscillations envelope component was not included in the GP model.

Considering the stellar radius from the TIC, the host star is a $\sim 3.98$ $\rm R_\odot$ red giant.
Under that assumption, the planet candidate has a radius of $\sim 1.243 \ R_\text{J}$, which, when combined with its period points to the possibility of the planet being an inflated hot Jupiter \parencite{Fortney_2007}.


\subsection{TIC 204650483.01}
\label{sub:tic204}

\begin{table}[!ptb]
    \centering
    \setlength{\tabcolsep}{14pt}
    \renewcommand{\arraystretch}{1.1}
    \begin{tabular}{lllll}
        \toprule
        \toprule
        Parameter                            & Prior                                & Posterior        & 84\%             & 16\%            \\
        \midrule
        $a_\text{meso}$ [ppm]                & $\mathcal{U}$(10, 1000)              & 150.44           & +20.78           & -17.52          \\ 
        $b_\text{meso}$ [$\mu$Hz]            & $\mathcal{U}$(1, 280)                & 7.47             & +2.05            & -1.66           \\ 
        $\sigma$ [ppm]                       & $\mathcal{U}$(10, 1000)              & 217.83           & +5.88            & -5.20           \\ 
        $P$ [days]                           & $\mathcal{U}$(9.5, 10.5)             & 9.9610           & +0.0060          & -0.0056         \\ 
        $t_0$ [BTJD]                         & $\mathcal{U}$(1574.5, 1576.5)        & 1575.6549        & +0.0073          & -0.0091         \\ 
        $R_\text{p} / R_\star$               & $\mathcal{LU}$(0.001, 0.1)           & 0.0321           & +0.0033          & -0.0019         \\ 
        $a / R_\star$                        & $\mathcal{LU}$(1.5, 20)              & 5.01             & +1.40            & -1.23           \\ 
        $i$ [deg]                            & $\mathcal{U}$(60.0, 90.0)            & 83.49            & +4.48            & -7.27           \\ 
        $e$                                  & $\mathcal{U}$(0.0, 0.5)              & 0.23             & +0.18            & -0.16           \\ 
        $\omega$ [deg]                       & $\mathcal{U}$(0.0, 360.0)            & 145.85           & +140.90          & -87.84          \\ 
        $u_1$                                & $\mathcal{N}$(0.6, 0.1)              & 0.63             & +0.09            & -0.09           \\ 
        $u_2$                                & $\mathcal{N}$(0.09, 0.1)             & 0.11             & +0.10            & -0.10           \\ 
        \bottomrule
    \end{tabular}
    \caption[Prior and posterior distributions for all parameters in the GP + transit model obtained in the characterization of the \textit{TESS} light curve of TIC 204650483.01]
    {Prior and posterior distributions for all parameters in the GP + transit model obtained in the characterization of the \textit{TESS} light curve of TIC 204650483.01.
    Figure~\ref{fig:tic204_param_hist} shows an alternative look at the posterior distributions of each parameter, through an histogram of all the posterior samples drawn during the fit.}
    \label{tab:tic204_params}
\end{table}
TIC 204650483.01 was the other not known candidate identified by my pipeline.
Initial analysis of the target's pixel files indicated that the origin of one of the transits might have been noise and not a physical source.
This suspicion was also motivated by the fact that our collaborators could not find the candidate's transit with their pipeline.

Nevertheless, given the promising characteristics of the remaining transit event, follow-up RVs were obtained with the \textit{CHIRON} spectrograph\footnote{The \textit{CHIRON} spectrograph is situated in the 1.5-meter telescope from the \textit{Small \& Moderate Aperture Research Telescope System} (\textit{SMARTS}), at the \textit{Cerro Tololo Inter-American Observatory} (\textit{CTIO}), in Chile (see \url{http://www.ctio.noao.edu/noao/content/CHIRON}).}, and the initial 10 RV measurements available indicate the presence of a planet.

Given the availability of RV measurements for this target, I perform two fits to the planet candidate.
The first fit is the already seen GP + transit model that fits only the light curve data and includes the characterization of the stellar signals together with the transit in the time domain.
For the second fit, I use the \textit{juliet} \texttt{Python} package \parencite{Espinoza_2019}, to fit a joint model to both transits and RVs simultaneously.
This joint model does not include the GP components to characterize the stellar signals, and instead considers only a simple white noise component for the transit and another for the RVs.

\begin{figure}[!ptb]
    \centering
    \includegraphics[width=0.91\textwidth, height=\textheight, keepaspectratio]{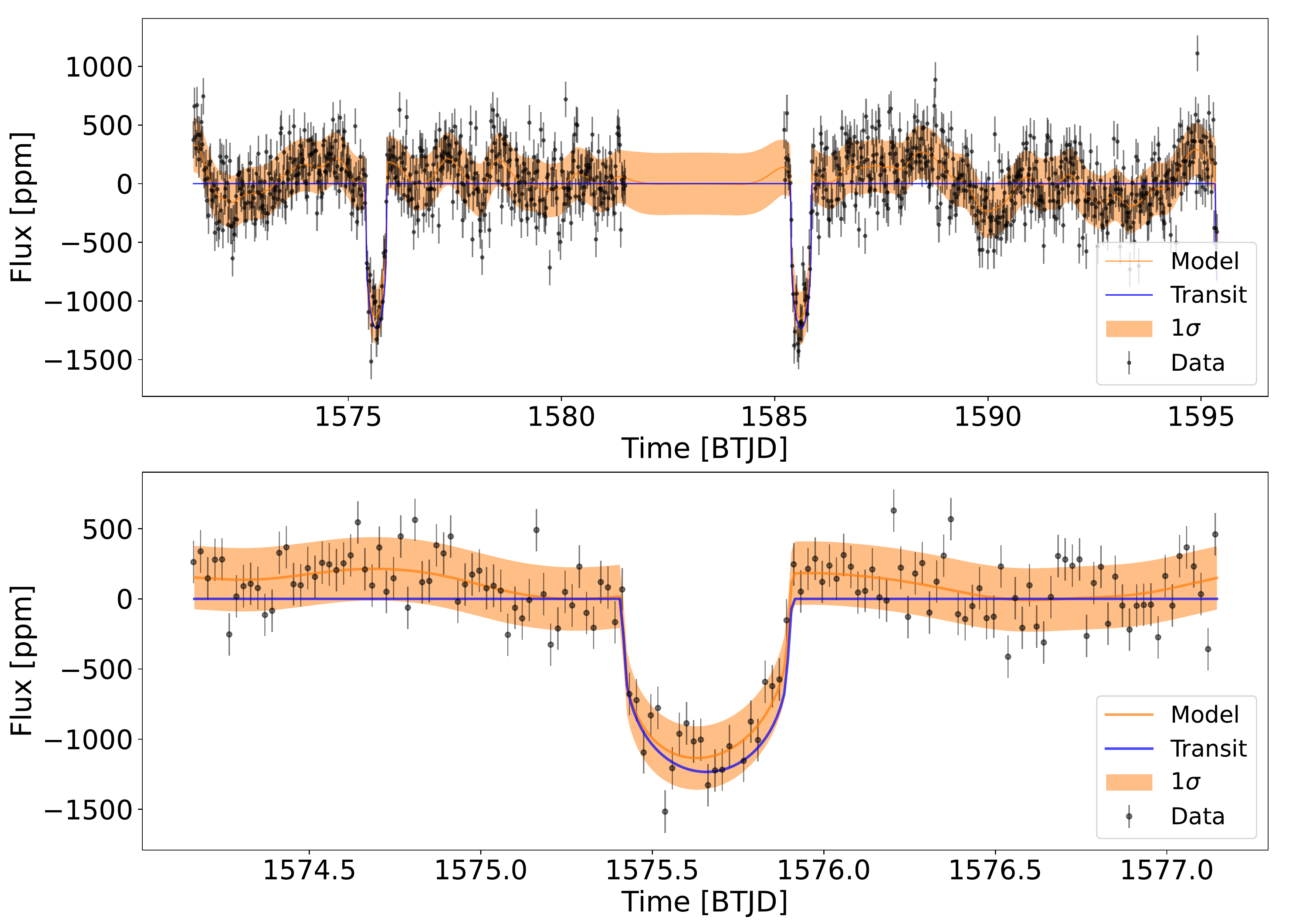}
    \includegraphics[width=0.91\textwidth, height=\textheight, keepaspectratio]{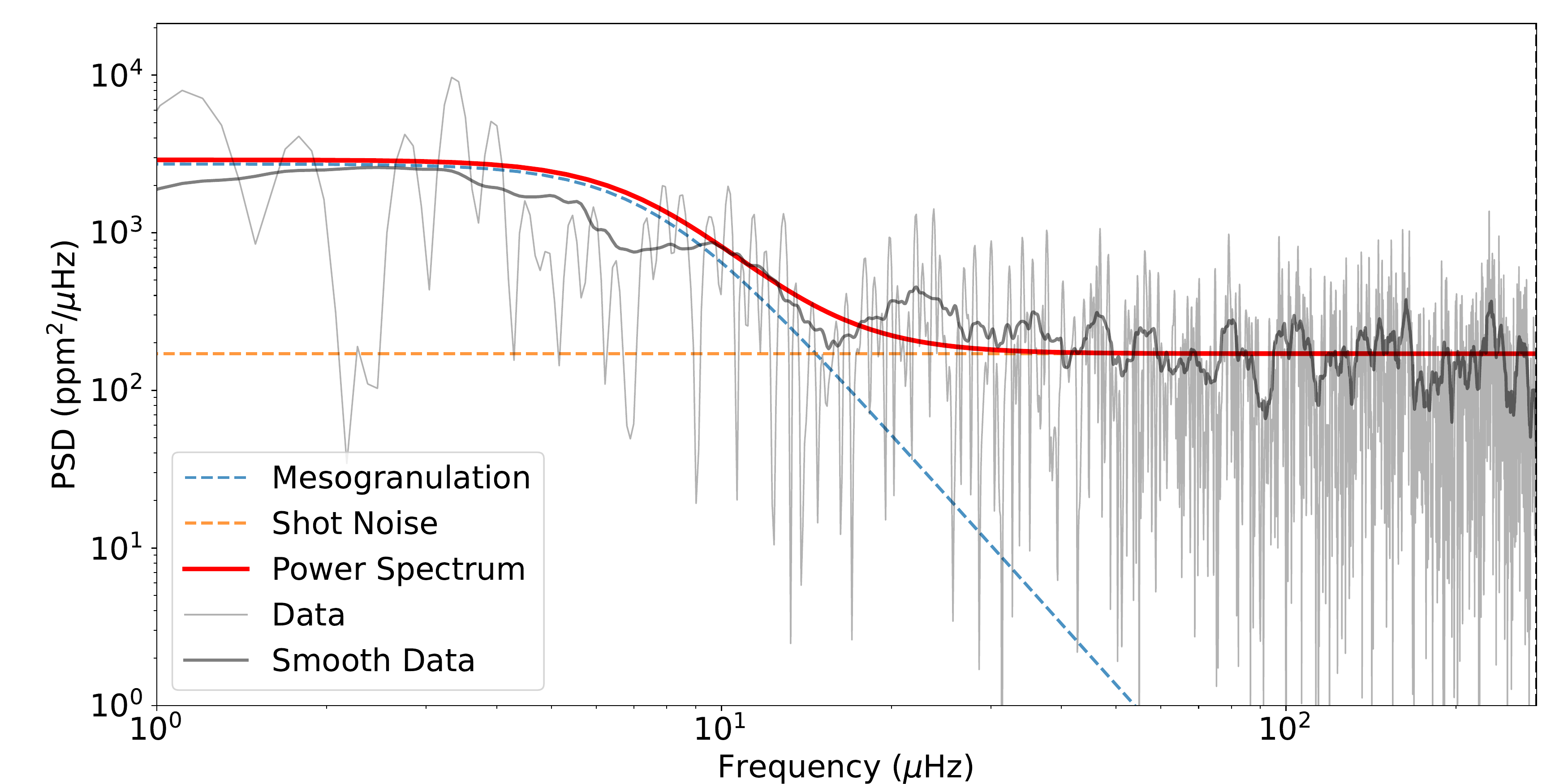}
    \caption[Light curve fit and corresponding power spectrum from the characterization of the \textit{TESS} light curve of planet candidate TIC 204650483.01]
    {\textit{Top}: Light curve fit of planet candidate TIC 204650483.01.
    Black points and error bars correspond to the light curve data.
    The GP predictive model is shown in orange, with the solid central line denoting the median of the distribution and the shaded area the 1-$\sigma$ interval.
    The transit component of the model is depicted by the blue solid line.
    The panel below it has a zoom in of the first transit of the light curve.
    \textit{Bottom}: Power spectrum of the light curve in the panel above.
    The PSD of the light curve is shown in light gray, with a slightly smoothed version overlapped in dark gray.
    The PSD of the sum of the GP components in the model in the panels above is shown as a solid red curve, with individual components identified by different colors (see legend).}
    \label{fig:tic204_fit}
\end{figure}

Starting with the results from the \textit{TESS} light curve characterization with the GP + transit model, Table~\ref{tab:tic204_params} shows the prior and posterior distributions of all model parameters.
Figure~\ref{fig:tic204_param_hist} shows an alternative look at the posterior distributions of each parameter, through an histogram of all the posterior samples drawn during the fit.
Figure~\ref{fig:tic204_fit} shows the light curve with the complete model in orange, as well as the transit component separately (in blue) in the top two panels.
Below is the PSD of the GP model, and its individual components, overplotted on the power spectrum of the light curve, with the contributions from the transit removed.
Similarly to the previous candidate, the oscillations envelope is also not visible in the power spectrum of the target, so I did not include this component in the GP model.

Considering the stellar radius from the TIC, the host is a $\sim 3.38$ $\rm R_\odot$ red-giant star and the candidate has a radius of $\sim 1.056 \ R_\text{J}$.

\begin{table}[!ptb]
    \centering
    \setlength{\tabcolsep}{14pt}
    \renewcommand{\arraystretch}{1.1}
    \begin{tabular}{lllll}
        \toprule
        \toprule
        Parameter                            & Prior                                & Posterior        & 84\%             & 16\%            \\
        \midrule
        \multicolumn{5}{c}{Model Parameters}                                                                                                \\
        \midrule
        $P$ [days]                                  & $\mathcal{U}$(9.5, 10.5)             & 9.9566           & +0.0030          & -0.0033         \\ 
        $t_0$ [BTJD]                                & $\mathcal{U}$(1574.5, 1576.5)        & 1575.6540        & +0.0058          & -0.0071         \\ 
        $\sqrt{e} \sin \omega$                      & $\mathcal{U}$(-1.0, 1.0)             & 0.26             & +0.21            & -0.24           \\ 
        $\sqrt{e} \cos \omega$                      & $\mathcal{U}$(-1.0, 1.0)             & -0.28            & +0.16            & -0.13           \\ 
        $r_1$                                       & $\mathcal{U}$(0.0, 1.0)              & 0.49             & +0.27            & -0.29           \\ 
        $r_2$                                       & $\mathcal{U}$(0.0, 1.0)              & 0.302            & +0.031           & -0.015          \\ 
        $a / R_\star$                               & $\mathcal{LU}$(1.0, 20.0)            & 5.20             & +1.06            & -1.29           \\ 
        $q_1$                                       & $\mathcal{U}$(0.0, 1.0)              & 0.63             & +0.24            & -0.25           \\ 
        $q_2$                                       & $\mathcal{U}$(0.0, 1.0)              & 0.44             & +0.29            & -0.25           \\ 
        $\sigma_\textit{TESS}$ [ppm]                & $\mathcal{LU}$(0.1, 1000.0)          & 210.69           & +7.11            & -7.16           \\ 
        $K$ [m$\cdot$s$^{-1}$]                      & $\mathcal{U}$(0.0, 200.0)            & 136.27           & +16.12           & -13.75          \\ 
        $\gamma_\textit{CHIRON}$ [m$\cdot$s$^{-1}$] & $\mathcal{U}$(41300.0, 41400.0)      & 41310.7          & +10.4            & -6.8            \\ 
        $\sigma_\textit{CHIRON}$ [m$\cdot$s$^{-1}$] & $\mathcal{LU}$(0.001, 100.0)         & 27.63            & +10.56           & -6.60           \\ 
        \midrule
        \multicolumn{5}{c}{Derived Properties}                                                                                              \\
        \midrule
        $u_1$                                       &                                      & 0.68             & +0.36            & -0.38           \\ 
        $u_2$                                       &                                      & 0.09             & +0.42            & -0.41           \\ 
        $e$                                         &                                      & 0.190            & +0.119           & -0.095          \\ 
        $\omega$ [deg]                              &                                      & -37.366          & +45.073          & -29.104         \\ 
        $a$ [AU]                                    &                                      & 0.082            & +0.017           & -0.020          \\
        $i$ [deg]                                   &                                      & 84.22            & +3.95            & -7.44           \\ 
        $R_\text{p} / R_\star$                      &                                      & 0.0309           & +0.0031          & -0.0014         \\ 
        $R_\text{p}$ [$R_\text{J}$]                 &                                      & 1.016            & +0.101           & -0.047          \\ 
        \bottomrule
    \end{tabular}
    \caption[Prior and posterior distributions for all parameters in the joint RV and transit model, as well as derived orbital and planetary properties obtained in the characterization of TIC 204650483.01]
    {Prior and posterior distributions for all parameters in the joint RV and transit model, as well as derived orbital and planetary properties obtained in the characterization of TIC 204650483.01.
    Besides the median of the posterior distributions, the 68\% confidence interval is also included, denoted by the 16\% and 84\% percentiles.}
    \label{tab:tic204_joint_params}
\end{table}
Considering now the joint fit to both transits and RVs, \textit{juliet} implements reparameterizations of some model parameters to improve the Bayesian sampling.
The eccentricity, $e$, and the argument of periapsis, $\omega$, are replaced by the more efficiently sampled parameters, $\sqrt{e} \cos \omega$ and $\sqrt{e} \sin \omega$, which keep an uniform sampling of the eccentricity \parencite{Eastman_2013}.
The quadratic limb-darkening coefficients, $u_1$ and $u_2$, are replaced by $q_1$ and $q_2$, a more efficiently sampled parameterization introduced in \textcite{Kipping_2013}.
Finally, the ratio of radii, $R_\text{p} / R_\star$, and the impact parameter, $b$, which is, itself, a reparametrization of the orbital inclination, are reparameterized by $r_1$ and $r_2$, two random values which can be uniformly sampled between 0 and 1 and that ensure that all sampled values (of $R_\text{p} / R_\star$ and $b$) are physically plausible, as shown by \textcite{Espinoza_2018}.

Besides the transit parameters, the joint fit now also includes the RV velocity semi-amplitude, $K$, the radial-velocity zero point, $\gamma$, and the radial-velocity jitter, $\sigma$.
The RV jitter term is distinguished from the light curve's one by their subscripts of \textit{CHIRON} and \textit{TESS}, respectively.

\begin{figure}[!ptb]
    \centering
    \includegraphics[width=\textwidth, height=\textheight, keepaspectratio]{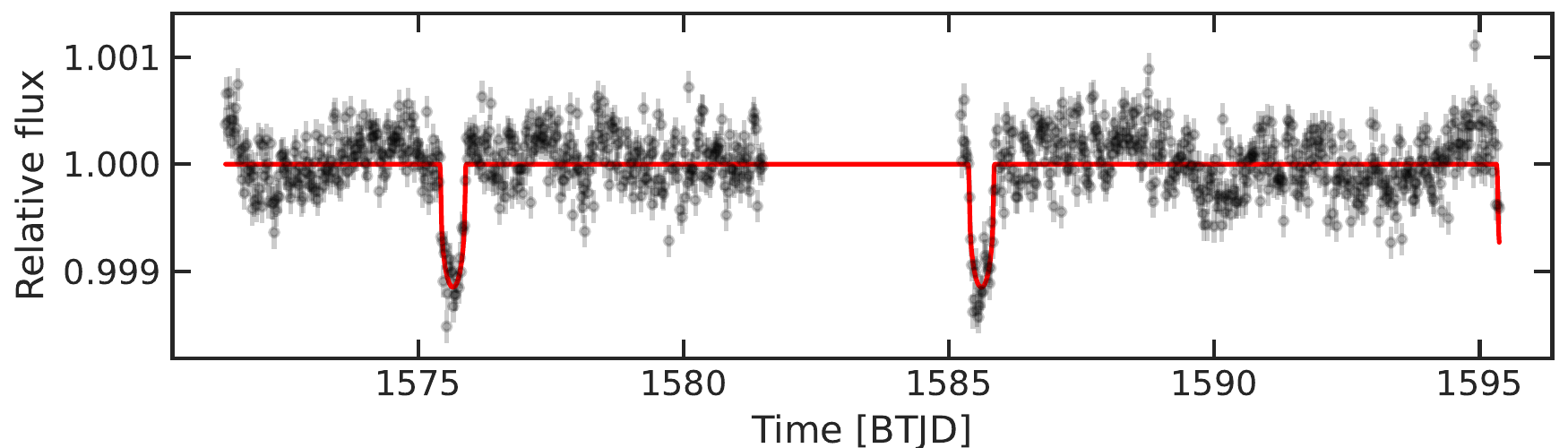}
    \includegraphics[width=\textwidth, height=\textheight, keepaspectratio]{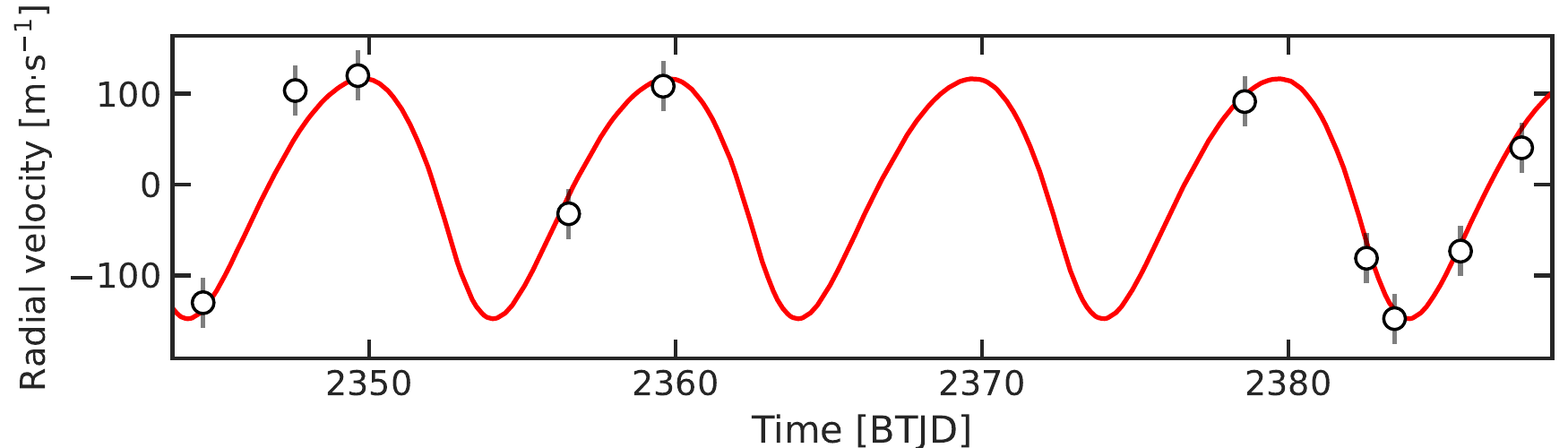}
    \includegraphics[width=\textwidth, height=\textheight, keepaspectratio]{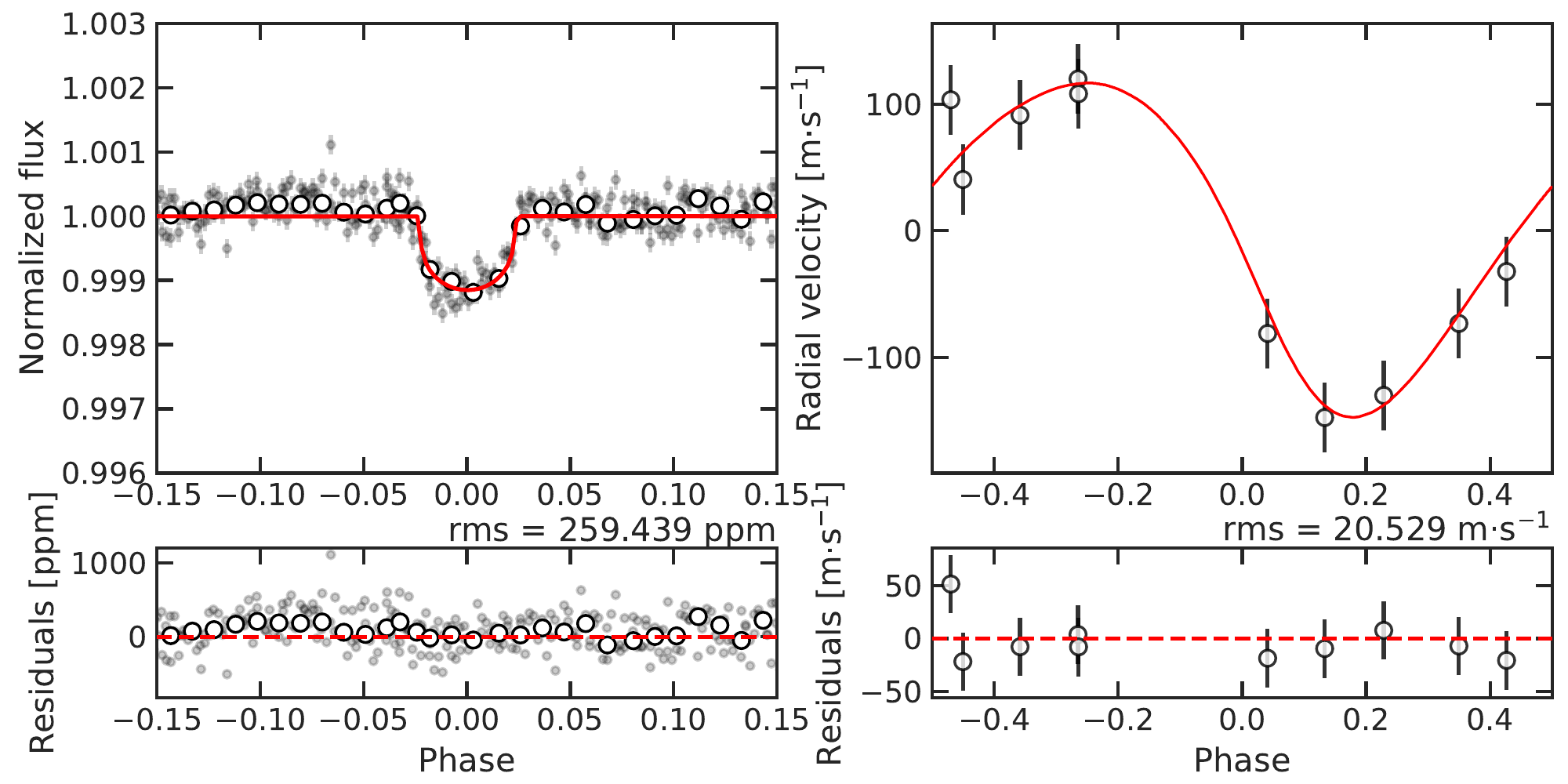}
    \caption[Results from the joint transit and RV fit of TIC 204650483.01]
    {Results from the joint transit and RV fit of TIC 204650483.01.
    The upper panel shows the transit model (in red) alongside the \textit{TESS} light curve of the target.
    Similarly, the middle panel shows the RV model (in red) alongside the RV measurements.
    The lower panels contain, from left to right, the phase-folded (on the planet's period) transit and RV models and data, respectively.
    Below each phase-folded panel are the residuals obtained from subtracting the model from the data points, with the red dashed line denoting a zero offset.
    The transit panels additionally include binned (10-point bins) data points.
    The root mean square of the residuals for each dataset is denoted above the panels, on the right.}
    \label{fig:tic204_joint}
\end{figure}

The results of the characterization are shown in Table~\ref{tab:tic204_joint_params}, which includes the prior and posterior distributions for all the parameters in the joint model, as well as the posterior distributions of all derived planetary and orbital parameters.
Figures~\ref{fig:tic204_model_param_hist} and \ref{fig:tic204_derived_param_hist} show an alternative look at the posterior distributions of each model parameter and derived property, respectively.
Figure~\ref{fig:tic204_joint} illustrates the results from the joint RV and transit fit done with \textit{juliet}.
The upper panels show both the transit light curve and the RV data with their respective models denoted by a solid red line.
The panels below show the same models phase-folded according to the planet's period, and their respective residuals.

The orbital period, transit epoch and planetary radius obtained from the joint fit all agree within 1-$\sigma$ with the values from the light curve characterization alone.
With the introduction of the RV data, the eccentricity can now be constrained, and the planet exhibits a slightly eccentric orbit with $e = 0.19_{-0.095}^{+0.119}$.
Despite the availability of RV measurements, I cannot estimate the planetary mass since it requires a value for the stellar mass, which is not available in the TIC.

An announcement and detailed characterization of this planet will be included in a publication led by me, targeting late 2021 (Pereira et al. in prep.).
This publication will also include a spectroscopic characterization of the host star, to determine more accurately its stellar properties. 
This, in turn, will provide an estimate of the stellar mass which will lead to an estimation of the planet's mass which will be combined with its radius to determine its mean density.
Considering that the light curves from the \textit{TESS} extended mission have been recently released for this target, I will also re-evaluate the presence of oscillations in the faster-cadence data, potentially enabling an asteroseismic characterization of this target.


\subsection{TIC 394918211.01}

\begin{table}[!ptb]
    \centering
    \setlength{\tabcolsep}{14pt}
    \renewcommand{\arraystretch}{1.1}
    \begin{tabular}{lllll}
        \toprule
        \toprule
        Parameter                            & Prior                                & Posterior        & 84\%             & 16\%            \\
        \midrule
        $a_\text{meso}$ [ppm]                & $\mathcal{U}$(10, 1000)              & 206.89           & +21.96           & -21.42          \\ 
        $b_\text{meso}$ [$\mu$Hz]            & $\mathcal{U}$(1, 280)                & 6.68             & +1.47            & -1.22           \\ 
        $\sigma$ [ppm]                       & $\mathcal{U}$(10, 1000)              & 703.44           & +10.11           & -10.17          \\ 
        $P$ [days]                           & $\mathcal{U}$(3.88, 4.88)            & 4.37798          & +0.00057         & -0.00055        \\ 
        $t_0$ [BTJD]                         & $\mathcal{U}$(1601.1, 1603.1)        & 1602.1205        & +0.0091          & -0.0092         \\ 
        $R_\text{p} / R_\star$                   & $\mathcal{LU}$(0.001, 0.1)           & 0.0389           & +0.0026          & -0.0017         \\ 
        $a / R_\star$                            & $\mathcal{LU}$(1.5, 20)              & 3.22             & +0.77            & -0.62           \\ 
        $i$ [deg]                            & $\mathcal{U}$(60.0, 90.0)            & 80.70            & +6.38            & -9.57           \\ 
        $e$                                  & $\mathcal{U}$(0.0, 0.5)              & 0.23             & +0.17            & -0.16           \\ 
        $\omega$ [deg]                       & $\mathcal{U}$(0.0, 360.0)            & 129.27           & +105.28          & -85.32          \\ 
        $u_1$                                & $\mathcal{N}$(0.6, 0.1)              & 0.58             & +0.09            & -0.09           \\ 
        $u_2$                                & $\mathcal{N}$(0.09, 0.1)             & 0.08             & +0.10            & -0.09           \\ 
        \bottomrule
    \end{tabular}
    \caption[Prior and posterior distributions for all parameters in our GP + transit model obtained in the characterization of the \textit{TESS} light curve of TIC 394918211.01]
    {Prior and posterior distributions for all parameters in our GP + transit model obtained in the characterization of the \textit{TESS} light curve of TIC 394918211.01.
    Figure~\ref{fig:tic204_param_hist} shows an alternative look at the posterior distributions of each parameter, through an histogram of all the posterior samples drawn during the fit.}
    \label{tab:tic394_params}
\end{table}
Finally, planet candidate TIC 394918211.01 was a candidate identified by our collaborators.
This target, although located in the southern hemisphere of \textit{TESS}, has a \textit{TESS} magnitude of 10.79, falling outside of the range considered by the southern sample defined here.
Using my pipeline, I confirmed the presence of a transit signal and classified it as a planet candidate.
An interesting difference between this candidate and all previous ones is that three sectors of data are available for the target from the first year of \textit{TESS} observations.

In addition to the \textit{TESS} light curve, 20 RV measurements have also been obtained for this target with the \textit{CORALIE} spectrograph \parencite{Queloz_2000}, with evidence pointing to the presence of an object of planetary nature.
In parallel to TIC 204650483.01, given the availability of RV measurements for this target, I perform two fits, one to the \textit{TESS} light curve only, using the GP + transit model and the other to both the transit and RVs simultaneously.

\begin{figure}[!ptb]
    \centering
    \includegraphics[width=0.91\textwidth, height=\textheight, keepaspectratio]{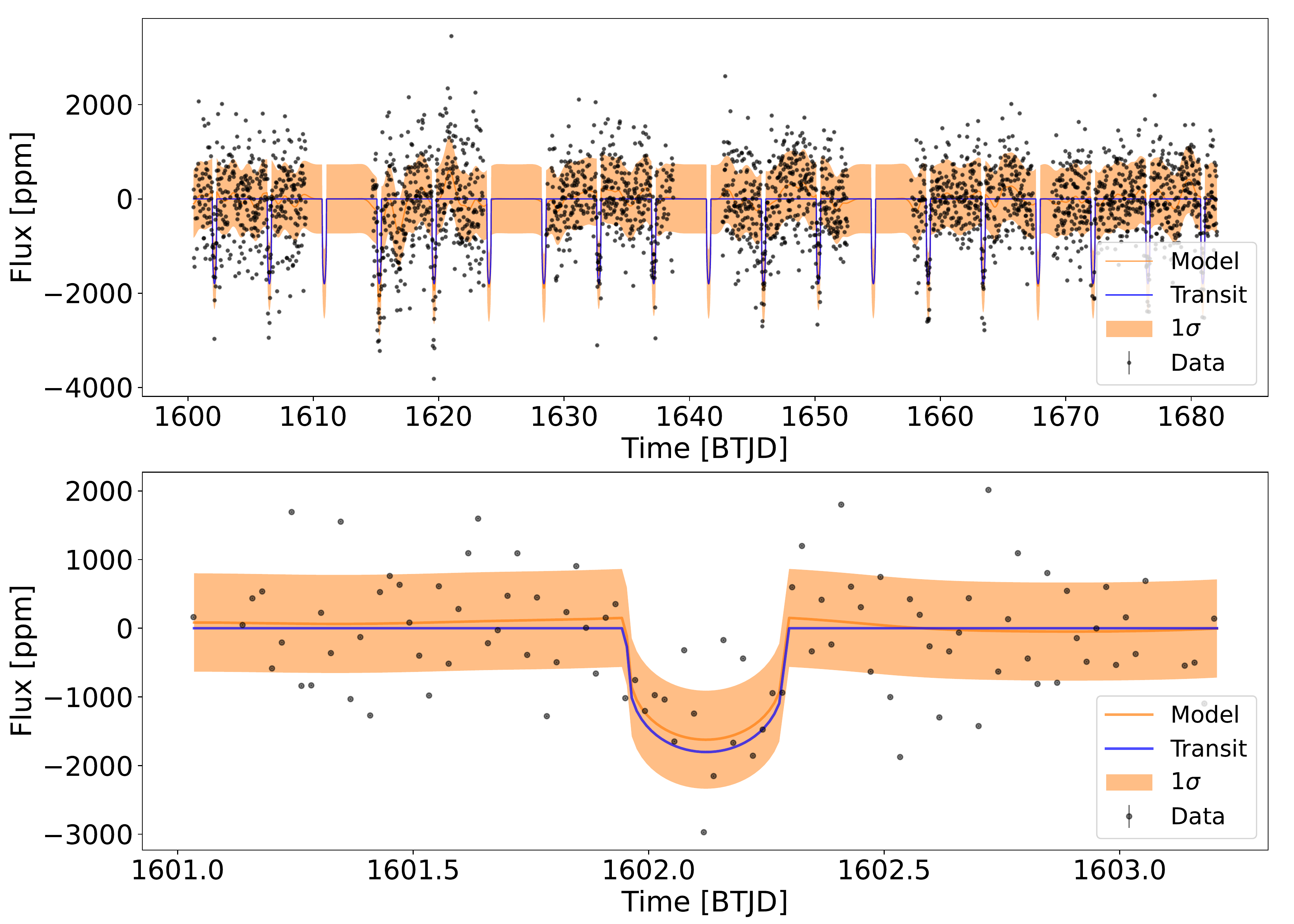}
    \includegraphics[width=0.91\textwidth, height=\textheight, keepaspectratio]{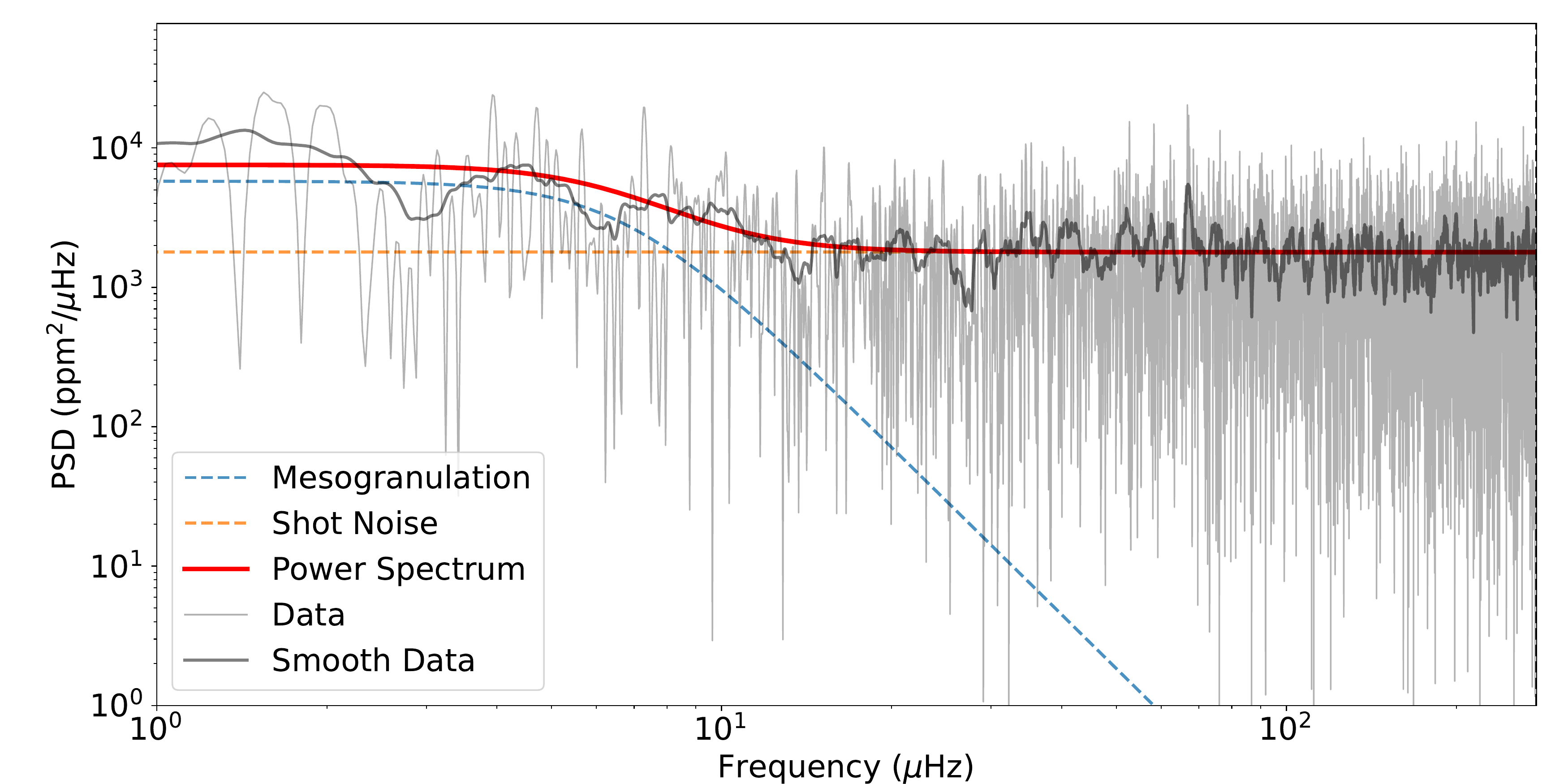}
    \caption[Light curve fit and corresponding power spectrum from the characterization of the \textit{TESS} light curve of planet candidate TIC 394918211.01]
    {\textit{Top}: Light curve fit of planet candidate TIC 394918211.01.
    Black points and error bars correspond to the light curve data.
    The GP predictive model is shown in orange, with the solid central line denoting the median of the distribution and the shaded area the 1-$\sigma$ interval.
    The transit component of the model is depicted by the blue solid line.
    The panel below it has a zoom in of the first transit of the light curve.
    
    \textit{Bottom}: Power spectrum of the light curve in the panel above.
    The PSD of the light curve is shown in light gray, with a slightly smoothed version overlapped in dark gray.
    The PSD of the sum of the GP components in the model in the panels above is shown as a solid red curve, with individual components identified by different colors (see legend).}
    \label{fig:tic394_fit}
\end{figure}

Table~\ref{tab:tic394_params} shows the prior and posterior distributions of all model parameters from the characterization of the \textit{TESS} light curve only, using the GP + transit model.
Figure~\ref{fig:tic204_param_hist} shows an alternative look at the posterior distributions of each parameter, through an histogram of all the posterior samples drawn during the fit.
Figure~\ref{fig:tic394_fit} shows the light curve with the complete model in orange, as well as the transit component separately (in blue) in the top two panels.
Below is the PSD of the GP model, and its individual components, overplotted on the power spectrum of the light curve, with the contributions from the transit removed.
Despite the larger dataset available for this target, due to its faintness, the oscillations envelope is once again not visible in the power spectrum of the target, which led us to not include this component in the GP model.
%
%

Considering the stellar radius from the TIC, the host star is a $\sim 3.41$ $\rm R_\odot$ red giant, with the planet candidate having a radius of $\sim 1.29 \ R_\text{J}$.
This estimated radius suggests that this planet candidate might be inflated \parencite{Fortney_2007}.


\begin{table}[!ptb]
    \centering
    \setlength{\tabcolsep}{14pt}
    \renewcommand{\arraystretch}{1.1}
    \begin{tabular}{lllll}
        \toprule
        \toprule
        Parameter                            & Prior                                & Posterior        & 84\%             & 16\%            \\
        \midrule
        \multicolumn{5}{c}{Model Parameters}                                                                                                \\
        \midrule
        $P$ [days]                           & $\mathcal{U}$(4.0, 5.0)              & 4.3782           & +0.0004          & -0.0005         \\ 
        $t_0$ [BTJD]                         & $\mathcal{U}$(1600.0, 1604.0)        & 1602.1195        & +0.0073          & -0.0062         \\ 
        $\sqrt{e} \sin \omega$               & $\mathcal{U}$(-1.0, 1.0)             & -0.05            & +0.21            & -0.19           \\ 
        $\sqrt{e} \cos \omega$               & $\mathcal{U}$(-1.0, 1.0)             & 0.09             & +0.16            & -0.18           \\ 
        $r_1$                                & $\mathcal{U}$(0.0, 1.0)              & 0.49             & +0.26            & -0.27           \\ 
        $r_2$                                & $\mathcal{U}$(0.0, 1.0)              & 0.381            & +0.027           & -0.016          \\ 
        $a / R_\star$                            & $\mathcal{LU}$(1.0, 20.0)            & 3.73             & +0.47            & -0.74           \\ 
        $q_1$                                & $\mathcal{U}$(0.0, 1.0)              & 0.53             & +0.28            & -0.26           \\ 
        $q_2$                                & $\mathcal{U}$(0.0, 1.0)              & 0.31             & +0.30            & -0.20           \\ 
        $\sigma_\textit{TESS}$ [ppm]         & $\mathcal{LU}$(0.1, 1000.0)          & 534.06           & +12.62           & -13.38          \\ 
        $K$ [m$\cdot$s$^{-1}$]               & $\mathcal{U}$(0.0, 200.0)            & 95.88            & +9.17            & -8.58           \\ 
        $\gamma_\textit{CORALIE}$            & $\mathcal{U}$(12400.0, 12650.0)      & 12555.3          & +6.2             & -6.6            \\ 
        $\sigma_\textit{CORALIE}$            & $\mathcal{LU}$(0.001, 100.0)         & 0.11             & +2.89            & -0.11           \\ 
        \midrule
        \multicolumn{5}{c}{Derived Properties}                                                                                              \\
        \midrule
        $u_1$                                &                                      & 0.44             & +0.33            & -0.28           \\ 
        $u_2$                                &                                      & 0.26             & +0.36            & -0.39           \\ 
        $e$                                  &                                      & 0.058            & +0.061           & -0.040          \\ 
        $\omega$ [deg]                       &                                      & -15.830          & +74.114          & -50.910         \\ 
        $a$ [AU]                             &                                      & 0.059            & +0.007           & -0.012          \\ 
        $i$ [deg]                            &                                      & 83.00            & +4.64            & -6.98           \\ 
        $R_\text{p} / R_\star$                   &                                      & 0.0387           & +0.0027          & -0.0016         \\ 
        $R_\text{p}$ [$R_\text{J}$]             &                                      & 1.284            & +0.089           & -0.052          \\ 
        \bottomrule
    \end{tabular}
    \caption[Prior and posterior distributions for all parameters in the joint RV and transit model, as well as derived orbital and planetary properties obtained in the characterization of TIC 394918211.01]
    {Prior and posterior distributions for all parameters in the joint RV and transit model, as well as derived orbital and planetary properties obtained in the characterization of TIC 394918211.01.
    Besides the median of the posterior distributions, the 68\% confidence interval is also included, denoted by the 16\% and 84\% percentiles.}
    \label{tab:tic394_joint_params}
\end{table}
For the joint fit of both transits and RVs, I consider the same model as in Section~\ref{sub:tic204}, with the same reparameterizations discussed earlier.
In this case, the RV jitter term is distinguished from the light curve's one by their subscripts of \textit{CORALIE} and \textit{TESS}, respectively.

\begin{figure}[!ptb]
    \centering
    \includegraphics[width=\textwidth, height=\textheight, keepaspectratio]{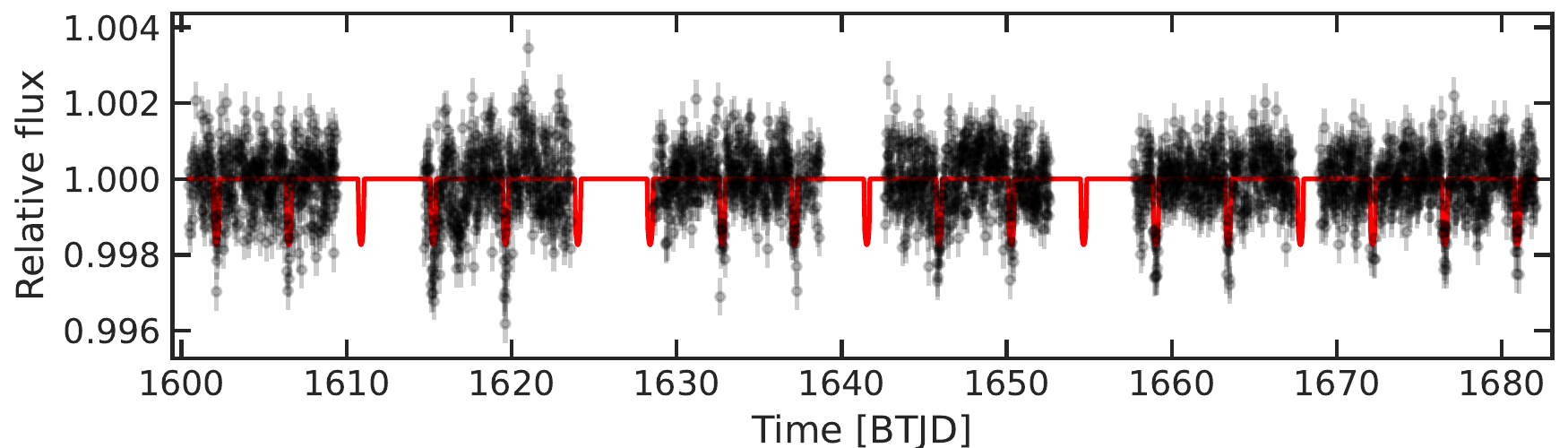}
    \includegraphics[width=\textwidth, height=\textheight, keepaspectratio]{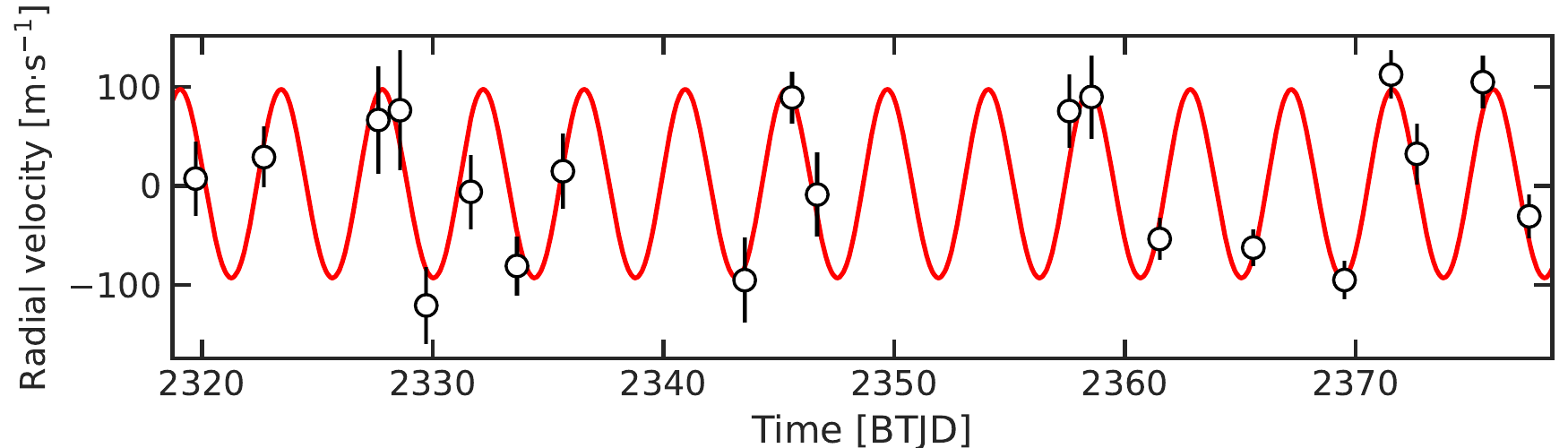}
    \includegraphics[width=\textwidth, height=\textheight, keepaspectratio]{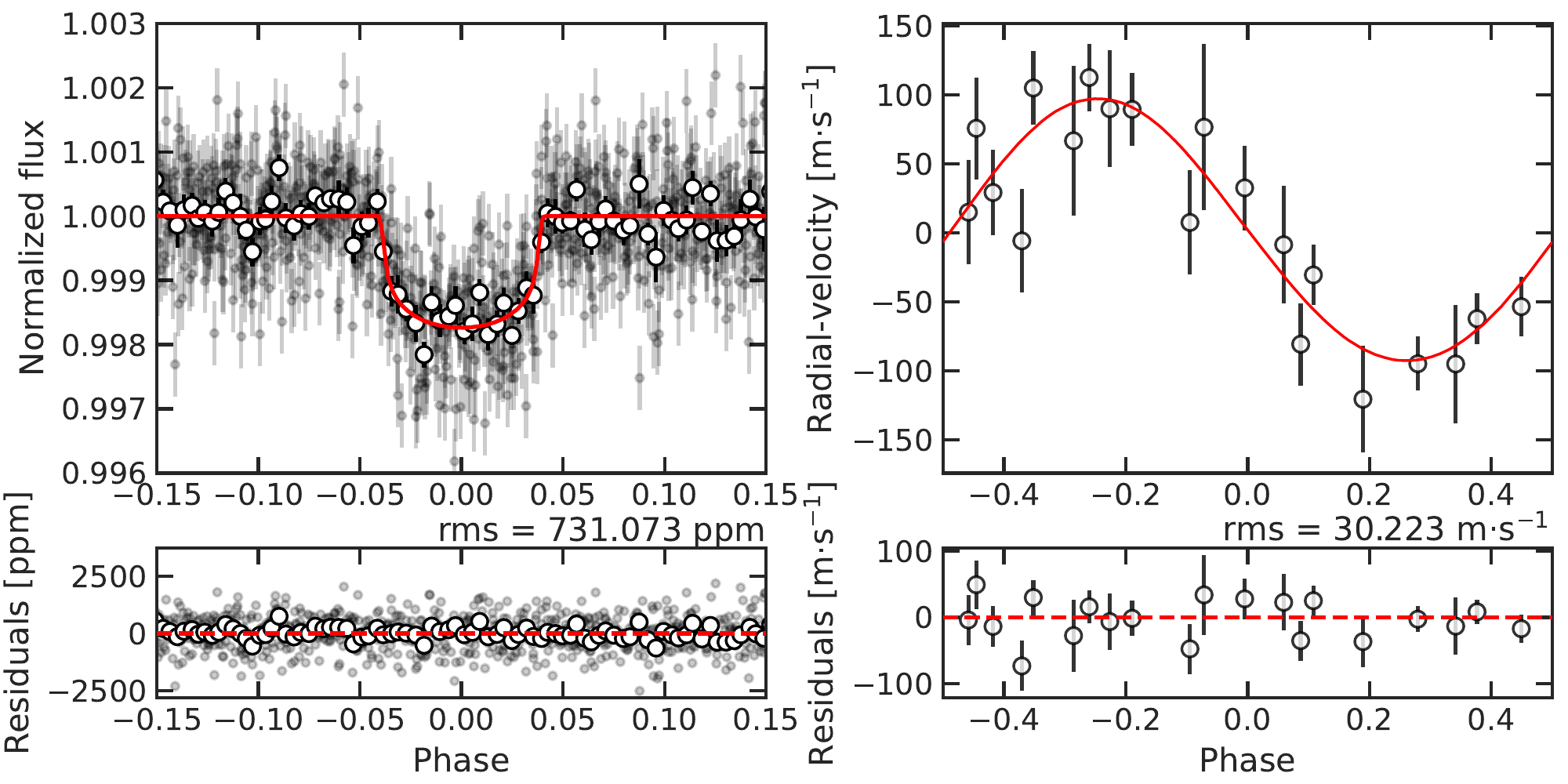}
    \caption[Results from the joint transit and RV fit of TIC 394918211.01]
    {Results from the joint transit and RV fit of TIC 394918211.01.
    The upper panel shows the transit model (in red) alongside the \textit{TESS} light curve of the target.
    Similarly, the middle panel shows the RV model (in red) alongside the RV measurements.
    The lower panels contain, from left to right, the phase-folded (on the planet's period) transit and RV models and data, respectively.
    Below each phase-folded panel are the residuals obtained from subtracting the model to the data points, with the red dashed line denoting a zero offset.
    The transit panels additionally include binned (10-point bins) data points.
    The root mean square of the residuals for each dataset is denoted above the panels, on the right.}
    \label{fig:tic394_joint}
\end{figure}

The results of the characterization are shown in Table~\ref{tab:tic394_joint_params}, which includes the prior and posterior distributions for all the parameters in the joint model, as well as the posterior distributions of all derived planetary and orbital parameters.
Figures~\ref{fig:tic394_model_param_hist} and \ref{fig:tic394_derived_param_hist} show an alternative look at the posterior distributions of each model parameter and derived property, respectively.
Figure~\ref{fig:tic394_joint} illustrates the results from the joint RV and transit fit done with \textit{juliet}.
The upper panels show both the transit light curve and the RV data with their respective models denoted by a solid red line.
The panels below show the same models phase-folded according to the planet's period, and their respective residuals.

Once again, the orbital period, transit epoch and planetary radius are in agreement within 1-$\sigma$ between both fits.
As with TIC 348835438.01, the radius estimate suggests that the planet might be inflated \parencite{Fortney_2007}, making this candidate another rare example of an inflated hot Jupiter orbiting an evolved star, and a prime example for the study of planetary inflation in giant planets orbiting evolved hosts (Section~\ref{sub:inflation}).
Unlike TIC 204650483.01, results for this candidate suggest only a very minor eccentricity, with a median distribution centered on $e = 0.058_{-0.040}^{+0.061}$.
Again, due to no stellar mass being available in the TIC, the planetary mass could not be estimated.


The publication announcing this discovery will be the same as the one containing TIC 204650483.01, led by me and expected in late 2021 (Pereira et al. in prep.).
Following the same lines as for TIC 204650483.01, the publication will include a spectroscopic characterization of the host star to determine more accurate stellar properties, again leading to an estimation of the its mass which will be combined with its radius to determine the planet's mean density.
There will also be a re-evaluation of the potential for asteroseismology with the new data from the \textit{TESS} extended mission.

This improved stellar characterization is all the more relevant for the study of planetary radius inflation, as precise planetary properties are essential to be able to constrain the processes involved \parencite{Grunblatt_2017,Thorngren_2021}.

%% file: chapters/search/summary.tex
\section{Summary}
\label{sec:search_summary}

Having introduced and analyzed the results from the characterization of all five planet candidates, I now draw some conclusions on the overall performance of the search.
Cross-checking the search's target list with that of known hosts indicates that only two known hosts are present in the sample.
Both these hosts, KELT-11 and TOI-197, were identified by the pipeline and correctly classified as planet candidate hosts, which reinforces the confidence in the results of the search.
Both these hosts are also subgiant stars, which ended up being included in the sample due to the expansion of the range of stellar radii considered, to account for uncertainties in the TIC values.

\begin{figure}[!ptb]
    \centering
    \includegraphics[width=\textwidth, height=\textheight, keepaspectratio]{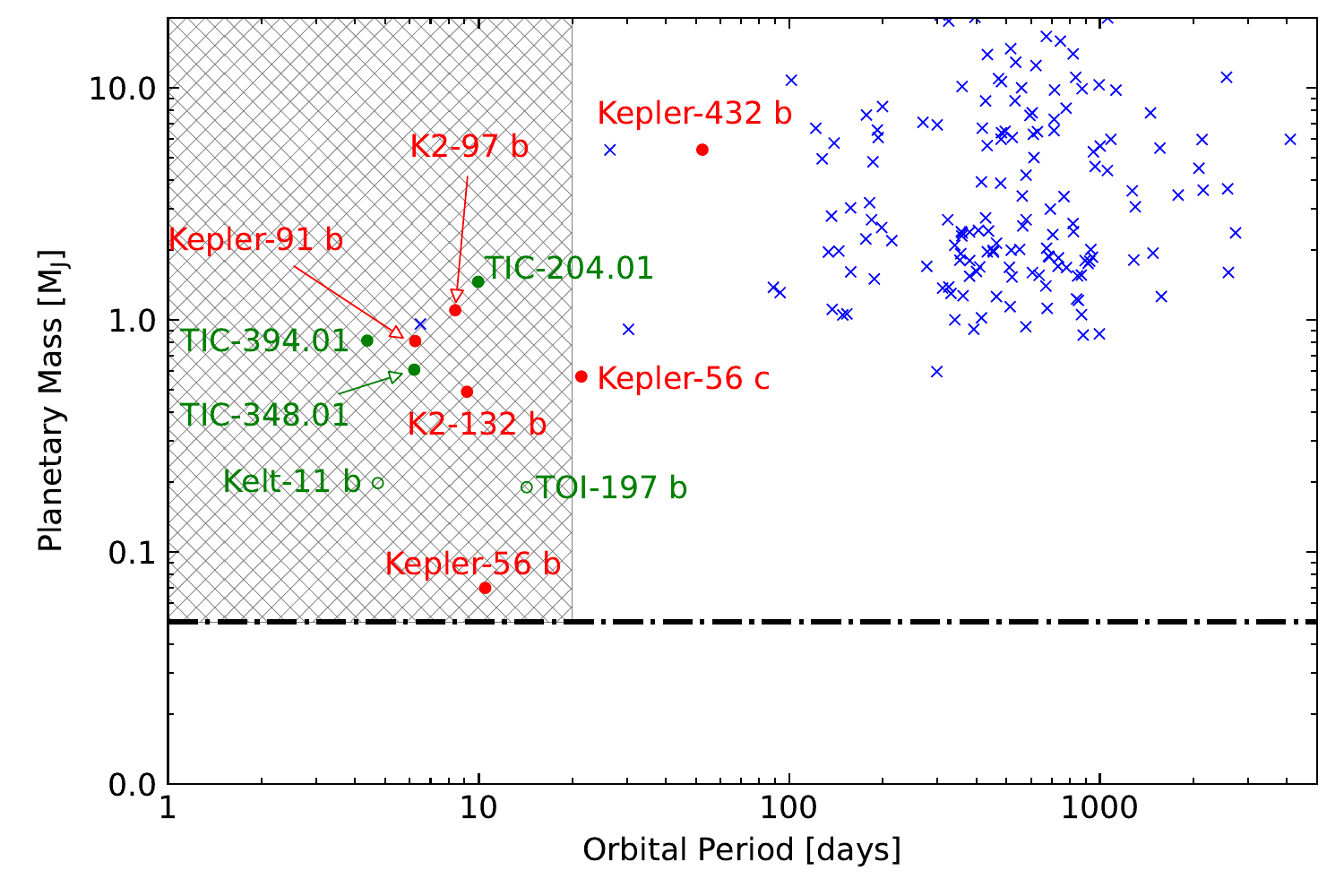}
    \caption[Updated mass-period diagram of known exoplanets orbiting red-giant branch stars, initially shown in Figure~\ref{fig:known_evolved_hosts}]
    {Updated mass-period diagram of known exoplanets orbiting red-giant branch stars, initially shown in Figure~\ref{fig:known_evolved_hosts}.
    Known planets orbiting red giants detected by the transit method are depicted as red circles and those detected in RV surveys as blue crosses (in which case the masses are lower limits).
    All planet candidates identified in our sample are shown in green, and fall within the shaded region.
    The two known planets orbiting subgiant hosts are shown as empty circles, whilst the three new candidates identified are shown as solid circles.
    The dashed-dotted line marks the mass of Neptune. 
    The shaded area approximately corresponds to the parameter space probed by \textit{TESS} considering one sector of data ($M > M_\text{Neptune}$; $P \leq 20$d).}
    \label{fig:updated_known_hosts}
\end{figure}

The remaining three, new planet candidates all have RV follow-up observations (although more RV measurements are still underway) and their planetary nature is tentatively confirmed.
Grunblatt et al. (in prep.) will publish the results on TIC 348835438.01, and I will publish results on TIC 204650483.01 and TIC 394918211.01 (Pereira et al. in prep.).
Returning to Figure~\ref{fig:known_evolved_hosts}, I can now show an updated version, Figure~\ref{fig:updated_known_hosts}, which includes all candidates from the search, both the known subgiant hosts (as green empty circles) and the three new planet candidates (as green solid circles).

%% file: chapters/conclusions.tex
\chapter{Conclusions}
\label{cha:conclusions}

At the start of this thesis, I introduced the general goal of improving the knowledge of the formation and evolution of planetary systems, focusing on giant planets orbiting giant stars. 
Specifically, I was interested in close-in giant planets orbiting red-giant stars, looking to better constrain their occurrence rate, which is still not fully understood \parencite{Grunblatt_2019} and to exploit their unique properties in the study of giant planet inflation \parencite{Lopez_2016,Grunblatt_2017}.

To pursue this goal, I specified three objectives.
In this final chapter, I recap all major results from my thesis in the context of each of these objectives.
Then, I finish by discussing some future prospects in the search for giant planets around giant stars.


\section{Thesis results}

\subsection{Transit characterization}

For the improved characterization of transiting light curves, in particular from evolved hosts, I developed and implemented a method using Gaussian processes, built on top of an existing framework, that could characterize transits and stellar signals simultaneously (Section~\ref{sub:gptransits}).
This change in methodology is particularly crucial for giant stars, as the granulation and oscillations have high amplitudes and similar timescales to that of the duration of transits of close-in giant planets orbiting evolved stars (Section~\ref{sec:stellar_signals}).

To that end, I defined an expression for the granulation signals in the time domain that is the exact counterpart of its frequency-domain equation, as well as an approximate expression for the oscillations envelope (Section~\ref{sec:modelling}).
This Gaussian process model was then combined with a quadratic limb-darkened model for the transit signal, building a unified model in the time domain.

To systematically test this method, I applied it to both \textit{TESS} simulated light curves (Section~\ref{sub:test_tess_lcs}) and \textit{Kepler} light curves (Section~\ref{sub:test_kepler}) without transits.
When testing with \textit{TESS} simulated light curves, I found that the method was capable of recovering the properties of the simulated stellar signals within uncertainties, when considering a model with mesogranulation and oscillations only.
Tests with \textit{Kepler} light curves showed that this method's implementation was capable of recovering the same stellar signals as the ones estimated using power-spectrum fitting methods, widely used in the literature.

In both tests, I show that the inferred value of $\nu_\text{max}$, a global asteroseismic property, was always accurate and precise, demonstrating the method's ability to do time-domain asteroseismology.

By adding transits to the \textit{TESS} simulated light curves, I then evaluated the performance of the method when recovering planetary properties, comparing it with a simpler transit model (Section~\ref{sub:test_transits}).
Overall, I found that the GP method determines more accurate and precise values, with particular improvements on the inferred planet radius (with respect to the stellar radius).

I have published an article with a description of the method and all results from Section~\ref{sec:applications} \parencite{Pereira_2019}.
The method's implementation is open-source and available through my Github repository (\url{https://github.com/Fill4/gptransits}).

\subsection{Transit search pipeline}

To explore the \textit{TESS} data, in particular its FFIs, and search for new planets, I assembled a data processing pipeline (Chapter~\ref{cha:search_pipeline}).
This pipeline was comprised of multiple open-source software packages, implemented and adjusted as necessary, with some additional data processing methods added.

The pipeline starts by extracting pixel information from the FFIs for any \textit{TESS} target identified by its TIC (which is a unique ID from the \textit{TESS} Input Catalog \parencite{Stassun_2019}) (Section~\ref{sec:photometry}).
Then, a light curve is extracted from the pixels, and is corrected for any known systematics.
The correction methodology further includes a set of time-stamped ranges that are removed from the data.
These were identified as noisy patterns present in a representative sample of light curves for each \textit{TESS} sector that were not removed by the initial correction procedure (Section~\ref{sec:corrections}).

Transits are searched for in the corrected light curves using two different tools, TLS and BLS, and the most significant transit found by each method is flagged for validation (Section~\ref{sec:transit_search}).
The pipeline processing is fully automated until this point, and for any targets which successfully completed all previous stages, a summary figure is created which contains relevant information for the transit validation (see Figure~\ref{fig:summary_plot_example}).

Finally, validation is performed both through a statistical evaluation of a signal detection efficiency (SDE) metric, calculated for each transit, as well as through the use of a tool for automated astrophysical false-positive evaluation of transits, \texttt{VESPA} (Section~\ref{sec:transit_validation}).

\subsection{\textit{TESS} southern search}

To search for planets in the \textit{TESS} data, I started by selecting a sample of 40,772 bright (\textit{TESS} magnitude < 10) low-luminosity red-giant branch stars ($R_\star \sim 2.5 - 10 \ \rm R_\odot$) observed in the southern hemisphere of \textit{TESS}.
All targets were then pushed through the pipeline, which extracted light curves from the \textit{TESS} FFIs, performed systematics correction and searched for transits, producing a summary plot for each one.

For the validation of transit signals, initial statistical false-positives were identified using the SDE metrics calculated by the pipeline for each target, resulting in a list of $\sim$3000 targets likely to have a physical transit signal.
This sample constitutes a reproducible, magnitude limited sample of \textit{TESS} southern hemisphere targets with statistically significant transit-like signals, representing an ideal starting point for population studies (i.e. occurrence rates), as discussed in Section~\ref{sec:sample_analysis}.

For my case, I carried out a visual analysis of all $\sim$3000 targets, short-listing 254 targets as having signals of physical origin.
This reduced sample was then re-analyzed both through an additional visual inspection, as well as through the application of \texttt{VESPA}, in order to try to attribute a specific astrophysical nature to each signal.

All things considered, I classified four targets with signals with a high probability of being of transits of planetary nature.
Of these four planet candidates, two were already known planets (the only known planets present in the sample), and two were new detections.
Below I list some of their properties:
\begin{itemize}[noitemsep, topsep=1ex]
    \item KELT-11 b: An already known planet \parencite{Pepper_2017} for which I determine a radius of $\sim 1.207 \ R_\text{J}$, in agreement within 2-$\sigma$ with the radius from \textcite{Pepper_2017}. 
    The same publication lists a mass of $\sim 0.195 \ M_\text{J}$ for the planet, orbiting a $\sim 2.72$ $\rm R_\odot$ subgiant star on a period of $\sim 4.737$ days.
    \item TOI-197.01: A \textit{TESS} confirmed planet orbiting a $\sim 2.943$ $\rm R_\odot$ subgiant star on a period of $\sim 14.277$ days \parencite{Huber_2019b}. 
    Similarly to the previous candidate, I find a radius of $\sim 0.753 \ R_\text{J}$, in agreement within 2-$\sigma$ with the published value. 
    Having access to RV follow-up, \textcite{Huber_2019b} also estimated a planetary mass of $\sim 0.190 \ M_\text{J}$.
    \item TIC 348835438.01: Planet candidate found independently by my pipeline and by colleagues in a collaboration, with confirmation to soon follow (Grunblatt et al., in prep.).
    The planet has a radius of $\sim 1.243 \ R_\text{J}$, orbiting a $\sim 3.98$ $\rm R_\odot$ red-giant star (according to the TIC) on a period of $\sim 6.202$ days. 
    The planet's radius suggests that it is possibly inflated.
    \item TIC 204650483.01: Planet candidate originally found by my pipeline.
    Although an initial analysis pointed to one of the transits being caused by noise, preliminary follow-up RVs clearly point to an object of planetary nature. 
    The planet has a radius of $\sim 1.016 \ R_\text{J}$ orbiting a $\sim 3.38$ $\rm R_\odot$ red-giant star (according to the TIC) on a period of $\sim 9.957$ days.
    Results also suggest that the planet has a slightly eccentric orbit with $e = 0.190$.
    A publication announcing this planet and including a detailed analysis of itself and its host star will be led by me (Pereira et al. in prep.).
\end{itemize}

In addition to the four candidates found during the search, I also introduced planet candidate TIC 394918211.01, initially found by colleagues in our collaboration and validated also by my pipeline.
This target, although located in the southern hemisphere of TESS, has a TESS magnitude of 10.79, falling outside of the range considered by my southern sample.
Through RV follow-up observations, I have found evidence pointing to the planetary nature of the candidate.

This planet candidate might be an inflated hot Jupiter with a radius of $\sim 1.284 \ R_\text{J}$, and orbits its host every $\sim 4.378$ days with a low eccentricity of $e = 0.058$.
The host is a $\sim 3.41$ $\rm R_\odot$ star, according to the TIC.

The announcement and detailed analysis of this planet and its host will also be led by me, in the same publication as TIC 204650483.01 (Pereira et al. in prep.).
The analysis will include a study of the possible origin and efficiency of the radius inflation of these planets.

All in all, three new planet candidates, tentatively confirmed as planets, have been found in the southern hemisphere of \textit{TESS}, improving the picture of close-in giant planets orbiting red-giant stars, as shown in Figure~\ref{fig:updated_known_hosts}.
Two of these candidates also appear to have inflated radii, potentially helping to unveil the origins and efficiency of the processes responsible for planetary radius inflation \parencite{Lopez_2016,Grunblatt_2017}, something I will explore with TIC 204650483.01 and TIC 394918211.01 (Pereira et al. in prep.).
Finally, the eventual confirmation of both new candidates found in our sample, TIC 348835438.01 and TIC 204650483.01, combined with our reproducible sample of $\sim$3000 \textit{TESS} southern hemisphere LLRGB targets with statistically significant transit signals, should provide a good starting point to an occurrence rate study of close-in giant planets orbiting evolved stars with \textit{TESS}, attesting to the sample's legacy value.

\section{Future Work}
\label{sub:future_work}

As just mentioned, with the results from the newly discovered planet candidates, I will look into planet radius inflation around evolved stars (Pereira et al. in prep).
At the same time, confirmation of the two planet candidates included in the magnitude-limited sample could lead to a study on the occurrence rate of close-in giant planets orbiting evolved stars with \textit{TESS}.
A straightforward path from here would be to expand on the sample of stars selected for this thesis.

Expanding to the northern sectors of \textit{TESS} could lead to a first all-sky occurrence rate study for these planets.
Considering the fact that, even with a decreased number of detections of oscillations \parencite{Mackereth_2021}, spectroscopic characterization of the hosts is still possible, the sample of stars could also be expanded to fainter targets, increasing the number of red-giant stars available approximately tenfold for every point in magnitude.
Moreover, given that the 10-min cadence of the \textit{TESS} expanded mission unlocks higher frequencies of oscillation, the target sample could also be further expanded into subgiant stars, widening the possibilities when probing the properties of planetary systems, from the main-sequence to the red-giant branch.

Looking further into the future, the \textit{PLAnetary Transits and Oscillations of stars} mission \parencite[\textit{PLATO};][]{Rauer_2014} should again provide a wide field-of-view.
Whilst data processing pipelines will surely need to be adapted, our characterization method should still prove useful at modeling any new planetary systems that are found.

%% file: chapters/appendix.tex
\input{chapters/characterization/parseval_norm}

\input{chapters/search/vespa_plots}

\input{chapters/search/histogram_plots}

%% file: chapters/characterization/parseval_norm.tex
\chapter{Parseval normalization of the \texttt{celerite} PSD}
\label{app:celerite_normalization}

The PSD in Equation~(\ref{eq:celerite_granulation_psd}), which we rewrite here,
\begin{equation}
    P(\omega) = \sqrt{\frac{2}{\pi}} \frac{S_0}{\left( \omega / \omega_0 \right)^4 + 1} .
	\label{eq:celerite_granulation_psd_appendix}
\end{equation}
shares the functional form of the PSD describing the granulation in \textcite{Kallinger_2014} (Equation~\ref{eq:granulation_psd_kallinger}),
\begin{equation}
    P(\nu) = \frac{2 \sqrt{2}}{\pi} \frac{a^2 / b}{\left(\nu / b \right)^4 + 1} ,
	\label{eq:kallinger_granulation_psd_appendix}
\end{equation}
However, Equation~(\ref{eq:celerite_granulation_psd_appendix}) is not normalized according to Parseval's theorem.

For a Parseval-normalized PSD, the variance of the light curve must equal $a^2$.
In order to normalize it, we need to find a constant $K$ that ensures that the previous condition is met.
From \textcite{Foreman-Mackey_2017}, the variance of a light curve described by the kernel in Equation~(\ref{eq:celerite_granulation_psd_appendix}) is
\begin{equation}
	k(\tau = 0) = \frac{S_0\omega_0}{\sqrt{2}} ,
\end{equation}
which gives
\begin{equation}
	a^2 = \frac{S_0\omega_0}{\sqrt{2}} .
	\label{eq:variance_appendix}
\end{equation}
Moreover, from \textcite{Foreman-Mackey_2017}, $\omega_0$ is expressed as
\begin{equation}
	\omega_0 = 2 \pi b .
	\label{eq:omega_appendix}
\end{equation}

Equating Equations~(\ref{eq:celerite_granulation_psd_appendix}) and (\ref{eq:kallinger_granulation_psd_appendix}), we get
\begin{equation}
	K \sqrt{\frac{2}{\pi}} \frac{S_0}{\left( \omega / \omega_0 \right)^4 + 1} = \frac{2\sqrt{2}}{\pi} \frac{a^2/b}{\left( \nu/b \right)^4 +1} ,
	\label{equality}
\end{equation}
and substituting for Equations~(\ref{eq:variance_appendix}) and (\ref{eq:omega_appendix}), $K$ becomes
\begin{equation}
	K = 2 \sqrt{2\pi}.
\end{equation}

With the value for $K$, we can then rewrite Equation~(\ref{eq:celerite_granulation_psd_appendix}) normalized according to Parseval's theorem as
\begin{equation}
	P(\omega) = \frac{4S_0}{\left( \omega / \omega_0 \right)^4 + 1} .
\end{equation}

%% file: chapters/search/vespa_plots.tex
\chapter{\texttt{VESPA} figures}
\label{app:vespa_plots}

\begin{figure}[!h]
    \centering
    \fbox{\includegraphics[width=\textwidth, height=\textheight, keepaspectratio]{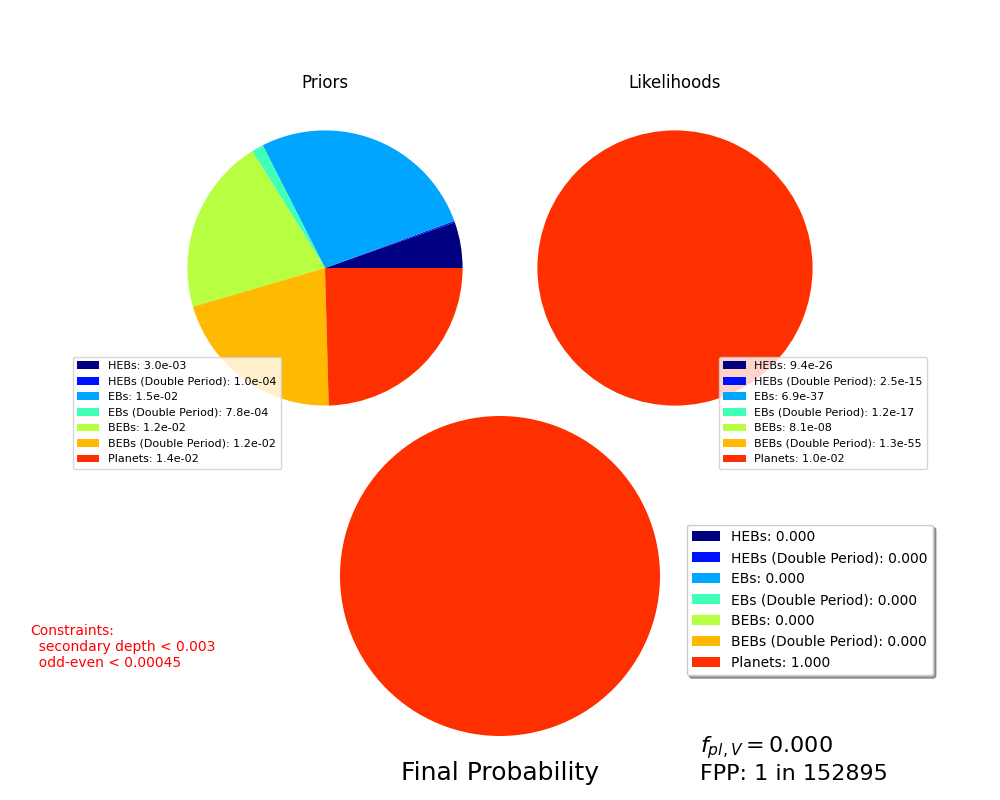}}
    \caption[False-positive probability summary plot from \texttt{VESPA}, for the transit of planet candidate TIC 55092869.01 (KELT-11 b)]
    {False-positive probability (FPP) summary plot from \texttt{VESPA}, for the transit of planet candidate TIC 55092869.01 (KELT-11 b).
    For a description of all the elements see Figure~\ref{fig:vespa_summary_example}.}
    \label{fig:vespa_summary_tic550}
\end{figure}

\begin{figure}[!h]
    \centering
    \fbox{\includegraphics[width=\textwidth, height=\textheight, keepaspectratio]{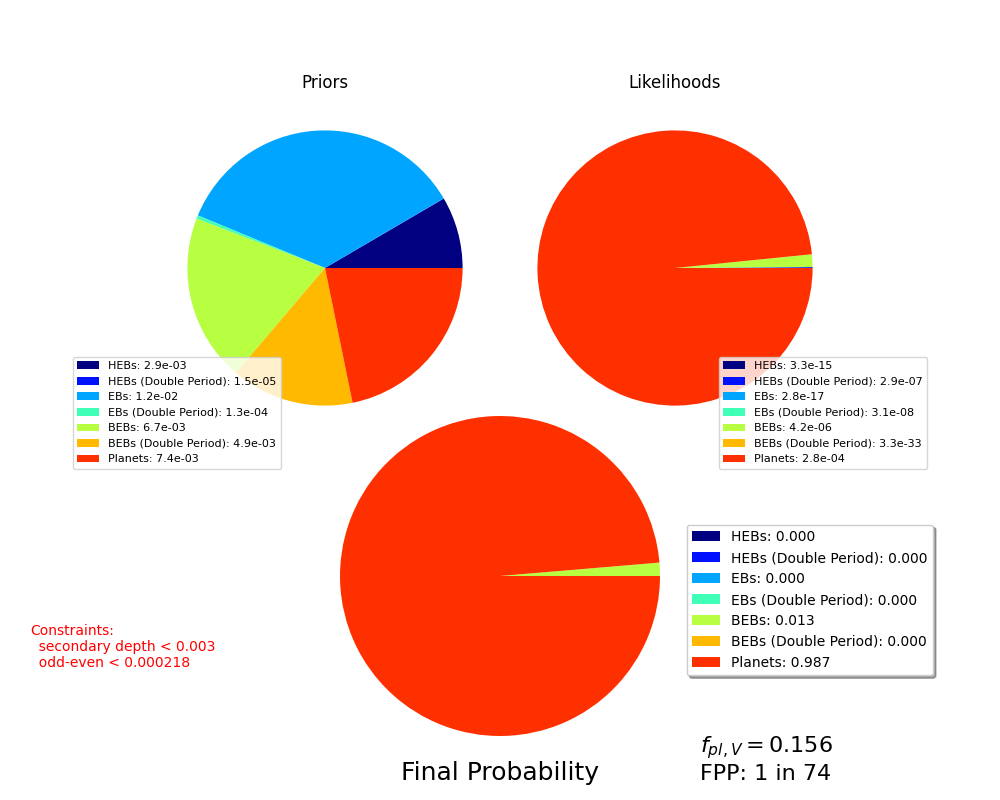}}
    \caption[False-positive probability summary plot from \texttt{VESPA}, for the transit of planet candidate TIC 441462736.01 (TOI-197 b)]
    {False-positive probability (FPP) summary plot from \texttt{VESPA}, for the transit of planet candidate TIC 441462736.01 (TOI-197 b).
    For a description of all the elements see Figure~\ref{fig:vespa_summary_example}.}
    \label{fig:vespa_summary_tic441}
\end{figure}

\begin{figure}[!h]
    \centering
    \fbox{\includegraphics[width=\textwidth, height=\textheight, keepaspectratio]{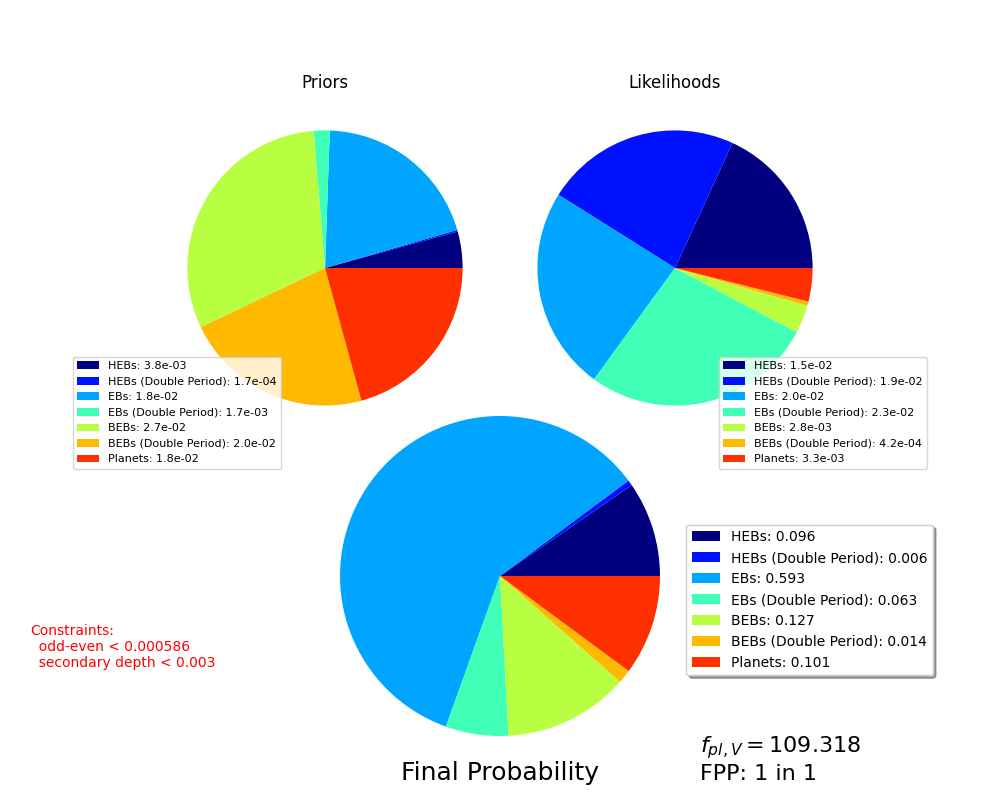}}
    \caption[False-positive probability summary plot from \texttt{VESPA}, for the transit of planet candidate TIC 348835438.01]
    {False-positive probability (FPP) summary plot from \texttt{VESPA}, for the transit of planet candidate TIC 348835438.01.
    For a description of all the elements see Figure~\ref{fig:vespa_summary_example}.}
    \label{fig:vespa_summary_tic348}
\end{figure}

\begin{figure}[!h]
    \centering
    \fbox{\includegraphics[width=\textwidth, height=\textheight, keepaspectratio]{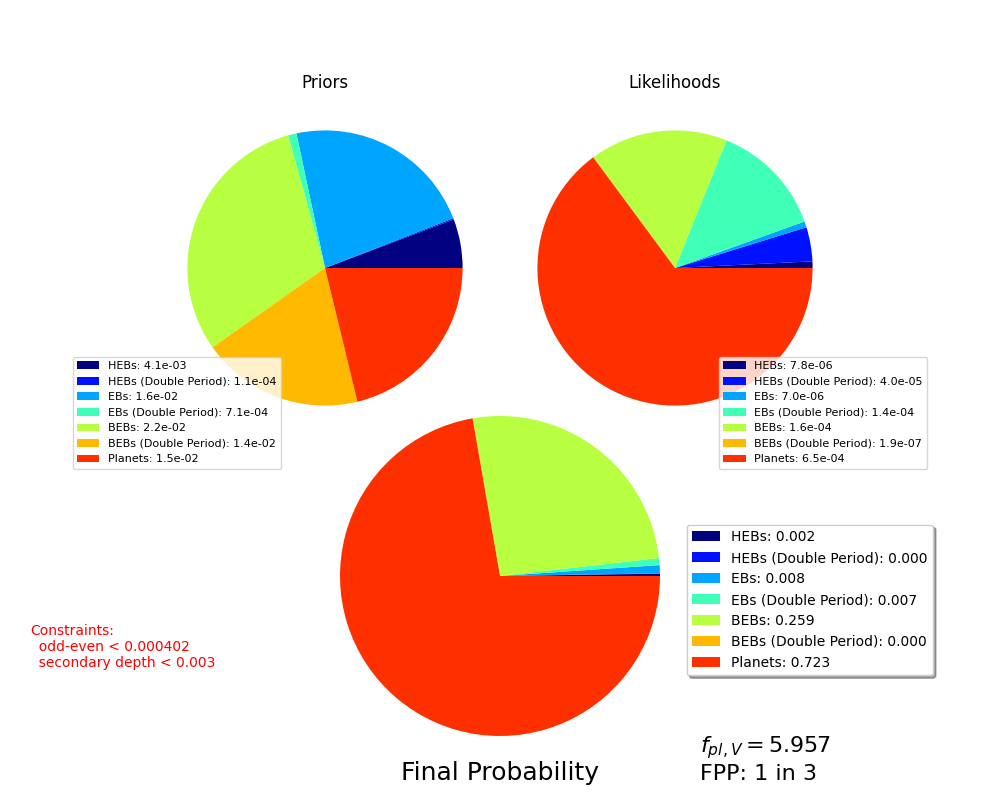}}
    \caption[False-positive probability summary plot from \texttt{VESPA}, for the transit of planet candidate TIC 204650483.01]
    {False-positive probability (FPP) summary plot from \texttt{VESPA}, for the transit of planet candidate TIC 204650483.01.
    For a description of all the elements see Figure~\ref{fig:vespa_summary_example}.}
    \label{fig:vespa_summary_tic204}
\end{figure}

\begin{figure}[!h]
    \centering
    \fbox{\includegraphics[width=\textwidth, height=\textheight, keepaspectratio]{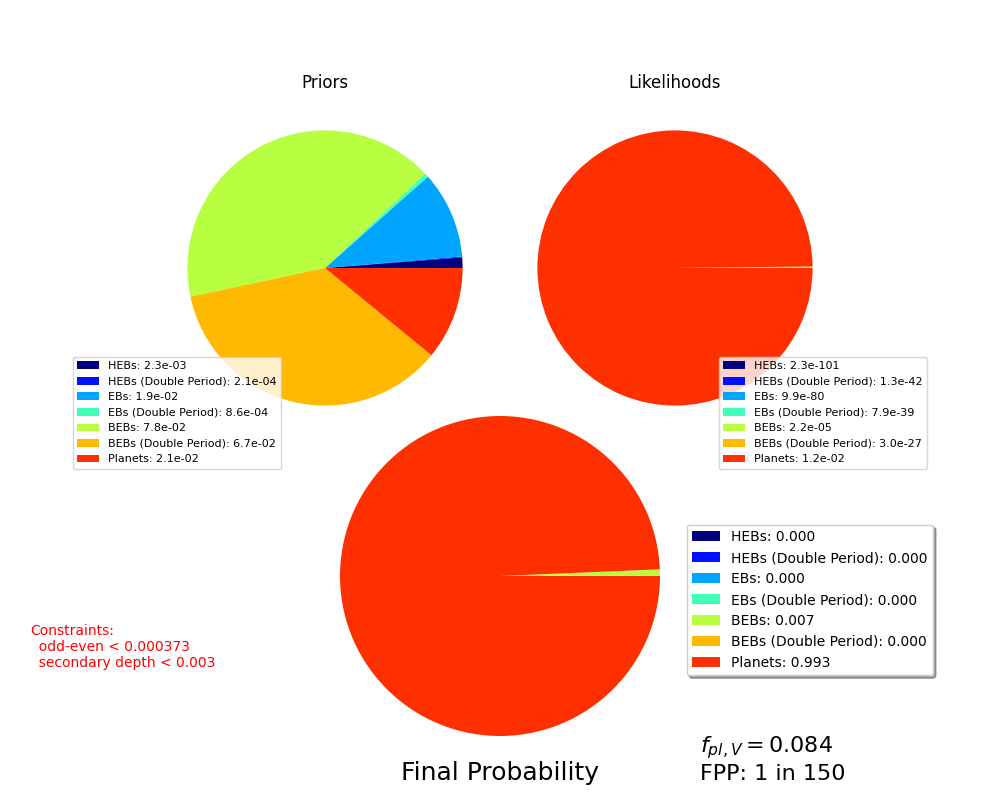}}
    \caption[False-positive probability summary plot from \texttt{VESPA}, for the transit of planet candidate TIC 394918211.01]
    {False-positive probability (FPP) summary plot from \texttt{VESPA}, for the transit of planet candidate TIC 394918211.01.
    For a description of all the elements see Figure~\ref{fig:vespa_summary_example}.}
    \label{fig:vespa_summary_tic394}
\end{figure}

%% file: chapters/search/histogram_plots.tex
\chapter{Posterior histograms}
\label{app:fit_histograms}

\begin{figure}[!h]
    \centering
    \includegraphics[width=\textwidth, height=\textheight, keepaspectratio]{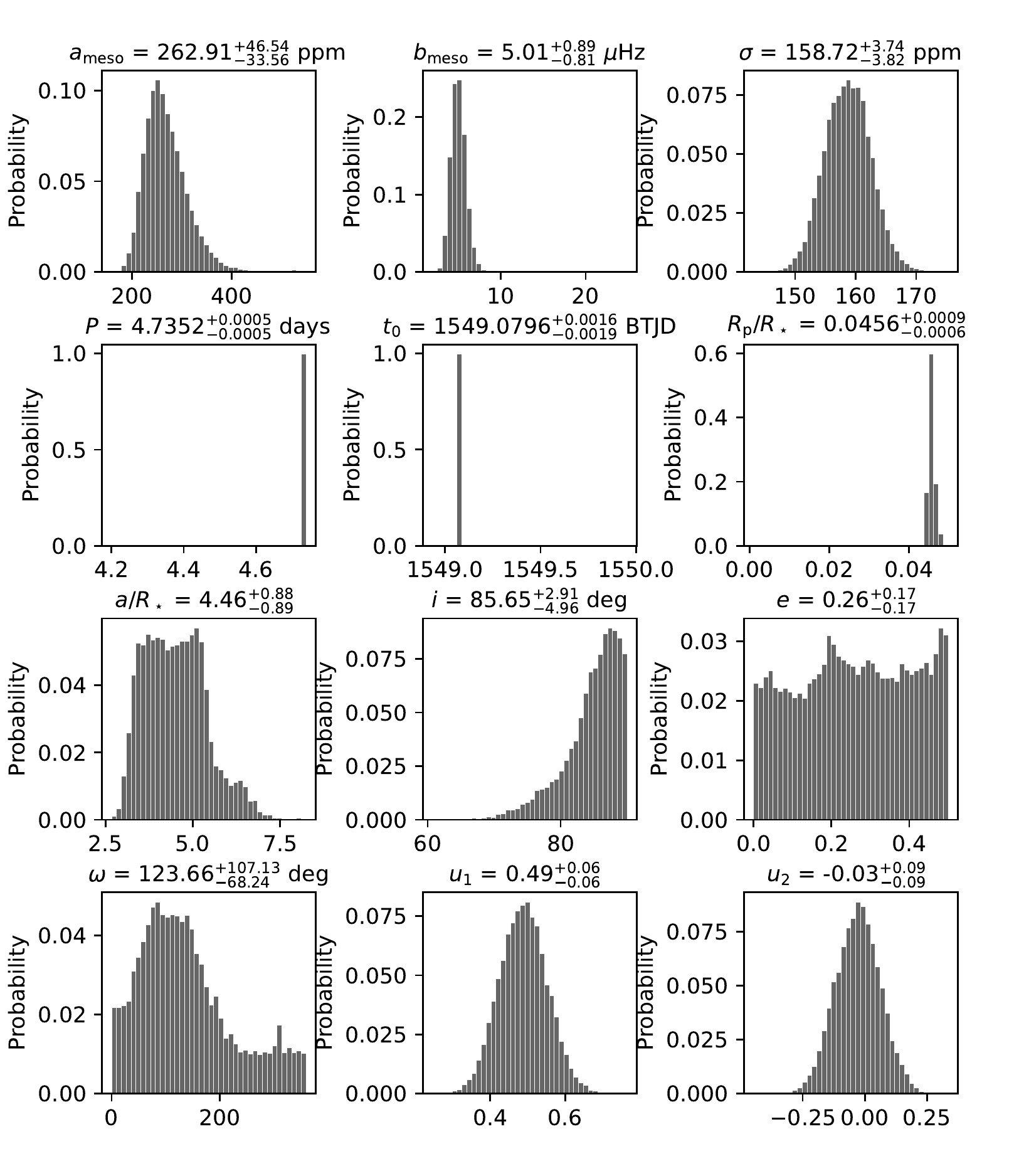}
    \caption[Posterior distributions for all parameters in the GP + transit model obtained in the characterization of the \textit{TESS} light curve of TIC 55092869.01 (KELT-11 b)]
    {Posterior distributions for all parameters in the GP + transit model obtained in the characterization of the \textit{TESS} light curve of TIC 55092869.01 (KELT-11 b).}
    \label{fig:tic550_param_hist}
\end{figure}

\begin{figure}[!h]
    \centering
    \includegraphics[width=\textwidth, height=\textheight, keepaspectratio]{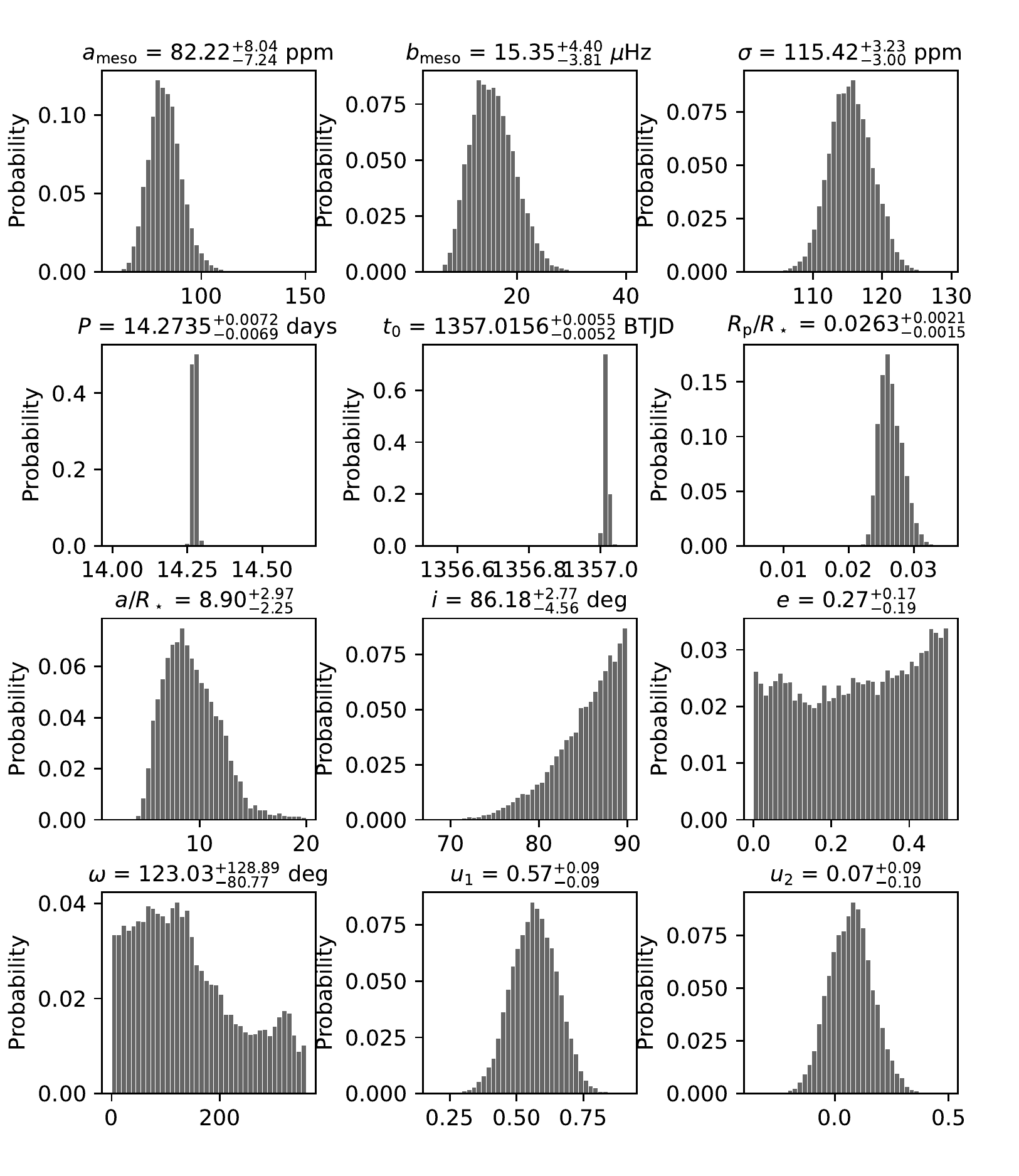}
    \caption[Posterior distributions for all parameters in the GP + transit model obtained in the characterization of the \textit{TESS} light curve of TIC 441462736.01 (TOI 197.01; HD 221416 b)]
    {Posterior distributions for all parameters in the GP + transit model obtained in the characterization of the \textit{TESS} light curve of TIC 441462736.01 (TOI 197.01; HD 221416 b).}
    \label{fig:tic441_param_hist}
\end{figure}

\begin{figure}[!h]
    \centering
    \includegraphics[width=\textwidth, height=\textheight, keepaspectratio]{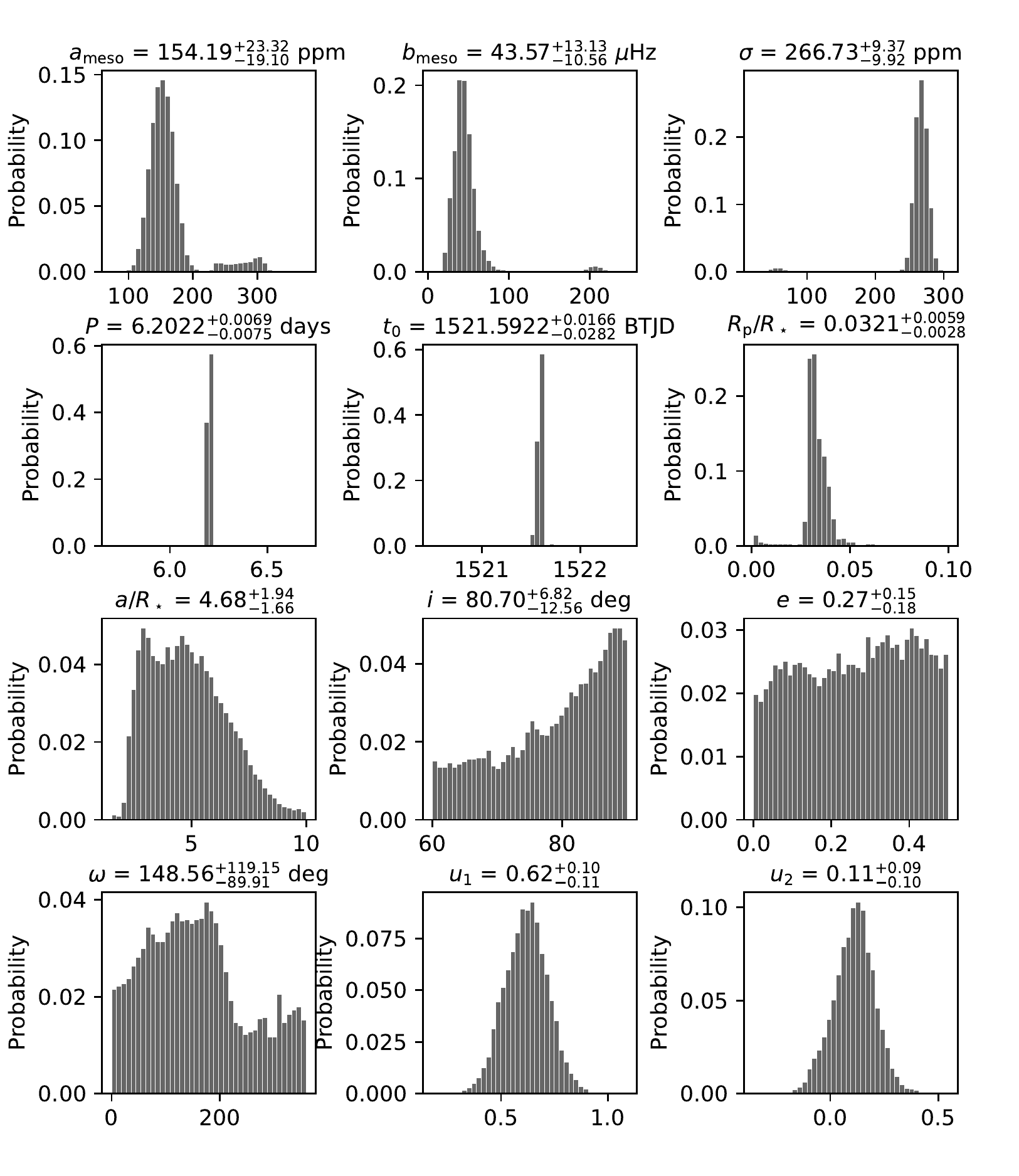}
    \caption[Posterior distributions for all parameters in the GP + transit model obtained in the characterization of the \textit{TESS} light curve of TIC 348835438.01]
    {Posterior distributions for all parameters in the GP + transit model obtained in the characterization of the \textit{TESS} light curve of TIC 348835438.01.}
    \label{fig:tic348_param_hist}
\end{figure}

\begin{figure}[!h]
    \centering
    \includegraphics[width=\textwidth, height=\textheight, keepaspectratio]{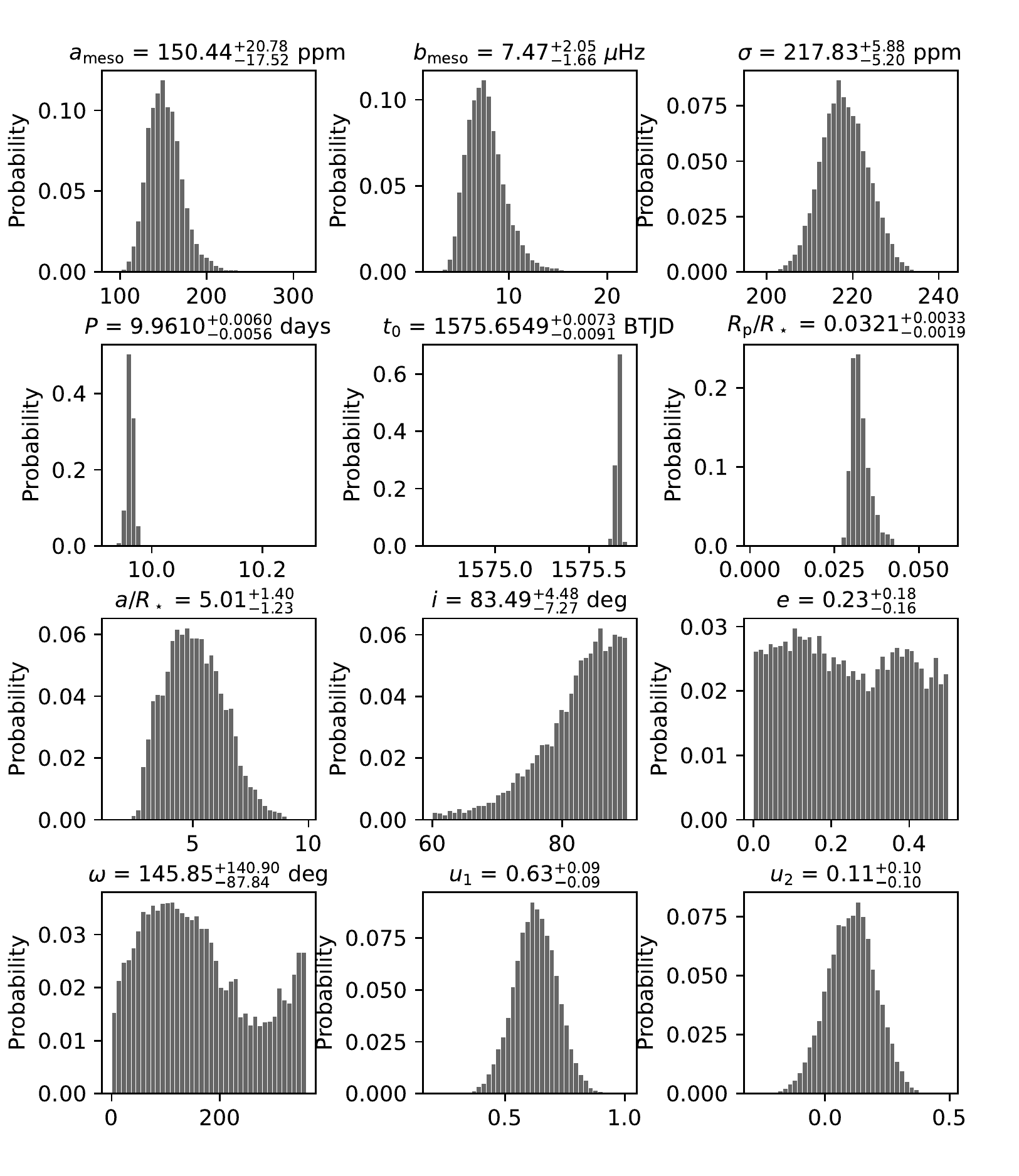}
    \caption[Posterior distributions for all parameters in the GP + transit model obtained in the characterization of the \textit{TESS} light curve of TIC 204650483.01]
    {Posterior distributions for all parameters in the GP + transit model obtained in the characterization of the \textit{TESS} light curve of TIC 204650483.01.}
    \label{fig:tic204_param_hist}
\end{figure}

\begin{figure}[!h]
    \centering
    \includegraphics[width=\textwidth, height=\textheight, keepaspectratio]{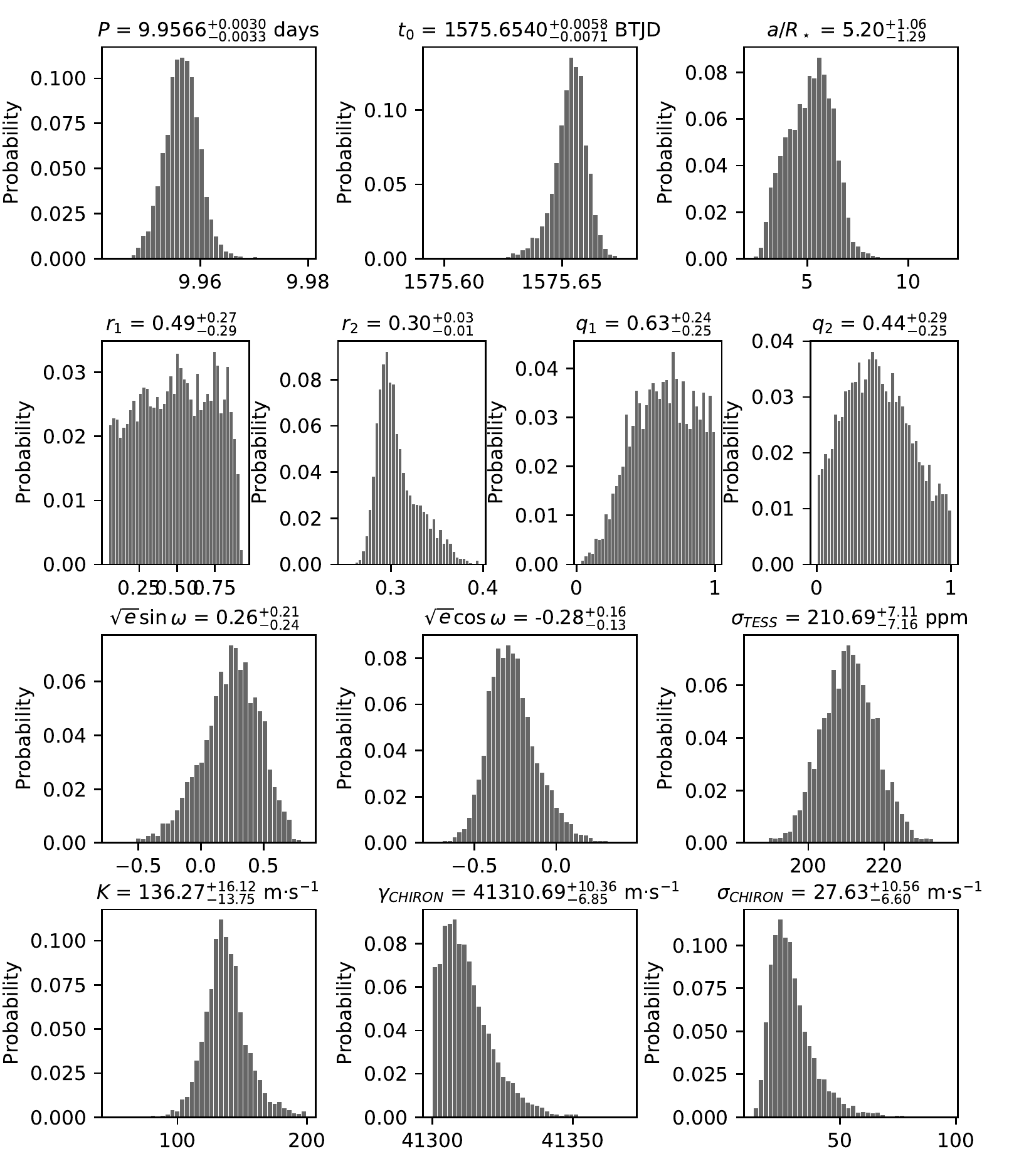}
    \caption[Posterior distributions for all model parameters in the joint RV and transit characterization of TIC 204650483.01]
    {Posterior distributions for all model parameters in the joint RV and transit characterization of TIC 204650483.01.}
    \label{fig:tic204_model_param_hist}
\end{figure}

\begin{figure}[!h]
    \centering
    \includegraphics[width=\textwidth, height=\textheight, keepaspectratio]{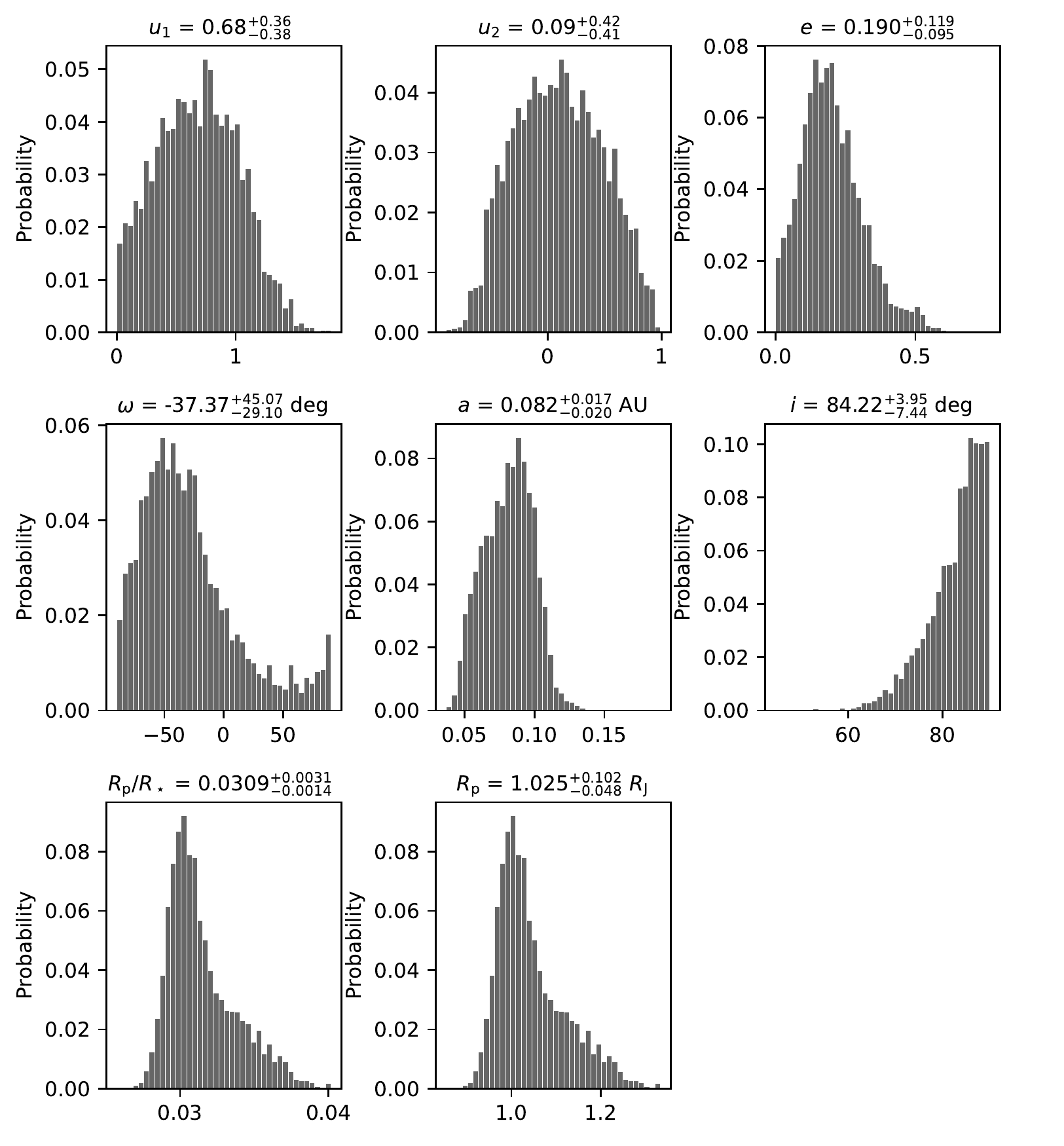}
    \caption[Posterior distributions for all derived parameters in the joint RV and transit characterization of TIC 204650483.01]
    {Posterior distributions for all derived parameters in the joint RV and transit characterization of TIC 204650483.01.}
    \label{fig:tic204_derived_param_hist}
\end{figure}

\begin{figure}[!h]
    \centering
    \includegraphics[width=\textwidth, height=\textheight, keepaspectratio]{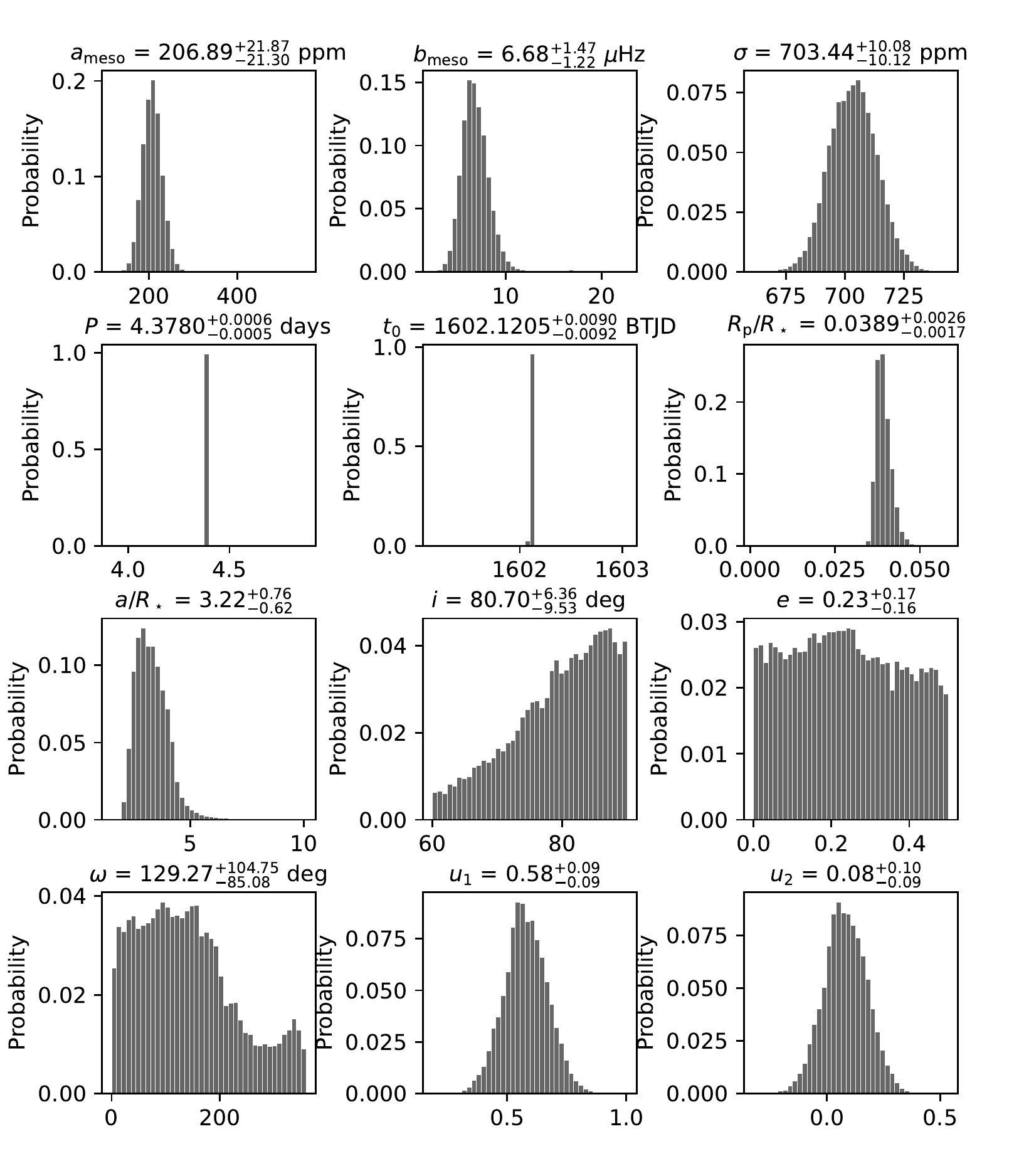}
    \caption[Posterior distributions for all parameters in the GP + transit model obtained in the characterization of the \textit{TESS} light curve of TIC 394918211.01]
    {Posterior distributions for all parameters in the GP + transit model obtained in the characterization of the \textit{TESS} light curve of TIC 394918211.01.}
    \label{fig:tic394_param_hist}
\end{figure}

\begin{figure}[!h]
    \centering
    \includegraphics[width=\textwidth, height=\textheight, keepaspectratio]{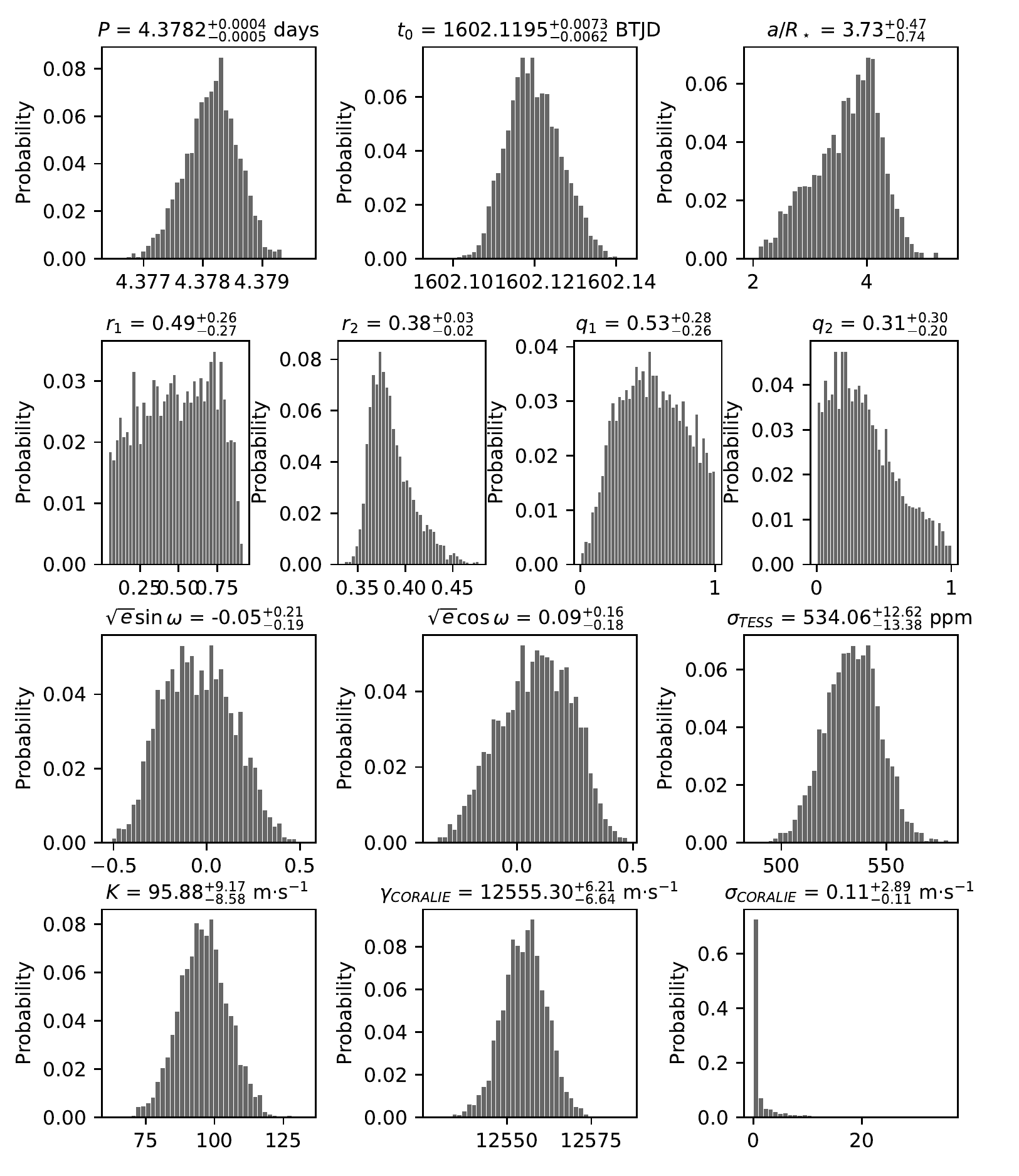}
    \caption[Posterior distributions for all model parameters in the joint RV and transit characterization of TIC 394918211.01]
    {Posterior distributions for all model parameters in the joint RV and transit characterization of TIC 394918211.01.}
    \label{fig:tic394_model_param_hist}
\end{figure}

\begin{figure}[!h]
    \centering
    \includegraphics[width=\textwidth, height=\textheight, keepaspectratio]{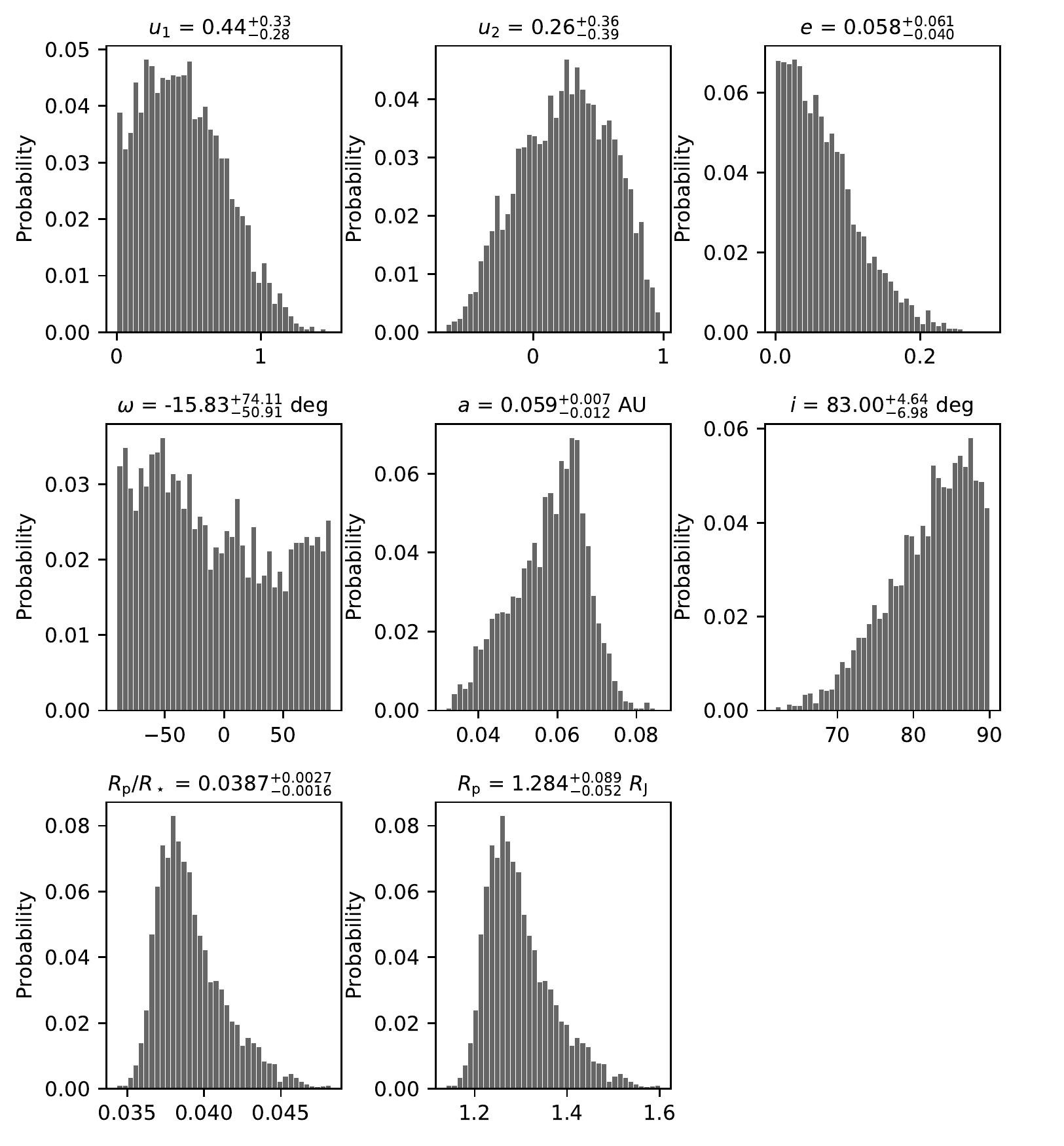}
    \caption[Posterior distributions for all derived parameters in the joint RV and transit characterization of TIC 394918211.01]
    {Posterior distributions for all derived parameters in the joint RV and transit characterization of TIC 394918211.01.}
    \label{fig:tic394_derived_param_hist}
\end{figure}